\documentclass{article}
\pdfoutput=0
\usepackage[english]{babel}
\usepackage{cite}
\usepackage{jheppub} 

\usepackage{gensymb}

\usepackage{placeins}

\usepackage{epsf}
\usepackage{amssymb}
\usepackage{amsmath}
\usepackage{amsfonts}
\usepackage{psfrag,epsfig,graphicx,graphics}

\usepackage{bm,slashed,multirow,
	soul,mathtools,xspace,array,comment}                
\allowdisplaybreaks
\usepackage{bbold}
\usepackage{color}
\usepackage{xcolor}
\usepackage{subfigure}
\usepackage[printonlyused]{acronym}
\usepackage{hyperref}

\usepackage{float}

\def\slashchar#1{\setbox0=\hbox{$#1$}
   \dimen0=\wd0
   \setbox1=\hbox{/} \dimen1=\wd1
   \ifdim\dimen0>\dimen1
      \rlap{\hbox to \dimen0{\hfil/\hfil}}
      #1
   \else
      \rlap{\hbox to \dimen1{\hfil$#1$\hfil}}
      /
   \fi}

\newcommand{\APP}{App.~}

\newcommand{\FIG}{Fig.~}
\newcommand{\FIGs}{Figs.~}
\newcommand{\SEC}{Sec.~}

\newcommand{\EQ}{Eq.~}
\newcommand{\EQs}{Eqs.~}

\newcommand{\cf}{cf.~}
\newcommand{\eg}{e.g.~}
\newcommand{\ie}{i.e.~}

\newcommand{\GeV}{\,\mbox{GeV}}

\def\bei{\begin{itemize}}
\def\ei{\end{itemize}}

\def\beeq{\begin{eqnarray}} 
\def\beqa{\begin{eqnarray}}
\def\bea{\begin{eqnarray}}

\def\eea{\end{eqnarray}}
\def\eqa{\end{eqnarray}}
\def\eeeq{\end{eqnarray}}

\def\eqar{\end{array}}
\def\beqar{\begin{array}}

\def\beas{\begin{eqnarray*}}
\def\beqas{\begin{eqnarray*}}

\def\eqas{\end{eqnarray*}}
\def\eeas{\end{eqnarray*}}

\def\beq{\begin{equation}} 
\def\be{\begin{equation}}

\def\ee{\end{equation}}
\def\eq{\end{equation}}
\def\eeq{\end{equation}}

\def\beqd{\begin{displaymath}}
\def\eeqd{\end{displaymath}}
\def\eqd{\end{displaymath}}

\def\beeq{\begin{eqnarray}} \def\eeeq{\end{eqnarray}}

\newcommand{\fin}{\end{document}}

\def\pv{\vec{p}_t}
\def\dv{\vec{\Delta}_t}

\def\meson{ \rho }

\newcommand{\mesonzp}{\meson_{p}^{0}}
\newcommand{\mesonzn}{\meson_{n}^{0}}
\newcommand{\mesonpp}{\meson_{p}^{+}}
\newcommand{\mesonmn}{\meson_{n}^{-}}

\def\fin{\end{document}}

\newcommand{\pt}{ \vec{p}_{t} }
\newcommand{\SgN}{ S_{\gamma N} }
\newcommand{\Msq}{ M_{\gamma \meson}^2 }

\title{Probing chiral-even and chiral-odd leading twist quark generalised parton distributions through the exclusive 
	photoproduction of a $ \gamma  \rho $ pair}

\author[1]{Goran Duplan\v{c}i\'{c},}
\author[2]{Saad Nabeebaccus,}
\author[1]{Kornelija Passek-Kumeri\v{c}ki,}
\author[3]{Bernard Pire,}
\author[4]{Lech Szymanowski,}
\author[2]{Samuel Wallon}

\affiliation[1]{Theoretical Physics Division, Rudjer Bo{\v s}kovi{\'c} Institute,
	HR-10002 Zagreb, Croatia}
\affiliation[2]{Universit\'e Paris-Saclay, CNRS/IN2P3, IJCLab, 91405 Orsay, France}
\affiliation[3]{CPHT, CNRS, Ecole polytechnique, Institut Polytechnique de Paris, 91128 Palaiseau, France}
\affiliation[4]{National Centre for Nuclear Research (NCBJ), 02-093 Warsaw, Poland}

\emailAdd{gorand@thphys.irb.hr}
\emailAdd{passek@irb.hr}
\emailAdd{saad.nabeebaccus@ijclab.in2p3.fr}
\emailAdd{bernard.pire@polytechnique.edu}
\emailAdd{Lech.Szymanowski@ncbj.gov.pl}
\emailAdd{samuel.wallon@ijclab.in2p3.fr}

\abstract{
We extend our studies of a new class of $2 \to 3$ exclusive processes using the collinear factorisation framework by considering the exclusive photoproduction of a $\gamma\,\rho$ pair, in the
kinematics where the pair has a large invariant mass, {and the outgoing $ \rho $-meson has a sufficiently large transverse momentum to not resonate with the nucleon}. We cover the whole kinematical range from medium energies in fixed target experiments to very large energies of colliders, by considering the experimental conditions of JLab~12-GeV, COMPASS, future EIC and LHC (in ultra-peripheral collisions) cases. One of the main interest in studying the present process is that it provides access to both chiral-even and chiral-odd GPDs, depending on the polarisation of the outgoing $\rho$-meson, both at leading twist. Our analysis covers both neutral and charged $ \rho $-mesons. We find that the order of magnitude of the obtained cross sections are sufficiently large for a dedicated experimental analysis to be performed, especially at JLab, for both longitudinally and transversely polarised $\rho$. Furthermore, we compute the linear photon beam polarisation asymmetry {which is} sizeable for a longitudinally polarised meson. These predictions are  obtained for both  asymptotic distribution amplitude (DA) and the \textit{holographic} DA.
}

\date{\today}
\begin{document}
	
	\maketitle

\preprint{}

\pagestyle{empty}
\newpage

\mbox{}

\pagestyle{plain}

\section{Introduction}

\label{sec:introduction}

In the present study, we extend our previous analysis in \cite{Duplancic:2022ffo} of generalised parton distributions (GPDs) in $2 \to 3$ exclusive processes, \eg{}\cite{Ivanov:2002jj,Enberg:2006he,ElBeiyad:2010pji,Pedrak:2017cpp,Pire:2019hos,Pedrak:2020mfm,Cosyn:2021dyr,Grocholski:2021man,Grocholski:2022rqj}, by considering
\begin{align}
	\label{eq:process}
	\gamma (q,\varepsilon _{q})+N(p_1, \lambda _{1}) \longrightarrow \gamma (k,\varepsilon _{k})+N'(p_2, \lambda _{2})+{\meson}(p_{\meson},\varepsilon_{ \meson })\,,
\end{align}
in the kinematical range from medium energies in fixed target experiments to very large energies of colliders, which corresponds to the experimental environment of JLab~12-GeV, COMPASS, future EIC and LHC in ultra-peripheral collisions (UPCs). The main motivation for considering the above photoproduction process is that it gives access to both leading twist chiral-even (CE) and chiral-odd (CO) GPDs, depending on the polarisation of the outgoing $  \meson  $ meson, described using its distribution amplitude (DA), also at the leading twist. In particular, this process allows us to learn more about the badly known chiral-odd GPDs. Factorisation for this process was recently proved in \cite{Qiu:2022bpq,Qiu:2022pla}, in which the hard scale is provided by the large relative transverse momentum of the outgoing $\gamma /  \meson  $ meson. The work presented here builds up on our earlier publications \cite{Boussarie:2016qop,Duplancic:2018bum} and our more recent work \cite{Duplancic:2022ffo}. One should note that the present paper extends the study performed in \cite{Boussarie:2016qop}, which focused on the neutral $  \rho^{0}  -$meson, to $  \rho^{0,\pm}  -$mesons of any possible charge.

 The paper is organised as follows: Details regarding the kinematics are recalled in \SEC\ref{sec:kinematics}. In \SEC\ref{sec:non-pert-inputs}, the non-perturbative inputs, namely the GPDs and the DAs, are presented. In \SEC\ref{sec:computation}, we show how the amplitude can be expressed in terms of tensorial structures, and how the computation reduces to that of basic building blocks. This section ends with a discussion of polarisation asymmetries. Results for the cross sections and linear polarisation asymmetries with respect to the incoming photon are the subject of \SEC\ref{sec:results}.  This section ends with an estimation of counting rates at various experiments.
 We end with conclusions in \SEC\ref{sec:conclusion}. In \APP\ref{app:CO-amplitudes}, the diagrams for the chiral-odd case are given in terms of building block integrals, for both the asymptotic and holographic DAs. {In \APP\ref{app:angle-outgoing-photon}, we discuss the effect, on the cross section, of the experimental constraints at JLab on the angle of the outgoing photon.} Finally, in \APP\ref{app:vanishing-circular-asymmetry}, the vanishing of the circular polarisation asymmetry wrt the incoming photon for the chiral-even is discussed.

\section{Kinematics}

\label{sec:kinematics}

From \eqref{eq:process}, one can define the following momenta,
\begin{align}
	P^{ \mu }=\frac{p_1^{ \mu }+p_{2}^{ \mu }}{2}\,,\quad  \Delta ^{ \mu }=p_2^{ \mu }-p_{1}^{ \mu }\,.
\end{align}
All momenta are decomposed in a Sudakov basis, such that a generic vector $ v $ can be written as
\begin{align}
v^{\mu}=a\,n^{ \mu }+b\,p^{ \mu }+v^{ \mu }_{\perp}\,,
\end{align}
with the two light-cone vectors $p$ and $n$ given by
\begin{align}
\label{sudakov2}
p^\mu = \frac{\sqrt{s}}{2}(1,0,0,1)\,,\qquad n^\mu = 
\frac{\sqrt{s}}{2}(1,0,0,-1) \,,\qquad p\cdot n = \frac{s}{2}\,.
\end{align}
For the transverse vectors, we use the following convention,
\begin{equation}
	\label{sudakov3}
	v_\bot^\mu = (0,v^x,v^y,0) \,, \qquad v_\bot^2 = -\vec{v}_t^2\,.
\end{equation}
The particle momenta for the process now read
\begin{align}
	\label{eqn:impini}
	p_1^\mu &= (1+\xi)\,p^\mu + \frac{M^2}{s(1+\xi)}\,n^\mu\,,\\[5pt]
	p_2^\mu &= (1-\xi)\,p^\mu + \frac{M^2+\vec{\Delta}^2_t}{s(1-\xi)}n^\mu + \Delta^\mu_\bot\,,\\[5pt]
	q^\mu &= n^\mu\,,\\[5pt]
	\label{eq:momentum-outgoing-photon}
	k^\mu &= \alpha \, n^\mu + \frac{(\vec{p}_t-\vec\Delta_t/2)^2}{\alpha s}\,p^\mu + p_\bot^\mu -\frac{\Delta^\mu_\bot}{2}\,,\\[5pt]
	\label{eq:p-rho}
	p_\meson^\mu &= \alpha_\meson \, n^\mu + \frac{(\vec{p}_t+\vec\Delta_t/2)^2+M^2_\meson}{\alpha_\meson s}\,p^\mu - p_\bot^\mu-\frac{\Delta^\mu_\bot}{2}\,,
\end{align}
where $M$ and $M_\meson$ are the masses of the nucleon and the $  \meson  -$meson respectively. The  square of the centre of mass energy of the photon-nucleon system is
then
\begin{align}
S_{\gamma N} = (q + p_1)^2 = (1+\xi)s + M^2\,,
\end{align}
while the square of the transferred momentum is
\begin{align}
	t = (p_2 - p_1)^2 = -\frac{1+\xi}{1-\xi}\vec{\Delta}_t^2 -\frac{4\xi^2M^2}{1-\xi^2}\,.
\end{align}
The invariant mass squared  of the $\gamma\meson$ system, $M^2_{\gamma\meson}$, provides the hard scale for factorisation. This is guaranteed by having a large \textit{relative} transverse momentum $  \vec{p}_{t}  $ between the outgoing photon and meson.

Collinear QCD factorisation implies that
\begin{equation}
 -u'= \left( p_{\meson}-q \right)^2\,,\qquad -t'= \left( k-q \right)^2\,,\qquad   M_{\gamma \meson}^2 =  \left(  p_{\meson}+k\right)^2\,,
\end{equation}
 are large, while 
 \begin{equation}
-t =  \left( p_2-p_1 \right)^2  \,,
 \end{equation}
   needs to be small. In practice, we employ the cuts
 \begin{align}
 	\label{eq:kinematical-cuts}
 -u',-t'&>1 \GeV ^2\,, \qquad   -t < 0.5  \GeV ^2\,.
 \end{align}
 
 We note that these cuts are sufficient to ensure that $ M^2_{\gamma\meson} > 2 $ GeV$ ^2 $. Furthermore, the above kinematical cuts ensure that the $\meson N'$ invariant mass, $ M_{\meson N'} $, is \textit{completely} out of the resonance region. Indeed, through a numerical analysis, taking into account \eqref{eq:kinematical-cuts}, we find that\footnote{In all rigor, we should also exclude the kinematical region where one of the pion decay products from the $\rho$ meson resonates with the final nucleon. Such a constraint is however difficult to fulfill at our level since we do not specify the kinematics of the $\rho$ decay.}
 \begin{align}
M_{\meson N'}^{2} > 3.11 \GeV^{2}\,,
 \end{align}
which is much larger than the mass squared of the $  \Delta  $ baryon, $ m_{ \Delta }^{2} \approx 1.52 \GeV^2 $. 
Thus, unlike the case of the charged pion in \cite{Duplancic:2022ffo}, we find that the larger mass of the $ \meson $-meson wrt to the pion mass is such that $ M_{\meson N'} ^{2}$ is pushed to larger values.

Neglecting  $\dv$ in front of $\pv$, as well as hadronic masses, we have that the approximate kinematics, as used in the hard part of the factorised amplitude, is
\begin{eqnarray}
	\label{skewness2}
	M^2_{\gamma\meson} \approx  \frac{\vec{p}_t^{\,2}}{\alpha\bar{\alpha}} ~, \qquad
	\alpha _{\meson} \approx 1-\alpha \equiv \bar{\alpha} ~,\qquad
	\xi =  \frac{\tau}{2-\tau} ~,
\end{eqnarray}
\begin{eqnarray}
	\label{eq:alpha}
	\tau \approx 
	\frac{M^2_{\gamma\meson}}{S_{\gamma N}-M^2}~,\qquad
	~-t'  \approx  \bar\alpha\, M_{\gamma\meson}^2  ~,\qquad -u'  \approx  \alpha\, M_{\gamma\meson}^2 \,.\quad \,
\end{eqnarray}

We choose as independent variables $(-t)$, $(-u')$ and $M_{\gamma\meson}^2$. 

{Regarding the polarisation vectors, we work in the \textit{axial} gauge, such that $ p \cdot  \varepsilon =0 $. In particular, this implies that the polarisation vector of the initial photon is given by
\begin{align}
	\varepsilon^\mu_q=\varepsilon^\mu_{q\perp} \,,
\end{align}
\ie{}it only has non-zero components in the transverse plane. For the polarisation vector of the outgoing photon, one obtains
\begin{align}
\varepsilon^\mu_k=\varepsilon^\mu_{k\perp} - \frac{\varepsilon_{k\perp} \cdot k_{\bot} }{p\cdot k}p^\mu\,.
\end{align}
Regarding the polarisation of the $ \meson $-meson, we have\footnote{Our conventions are such that $ \varepsilon ^{ \mu }_{ \rho }(p_{ \rho },L) =(0,0,0,-1)$ in the $ \meson $-meson rest frame.}
\begin{align}
	\label{eq:pol-rho-long}
\varepsilon ^{ \mu }_{ \rho }(p_{ \rho },L)= \frac{1}{M_{ \rho }}p_{ \rho }^{ \mu }-\frac{M_{ \rho }}{ \left( p \cdot p_{ \rho } \right) }p^{ \mu }\,,
\end{align}
for the longitudinally polarised case. For the transversely polarised $ \meson $-meson, we exploit the transversity relation $ p_{\meson} \cdot  \varepsilon _{\meson}(p_{\meson},T) =0 $ to write the polarisation vector as
\begin{align}
	\label{eq:pol-rho-trans}
 \varepsilon ^{ \mu }_{ \rho }(p_{ \rho },T)=  \varepsilon ^{ \mu }_{ \rho \perp}-\frac{\varepsilon_{ \rho \perp} \cdot p_{ \rho }}{p \cdot p_{ \rho }}p^{ \mu }\,,
\end{align}
where we have chosen the basis for which $ p \cdot  \varepsilon _{ \rho }(p_{\meson},T) =0$. The sum over {all} 3 polarisations gives
\begin{equation}
	\sum_{i=L,\,T} \varepsilon_{\meson}^\mu(p_{\meson},i)\, \varepsilon_{\meson}^{\nu\ast}(p_{\meson},i) = -g^{\mu\nu} + \frac{p_{\meson}^\mu
		p_{\meson}^\nu}{m_\meson^2}.
\end{equation}
Using \eqref{eq:pol-rho-long} and \eqref{eq:pol-rho-trans}, restricting the sum to only \textit{transverse} polarisations  leads to
\begin{align}
	\label{eq:sum-pol-rhoT}
	\sum_{T}\varepsilon_{\meson}^\mu(p_{\meson},T)\, \varepsilon_{\meson}^{\nu\ast}(p_{\meson},T)= -g^{ \mu  \nu }_{\perp}+\frac{p^{ \mu }p_{\meson \perp}^{ \nu }+p^{  \nu  }p_{\meson \perp}^{  \mu  }}{p \cdot p_{ \meson}}-\frac{p_{ \meson\perp}^2}{ \left( p \cdot p_{\meson} \right)^2 }p^{ \mu }p^{ \nu }\,,
\end{align}
where $ p_{ \rho \perp} \equiv -p_{\perp}-\frac{ \Delta _{\perp}}{2} $, see \eqref{eq:p-rho}.

}

Further details on the kinematics can be found in our previous works \cite{Boussarie:2016qop,Duplancic:2018bum,Duplancic:2022ffo}.

\section{Non-perturbative inputs}

\label{sec:non-pert-inputs}

For the self-consistency of the paper, we choose to recall the basic non-perturbative ingredients needed for computing the amplitude.

\subsection{Generalised parton distributions}

\label{sec:GPDs}

For our case, both the $p \to n$ and $n \to p$ quark transition GPDs are needed. By 
isospin symmetry, they are identical and are related to the proton GPD by the relation \cite{Mankiewicz:1997aa}
\beqa
\label{TransitionGPD}
\langle n | \bar{d} \, \Gamma \, u | p \rangle = \langle p | \bar{u} \, \Gamma 
\, d | n \rangle =
\langle p | \bar{u} \, \Gamma \, u | p \rangle  - \langle p |
\bar{d} \, \Gamma \, d | p \rangle\,.
\eqa
Therefore, we only use the proton GPDs in practice. The chiral-even GPDs of a parton $q$ (where $q = u,\ d$) in the nucleon target   are  defined by~\cite{Diehl:2003ny}:
\beqa
\label{defGPDEvenV}
&&\langle p(p_2,\lambda_2)|\, \bar{q}\left(-\frac{y}{2}\right)\,\gamma^+q \left(\frac{y}{2}\right)|p(p_1,\lambda_1) \rangle \\ \nonumber 
&&= \int_{-1}^1dx\ e^{-\frac{i}{2}x(p_1^++p_2^+)y^-}\bar{u}(p_2,\lambda_2)\, \left[ \gamma^+ H^{q}(x,\xi,t)   +\frac{i}{2m}\sigma^{+ \,\alpha}\Delta_\alpha  \,E^{q}(x,\xi,t) \right]
u(p_1,\lambda_1)\,,
\eqa
for the chiral-even vector GPDs, and
\beqa
\label{defGPDEvenA}
&&\langle p(p_2,\lambda_2)|\, \bar{q}\left(-\frac{y}{2}\right)\,\gamma^+ \gamma^5 q\left(\frac{y}{2}\right)|p(p_1,\lambda_1)\rangle \\ \nonumber
&&= \int_{-1}^1dx\ e^{-\frac{i}{2}x(p_1^++p_2^+)y^-}\bar{u}(p_2,\lambda_2)\, \left[ \gamma^+ \gamma^5 \tilde H^{q}(x,\xi,t)   +\frac{1}{2m}\gamma^5 \Delta^+  \,\tilde E^{q}(x,\xi,t) \right]
u(p_1,\lambda_1)\,.
\eqa
for chiral-even axial GPDs. In the above, $\lambda_1$ and $\lambda_2$ are the light-cone helicities of the nucleons with momenta $p_1$ and $p_2$. In our analysis, 
the contributions from $ E^{q} $ and $  \tilde{E}^{q}  $  are neglected, since they are suppressed by kinematical factors at the cross section level, see \eqref{squareCEresult}.

The transversity (chiral-odd) GPD of a quark $q$   is defined by
\beqa
&&\langle p(p_2,\lambda_{2})|\, \bar{q}\left(-\frac{y}{2}\right)i\,\sigma^{+j} q \left(\frac{y}{2}\right)|p(p_1,\lambda_{1})\rangle \\ \nonumber
&&= \int_{-1}^1dx\ e^{-\frac{i}{2}x(p_1^++p_2^+)y^-}\bar{u}(p_2,\lambda_{2})\, \left[i\,\sigma^{+j}H_T^{q}(x,\xi,t)
+\dots
\right]u(p_1,\lambda_{1})\,,
\label{defCOGPD}
\eqa
where $\dots$ denote the remaining three chiral-odd GPDs whose contributions are omitted in the present analysis.

 The GPDs are parametrised in terms of double distributions \cite{Radyushkin:1998es}. The details can be found in \cite{Boussarie:2016qop,Duplancic:2018bum}, and we do not repeat them here. The $t$-dependence of the GPDs is modelled by a simplistic dipole ansatz, discussed in \APP E of \cite{Duplancic:2022ffo}.

In our current study, which is performed at leading order in $\alpha_s$, we neglect any evolution of the GPDs/PDFs, and take a fixed factorisation scale of $  \mu _{F}^2=10 \GeV^2 $. As in \cite{Boussarie:2016qop,Duplancic:2018bum,Duplancic:2022ffo}, the PDF datasets that we use to construct the GPDs are
\begin{itemize}
	\item 
	For $x q(x)$,  the GRV-98 parameterisation~\cite{Gluck:1998xa}, as 
	made available from the Durham database. 
		
	\item
	For $x \Delta q(x)\,,$  the  GRSV-2000 
	parameterisation~\cite{Gluck:2000dy}, also available from the Durham 
	database. Two scenarios are proposed within this parameterisation:
\begin{itemize}
	\item The \textit{standard} scenario, for which the light sea quark and anti-quark distributions are \textit{flavour-symmetric},
	\item The \textit{valence} scenario, which corresponds to \textit{flavour-asymmetric} light sea quark densities.
\end{itemize}
		The above two scenarios can be used to obtain an order of magnitude estimate of the theoretical uncertainties.\footnote{Using more recent tables for the PDFs leads to variations that are smaller than the above-mentioned theoretical uncertainties. This effect was studied in \cite{Boussarie:2016qop} (see \eg{}\FIG 8).}
	\end{itemize}

\subsection{Distribution amplitudes}

\label{sec:DAs}

The chiral-even light-cone DA for the longitudinally polarised $ \meson_{L} $ meson is defined, at the leading 
twist 2, by the matrix element~\cite{Ball:1996tb},
\begin{equation}
	\langle 0|\bar{q}(0)\gamma^\mu T^{i} q(x)|\meson^{i}_{L}(p_\meson,\varepsilon_{\meson}) \rangle = p_\meson^\mu f_{\meson}^{\parallel}\int_0^1dz\ e^{-izp_\meson\cdot x}\ \phi_{\parallel}(z),
	\label{defDArhoL}
\end{equation}
with $f^{\parallel}_{\meson}=216\,\mbox{MeV}$ and $ i= 0,\pm $.\footnote{The wave functions are  $|\meson^0\rangle =\frac{1}{\sqrt{2}}(|u\bar u\rangle -|d\bar d\rangle)$, $|\meson^+\rangle =|u\bar d\rangle $ and $|\meson^-\rangle =|d\bar u\rangle $ for the $\meson^0$-, $ \meson^+ $- and $ \meson^{-} $-mesons respectively.} In the above, $ q = (u\; d) $ is a 2D vector in flavour space, and the matrices $ T^{i} $ (in flavour space) are defined by
\begin{align}
T^{0}=\frac{1}{\sqrt{2}}\begin{pmatrix}
1 & 0 \\
0 & -1
\end{pmatrix}\;,
\qquad 
T^{+}=\begin{pmatrix}
	0 & 0 \\
	1 & 0
\end{pmatrix}\;,
\qquad
T^{-}=\begin{pmatrix}
	0 & 1 \\
	0 & 0
\end{pmatrix}\;.
\end{align}
The chiral-odd light-cone DA for the transversely polarised meson vector $\meson_T$  is defined as
\begin{equation}
	\langle 0|\bar{q}(0)\sigma^{\mu\nu}T^{i}q(x)|\meson^{i}_{T}(p_\meson,\varepsilon_{\meson}) \rangle = i(\varepsilon^\mu_{\meson }\, p^\nu_\meson - \varepsilon^\nu_{\meson }\, p^\mu_\meson)f_\meson^\bot\int_0^1dz\ e^{-izp_\meson\cdot x}\ \phi_\bot(z),
	\label{defDArhoT}
\end{equation}
where $\varepsilon^\mu_{\meson}$ is the $\meson$-meson transverse polarisation and  $f_\meson^\bot$ = 160 MeV. 

For the computation, we use the asymptotic form of the distribution amplitude, $ \phi^{\rm as} $, as well as an alternative form, which is often  called `holographic' DA, $ \phi^{\rm hol} $. They are given by
\begin{align}
\label{DA-asymp}
\phi^{\rm as}(z)&= 6 z (1-z)\,,\\[5pt]
\label{DA-hol}
 \phi^{\rm hol}(z)&= \frac{8}{ \pi } \sqrt{z (1-z)}\,,
\end{align}
where both are normalised to 1. With the above two forms of the DA, the integration over $ z $ can be performed analytically. For the chiral-even case, including the building block integrals, the results can be found in \APP D of \cite{Duplancic:2018bum} for the asymptotic DA case, and in \APP C of \cite{Duplancic:2022ffo} for the holographic DA case. For the chiral-odd case, the results can be found in \APP\ref{app:CO-amplitudes}.

\section{The computation}

\label{sec:computation}

\subsection{Gauge invariant decomposition of the hard amplitude}

\label{sec:amplitude}

{In the framework of collinear factorisation, we set $   \vec{\Delta}  _{t} = 0 $ in the hard amplitude, which implies that $  \left( -t \right) =  \left( -t \right)_{ \mathrm{min} }   $, where
\begin{align}
	\left( -t \right) _{ \mathrm{min} }=\frac{4\xi^2M^2}{1-\xi^2}\,.
\end{align}
}

For the sake of completeness, we remind the reader of the properties of the diagrams contributing to the coefficient function, which significantly simplify the calculation. The hard part is described at leading order in $\alpha_s$ by 20 
Feynman diagrams. As discussed in \cite{Duplancic:2018bum,Duplancic:2022ffo}, half of the diagrams are related by $ C- $parity transformations.\footnote{This corresponds to a $C-$parity transformation ($z \leftrightarrow 1-z$ and $x \leftrightarrow -x$) \textit{after} the electric charges have been factored out, such that effectively, $q$ and $\bar{q}$ have a charge of 1.}

\def\diagici{2.65cm}
\begin{figure}[h]
	\begin{center}
		\psfrag{z}{\begin{small} $z$ \end{small}}
		\psfrag{zb}{\raisebox{0cm}{ \begin{small}$\bar{z}$\end{small}} }
		\psfrag{gamma}{\raisebox{+.1cm}{ $\,\gamma$} }
		\psfrag{pi}{$\,\pi$}
		\psfrag{rho}{$\,\pi$}
		\psfrag{TH}{\hspace{-0.2cm} $T_H$}
		\psfrag{tp}{\raisebox{.5cm}{\begin{small}     $t'$       \end{small}}}
		\psfrag{s}{\hspace{.6cm}\begin{small}$s$ \end{small}}
		\psfrag{Phi}{ \hspace{-0.3cm} $\phi$}
		\hspace{-.4cm}
		\begin{picture}(430,170)
			\put(0,20){\includegraphics[width=15.2cm]{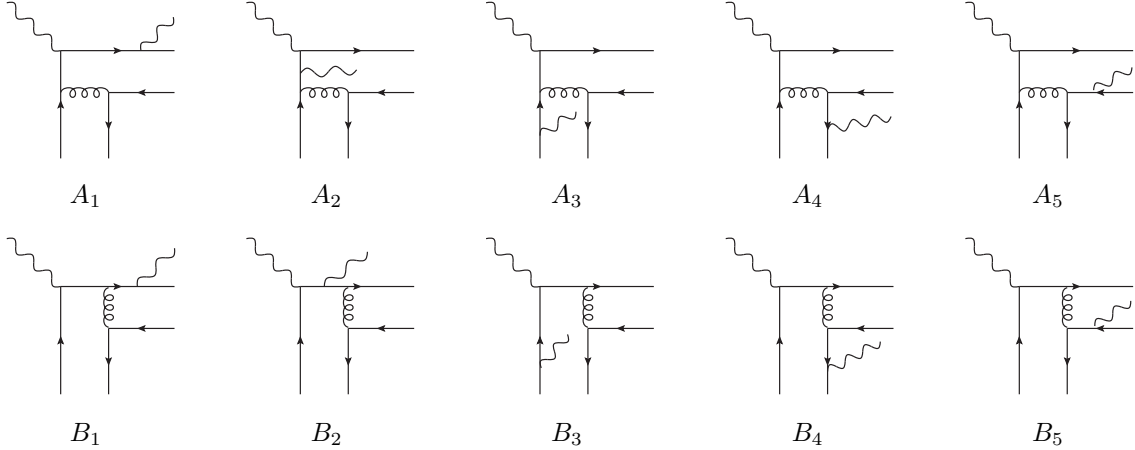}}
			\put(28,95){$A_1$}
			\put(119,95){$A_2$}
			\put(210,95){$A_3$}
			\put(301,95){$A_4$}
			\put(392,95){$A_5$}
			\put(28,5){$B_1$}
			\put(119,5){$B_2$}
			\put(210,5){$B_3$}
			\put(301,5){$B_4$}
			\put(392,5){$B_5$}
		\end{picture}
		\caption{Half of the Feynman diagrams contributing to the hard part of the amplitude.}
		\label{fig:diagrams}
	\end{center}
\end{figure}

The sets of diagrams (without including charge factors) are denoted as $(\cdots  
)$. We denote the $ A $ and $ B $ diagrams  by the order in which the incoming photon
and virtual gluon join one of the quark lines. The numbers (1 to 5) denote the five different ways
of attaching the outgoing photon to the quark lines. The remaining set of diagrams, $ C $ and $ D $, is
obtained by exchanging the role of the two quarks in the $ t- $channel. In practice, one obtains 4 separately QED gauge invariant sets of diagrams, namely $(AB)_{123}$, $(AB)_{45}$, $(CD)_{12}$ and $(CD)_{345}$ \cite{Duplancic:2018bum,Duplancic:2022ffo}. \FIG\ref{fig:diagrams} shows the first two sets.

Defining the charges $ Q_{q} $ through $e_q=Q_q |e|$, by QED gauge invariance, one can write any 
amplitude for photon meson production 
as the sum of three separate gauge invariant terms, in the form
\beqa
\label{generic-decomposition}
{\cal M}=
(Q_1^2 + Q_2^2) {\cal M}_{\rm sum} + (Q_1^2 - Q_2^2) {\cal M}_{\rm diff} + 2 
Q_1 \, Q_2 {\cal M}_{\rm prod} \,,
\eqa
where $Q_1$ is the charge of the quark entering the DA and $Q_2$ is
the charge of the  quark leaving the DA, in each diagram.

\subsection{Chiral-even case}
\label{sec:amplitude-chiral-even}

The parity properties of the $q \bar{q}$ correlators appearing
in the DA and in the GPDs allows the separation of the contributions for parity $(+)$, denoted as $S$ and parity $(-)$, 
denoted as $P$.
Only two structures occur in the hard part, namely $SS$ (no $\gamma^5$ 
matrices, vector GPD case) and $SP$ (one $\gamma^5$, axial GPD case).\footnote{Note that the $ SS $ structure is equivalent to the $ PP $ structure that enters the amplitude for the charged pion \cite{Duplancic:2022ffo} since the two $ \gamma ^{5} $ matrices can be combined through anti-commutation relations. {They are of course associated with different GPDs in each case.}}

A careful examination of the $C-$parity transformation which relates the two 
sets of 10 diagrams  gives the following results. For the vector contribution, 
the sum of 
diagrams reads
\beqa
\label{sumVmeson}
&&{\cal M}_ \meson ^V \\
&&=
Q_1^2 [(AB)_{123}]_{SS}  \otimes {f} + Q_1 Q_2 [(AB)_{45}]_{SS} \otimes 
{f}
+ Q_2^2 [(AB)_{123}]_{SS}^{(C)}  \otimes {f} + Q_1 Q_2 
[(AB)_{45}]_{SS}^{(C)} \otimes {f}\,,\nonumber
\eqa
while for the axial contribution, one gets
\beqa
\label{sumAmeson}
&&{\cal M}_ \meson ^A \\
&&=
Q_1^2 [(AB)_{123}]_{SP} \otimes \tilde{f} + Q_1 Q_2 [(AB)_{45}]_{SP} \otimes \tilde{f}
- Q_2^2 [(AB)_{123}]_{SP}^{(C)}  \otimes \tilde{f}- Q_1 Q_2 [(AB)_{45}]_{SP}^{(C)} 
\otimes \tilde{f}\, . \nonumber
\eqa
In the above two formulae, $f$ denotes a GPD of the set  $H, E$ appearing in the decomposition of the 
vector correlator  (\ref{defGPDEvenV}), while 
$\tilde{f}$ denotes a GPD of the set  $\tilde{H}, \tilde{E}$ appearing in the 
decomposition of the axial correlator  (\ref{defGPDEvenA}). The symbol $\otimes$ represents the integration over $x$. The integration over $z$ 
for the $  \meson  $-meson DA is implicit, since the DA is symmetric 
over $z \leftrightarrow 1-z$. {The superscript $ (C) $ denotes $ x \to -x $ and $ z \to  \left( 1-z \right)  $}.

The above decomposition is convenient since the integration over $z$ is performed \textit{analytically}, while the integration over $x$ is performed \textit{numerically}. This allows us to evaluate the amplitude in blocks which can be used for computing various observables. 
 \EQs\eqref{sumVmeson} and \eqref{sumAmeson} are obtained by making the identification
\begin{align}
	\left[  \left( CD \right)_{345}  \right] _{SP}&=	- \left[  \left( AB \right)_{123}  \right] _{SP}^{(C)}\;,\\[5pt]
	\left[  \left( CD \right)_{12}  \right] _{SP}&=	- \left[  \left( AB \right)_{45}  \right] _{SP}^{(C)}\;,\\[5pt]
	\left[  \left( CD \right)_{345}  \right] _{SS}&=	 \left[  \left( AB \right)_{123}  \right] _{SS}^{(C)}\;,\\[5pt]
	\left[  \left( CD \right)_{12}  \right] _{SS}&=	 \left[  \left( AB \right)_{45}  \right] _{SS}^{(C)}\;.
\end{align}

We introduce a few convenient notations.
A superscript $s$ (resp. $a$) refers to the symmetric (resp. 
antisymmetric) structures of the hard amplitude and of the GPD wrt $ x $, \ie
\beqa
\label{Def:a-s}
f(x) = \frac{1}{2} (f(x) + f(-x)) +  \frac{1}{2} (f(x) - f(-x))  = f^s(x) + 
f^a(x)\,.
\eqa
This thus leads to
\beqa
\label{generic-decomposition-rho-vector}
{\cal M}^V_ \meson  &=&
(Q_1^2 + Q_2^2)\, [(AB)_{123}]^s_{SS} \otimes {f}^s 
+ (Q_1^2 - Q_2^2) \,[(AB)_{123}]^a_{SS} \otimes {f}^a \nonumber \\
&& + 2 Q_1 \, Q_2 \,
[(AB)_{45}]^s_{SS} \otimes {f}^s\,,
\eqa
for the vector GPD contribution, and
\beqa
\label{generic-decomposition-rho-axial}
{\cal M}^A_ \meson &=&
(Q_1^2 + Q_2^2) \,[(AB)_{123}]^a_{SP} \otimes \tilde{f}^a 
+ (Q_1^2 - Q_2^2) \,[(AB)_{123}]^s_{SP} \otimes \tilde{f}^s \nonumber \\
&&+ 2 Q_1 \, Q_2 \,
[(AB)_{45}]^a_{SP} \otimes \tilde{f}^a\;,
\eqa
for the axial GPD contribution, \ie{}$SP$. In the above formulae, 
$Q_1=Q_u$ and $Q_2=Q_d$ corresponds to a $\meson^+,$  $Q_1=Q_d$ and $Q_2=Q_u$ corresponds to a 
$\meson^-$, and $ Q_1 = Q_2 = Q_{u,d} $ corresponds to $ \meson^{0} $.

In the case of $\rho^0$ meson 
production~\cite{Boussarie:2016qop}, which is $C (-)$, the exchange in the $t$-channel is fixed to be $C(-)$. In \EQ\eqref{generic-decomposition-rho-vector}, this implies that only the symmetric part of the vector GPD contributes, while in \EQ\eqref{generic-decomposition-rho-axial}, only the anti-symmetric part of the axial GPD contributes. On the other hand, $\meson^+$ production (and similarly for $\meson^-$) 
involves
both $C$-parity exchanges in $t-$channel, which explains why both symmetrical and 
antisymmetrical parts of  the GPDs are involved in 
\EQs\eqref{generic-decomposition-rho-vector} and \eqref{generic-decomposition-rho-axial}. 

The detailed evaluation of one diagram was already illustrated in \cite{Duplancic:2018bum}, and therefore, we do not repeat it here.

\subsubsection{Tensor structure}

For convenience, we introduce the common normalisation coefficients\footnote{{Note that the sign has been corrected here wrt our previous publication \cite{Boussarie:2016qop}. We note however that this does not affect the cross section, which correspond to the square of the amplitude and is therefore insensitive to the sign.}}
\begin{align}
	\label{coefCE}
		C^{\parallel} &= {-} \frac{4}{9}\,f_{\meson}^{\parallel} \, \alpha_{em}\,\alpha_s\,\pi^2 \,.
\end{align}   
Note that we include the charge factors $Q_u$ and $Q_d$  inside
the hard matrix element, using the decompositions obtained in 
\EQs\eqref{generic-decomposition-rho-vector} and
\eqref{generic-decomposition-rho-axial}.

For the $SS$ sector, 
two tensor structures appear, namely
\beqa
\label{def:TA-TB}
T_A &=& (\varepsilon_{q\perp} \cdot \varepsilon_{k\perp}^*)\,, \nonumber \\                                                  
T_B &=& (\varepsilon_{q\perp} \cdot p_\perp) (p_\perp \cdot                      
\varepsilon_{k\perp}^*)\,,
\eqa
while for the $SP$ sector, the two following structures appear
\beqa
\label{def:TA5-TB5}
T_{A_5} &=& (p_\perp \cdot                                      
\varepsilon_{k\perp}^*) \,  \epsilon^{n \,p \,\varepsilon_{q\perp}\, p_\perp}\,, 
\nonumber \\
T_{B_5} &=& -(p_\perp \cdot \varepsilon_{q\perp})\, \epsilon^{n \,p 
	\varepsilon_{k\perp}^*\, p_\perp}\,.
\eqa

\subsubsection{Organisation of the chiral-even amplitude}
\label{sec:organising-amplitude}

The scattering amplitude of the process \eqref{eq:process}, in the factorised 
form,
is expressed in terms of  form factors ${\cal H}_\meson$, ${\cal E}_\meson$, $\tilde 
{\cal H}_\meson,$ $\tilde {\cal E}_\meson$, analogous to Compton form factors in DVCS, 
and reads
\begin{eqnarray}
	\mathcal{M}^{\parallel}_\meson \equiv 
	\frac{1}{n\cdot p}\bar{u}(p_2,\lambda_2) \!\! \left[   \slashed {n}  {\cal 
		H}_\meson(\xi,t) +\frac{i\,\sigma^{n\,\alpha}\Delta_\alpha}{2m}  {\cal 
		E}_\meson(\xi,t) +    \slashed{n} \gamma^5  \tilde {\cal H}_\meson(\xi,t)
	+ \frac{n\cdot \Delta}{2m} \,\gamma^5\, \tilde {\cal E}_\meson(\xi,t)
	\right] \!\! u(p_1,\lambda_1). \!\!\!\!\!\nonumber
	\\
	\label{CEGPD}
\end{eqnarray}

We isolate the tensor structures of the form factors as
\begin{eqnarray}
	\label{dec-tensors-quarks}
	\mathcal{H}_\meson(\xi , t) &=&  \mathcal{H}_{\meson A} (\xi , t) T_{A} + 
	\mathcal{H}_{\meson B} (\xi , t) T_{B} \,,\nonumber\\
	\mathcal{\tilde{H}}_\meson(\xi , t) &=& \mathcal{\tilde{H}}_{\meson A_5} (\xi , t) T_{A_5} + 
	\mathcal{\tilde{H}}_{\meson B_5} (\xi , t) T_{B_5} \,.
\end{eqnarray}

These coefficients can be expressed in terms  of the sum over diagrams of the 
integral of the product of their traces, of GPDs and DAs, as defined and given 
explicitly in  
\APP{}C of \cite{Duplancic:2022ffo} for the case of the holographic DA, and \APP{}D of \cite{Duplancic:2018bum} for the asymptotic DA case.
We introduce dimensionless coefficients $N$ and $\tilde{N}$ as follows:
\beqa
\label{form-factors-NA-NB}
{\mathcal{H}}_{\meson A} = \frac{1}s C^{\parallel} {N}_{\meson A} \,, \quad
{\mathcal{H}}_{\meson B} = \frac{1}{s^2} C^{\parallel} {N}_{\meson B}\,.
\eqa
and
\beqa
\label{form-factors-tilde-NA5-NB5}
\tilde{\mathcal{H}}_{\meson A_5} = \frac{1}{s^3}C^{\parallel} \tilde N_{\meson A_5} \,, \quad
\tilde{\mathcal{H}}_{\meson B_5} = \frac{1}{s^3}C^{\parallel} \tilde N_{\meson B_5}\,,
\eqa
In order to emphasise the gauge invariant structure and to organise the 
numerical study, we factorise out the charge coefficients, and put an explicit
index $q$ for the flavour of the quark GPDs $f^q$ and $\tilde{f}^q$. In 
accordance
with the decompositions (\ref{generic-decomposition-rho-vector})
and (\ref{generic-decomposition-rho-axial}) we thus introduce\footnote{{Effectively, what changes here for the $ \meson $-meson case from the pion case is that the association of the coefficients in \eqref{gauge-tildeNA5} to \eqref{gauge-NB} to the GPDs is swapped, \ie{}vector for axial and vice-versa. This explains why the `tildes' are swapped wrt to the pion case.}}
\beqa
\label{gauge-NA}
{N}^q_{\meson A}(Q_1,Q_2)&= 
(Q_1^2 + Q_2^2) {N}^q_A[(AB)_{123}]^s
+ (Q_1^2 - Q_2^2) {N}^q_A[(AB)_{123}]^a
+ 2 Q_1 \, Q_2 \,
{N}^q_A[(AB)_{45}]^s\,,  \qquad
\\
\label{gauge-NB}
{N}^q_{\meson B}(Q_1,Q_2) &=
(Q_1^2 + Q_2^2) {N}^q_B[(AB)_{123}]^s
+ (Q_1^2 - Q_2^2) {N}^q_B[(AB)_{123}]^a
+ 2 Q_1 \, Q_2 \,
{N}^q_B[(AB)_{45}]^s\,,  \qquad
\eqa
and
\beqa
\label{gauge-tildeNA5}
\tilde N^q_{\meson A_5}(Q_1,Q_2) &=
(Q_1^2 + Q_2^2) 
\tilde{N}^q_{A_5}[(AB)_{123}]^a
+
(Q_1^2 - Q_2^2) \tilde{N}^q_{A_5}[(AB)_{123}]^s
+ 2 Q_1 \, Q_2 \,\tilde{N}^q_{A_5}[(AB)_{45}]^a\,, \qquad
\\
\label{gauge-tildeNB5}
\tilde N^q_{\meson B_5}(Q_1,Q_2) &=
(Q_1^2 + Q_2^2) 
\tilde{N}^q_{B_5}[(AB)_{123}]^a
+
(Q_1^2 - Q_2^2) \tilde{N}^q_{B_5}[(AB)_{123}]^s
+ 2 Q_1 \, Q_2 \,\tilde{N}^q_{B_5}[(AB)_{45}]^a\,. \qquad
\eqa
The above 4 terms, which have a superscript `$ q $', are not to be confused with the coefficients that appear in \eqref{form-factors-NA-NB} and \eqref{form-factors-tilde-NA5-NB5}. Instead, the 4 terms are used in (\ref{Nrho0p} - \ref{NTilderho-}) below to construct the coefficients.

For the specific case of our four processes, namely $\gamma \meson^{0}$ production on 
a proton (denoted by $ \mesonzp $), $\gamma \meson^{0}$ production on a neutron (denoted by $ \mesonzn $), $\gamma \meson^{+}$ production on 
a proton and $\gamma \meson^{-}$ production on a neutron,  taking into account the 
structure (\ref{TransitionGPD}) of the transition GPDs structure
we thus need to compute the coefficients
\begin{alignat}{4}
	\label{Nrho0p}
	&{N}_{\mesonzp A}&\,=\,&\frac{1}{\sqrt{2}} \left[ {N}^u_{\meson A}(Q_u,Q_u) - {N}^d_{\meson A}(Q_d,Q_d) \right]  \,,&\quad&
	{N}_{\mesonzp B}&\,=\,&\frac{1}{\sqrt{2}} \left[ {N}^u_{\meson B}(Q_u,Q_u) - {N}^d_{\meson B}(Q_d,Q_d) \right] \,, \\[5pt]
	\label{Nrho0n}
	&{N}_{\mesonzn A}&\,=\,&\frac{1}{\sqrt{2}} \left[ {N}^u_{\meson A}(Q_d,Q_d) - {N}^d_{\meson A}(Q_u,Q_u) \right]  \,,&\quad&
	{N}_{\mesonzn B}&\,=\,&\frac{1}{\sqrt{2}} \left[ {N}^u_{\meson B}(Q_d,Q_d) - {N}^d_{\meson B}(Q_u,Q_u) \right] \,, \\[5pt]
	\label{Nrho+}
	&{N}_{\meson^+ A}&\,=\,&{N}^u_{\meson A}(Q_u,Q_d) - {N}^d_{\meson A}(Q_u,Q_d) \,, &\quad&
	{N}_{\meson^+ B}&\,=\,&{N}^u_{\meson B}(Q_u,Q_d) - {N}^d_{\meson B}(Q_u,Q_d)\,, \\[5pt]
	\label{Nrho-}
	&{N}_{\meson^- A}&\,=\,&{N}^u_{\meson A}(Q_d,Q_u) - {N}^d_{\meson A}(Q_d,Q_u)\,, &\quad&
	{N}_{\meson^- B}&\,=\,&{N}^u_{\meson B}(Q_d,Q_u) - {N}^d_{\meson B}(Q_d,Q_u)\,,
\end{alignat}
corresponding to the case with vector GPDs, as well as
\begin{alignat}{4}
	\label{NTilderho0p}
	&\tilde{N}_{\mesonzp A_5}&\,=\,&\frac{1}{\sqrt{2}} \left[ \tilde{N}^u_{\meson A_5}(Q_u,Q_u) - \tilde{N}^d_{\meson A_5}(Q_d,Q_d) \right]  \,,&\quad&
	\tilde{N}_{\mesonzp B_5}&\,=\,&\frac{1}{\sqrt{2}} \left[ \tilde{N}^u_{\meson B_5}(Q_u,Q_u) - \tilde{N}^d_{\meson B_5}(Q_d,Q_d) \right] \,, \\[5pt]
	\label{NTilderho0n}
	&\tilde{N}_{\mesonzn A_5}&\,=\,&\frac{1}{\sqrt{2}} \left[ \tilde{N}^u_{\meson A_5}(Q_d,Q_d) - \tilde{N}^d_{\meson A_5}(Q_u,Q_u) \right]  \,,&\quad&
	\tilde{N}_{\mesonzn B_5}&\,=\,&\frac{1}{\sqrt{2}} \left[ \tilde{N}^u_{\meson B_5}(Q_d,Q_d) - \tilde{N}^d_{\meson B_5}(Q_u,Q_u) \right] \,, \\[5pt]
	\label{NTilderho+}
	&\tilde{N}_{\meson^+ A_5}&\,=\,&\tilde{N}^u_{\meson A_5}(Q_u,Q_d) - \tilde{N}^d_{\meson A_5}(Q_u,Q_d) \,, &\quad&
	\tilde{N}_{\meson^+ B_5}&\,=\,&\tilde{N}^u_{\meson B_5}(Q_u,Q_d) - \tilde{N}^d_{\meson B_5}(Q_u,Q_d)\,, \\[5pt]
	\label{NTilderho-}
	&\tilde{N}_{\meson^- A_5}&\,=\,&\tilde{N}^u_{\meson A_5}(Q_d,Q_u) - \tilde{N}^d_{\meson A_5}(Q_d,Q_u)\,, &\quad&
	\tilde{N}_{\meson^- B_5}&\,=\,&\tilde{N}^u_{\meson B_5}(Q_d,Q_u) - \tilde{N}^d_{\meson B_5}(Q_d,Q_u)\,,
\end{alignat}
which correspond to the case of axial GPDs.
Therefore, for each flavour $u$ and $d$, 
knowing the 12 numerical coefficients 
\beqa
\label{list-N-to-be-computed}
&& \tilde{N}^q_{A_5}[(AB)_{123}]^s, \ \tilde{N}^q_{A_5}[(AB)_{123}]^a, \ \tilde{N}^q_{A_5}[(AB)_{45}]^a, 
\nonumber \\
&& \tilde{N}^q_{B_5}[(AB)_{123}]^s, \ \tilde{N}^q_{B_5}[(AB)_{123}]^a, \ \tilde{N}^q_{B_5}[(AB)_{45}]^a, 
\nonumber \\ 
&& {N}^q_A[(AB)_{123}]^s, \ {N}^q_A[(AB)_{123}]^a, \ 
{N}^q_A[(AB)_{45}]^s, \nonumber \\
&& {N}^q_B[(AB)_{123}]^s, \ {N}^q_B[(AB)_{123}]^a, \  
{N}^q_B[(AB)_{45}]^s,
\eqa
for two given GPDs $f$ and $\tilde{f}$ (in practice $H$ and 
$\tilde{H}$, see next subsection), one can reconstruct the scattering amplitudes of the two processes.
These 12 coefficients can be expanded in terms of 5 building block integrals which we label as $I_b$, $I_c$, $I_h$, $I_i$ and $I_e$ for the asymptotic DA case, and 2 extra building blocks labelled as $\chi_b$, $\chi_c$
for the case of the holographic DA. The building block integrals can be found in \APP{}C of \cite{Duplancic:2022ffo}, and in \APP{}D of \cite{Duplancic:2018bum}.

\subsubsection{Cross section}
\label{sec:cross section}

In the forward limit $\Delta_{\bot} = 0 = P_{\bot}$, one can show that the 
square of the $\mathcal{M}^{\parallel}_\meson$ from \eqref{CEGPD} reads, after summing over nucleon helicities
\begin{eqnarray}
	\label{squareCEresult}
	\mathcal{M}^{\parallel}_{\meson} \mathcal{M}_{\meson}^{\parallel*} &\equiv  &
	\sum_{\lambda_2,\, \lambda_1}
	\mathcal{M}^{\parallel}_{\meson} (\lambda_1,\lambda_2)\,
	\mathcal{M}_{\meson}^{\parallel*}(\lambda_1,\lambda_2) \\  
	&=&   8(1-\xi^2) 
	\left(  {\cal H}_{\meson}(\xi,t)  {\cal H}^{*}_\meson(\xi,t)    +  \tilde {\cal 
		H}_\meson(\xi,t) \tilde {\cal H}^{*}_\meson(\xi,t)  \right) \nonumber \\
	&&+8\,\frac{\xi^4}{1-\xi^2}   
	\left(  {\cal E}_\meson(\xi,t)
	{\cal E}^{*}_\meson(\xi,t)
	+  \tilde {\cal E}_\meson(\xi,t)
	\tilde {\cal E}^{*}_\meson(\xi,t)
	\right)\nonumber
	\\ 
	&&-8\, \xi^2   \left(  {\cal H}_\meson(\xi,t) {\cal E}^{ *}_\meson(\xi,t) + 
	{\cal H}^{*}_\meson(\xi,t) {\cal E}_\meson(\xi,t)
	+
	\tilde {\cal H}_\meson(\xi,t)\tilde {\cal E}^{*}_\meson(\xi,t) 
	+
	\tilde {\cal H}^{*}_\meson(\xi,t)\tilde {\cal E}_\meson(\xi,t)
	\right) .\nonumber
\end{eqnarray}
For moderately small values of $\xi$, this becomes
\begin{eqnarray}
	\label{squareCEresultsmallxi}
	\mathcal{M}^{\parallel}_\meson \mathcal{M}^{\parallel*}_\meson &\simeq&   8
	\left(  {\cal H}_\meson(\xi,t) \, {\cal H}^{*}_\meson(\xi,t)    +  \tilde {\cal 
		H}_\meson(\xi,t)\, \tilde {\cal H}^{*}_\meson(\xi,t)  \right).
\end{eqnarray}
Hence we will restrict ourselves to the GPDs $H$, $\tilde{H}$ to perform our 
estimates of the cross section\footnote{In practice, we keep the first line in 
	the RHS of \eqref{squareCEresult}.}. We note that this approximation remains valid for the linear polarisation asymmetry wrt the incoming photon, as the above equation still contains the helicities of the incoming and outgoing photons.

We now perform the sum/averaging over the polarisations of the incoming and outgoing photons,
\begin{eqnarray}
	\label{FF-squared-H}
	|{\mathcal{H}}_\meson(\xi , t)|^2 & \equiv & \sum_{\lambda_k, \lambda_q} 
	{\mathcal{H}}_\meson(\xi , t, \lambda_k, \lambda_q) \, 
	{\mathcal{H}}_\meson(\xi , t, \lambda_k, \lambda_q) \\ \nonumber
	&=& 2|{\mathcal{H}}_{\meson A} (\xi , t)|^2 + p_\bot^4 | {\mathcal{H}}_{\meson B}
	(\xi , t)|^2 + p_\bot^2 \left[ {\mathcal{H}}_{\meson A} (\xi , 
	t){\mathcal{H}}^{\ast}_{\meson B} (\xi , t) + {\mathcal{H}}^{\ast}_{\meson A} (\xi , 
	t){\mathcal{H}}_{\meson B} (\xi , t) \right], \\ \nonumber \\
	|\tilde{\mathcal{H}}_\meson(\xi , t)|^2 &\equiv & \sum_{\lambda_k, \lambda_q} 
	 \tilde{\mathcal{H}} _\meson(\xi , t, \lambda_k, \lambda_q) \,  \tilde{\mathcal{H}} _\meson^*(\xi , t, \lambda_k, 
	\lambda_q)
	\\ \nonumber
	\label{FF-squared-HTilde}
	&=& \frac{s^2 p_\bot^4}{4} \left(|  \tilde{\mathcal{H}} _{\meson A_5} (\xi , t)|^2 + | 
	 \tilde{\mathcal{H}} _{\meson B_5} (\xi , t) |^2\right). \nonumber 
\end{eqnarray}
Finally, we define the averaged amplitude squared $|\mathcal{\overline{M}}^{\parallel}_\meson|^2,$ 
which includes 
the factor 1/4 coming from the averaging over the 
polarisations of the initial particles. Collecting all prefactors, which read 
\beq
\label{coefficients}
\frac{1}{s^2}   8 (1-\xi^2) |C^{\parallel}|^2 \frac{1}{2^2}\,,
\eq
we have that
\beqa
\label{all-rho}
&&|\mathcal{\overline{M}}^{\parallel}_{\meson}|^2 = \frac{2}{s^2}   (1-\xi^2)  |C^{\parallel}|^2 
\left\{ 
2 \left|{N}_{\meson A} \right|^2 
+ \frac{p_\perp^4}{s^2} \left|{N}_{\meson B} \right|^2 
\right.\\
&&
\left.
+ \frac{p_\perp^2}s \left({N}_{\meson A} 
{N}_{\meson B}^* + c.c. \right)
+ \frac{p_\perp^4}{4 s^2} 
\left|\tilde N_{\meson A_5} \right|^2 
+ \frac{p_\perp^4}{4 s^2} \left|\tilde N_{\meson B_5} \right|^2
\right\}.\nonumber
\eqa
Here $\meson$ corresponds to $\mesonzp$, $ \mesonzn $, $ \mesonpp $ or $ \mesonmn $, where the subscript denotes the target. The corresponding coefficients
${N}_{\meson A}$, ${N}_{\meson B}$, $\tilde N_{\meson A_5}$, $\tilde N_{\meson B_5}$
 are given by (\ref{Nrho0p} - \ref{NTilderho-}).

The differential cross section as a function of $t$, $M^2_{\gamma\meson},$ $-u'$ 
then reads
\begin{equation}
	\label{difcrosec}
	\left.\frac{d\sigma^{\parallel}}{dt \,du' \, dM^2_{\gamma\meson}}\right|_{\ -t=(-t)_{ \mathrm{min} }} = 
	\frac{|\mathcal{\overline{M}}^{\parallel}_\meson|^2}{32S_{\gamma 
			N}^2M^2_{\gamma\meson}(2\pi)^3}\,.
\end{equation}

\subsection{Chiral-odd case}

As before, one can group diagrams based on their charges. Using the same notations as in \SEC\ref{sec:amplitude-chiral-even}, exploiting the $ C- $parity symmetry of the process, one obtains
\begin{align}
\label{sumTmeson}
{\cal M}_ \meson ^\perp=Q_1^2 [(AB)_{123}]_{TT}  \otimes {f}_{T} + Q_1 Q_2 [(AB)_{45}]_{TT} \otimes 
{f}_{T}
+ Q_2^2 [(AB)_{123}]_{TT}^{(C)}  \otimes {f}_{T} + Q_1 Q_2 
[(AB)_{45}]_{TT}^{(C)} \otimes {f}_{T}\,,
\end{align}
where
\begin{align}
	\label{eq:CD345TT}
	\left[  \left( CD \right)_{345}  \right] _{TT}&= \left[  \left( AB \right)_{123}  \right] _{TT}^{(C)}\;,\\[5pt]
		\label{eq:CD12TT}
\left[  \left( CD \right)_{12}  \right] _{TT}&=	 \left[  \left( AB \right)_{45}  \right] _{TT}^{(C)}\;,
\end{align}
have been used. In the above, $ f_{T} $ represents a generic `tensor' chiral-odd GPD (in practice, $ H_{T} $). Only 8 diagrams out of the 20 diagrams are non-vanishing in the chiral-odd case. They are $ A_3 $, $ B_1 $, $ A_4 $ and $ B_5 $, and the corresponding ones given by the symmetry transformation in \eqref{eq:CD345TT} and \eqref{eq:CD12TT}. Writing the GPDs in terms of symmetric and anti-symmetric parts, we have
\begin{eqnarray}
\label{generic-decomposition-rho-tensor}
{\cal M}^\perp_ \meson  &=&
(Q_1^2 + Q_2^2)\, [(AB)_{123}]^s_{TT} \otimes {f}_{T}^s 
+ (Q_1^2 - Q_2^2) \,[(AB)_{123}]^a_{TT} \otimes {f}_{T}^a \nonumber \\
&& + 2 Q_1 \, Q_2 \,
[(AB)_{45}]^s_{TT} \otimes {f}^s_{T}\,.
\end{eqnarray}
The detailed evaluation of one diagram was performed in \cite{Boussarie:2016qop}, and we do not repeat this here.

\subsubsection{Tensor structure}

It is convenient to introduce the common normalisation factor\footnote{{Note that the sign has been corrected here wrt our previous publication \cite{Boussarie:2016qop}. We note however that this does not affect the cross section, which correspond to the square of the amplitude and is therefore insensitive to the sign.}}
\begin{align}
C^{\perp} &=  {\frac{4}{9}}\,f_{\meson}^{\perp} \, \alpha_{em}\,\alpha_s\,\pi^2 \,.
\end{align}
Note that we include the charge factors $Q_u$ and $Q_d$  inside
the hard matrix element, using the decomposition obtained in 
\eqref{generic-decomposition-rho-tensor}.

In this case, two tensor structures appear, namely
\begin{eqnarray} 
	\label{def-TiAperp}
	T_{A\perp}^i & = & \frac{-8s}{\bar{\alpha}}\left\{ \alpha\varepsilon_{k\perp}^{i*}\left[\left(p_{\perp}\cdot\varepsilon_{q\perp}\right)\left(p_{\perp}\cdot\varepsilon_{\meson\perp}^*\right)+\alpha\bar{\alpha}\xi s\left(\varepsilon_{q\perp}\cdot\varepsilon_{\meson\perp}^*\right)\right]\right.\\
	\nonumber
	&  & \!\!-\bar{\alpha}\varepsilon_{\meson\perp}^{i*}\left[\alpha\left(\alpha-2\right)\xi s\left(\varepsilon_{q\perp}\cdot\varepsilon_{k\perp}^*\right)-\left(p_{\perp}\cdot\varepsilon_{q\perp}\right)\left(p_{\perp}\cdot\varepsilon_{k\perp}^*\right)\right]\\ \nonumber
	&  &\!\! +p_{\perp}^{i}\left[\left(p_{\perp}\cdot\varepsilon_{\meson\perp}^*\right)\left(\varepsilon_{q\perp}\cdot\varepsilon_{k\perp}^*\right)-\bar{\alpha}\left(\varepsilon_{k\perp}^*\cdot\varepsilon_{\meson\perp}^*\right)\left(p_{\perp}\cdot\varepsilon_{q\perp}\right)\right]\nonumber\\ 
	&  &\!\! \left.+\varepsilon_{q\perp}^i\left[-\left(p_{\perp}\cdot\varepsilon_{\meson\perp}^*\right)\left(p_{\perp}\cdot\varepsilon_{k\perp}^*\right)+\alpha\bar{\alpha}\left(\alpha-2\right)\xi s\left(\varepsilon_{k\perp}^*\cdot\varepsilon_{\meson\perp}^*\right)\right]\right\} \,,\nonumber
\end{eqnarray} 
\begin{eqnarray} 
	\label{def-TiBperp}
	T_{B\perp}^i & = & \frac{8s}{\alpha\bar{\alpha}}\left\{ \bar{\alpha}\varepsilon_{\meson\perp}^{i*}\left[\left(p_{\perp}\cdot\varepsilon_{q\perp}\right)\left(p_{\perp}\cdot\varepsilon_{k\perp}^*\right)-\alpha\left(2\alpha-1\right)\xi s\left(\varepsilon_{q\perp}\cdot\varepsilon_{k\perp}^*\right)\right]\right.\\
	&  & +\alpha\varepsilon_{k\perp}^{i*}\left[\bar{\alpha}\left(2\alpha-1\right)\xi s\left(\varepsilon_{q\perp}\cdot\varepsilon_{\meson\perp}^*\right)+\left(p_{\perp}\cdot\varepsilon_{q\perp}\right)\left(p_{\perp}\cdot\varepsilon_{\meson\perp}^*\right)\right]\nonumber\\ 
	&  & +\varepsilon_{q\perp}^i\left[-\left(p_{\perp}\cdot\varepsilon_{\meson\perp}^*\right)\left(p_{\perp}\cdot\varepsilon_{k\perp}^*\right)-\alpha\bar{\alpha}\xi s\left(\varepsilon_{k\perp}^*\cdot\varepsilon_{\meson\perp}^*\right)\right] \nonumber \\ 
	&  & \left.+p_{\perp}^{i}\left[-\alpha\left(p_{\perp}\cdot\varepsilon_{\meson\perp}^*\right)\left(\varepsilon_{q\perp}\cdot\varepsilon_{k\perp}^*\right)-\bar{\alpha}\left(\varepsilon_{q\perp}\cdot\varepsilon_{\meson\perp}^*\right)\left(p_{\perp}\cdot\varepsilon_{k\perp}^*\right)\right]\right\}\,.
	\nonumber
\end{eqnarray}
When summing over the polarisations $ \varepsilon_{\meson\perp} $ of the $ \meson $-meson in order to compute the square of the amplitude, only
\begin{align}
	\sum_{T}\varepsilon_{\meson\perp}^\mu\, \varepsilon_{\meson\perp}^{\nu\ast}= -g^{ \mu  \nu }_{\perp}\,,
\end{align}
is needed, since we have chosen the basis as defined in  \SEC\ref{sec:kinematics}.

\subsubsection{Organisation of the chiral-odd amplitude}

Following the same steps as the chiral-even case, we can write the chiral-odd amplitude in terms of form factors ${\cal H}^{j}_{T \meson}, \tilde {\cal H}^{j}_{T \meson},$  ${\cal E}^{j}_{T \meson}, \tilde {\cal E}^{j}_{T \meson}$, analogous to Compton form factors in DVCS,
\begin{eqnarray}
	\label{eq:chiral-odd-amplitude}
 \mathcal{M}^{\perp}_{\meson} &\equiv& \frac{1}{n\cdot p}\,\bar{u}(p_2,\lambda_{2})\, \left[i\,\sigma^{nj}{\cal H}_{T \meson\, j}(\xi,t)
+ \frac{P\cdot n\;\Delta^j-\Delta\cdot n\;P^j}{m^2} \tilde {\cal H}_{T \meson\, j}(\xi,t) \right.
\nonumber \\
&& \left.
+ \frac{\gamma\cdot n\;\Delta^j-\Delta\cdot n\;\gamma^j}{2m}{\cal E}_{T \meson\, j}(\xi,t)
+ \frac{\gamma\cdot n\;P^j-P\cdot n\;\gamma^j}{m} \tilde {\cal E}_{T \meson\, j}(\xi,t) \right]u(p_1,\lambda_{1})\,,
\label{COGPD}
\end{eqnarray}
where $ j $ corresponds to a transverse vector index. From the form factors, one can isolate the following tensor structures
\begin{align}
	\mathcal{H}^{j}_{T }(\xi , t) = \mathcal{H}_{{T \meson}\,A} (\xi , t) T_{A\perp}^j + \mathcal{H}_{T\meson\,B} (\xi , t) T_{B\perp}^j.
\end{align}
We further express the above coefficients in terms of dimensionless ones through
\beqa
\label{form-factors-NT}
\mathcal{H}_{T \meson\,A} = \frac{1}{s^3}C^{\perp} N_{T\meson\, A} \,, \\
\mathcal{H}_{T \meson\,B} = \frac{1}{s^3}C^{\perp} N_{T\meson\, B}\,.
\eqa
Proceeding as in \SEC\ref{sec:organising-amplitude}, the electric charges are factorised, and we introduce an explicit index $ q $ to denote the flavour of the quark GPDs $ f_{T}^{q} $ and $ \tilde{f}_{T}^{q} $. Thus, using the decomposition in \eqref{generic-decomposition-rho-tensor}, we have that
\beqa
\label{gauge-NAT}
{N}^q_{T\meson\, A}(Q_1,Q_2)&= 
(Q_1^2 + Q_2^2) {N}^q_{T\,A}[(AB)_{123}]^s
+ (Q_1^2 - Q_2^2) {N}^q_{T\,A}[(AB)_{123}]^a
+ 2 Q_1 \, Q_2 \,
{N}^q_{T\,A}[(AB)_{45}]^s\,,  \qquad
\\
\label{gauge-NBT}
{N}^q_{T\meson\, B}(Q_1,Q_2) &=
(Q_1^2 + Q_2^2) {N}^q_{T\,B}[(AB)_{123}]^s
+ (Q_1^2 - Q_2^2) {N}^q_{T\,B}[(AB)_{123}]^a
+ 2 Q_1 \, Q_2 \,
{N}^q_{T\,B}[(AB)_{45}]^s\,.  \qquad
\eqa
Just like in the chiral-even case, the above 2 terms, which have a superscript `$ q $', are not to be confused with the coefficients that appear in \eqref{form-factors-NT}. Instead, the 2 terms are used in (\ref{NrhoT0p} - \ref{NrhoT-}) below to construct the coefficients.

For the specific case of our four processes, namely $\gamma \meson^{0}$ production on 
a proton (denoted by $ \mesonzp $), $\gamma \meson^{0}$ production on a neutron (denoted by $ \mesonzn $), $\gamma \meson^{+}$ production on 
a proton and $\gamma \meson^{-}$ production on a neutron,  taking into account the 
structure (\ref{TransitionGPD}) of the transition GPDs structure, we find that the following coefficients need to be computed,
\begin{alignat}{4}
	\label{NrhoT0p}
	&{N}_{T\mesonzp A}&\,=\,&\frac{1}{\sqrt{2}} \left[ {N}^u_{T\meson A}(Q_u,Q_u) - {N}^d_{T\meson A}(Q_d,Q_d) \right]  \,,&\quad&
	{N}_{T\mesonzp B}&\,=\,&\frac{1}{\sqrt{2}} \left[ {N}^u_{T\meson B}(Q_u,Q_u) - {N}^d_{T\meson B}(Q_d,Q_d) \right] \,, \\[5pt]
	\label{NrhoT0n}
	&{N}_{T\mesonzn A}&\,=\,&\frac{1}{\sqrt{2}} \left[ {N}^u_{T\meson A}(Q_d,Q_d) - {N}^d_{T\meson A}(Q_u,Q_u) \right]  \,,&\quad&
	{N}_{T\mesonzn B}&\,=\,&\frac{1}{\sqrt{2}} \left[ {N}^u_{T\meson B}(Q_d,Q_d) - {N}^d_{T\meson B}(Q_u,Q_u) \right] \,, \\[5pt]
	\label{NrhoT+}
	&{N}_{T\meson^+ A}&\,=\,&{N}^u_{T\meson A}(Q_u,Q_d) - {N}^d_{T\meson A}(Q_u,Q_d) \,, &\quad&
	{N}_{T\meson^+ B}&\,=\,&{N}^u_{T\meson B}(Q_u,Q_d) - {N}^d_{T\meson B}(Q_u,Q_d)\,, \\[5pt]
	\label{NrhoT-}
	&{N}_{T\meson^- A}&\,=\,&{N}^u_{T\meson A}(Q_d,Q_u) - {N}^d_{T\meson A}(Q_d,Q_u)\,, &\quad&
	{N}_{T\meson^- B}&\,=\,&{N}^u_{T\meson B}(Q_d,Q_u) - {N}^d_{T\meson B}(Q_d,Q_u)\,.
\end{alignat}
In practice, we deduce that the following 6 numerical coefficients need to be computed, for each flavour $ u $ and $ d $,
\beqa
\label{list-NT-to-be-computed}
&& {N}^q_{T\,A}[(AB)_{123}]^s, \ {N}^q_{T\,A}[(AB)_{123}]^a, \ 
{N}^q_{T\,A}[(AB)_{45}]^s, \nonumber \\
&& {N}^q_{T\,B}[(AB)_{123}]^s, \ {N}^q_{T\,B}[(AB)_{123}]^a, \  
{N}^q_{T\,B}[(AB)_{45}]^s\,,
\eqa
for a given chiral-odd GPD $ f_{T} $ (in practice, $ H_{T} $). This is sufficient to reconstruct the amplitude for all the 4 processes we are interested in. The 6 coefficients can be expressed in terms of 3 building block integrals, which we label as $ I_{e} $, $ I_{i} $ and $ I_{d} $ for the asymptotic DA case, and 1 extra building block integral labelled $ \chi_{a} $ for the case of the holographic DA. The expressions for the coefficients in \eqref{list-NT-to-be-computed} in terms of the building block integrals are given in \APP\ref{app:CO-amplitudes}.

\subsubsection{Cross section}

In the forward limit $\Delta_{\bot} = 0 = P_{\bot}$, one can show that the square of $\mathcal{M}_{\bot}$ in \eqref{eq:chiral-odd-amplitude} reads, after summing over nucleon helicities,
\begin{eqnarray}
	\label{squareCOresult}
	 \mathcal{M}^{\perp}_{\meson} \mathcal{M}^{\perp *}_{\meson}  &\equiv & 
	\sum_{\lambda_1,\, \lambda_2}
	\mathcal{M}^{\perp}_{\meson} (\lambda_1,\lambda_2)\,
	\mathcal{M}^{\perp *}_{\meson}(\lambda_1,\lambda_2) 
	\\ \nonumber 
	&=& 8\left[\!   
	-(1-\xi^2) {\cal H}_{T\meson}^{ i}(\xi,t)
	{\cal H}_{T\meson}^{ j \,*}(\xi,t)
	- \frac{\xi^2}{1-\xi^2} [ \xi \,
	{\cal E}_{T\meson}^{ i}(\xi,t) 
	- \tilde {\cal E}_{T\meson}^{ i}(\xi,t)       ]  [ \xi \,
	{\cal E}_{T\meson}^{j *}(\xi,t) - 
	\tilde {\cal E}_{T\meson}^{j *}(\xi,t) ] 
	\right. \\ \nonumber
	&& \left.
	+\, \xi  \left\{ 
	{\cal H}_{T\meson}^{i}(\xi,t)[\xi \,
	{\cal E}_{T\meson}^{j}(\xi,t) - 
	\tilde {\cal E}_{T\meson}^{ j}(\xi,t)  ]^* +
	{\cal H}_{T\meson}^{i *}(\xi,t)[\xi \, 
	{\cal E}_{T\meson}^{j}(\xi,t) 
	- \tilde {\cal E}_{T\meson}^{j}(\xi,t)  ]
	\right\}\right] g_{\perp ij}.
\end{eqnarray}
For moderately small values of $\xi$, it reduces to
\begin{eqnarray}
	\label{squareCOresultsmallxi}
	\mathcal{M}^{\perp}_{\meson} \mathcal{M}^{\perp *}_{\meson} 
	&=& -8\, {\cal H}_{T\meson}^{ i}(\xi,t)\,
	{\cal H}_{T\meson}^{ j \,*}(\xi,t) \,  g_{\perp ij}\,.
\end{eqnarray}
Hence, we will restrict ourselves to $H_{T\meson}$ to perform our estimates of the cross section.\footnote{In practice, we keep the first term
	in the RHS of \EQ\eqref{squareCOresult}.} Performing the sum over the transverse polarisations of the $  \meson $-meson, and the incoming and outgoing photons, one obtains
\begin{alignat}{2}
	\label{FF-squared-HT}
-g_{\perp i\, j} \sum_{\lambda_k \lambda_q \lambda_\meson} \mathcal{H}_{T\meson}^{i}(\xi , t, \lambda_k, \lambda_q, \lambda_\meson) \mathcal{H}_{T\meson}^{j*}(\xi , t, \lambda_k, \lambda_q, \lambda_\meson) 
	&\,=\,& 512\xi^2 s^4 \left(\alpha^4 | \mathcal{H}_{T\meson\, A} (\xi , t)|^2 +  |\mathcal{H}_{T\meson\, B} (\xi , t)|^2 \right).
\end{alignat}
We can now compute the averaged amplitude squared $ |{\cal M}_{\meson}^{\perp}|^2 $, which includes a factor of $ 1/4 $ coming from the averaging of the polarisations of the incoming particles. Collecting all prefactors,
\begin{align}
512 \xi ^2 s^4 \times	\frac{1}{s^6}|C^{\perp}|^2 \times \frac{1}{4}\times 8  \left( 1-\xi^2 \right) \,,
\end{align}
we have that
\beqa
\label{all-rhoT}
&&|\mathcal{\overline{M}}_{\meson}^{\perp}|^2 = \frac{1024}{s^2}   \xi^2 (1-\xi^2)  |C^{\perp}|^2 
 \left[ \alpha^4 |{N}_{T\meson A} (\xi , t)|^2 +  |{N}_{T\meson B} (\xi , t)|^2 \right] \,.
\eqa
Here $\meson$ corresponds to $\mesonzp$, $ \mesonzn $, $ \mesonpp $ or $ \mesonmn $, where the subscript denotes the target. The corresponding coefficients
${N}_{T\meson A}$ and ${N}_{T\meson B}$ are given by (\ref{NrhoT0p} - \ref{NrhoT-}).

As for the longitudinally polarised $ \meson $-meson case, the differential cross section as a function of $ \left( -t \right) $, $M^2_{\gamma\meson},$ $ \left( -u' \right) $ 
then reads
\begin{equation}
	\label{difcrosecT}
	\left.\frac{d\sigma^{\perp}}{d \left( -t \right)  \,d \left( -u' \right)  \, dM^2_{\gamma\meson}}\right|_{\ -t=(-t)_{ \mathrm{min} }} = 
	\frac{|\mathcal{\overline{M}}^{\perp}_\meson|^2}{32S_{\gamma 
			N}^2M^2_{\gamma\meson}(2\pi)^3}\,.
\end{equation}

\subsection{Polarisation asymmetry}

\label{sec:polarisation-asymmetry}

\subsubsection{Chiral-even case}

\label{sec:CE-pol-asym}

In the chiral-even case, as discussed in \APP\ref{app:vanishing-circular-asymmetry}, the circular polarisation asymmetry vanishes as a result of conservation of parity $ P $ for an unpolarised target, which is the case we consider here.\footnote{The circular double spin asymmetry does not vanish and may be an interesting observable for a polarised target experiment.} Therefore, we compute the \textit{linear} polarisation asymmetry (LPA) wrt the incoming photon, which is defined by
\begin{align}
	\label{eq:LPA}
	\mathrm{LPA} =\frac{	\int d \sigma _{x}-\int d \sigma _{y}}{\int d \sigma _{x}+\int d \sigma _{y}}\,,
\end{align}
where $ d \sigma _{x (y)} $ corresponds to the differential cross section with the incoming photon linearly polarised along the $ x (y) $-direction. The integral symbol in \EQ\eqref{eq:LPA} corresponds to phase space integration and hence, the LPA can be calculated at the fully differential (by dropping the integral altogether), single differential or integrated levels.

The LPA is usually calculated in the lab frame, which corresponds to fixing the directions of the polarisation vectors. However, for convenience in performing the computation, we first take the polarisation vector in the $ x $-direction to be along $ p_{\perp} $, which changes on an event-by-event basis. Then, the polarisation vector in the $ y $-direction is chosen such that the $ x,\,y,\,z $-directions form a right-handed basis. Thus,
\begin{align}
	\label{eq:polx}
	\varepsilon_{x}^{ \mu }(q)&\equiv \frac{p_{\perp}^{ \mu }}{| \vec{p}_{t}| }\,,\\
	\label{eq:poly}
	\varepsilon_{y}^{ \mu }(q)&\equiv - \frac{2}{s | \vec{p} _{t}|} \epsilon ^{p n p_{\perp} \mu }\,.
\end{align}
The LPA corresponding to this choice of polarisation vectors is denoted by $  \mathrm{LPA}_{ \mathrm{max} }  $, since the directions of the polarisation vectors are such that the LPA is maximised. The LPA in the lab frame, $  \mathrm{LPA}_{ \mathrm{Lab} }  $,  can then be related to $  \mathrm{LPA}_{ \mathrm{max} }  $ via a simple modulation of $ \cos 2 \theta $, where $ \theta  $ corresponds to the angle between $ p_{\perp} $ and the $ x $-direction defined by the lab frame. Thus,
\begin{align}
	\mathrm{LPA} _{ \mathrm{Lab} }= \mathrm{LPA}_{ \mathrm{max} } \cos 2 \theta \,.
\end{align}
The proof of this result, including the derivation of relevant expressions for the LPA, can be found in \APP F in \cite{Duplancic:2022ffo}. When showing the results in \SEC\ref{sec:results}, we therefore choose to show plots for $  \mathrm{LPA}_{ \mathrm{max} }  $, as the modification due to $ \cos 2 \theta $ is trivial. 

We now turn to the calculation of $  \mathrm{LPA}_{ \mathrm{max} }  $. Amplitudes corresponding to specific linear polarisation states in \eqref{eq:polx} and \eqref{eq:poly} can be defined as
\begin{align}
	{\cal M } _{x} = \varepsilon_{x}^{ \mu }(q) {\cal M } _{ \mu }\,,\qquad 	 {\cal M } _{y} = \varepsilon_{y}^{ \mu }(q) {\cal M } _{ \mu }\,.
\end{align}
For convenience, let us decompose the amplitude as (\cf\EQs\eqref{def:TA-TB} to \eqref{dec-tensors-quarks})
\begin{align}
	\label{eq:decomposition-tensors}
	{\cal M } =C_{A}T_{A}+C_{B}T_{B}+C_{A_{5}}T_{A_{5}}+C_{B_{5}}T_{B_{5}}\,,
\end{align}
\ie{}in terms of the tensor structures $ T_{A},\,T_{B},\,T_{A_5},\,T_{B_5} $ defined in \eqref{def:TA-TB} and \eqref{def:TA5-TB5}. We note that the coefficients of the tensor structures include the spinors of the nucleons, as well as Dirac matrices associated with the definition of the GPDs. More explicitly, using \eqref{CEGPD} and \eqref{dec-tensors-quarks}, 
\begin{align}
	\label{eq:CA}
	C_A&\equiv\frac{1}{n\cdot p}\bar{u}(p_2,\lambda_2) \slashed{n}  u(p_1,\lambda_1)\mathcal{{H}}_{\meson A} (\xi , t)\,,\\
	C_B&\equiv\frac{1}{n\cdot p}\bar{u}(p_2,\lambda_2) \slashed{n}  u(p_1,\lambda_1)\mathcal{{H}}_{\meson B} (\xi , t)\,,\\
	C_{A_5}&\equiv\frac{1}{n\cdot p}\bar{u}(p_2,\lambda_2) \slashed{n} \gamma^5 u(p_1,\lambda_1)\mathcal{\tilde{H}}_{\meson A_5} (\xi , t)\,,\\
	\label{eq:CB5}
	C_{B_5}&\equiv\frac{1}{n\cdot p}\bar{u}(p_2,\lambda_2) \slashed{n} \gamma^5 u(p_1,\lambda_1)\mathcal{\tilde{H}}_{\meson B_5} (\xi , t)\,.
\end{align}
By squaring the amplitude, and summing over the polarisation $  \lambda _{k} $ of the outgoing photon, we obtain
\begin{align}
	\label{eq:pol-amp-sq-x}
	\sum_{ \lambda _{k}}| {\cal M }_{x}|^2 &=|C_{A}|^2 + | \pt|^{4} |C_{B}|^2+\frac{s^2}{4}| \pt|^4 |C_{B_{5}}|^2-2 |\pt|^2  \mathrm{Re}(C_{A}^{*}C_{B}) \,,\\[5pt]
	\label{eq:pol-amp-sq-y}
	\sum_{ \lambda _{k}}| {\cal M }_{y}|^2 &=|C_{A}|^2 +\frac{s^2}{4}| \pt|^4 |C_{A_{5}}|^2 \,.
\end{align}
From the above polarised amplitude squared, one can compute the $  \mathrm{LPA}_{ \mathrm{max} }  $ at various levels (from fully differential to integrated). 

\subsubsection{Chiral-odd case}

\label{sec:CO-pol-asym}

	For the case of the transversely-polarised $ \meson $-meson, we find, through a direct computation, that both the circular and linear polarisation asymmetries \textit{always} vanish in the limit of $  \Delta _{\perp}=0 $. This is the consequence of the fact that, after squaring the amplitude and summing/averaging over all polarisations except $  \varepsilon _{q} $, one obtains, after setting $  \Delta _{\perp}=0 $,
	\begin{align}
		\label{eq:asymmetry-CO}
		\sum_{\substack{\lambda_1,\, \lambda_2,\\  \lambda _{k},\, \lambda _{\meson}}}
		\mathcal{M}^{\perp}_{\meson} \,
		\mathcal{M}^{\perp *}_{\meson} = -\frac{512}{s^2}   \xi^2 (1-\xi^2)  |C^{\perp}|^2 
		\left[ \alpha^4 |{N}_{T\meson A} (\xi , t)|^2 +  |{N}_{T\meson B} (\xi , t)|^2 \right]  \left(  \varepsilon _{q}^{*} \cdot  \varepsilon _{q} \right) \,.
	\end{align}
	First, we note that upon summing over the transverse polarisations of the incoming photon, one recovers the averaged amplitude squared in \eqref{all-rhoT}. Second, the term $ \left(  \varepsilon _{q}^{*} \cdot  \varepsilon _{q} \right)  $ is trivially $ -1 $, and thus can never give rise to any polarisation asymmetry.
	
	Comparing with the chiral-even case in \SEC\ref{sec:CE-pol-asym}, we find that repeating the same steps leads to two types of terms, namely the same one that appears in \eqref{eq:asymmetry-CO} $ \left(  \varepsilon _{q}^{*} \cdot  \varepsilon _{q} \right)  $, and also $ \left(  \varepsilon _{q}^{*} \cdot p_{\perp}\right) \left( \varepsilon _{q}  \cdot p_{\perp} \right)  $. It is terms of the latter type that lead to linear polarisation asymmetries. For instance, in \eqref{eq:pol-amp-sq-x} and \eqref{eq:pol-amp-sq-y}, we note that $ |C_{A}|^{2} $ comes from terms of the first type (from the square of the tensor $ T_A $ after summing over the polarisation of the outgoing photon, see \eqref{def:TA5-TB5}), and it is easy to see that $ |C_{A}|^{2} $ indeed cancels in the computation of the LPA.
	
	Before ending this section, we stress that the result in \eqref{eq:asymmetry-CO} is obtained by working in the limit of $  \Delta _{\perp}=0 $. In general, for non-zero $  \Delta _{\perp}$, the analogue of \eqref{eq:asymmetry-CO} contains all possible contractions involving $  \varepsilon _{q} $, namely 
 $ \left(  \varepsilon _{q}^{*} \cdot  \varepsilon _{q} \right)  $,  $ \left(  \varepsilon _{q}^{*} \cdot p_{\perp}\right) \left( \varepsilon _{q}  \cdot p_{\perp} \right)  $, $ \left(  \varepsilon _{q}^{*} \cdot  \Delta _{\perp}\right) \left( \varepsilon _{q}  \cdot p_{\perp} \right)  $, $ \left(  \varepsilon _{q}^{*} \cdot  p _{\perp}\right) \left( \varepsilon _{q}  \cdot   \Delta  _{\perp} \right)  $ and $ \left(  \varepsilon _{q}^{*} \cdot  \Delta _{\perp}\right) \left( \varepsilon _{q}  \cdot  \Delta _{\perp} \right)  $, which gives rise to polarisation asymmetries. On the other hand, the result depends on transversity GPDs other than $ H_{T} $, whose contributions are beyond the scope of work. We therefore postpone the analysis of polarisation asymmetries for non-zero $  \Delta _{\perp} $ for a future publication.

\section{Results}

\label{sec:results}

\subsection{Conventions for plots}

\label{sec:conventions-plots}
For consistency, we use the same conventions for the plots as in our previous study \cite{Duplancic:2022ffo}.
We typically include 4 cases, considering 2 models for the DA (asymptotic or holographic), and 2 GPD models (valence or standard scenario). The conventions used throughout this section are:
\begin{itemize}
	\item Solid line: asymptotic DA, valence scenario
	\item Dashed line: Holographic DA, valence scenario
	\item Dotted line: asymptotic DA, standard scenario
	\item Dot-dashed line: Holographic DA, standard scenario
\end{itemize}
Being dashed implies the use of the holographic DA, while being dotted implies the use of the standard scenario for the GPD.

We present results for JLab kinematics in \SEC\ref{sec:jlab-kinematics}, COMPASS kinematics in \SEC\ref{sec:COMPASS-kinematics} and EIC as well as LHC in UPCs kinematics in \SEC\ref{sec:EIC-LHC-UPC-kinematics}. In each subsection, we present results for fully-differential cross sections first, then single-differential cross sections (\ie{}integrated over $  \left( -t \right)  $ and $  \left( -u' \right)  $), followed by integrated cross sections as a function of $ \SgN $, and finally the linear polarisation asymmetries wrt the incoming photon. Each figure has 4 plots, with:
\begin{itemize}
\item top left corresponding to $\gamma  \meson^{0} $ photoproduction on proton target (denoted by $ \mesonzp $),
\item top right to $ \gamma \meson^{0} $ photoproduction on neutron target (denoted by $ \mesonzn $),
\item  bottom left to $ \gamma \meson^{+} $ photoproduction on a proton target (denoted by $ \mesonpp $),
\item  bottom right to $ \gamma \meson^{-} $ photoproduction on a neutron target (denoted by $ \mesonmn $).
\end{itemize}  Finally, the figures are presented in such an order that the chiral-even case (longitudinally polarised $ \meson $-meson) always appears before the chiral-odd case (transversely polarised $ \meson $-meson). We note that since the polarisation asymmetry is always vanishing for the chiral-odd case, only plots for the linear polarisation asymmetry corresponding to the chiral-even case are shown. Furthermore, these correspond to $  \mathrm{LPA}_{ \mathrm{max} }  $, see \SEC\ref{sec:CE-pol-asym}.

\subsection{Description of the numerics}

The GPDs are computed as tables in $ x $, for different $  \xi  $. For the amplitudes, we compute tables at different $ (-u') $ and $ \Msq $, at a particular value of $ \SgN $. To compute the fully differential cross section (and hence amplitudes), $ (-t) $ is fixed to its minimum value $ (-t)_{ \mathrm{min} } $, see \eqref{difcrosec}. The  $ t $-dependence of the cross section is then modelled by a simplistic ansatz, namely a factorised dipole form	
\beq
\label{dipole}
F_H(t)= \frac{ \left( t_{ \mathrm{min} }-C \right)^{2} }{(t-C)^2}\,,
\eq
with $C=0.71~{\rm GeV}^2.$

We compute the cross section covering the \textit{full} phase space in the region $20 \GeV^2 <\SgN< 20000 \GeV^2 $, since this covers the full kinematical range of JLab, COMPASS, EIC, and most of the relevant kinematical range for UPCs at LHC, see \SEC\ref{sec:UPC-LHC}. We compute 7 sets of amplitude tables in total:
\begin{itemize}
	\item $ \SgN = 20 \GeV^2  $, $ 2.1 \leq \Msq \leq 10 \GeV^2$ with a uniform step of 0.1~GeV$^2$
	\item $ \SgN = 200 \GeV^2  $, $ 2.1 \leq \Msq \leq 51.4 \GeV^2$ with a uniform step of 0.2~GeV$^2$
	\item $ \SgN = 200 \GeV^2  $, $ 2.1 \leq \Msq \leq 110.5 \GeV^2$ with a uniform step of 1.1~GeV$^2$
	\item $ \SgN = 2000 \GeV^2  $, $ 2.1 \leq \Msq \leq 51.4 \GeV^2$ with a uniform step of 0.2~GeV$^2$
	\item $ \SgN = 2000 \GeV^2  $, $ 2.1 \leq \Msq \leq 1041.1 \GeV^2$ with a uniform step of 10.5~GeV$^2$
	\item $ \SgN = 20000 \GeV^2  $, $ 2.1 \leq \Msq \leq 51.4 \GeV^2$ with a uniform step of 0.2~GeV$^2$
	\item $ \SgN = 20000 \GeV^2  $, $ 2.1 \leq \Msq \leq 10396.6 \GeV^2$ with a uniform step of 105~GeV$^2$
\end{itemize}
The first, third, fifth and seventh sets cover the full range of the phase space, while the second, fourth and sixth sets are needed to resolve the peak in $M_{\gamma \meson}^2$ (importance sampling), like for the charged pion case \cite{Duplancic:2022ffo}. This is particularly important for the chiral-even case, \ie{}for the longitudinally polarised $ \meson $-meson.

For each amplitude table, the whole range of $ (-u') $ is covered. More details regarding the boundaries of the kinematic variables can be found in \APP{}E of \cite{Duplancic:2022ffo}, and in \APP{}E of \cite{Duplancic:2018bum}.
At each value of $ \SgN = 200,\,2000,\,20000 \GeV^2 $, two separate datasets were needed, one to cover the whole range of the phase space, and the other to ensure that peaks in the distribution of $ \Msq $ were well-resolved in the chiral-even case. This is not needed for the $ \SgN=20 \GeV^2 $ case, as the peak is moderate in that case. We refer to \SEC 5.2.1 in \cite{Duplancic:2022ffo} for details regarding the importance sampling procedure.

In practice, we compute amplitude tables in $ (-u') $ for each of value of $ \Msq $. The steps we take are
\begin{itemize}
	\item 
	we calculate, for each of the above types of GPDs (in the present paper $H$, 
	$\tilde{H}$ and $ H_{T} $), sets of $u$- and $d$- quarks GPDs indexed by $M^2_{\gamma\meson}$, \ie{}ultimately by $\xi$ given by 
	\beqa
	\label{rel-xi-M2-S}
	\xi = \frac{M^2_{\gamma \meson}}{2(S_{\gamma N}-M^2)-M^2_{\gamma \meson}}\,.
	\eqa
	The GPDs are computed as tables of 1000 values for $x$ ranging from $-1$ to 
	$1$, unless importance sampling is needed, in which case 1000 more values around the peak is added, see 5.2.1 in \cite{Duplancic:2022ffo}.
	
	\item we compute the building block integrals which do not depend on 
	$-u'$. In the asymptotic DA case, this corresponds to $ I_e $ (see \APP{}D in  \cite{Duplancic:2018bum} for the notation), while in the holographic DA case, this corresponds to both $ I_e $ and  $  \chi _{c} $, see \APP{}C of \cite{Duplancic:2022ffo}.
	
	\item
	we choose 100 values of  $(-u')$, linearly varying from $(-u')_{ \mathrm{min} }=1~{\rm GeV}^2$ 
	up to its maximum possible value
	$(-u')_{ \mathrm{maxMax} }$ (see \APP{}E in \cite{Duplancic:2018bum} for how this is computed). Again, if importance sampling is needed (when the cross section varies rapidly at the boundaries), an extra set of 100 values of $ (-u') $ is added at each boundary.
	
	\item 
	at each value of $(-u')$, we compute, for each GPD and each flavour $u$ and $d$, the remaining building 
	block integrals, which are $I_b$, $I_c$, $ I_d $, $I_h$, $I_i$ in the asymptotic DA case, and $ \chi_{a} $ and $  \chi _{b} $ in the holographic DA case.\footnote{We note that the chiral-odd case requires the computation of the extra building blocks $ I_d $ and $ \chi_{a} $, which are not needed in the chiral-even case.}
	
	\item 
	this gives, for each of these couples of values of ($M^2_{\gamma \meson}, -u')$
	and each flavour, a set of 12 coefficients listed in 
	\EQ\eqref{list-N-to-be-computed} for the CE case, and \EQ\eqref{list-NT-to-be-computed} for the CO case.
	
	\item 
	one can then get the desired cross sections using \EQs\eqref{difcrosec} (CE case) and \eqref{difcrosecT} (CO case).
	
\end{itemize}

To optimise the computation, we use a mapping procedure, described in \SEC 5.2 of \cite{Duplancic:2022ffo}, which allows us to obtain amplitude tables corresponding to lower values of $ \SgN $ from a single table (which can correspond to any one of the 7 sets of amplitude tables mentioned above). This allows for a significant decrease in computing time, from the order of months to only a few days.

\subsection{JLab kinematics}

\label{sec:jlab-kinematics}

The electron beam at JLab hits a fixed target consisting of protons and neutrons, at an energy of $ 12\,\GeV $. The electron-nucleon centre-of-mass energy, $ S_{eN} $, is thus roughly 23 $ \GeV^2 $. Therefore, for most of the plots in this section, we use $ S_{\gamma N}=20\,\GeV^2 $ as a representative value for JLab kinematics. This allows us to probe GPDs for the range of skewnesses of $0.04 \leq \xi \leq 0.33$.

At this point, we would like to point out that a programming mistake, related to the sign of the interference term in the squared amplitude, \cf\eqref{all-rho}, was made in the previous publication~\cite{Boussarie:2016qop}. Thus, the plots corresponding to the $ \meson^{0} $-meson case are slightly different.

\subsubsection{Fully differential cross section}

\label{sec:jlab-fully-diff-X-section}

\begin{figure}[t!]
	\psfrag{HHH}{\hspace{-1.5cm}\raisebox{-.6cm}{\scalebox{.8}{$-u' ({\rm 
					GeV}^{2})$}}}
	\psfrag{VVV}{\raisebox{.3cm}{\scalebox{.9}{$\hspace{-.4cm}\displaystyle\left.\frac{d 
					\sigma^{\mathrm{even}}_{\gamma\mesonzp}}{d M^2_{\gamma \mesonzp} d(-u') d(-t)}\right|_{(-t)_{\rm min}}({\rm pb} \cdot {\rm GeV}^{-6})$}}}
	\psfrag{TTT}{}
	{
		{\includegraphics[width=18pc]{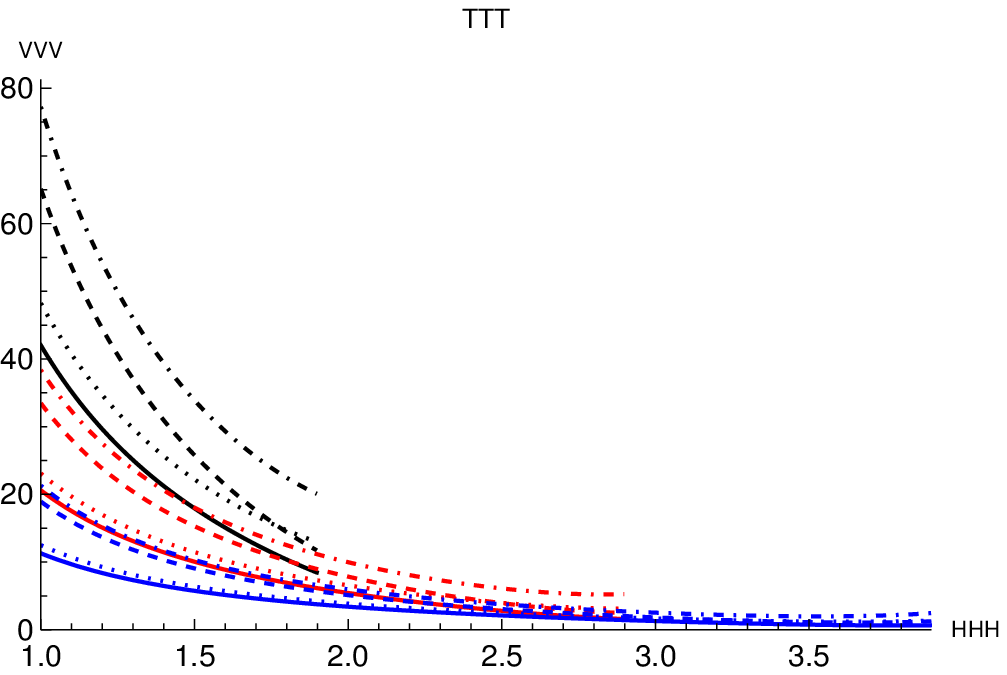}}
		\psfrag{VVV}{\raisebox{.3cm}{\scalebox{.9}{$\hspace{-.4cm}\displaystyle\left.\frac{d 
						\sigma^{\mathrm{even}}_{\gamma\mesonzn}}{d M^2_{\gamma \mesonzn} d(-u') d(-t)}\right|_{(-t)_{\rm min}}({\rm pb} \cdot {\rm GeV}^{-6})$}}}
		{\includegraphics[width=18pc]{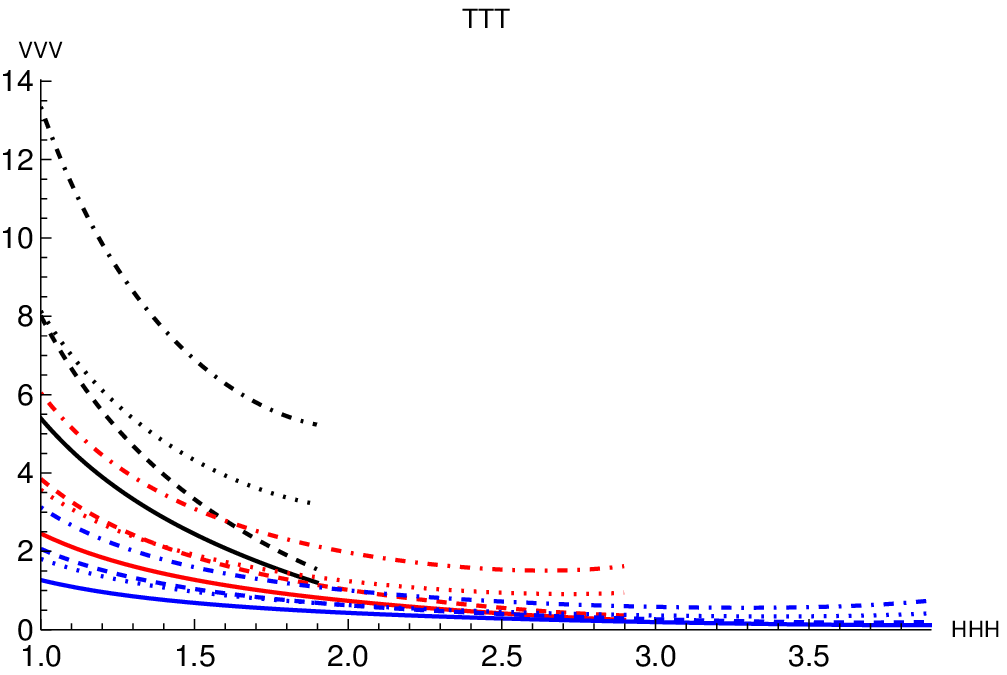}}
	\\[25pt]
	{	\psfrag{VVV}{\raisebox{.3cm}{\scalebox{.9}{$\hspace{-.4cm}\displaystyle\left.\frac{d 
						\sigma^{\mathrm{even}}_{\gamma\mesonpp}}{d M^2_{\gamma \mesonpp} d(-u') d(-t)}\right|_{(-t)_{\rm min}}({\rm pb} \cdot {\rm GeV}^{-6})$}}}
		\includegraphics[width=18pc]{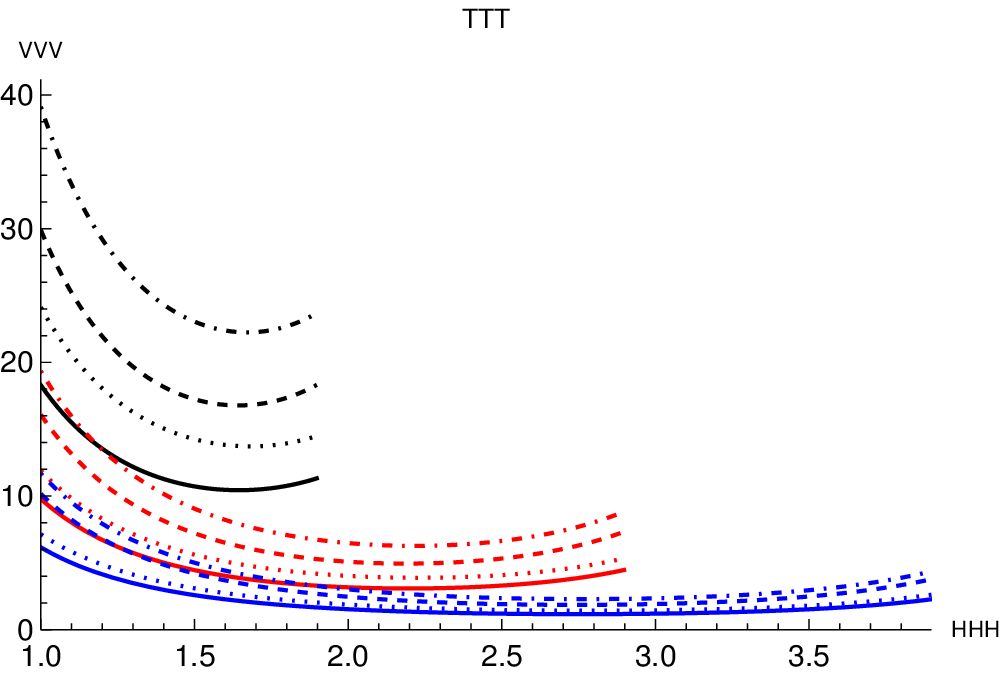}}
	\psfrag{VVV}{\raisebox{.3cm}{\scalebox{.9}{$\hspace{-.4cm}\displaystyle\left.\frac{d 
					\sigma^{\mathrm{even}}_{\gamma\mesonmn}}{d M^2_{\gamma \mesonmn} d(-u') d(-t)}\right|_{(-t)_{\rm min}}({\rm pb} \cdot {\rm GeV}^{-6})$}}}
	{\includegraphics[width=18pc]{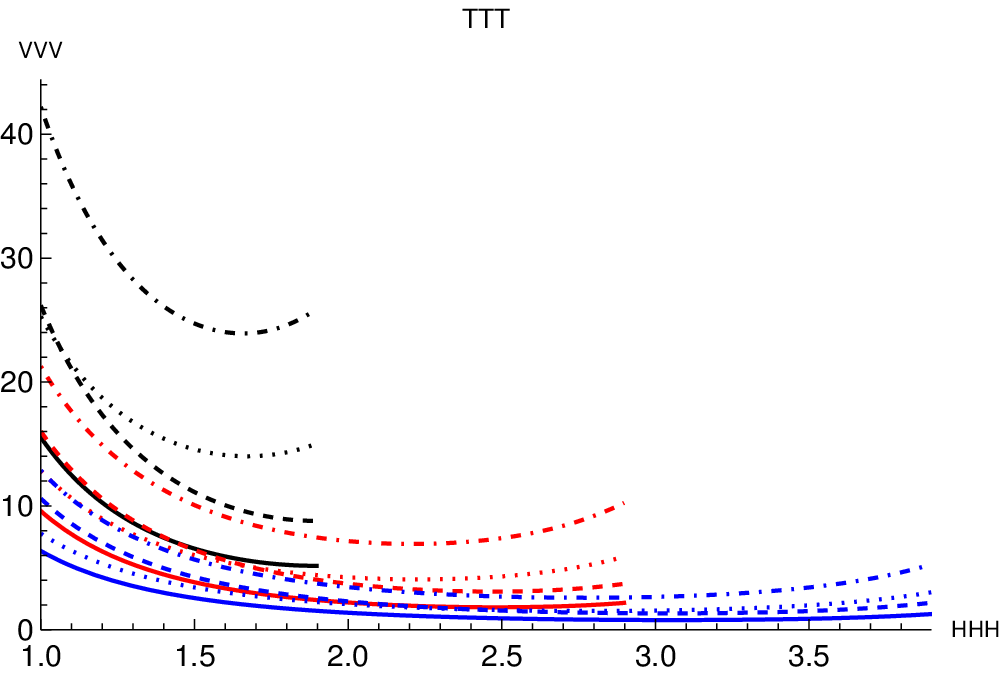}}}
	\vspace{0.2cm}
	\caption{\small The fully differential cross section for {longitudinally polarised} $ \mesonzp,\,\mesonzn,\,\mesonpp,\,\mesonmn$ is shown as a function of $  \left( -u' \right)  $ on the top left, top right, bottom left and bottom right plots respectively for different values of $ M_{\gamma \meson}^2 $. The black, red and blue curves correspond to $ M_{\gamma \meson}^{2}=3,\,4,\,5\, $ GeV$ ^2 $ respectively. The dashed (non-dashed) lines correspond to holographic (asymptotic) DA, while the dotted (non-dotted) lines correspond to the standard (valence) scenario. $ S_{\gamma N} $ is fixed at 20 GeV$ ^2 $.}
	\label{fig:jlab-fully-diff-diff-M2}
\end{figure}

\begin{figure}[t!]
	\psfrag{HHH}{\hspace{-1.5cm}\raisebox{-.6cm}{\scalebox{.8}{$-u' ({\rm 
					GeV}^{2})$}}}
	\psfrag{VVV}{\raisebox{.3cm}{\scalebox{.9}{$\hspace{-.4cm}\displaystyle\left.\frac{d 
					\sigma^{\mathrm{odd}}_{\gamma\mesonzp}}{d M^2_{\gamma\mesonzp} d(-u') d(-t)}\right|_{(-t)_{\rm min}}({\rm pb} \cdot {\rm GeV}^{-6})$}}}
	\psfrag{TTT}{}
	{
		{\includegraphics[width=18pc]{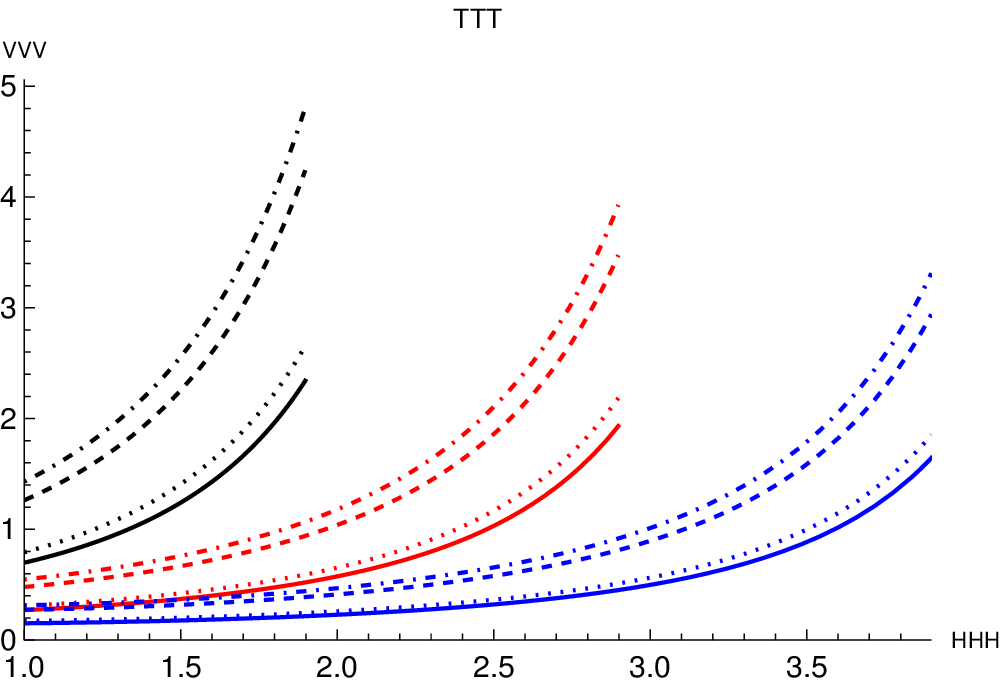}}
		\psfrag{VVV}{\raisebox{.3cm}{\scalebox{.9}{$\hspace{-.4cm}\displaystyle\left.\frac{d 
						\sigma^{\mathrm{odd}}_{\gamma\mesonzn}}{d M^2_{\gamma \mesonzn} d(-u') d(-t)}\right|_{(-t)_{\rm min}}({\rm pb} \cdot {\rm GeV}^{-6})$}}}
		{\includegraphics[width=18pc]{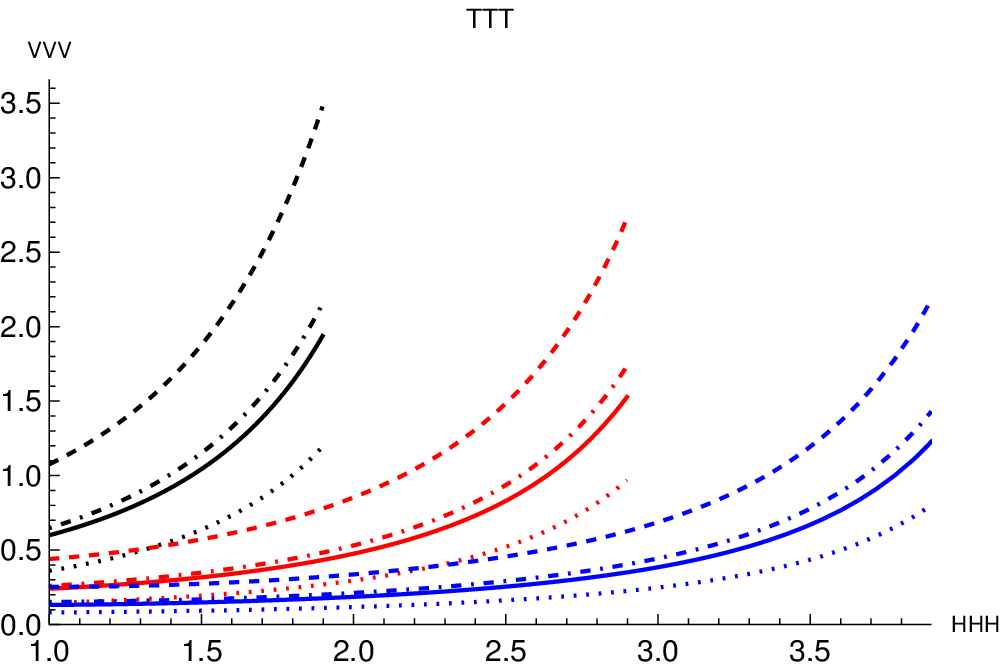}}
		\\[25pt]
		{	\psfrag{VVV}{\raisebox{.3cm}{\scalebox{.9}{$\hspace{-.4cm}\displaystyle\left.\frac{d 
							\sigma^{\mathrm{odd}}_{\gamma\mesonpp}}{d M^2_{\gamma \mesonpp} d(-u') d(-t)}\right|_{(-t)_{\rm min}}({\rm pb} \cdot {\rm GeV}^{-6})$}}}
			\includegraphics[width=18pc]{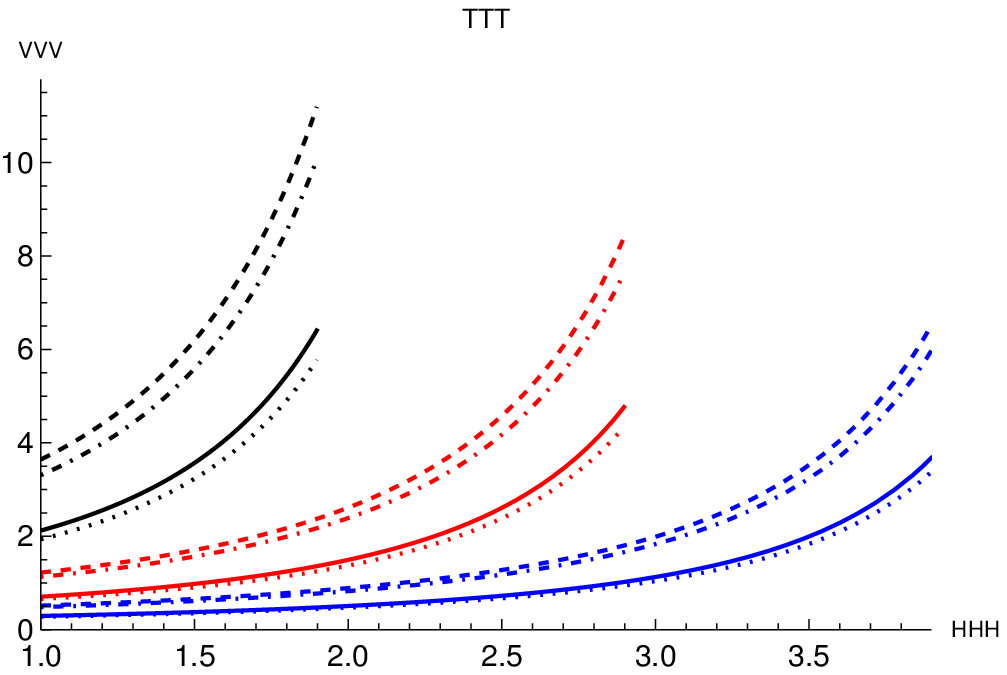}}
		\psfrag{VVV}{\raisebox{.3cm}{\scalebox{.9}{$\hspace{-.4cm}\displaystyle\left.\frac{d 
						\sigma^{\mathrm{odd}}_{\gamma\mesonmn}}{d M^2_{\gamma\mesonmn} d(-u') d(-t)}\right|_{(-t)_{\rm min}}({\rm pb} \cdot {\rm GeV}^{-6})$}}}
		{\includegraphics[width=18pc]{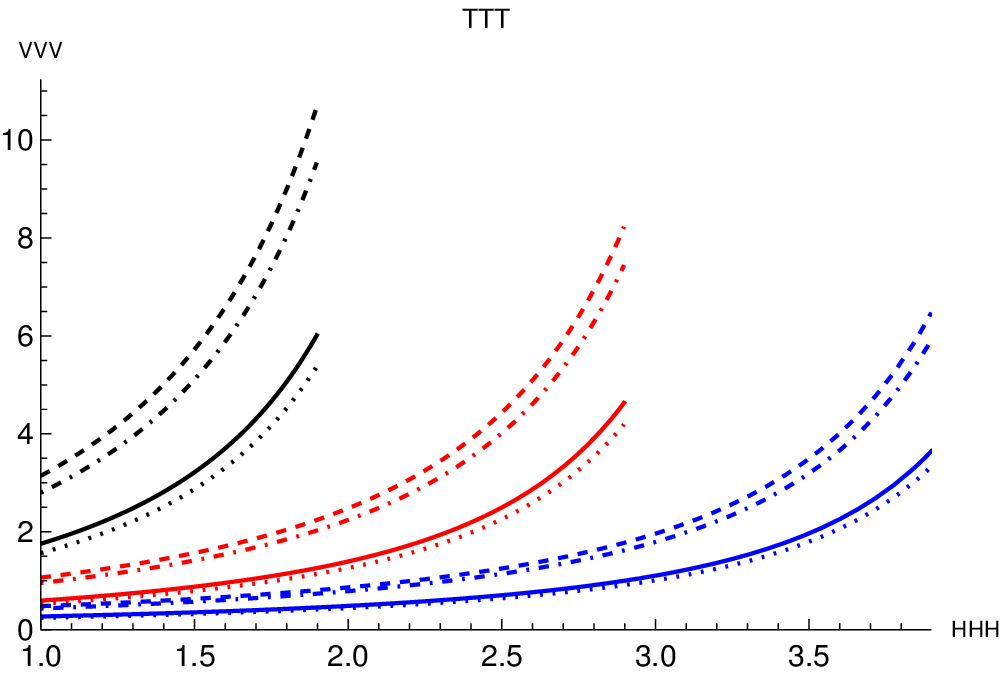}}}
	\vspace{0.2cm}
	\caption{\small The fully differential cross section for {transversely polarised} $ \mesonzp,\,\mesonzn,\,\mesonpp,\,\mesonmn$ is shown as a function of $  \left( -u' \right)  $ on the top left, top right, bottom left and bottom right plots respectively  for different values of $ M_{\gamma \meson}^2 $. The black, red and blue curves correspond to $ M_{\gamma \meson}^{2}=3,\,4,\,5\, $ GeV$ ^2 $ respectively. The dashed (non-dashed) lines correspond to holographic (asymptotic) DA, while the dotted (non-dotted) lines correspond to the standard (valence) scenario. $ S_{\gamma N} $ is fixed at 20 GeV$ ^2 $.}
	\label{fig:co-jlab-fully-diff-diff-M2}
\end{figure}

\begin{figure}[t!]
	\psfrag{HHH}{\hspace{-1.5cm}\raisebox{-.6cm}{\scalebox{.8}{$-u' ({\rm 
					GeV}^{2})$}}}
	\psfrag{VVV}{\raisebox{.3cm}{\scalebox{.9}{$\hspace{-.4cm}\displaystyle\left.\frac{d 
					\sigma^{\mathrm{even}}_{\gamma\mesonzp}}{d M^2_{\gamma\mesonzp} d(-u') d(-t)}\right|_{(-t)_{\rm min}}({\rm pb} \cdot {\rm GeV}^{-6})$}}}
	\psfrag{TTT}{}
	{\includegraphics[width=18pc]{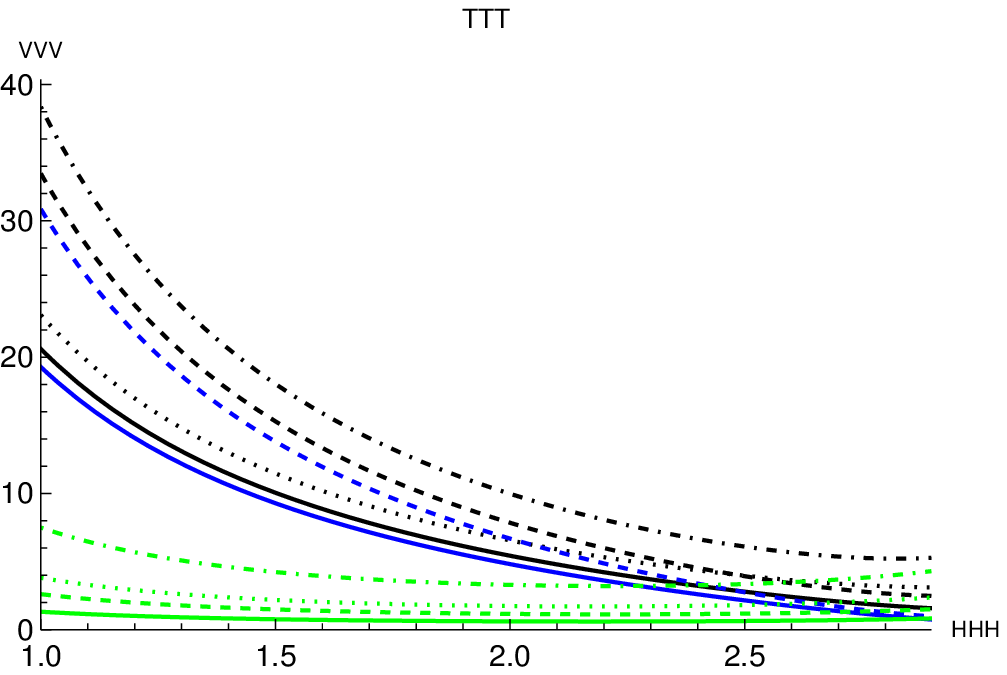}}
	\psfrag{VVV}{\raisebox{.3cm}{\scalebox{.9}{$\hspace{-.4cm}\displaystyle\left.\frac{d 
					\sigma^{\mathrm{even}}_{\gamma\mesonzn}}{d M^2_{\gamma \mesonzn} d(-u') d(-t)}\right|_{(-t)_{\rm min}}({\rm pb} \cdot {\rm GeV}^{-6})$}}}
	{\includegraphics[width=18pc]{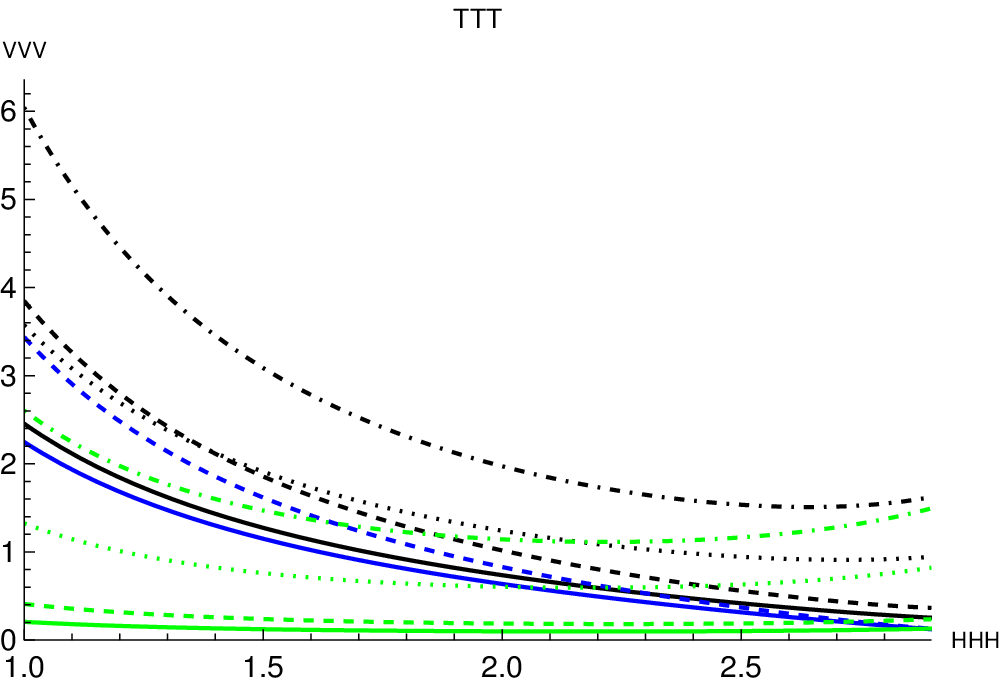}}
	\psfrag{VVV}{\raisebox{.3cm}{\scalebox{.9}{$\hspace{-.4cm}\displaystyle\left.\frac{d 
					\sigma^{\mathrm{even}}_{\gamma\mesonpp}}{d M^2_{\gamma\mesonpp} d(-u') d(-t)}\right|_{(-t)_{\rm min}}({\rm pb} \cdot {\rm GeV}^{-6})$}}}
	\\[25pt]
	{\includegraphics[width=18pc]{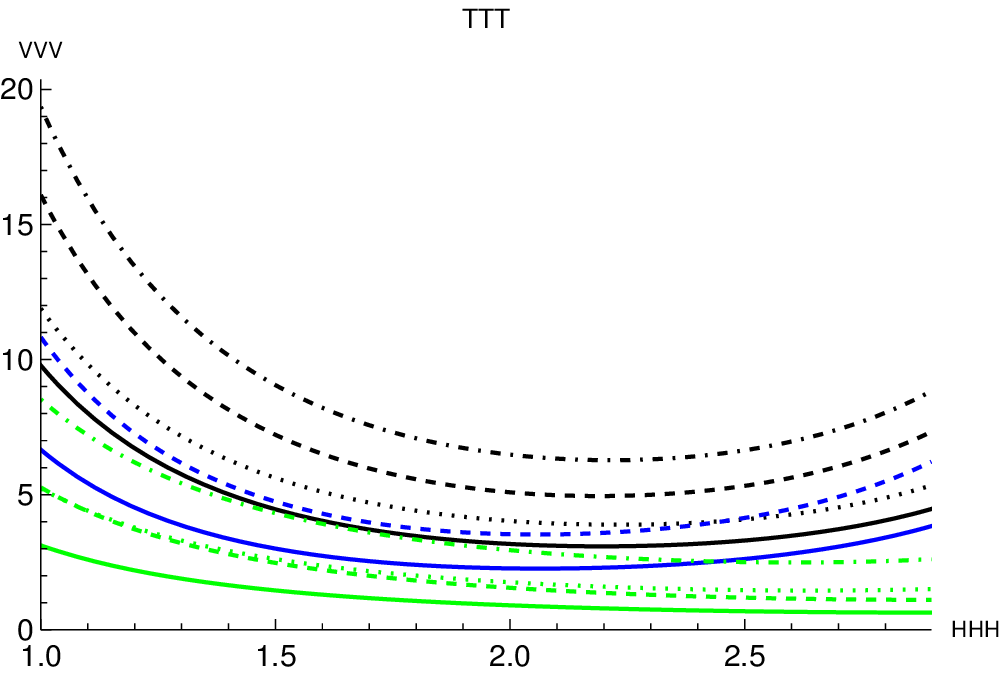}}
	\psfrag{VVV}{\raisebox{.3cm}{\scalebox{.9}{$\hspace{-.4cm}\displaystyle\left.\frac{d 
					\sigma^{\mathrm{even}}_{\gamma\mesonmn}}{d M^2_{\gamma\mesonmn} d(-u') d(-t)}\right|_{(-t)_{\rm min}}({\rm pb} \cdot {\rm GeV}^{-6})$}}}
	{\includegraphics[width=18pc]{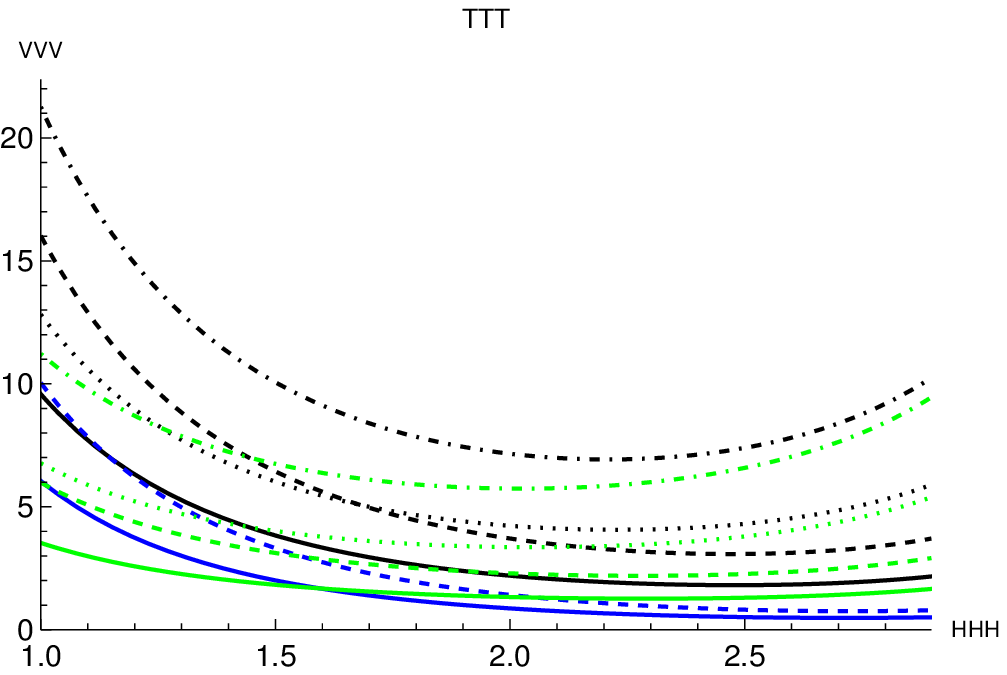}}
	\vspace{0.2cm}
	\caption{\small The fully differential cross section for {longitudinally polarised} $ \mesonzp,\,\mesonzn,\,\mesonpp,\,\mesonmn$ is shown as a function of $  \left( -u' \right)  $ on the top left, top right, bottom left and bottom right plots respectively. The blue and green curves correspond to contributions from the vector and axial GPDs respectively. The black curves correspond to the total contribution, \ie{}vector and axial GPD contributions combined. As before, the dashed (non-dashed) lines correspond to holographic (asymptotic) DA, while the dotted (non-dotted) lines correspond to the standard (valence) scenario. We fix $ S_{\gamma N}= 20\,  \mathrm{GeV}^{2}  $ and $ M_{\gamma \meson}^{2}= 4\,  \mathrm{GeV}^{2}  $. Note that the vector contributions consist of only two curves in each case, since they are insensitive to either valence or standard scenarios.}
	\label{fig:jlab-fully-diff-VandA}
\end{figure}

\begin{figure}[t!]
	\psfrag{HHH}{\hspace{-1.5cm}\raisebox{-.6cm}{\scalebox{.8}{$-u' ({\rm 
					GeV}^{2})$}}}
	\psfrag{VVV}{\raisebox{.3cm}{\scalebox{.9}{$\hspace{-.4cm}\displaystyle\left.\frac{d 
					\sigma^{\mathrm{even}}_{\gamma\mesonzp}}{d M^2_{\gamma\mesonzp} d(-u') d(-t)}\right|_{(-t)_{\rm min}}({\rm pb} \cdot {\rm GeV}^{-6})$}}}
	\psfrag{TTT}{}
	{
		{\includegraphics[width=18pc]{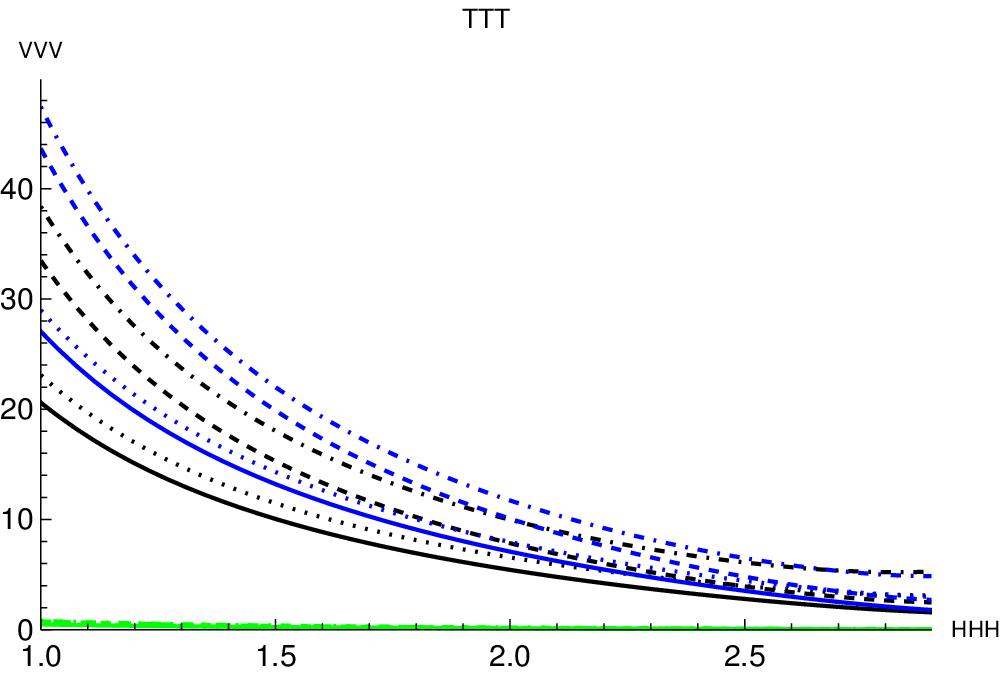}}
		\psfrag{VVV}{\raisebox{.3cm}{\scalebox{.9}{$\hspace{-.4cm}\displaystyle\left.\frac{d 
						\sigma^{\mathrm{even}}_{\gamma\mesonzn}}{d M^2_{\gamma \mesonzn} d(-u') d(-t)}\right|_{(-t)_{\rm min}}({\rm pb} \cdot {\rm GeV}^{-6})$}}}
		{\includegraphics[width=18pc]{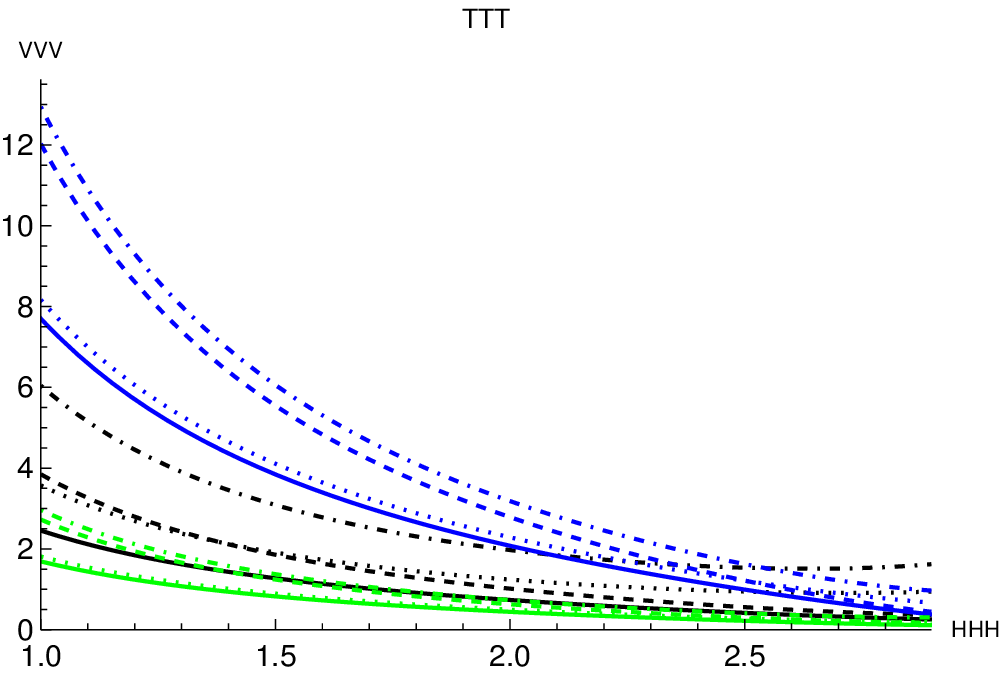}}}
	\\[25pt]
	\psfrag{VVV}{\raisebox{.3cm}{\scalebox{.9}{$\hspace{-.4cm}\displaystyle\left.\frac{d 
					\sigma^{\mathrm{even}}_{\gamma\mesonpp}}{d M^2_{\gamma\mesonpp} d(-u') d(-t)}\right|_{(-t)_{\rm min}}({\rm pb} \cdot {\rm GeV}^{-6})$}}}
	{
		{\includegraphics[width=18pc]{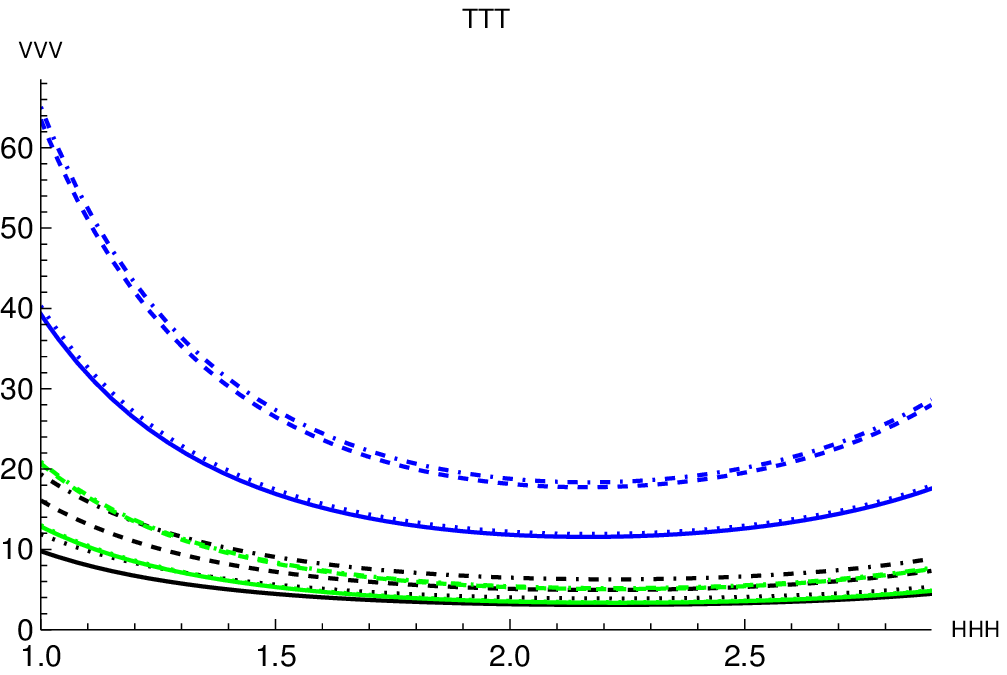}}
		\psfrag{VVV}{\raisebox{.3cm}{\scalebox{.9}{$\hspace{-.4cm}\displaystyle\left.\frac{d 
						\sigma^{\mathrm{even}}_{\gamma\mesonmn}}{d M^2_{\gamma\mesonmn} d(-u') d(-t)}\right|_{(-t)_{\rm min}}({\rm pb} \cdot {\rm GeV}^{-6})$}}}
		{\includegraphics[width=18pc]{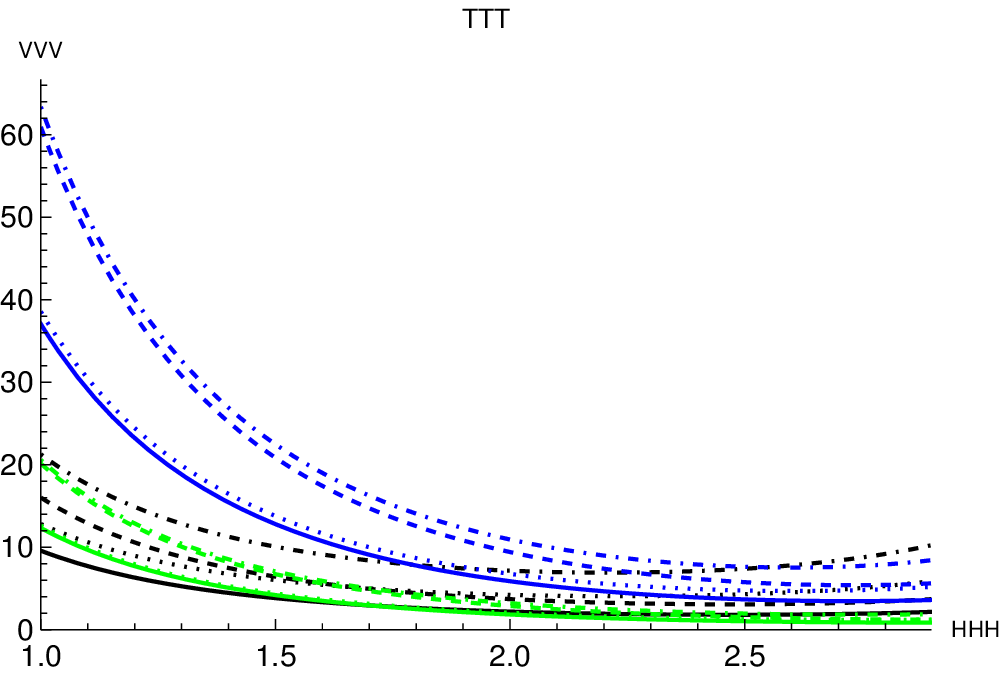}}}
	\vspace{0.2cm}
	\caption{\small The fully differential cross section for {longitudinally polarised} $ \mesonzp,\,\mesonzn,\,\mesonpp,\,\mesonmn$ is shown as a function of $  \left( -u' \right)  $ on the top left, top right, bottom left and bottom right plots respectively. The blue and green curves correspond to contributions from the $u$-quark ($ H_{u} $ and $  \tilde{H} _{u} $) and $d$-quark ($ H_{d} $ and $  \tilde{H} _{d} $) GPDs respectively. The black curves correspond to the total contribution. Otherwise, conventions are the same as in previous plots. We fix $ S_{\gamma N}= 20\,  \mathrm{GeV}^{2}  $ and $ M_{\gamma \meson}^{2}= 4\,  \mathrm{GeV}^{2}  $.}
	\label{fig:jlab-fully-diff-uandd}
\end{figure}

The effect of different values of $ M_{\gamma \meson}^2 $ on the cross section  is shown in \FIG\ref{fig:jlab-fully-diff-diff-M2} for the longitudinally polarised $ \meson $-meson case. The values chosen for $ M_{\gamma \meson}^2 $ are 3, 4 and 5 GeV$ ^2 $. As $ M_{\gamma \meson}^2 $ grows, the range of allowed $ (-u') $ values increases. On the other hand, the value of the cross section itself decreases. When integrating over $(-u')$, these two competing effects will become clearer later when we show the single differential plots in \SEC\ref{sec:sing-diff-X-section-JLab} as a function of $ M_{\gamma \meson}^2 $, leading to a peak in the distribution at low values of $ M_{\gamma \meson}^2 $. In general, the GPD model corresponding to the standard scenario leads to a larger value for the cross section. The maximum value of $  \left( -u' \right)  $ allowed by the kinematics, attained when $-t=(-t)_{\rm max}\,,$ is given by (see \APP{}E of \cite{Duplancic:2018bum})
\beqa
\label{Def:mupmaxMax}
(-u')_{\rm maxMax} = (-t)_{\rm max} -m_\meson^2 + M^2_{\gamma \meson}   - (-t')_{\rm 
	min}\,.
\eqa
In the case of the $ \meson $-meson, this has the effect of cutting the upper end of the $  \left( -u' \right)  $ range at a smaller value, compared to the pion case, see \FIG{}3 of \cite{Duplancic:2018bum}. In general, using a holographic DA gives a higher cross section than using an asymptotic DA. {We observe that the model used for the GPD (valence vs standard) has a small effect for the photoproduction of $ \gamma\mesonpp $ and $ \gamma\mesonzp $, compared to $ \gamma \mesonmn $ and $ \gamma \mesonzn $.} Finally, we note that the case of $ \gamma \meson^{0} $ photoproduction on a proton target has the largest cross section, followed by the two charged $ \meson $-meson cases, and lastly the $\gamma \meson^{0} $ photoproduction on a neutron target.

The corresponding figure for the differential cross section as a function of $  \left( -u' \right)  $ for the chiral-odd case is shown in \FIG\ref{fig:co-jlab-fully-diff-diff-M2}. In this case, the cross section increases with $  \left( -u' \right)  $, as opposed to the chiral-even case. Although the chiral-odd cross section seems smaller than the chiral-even one at first sight, the maximum value over the range of $  \left( -u' \right)  $ plays a key role when one performs the phase space integration over $  \left( -u' \right)  $ and $  \left( -t \right)  $ (as can be understood from the phase space figures in \APP{}D of \cite{Boussarie:2016qop}). This explains why the single differential cross sections as a function of $ \Msq $ are not heavily suppressed for the chiral-odd case when compared with the chiral-even one, see \FIGs\ref{fig:jlab-sing-diff} and \ref{fig:co-jlab-sing-diff}. We observe that the case of photoproduction of $ \gamma \meson^{0}_{n} $ has the strongest dependence on the GPD model used. This can be traced back to the larger sensitivity of the $ d $-quark transversity GPD vs the $ u $-quark one, as can be seen from \FIG{}4 in \cite{Boussarie:2016qop}. As with the chiral-even case, using a holographic DA gives the larger cross section. 

An interesting observation is that the plots for $ \gamma \mesonpp $ and $ \gamma \mesonmn $ are very similar. In fact, a closer look indicates that the difference between them becomes negligible when $  \left( -u' \right)  $ becomes larger. This effect can be traced back to \eqref{gauge-NAT}, \eqref{gauge-NBT}, \eqref{NrhoT+} and \eqref{NrhoT-}. One then finds that the only difference between the amplitudes of $ \mesonpp $ and $ \mesonmn $ comes from the terms $ (Q_1^2 - Q_2^2) {N}^q_{T\,A}[(AB)_{123}]^a $ and $ (Q_1^2 - Q_2^2) {N}^q_{T\,B}[(AB)_{123}]^a $, since they are anti-symmetric wrt the exchange of $ Q_1 $ and $ Q_2 $. Furthermore, from \eqref{eq:NTA-a-A3}, one finds that $ {N}^q_{T\,A}[(AB)_{123}]^a = 0 $. In the cross section \eqref{all-rhoT}, one also observes that the coefficient $ |{N}_{T\meson A}|^2 $ has a factor of $  \alpha ^4 $ in front compared to the $ |{N}_{T\meson B}|^2 $, which includes the contribution $ {N}^q_{T\,B}[(AB)_{123}]^a $ that causes the difference between $ \gamma \mesonpp $ and $ \gamma \mesonmn $. Since $  \alpha \propto  \left( -u' \right)  $ (see \eqref{eq:alpha}), this explains why the difference between $ \mesonpp $ and $ \mesonmn $ becomes negligible as $  \left( -u' \right)  $ increases.

The relative contributions of the vector and axial GPDs to the cross section for the longitudinally-polarised $ \meson $-meson are shown in \FIG\ref{fig:jlab-fully-diff-VandA}. The kinematical variables chosen for the plots are $ \SgN = 20 \GeV^2 $ and $ \Msq = 4 \GeV^2 $. The first point to note is that the vector contribution does not depend on the valence or standard scenarios, since they only enter the modelling of the axial GPDs. Hence, only two blue curves appear on each plot in the figure, corresponding to the DA model. Moreover, we note that the total contribution (black curve) corresponds simply to the sum of the vector (blue) and axial (green) contributions, since there is no interference between them, see \eqref{squareCEresult}. We thus find that the largest contribution to the cross section for the neutral meson case comes from the vector GPDs $ H $, while the contribution from the axial GPDs become more important for the charged $ \meson $-meson cases. This effect was also observed in the case of charged pions (see \FIG{}4 in \cite{Duplancic:2022ffo} and \FIG{}4 in \cite{Duplancic:2018bum}).

\begin{figure}[t!]
	\psfrag{HHH}{\hspace{-1.5cm}\raisebox{-.6cm}{\scalebox{.8}{$-u' ({\rm 
					GeV}^{2})$}}}
	\psfrag{VVV}{\raisebox{.3cm}{\scalebox{.9}{$\hspace{-.4cm}\displaystyle\left.\frac{d 
					\sigma^{\mathrm{odd}}_{\gamma\mesonzp}}{d M^2_{\gamma\mesonzp} d(-u') d(-t)}\right|_{(-t)_{\rm min}}({\rm pb} \cdot {\rm GeV}^{-6})$}}}
	\psfrag{TTT}{}
	{
		{\includegraphics[width=18pc]{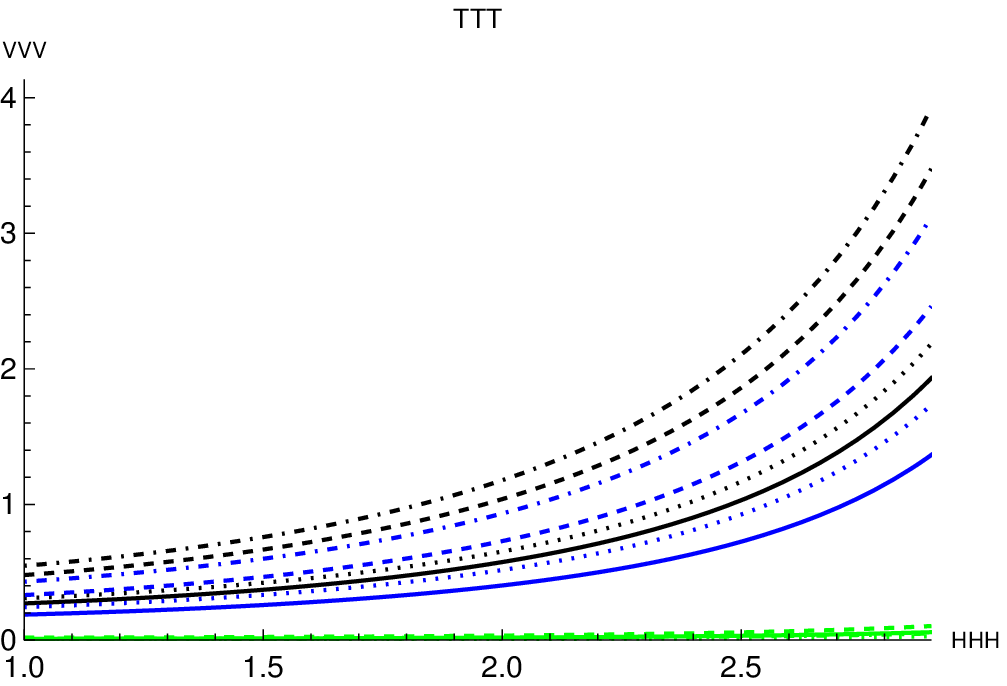}}
		\psfrag{VVV}{\raisebox{.3cm}{\scalebox{.9}{$\hspace{-.4cm}\displaystyle\left.\frac{d 
						\sigma^{\mathrm{odd}}_{\gamma\mesonzn}}{d M^2_{\gamma \mesonzn} d(-u') d(-t)}\right|_{(-t)_{\rm min}}({\rm pb} \cdot {\rm GeV}^{-6})$}}}
		{\includegraphics[width=18pc]{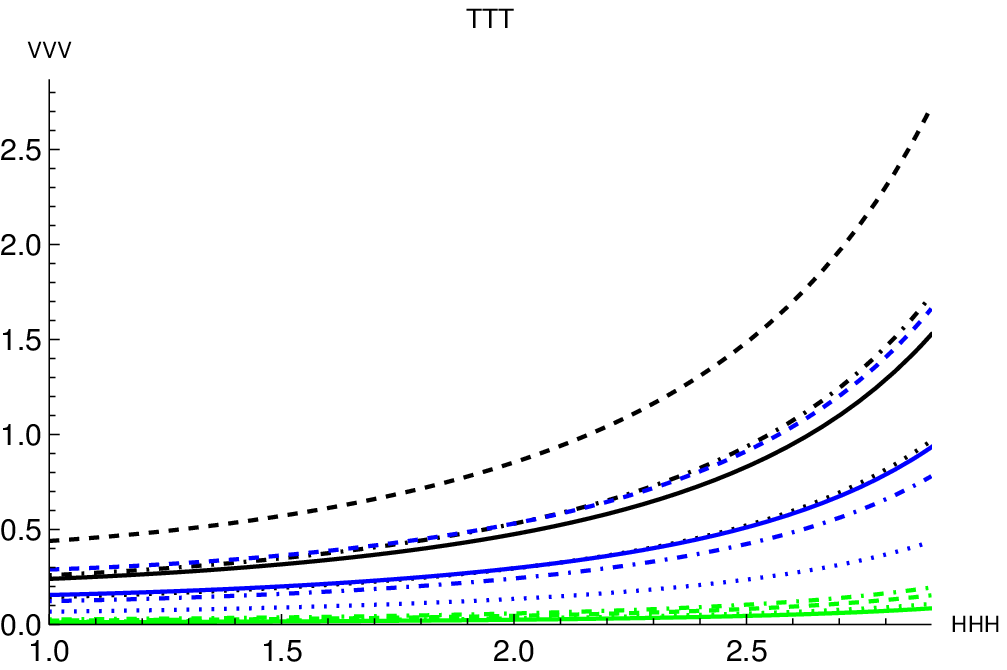}}}
	\\[25pt]
	\psfrag{VVV}{\raisebox{.3cm}{\scalebox{.9}{$\hspace{-.4cm}\displaystyle\left.\frac{d 
					\sigma^{\mathrm{odd}}_{\gamma\mesonpp}}{d M^2_{\gamma\mesonpp} d(-u') d(-t)}\right|_{(-t)_{\rm min}}({\rm pb} \cdot {\rm GeV}^{-6})$}}}
	{
		{\includegraphics[width=18pc]{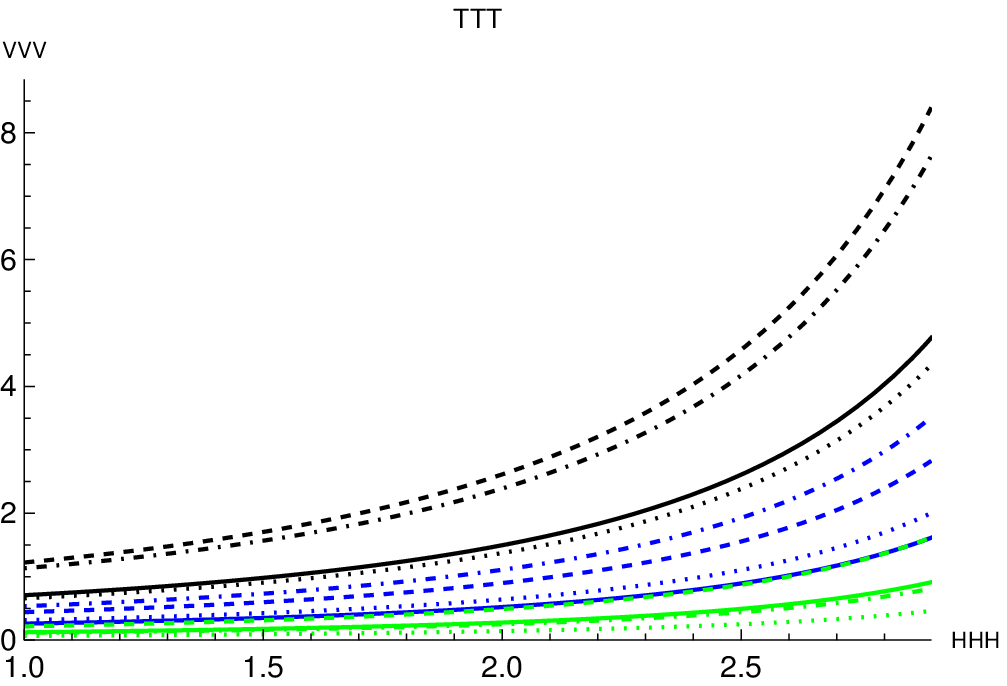}}
		\psfrag{VVV}{\raisebox{.3cm}{\scalebox{.9}{$\hspace{-.4cm}\displaystyle\left.\frac{d 
						\sigma^{\mathrm{odd}}_{\gamma\mesonmn}}{d M^2_{\gamma\mesonmn} d(-u') d(-t)}\right|_{(-t)_{\rm min}}({\rm pb} \cdot {\rm GeV}^{-6})$}}}
		{\includegraphics[width=18pc]{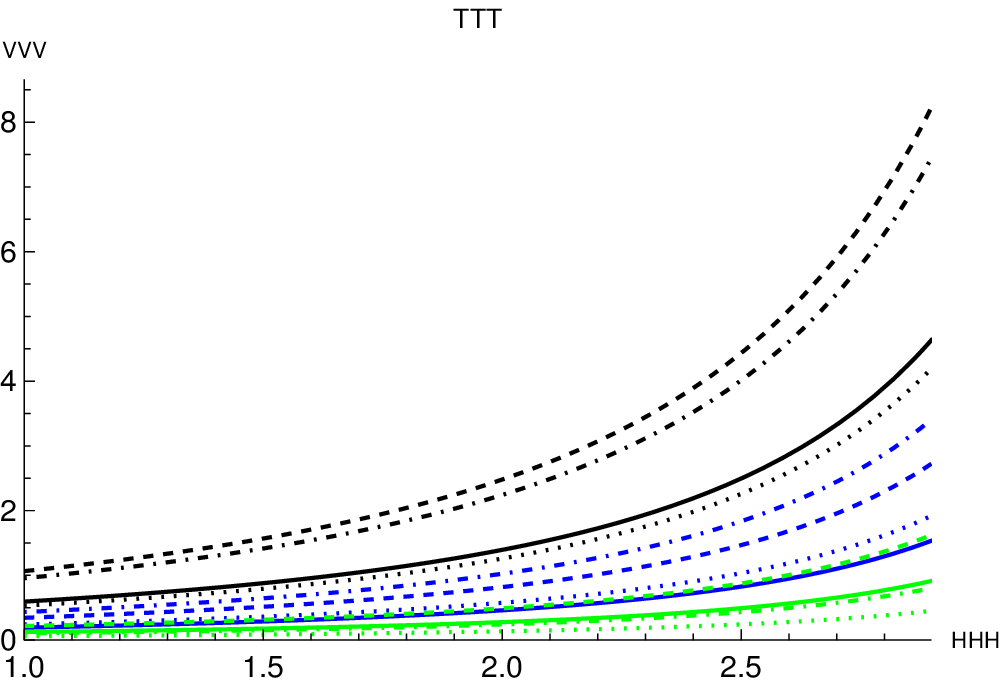}}}
	\vspace{0.2cm}
	\caption{\small The fully differential cross section for {transversely polarised} $ \mesonzp,\,\mesonzn,\,\mesonpp,\,\mesonmn$ is shown as a function of $  \left( -u' \right)  $ on the top left, top right, bottom left and bottom right plots respectively. The blue and green curves correspond to contributions from the $u$-quark ($ H_{u} $ and $  \tilde{H} _{u} $) and $d$-quark ($ H_{d} $ and $  \tilde{H} _{d} $) GPDs respectively. The black curves correspond to the total contribution. Otherwise, conventions are the same as in previous plots. We fix $ S_{\gamma N}= 20\,  \mathrm{GeV}^{2}  $ and $ M_{\gamma \meson}^{2}= 4\,  \mathrm{GeV}^{2}  $.}
	\label{fig:co-jlab-fully-diff-uandd}
\end{figure}

To conclude this subsection, the relative contributions of the $ u $- and $d$-quark GPDs to the cross section are shown in \FIG\ref{fig:jlab-fully-diff-uandd} for the longitudinally polarised $ \meson $-meson case and \FIG\ref{fig:co-jlab-fully-diff-uandd} for the transversely polarised $ \meson $-meson case. To generate the plots, $ \SgN = 20 \GeV^2 $ and $ \Msq = 4 \GeV^2 $ were used. Here, unlike in \FIG\ref{fig:jlab-fully-diff-VandA}, there are important interference terms between the $u$-quark and $d$-quark contributions, and therefore, the total contribution (black) is \textit{not} simply a sum of the individual quark GPD contributions. An interesting point to note is that the interference terms (which are not shown on the plots) are very sensitive to the axial GPDs $ \tilde{H} $

\FloatBarrier

\subsubsection{Single differential cross section}

\label{sec:sing-diff-X-section-JLab}

\begin{figure}[t!]
	\psfrag{HHH}{\hspace{-1.5cm}\raisebox{-.6cm}{\scalebox{.8}{$ M_{\gamma \meson}^2 ({\rm 
					GeV}^{2})$}}}
	\psfrag{VVV}{\raisebox{.3cm}{\scalebox{.9}{$\hspace{-.4cm}\displaystyle\frac{d 
					\sigma^{\mathrm{even}}_{\gamma\mesonzp}}{d M^2_{\gamma\mesonzp}}({\rm pb} \cdot {\rm GeV}^{-2})$}}}
	\psfrag{TTT}{}
	{
		{\includegraphics[width=18pc]{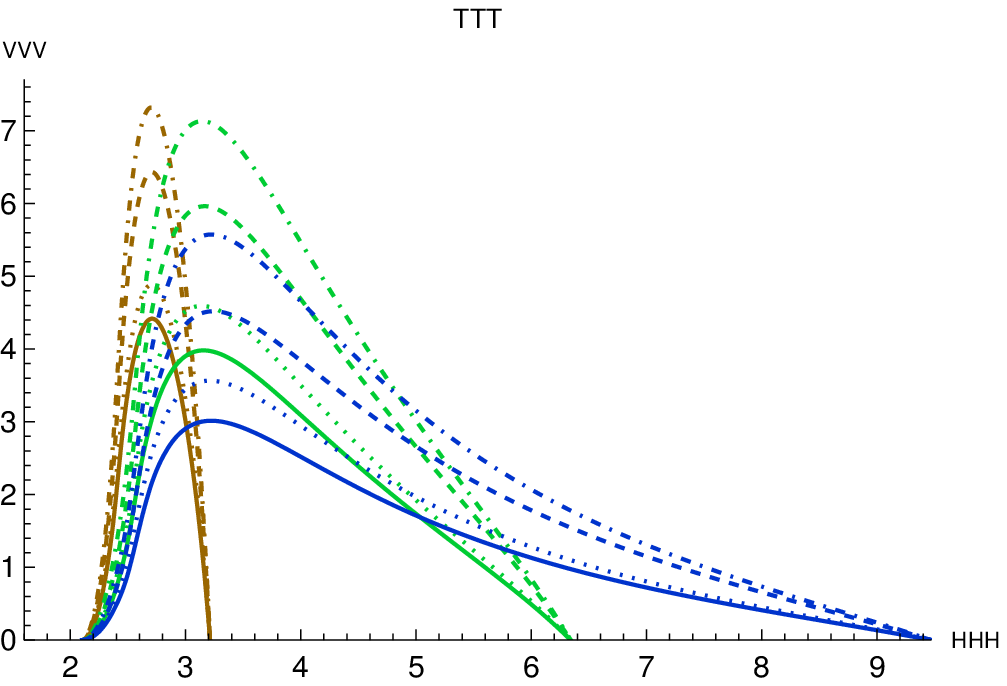}}
		\psfrag{VVV}{\raisebox{.3cm}{\scalebox{.9}{$\hspace{-.4cm}\displaystyle\frac{d 
						\sigma^{\mathrm{even}}_{\gamma\mesonzn}}{d M^2_{\gamma \mesonzn}}({\rm pb} \cdot {\rm GeV}^{-2})$}}}
		{\includegraphics[width=18pc]{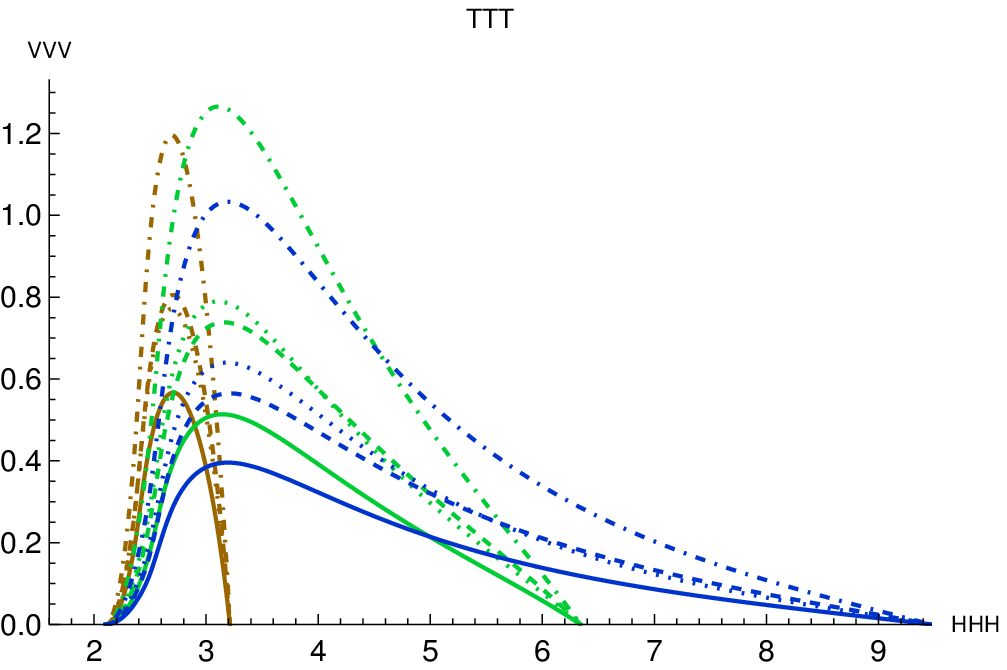}}}
	\\[25pt]
	\psfrag{VVV}{\raisebox{.3cm}{\scalebox{.9}{$\hspace{-.4cm}\displaystyle\frac{d 
					\sigma^{\mathrm{even}}_{\gamma\mesonpp}}{d M^2_{\gamma\mesonpp}}({\rm pb} \cdot {\rm GeV}^{-2})$}}}
	\psfrag{TTT}{}
	{
		{\includegraphics[width=18pc]{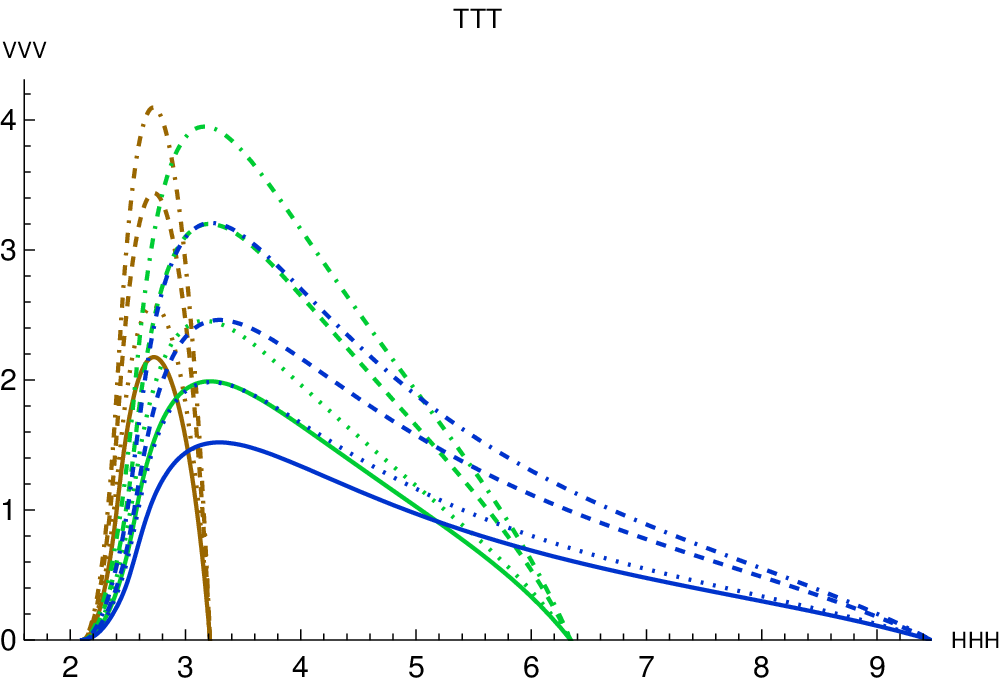}}
		\psfrag{VVV}{\raisebox{.3cm}{\scalebox{.9}{$\hspace{-.4cm}\displaystyle\frac{d 
						\sigma^{\mathrm{even}}_{\gamma\mesonmn}}{d M^2_{\gamma\mesonmn}}({\rm pb} \cdot {\rm GeV}^{-2})$}}}
		{\includegraphics[width=18pc]{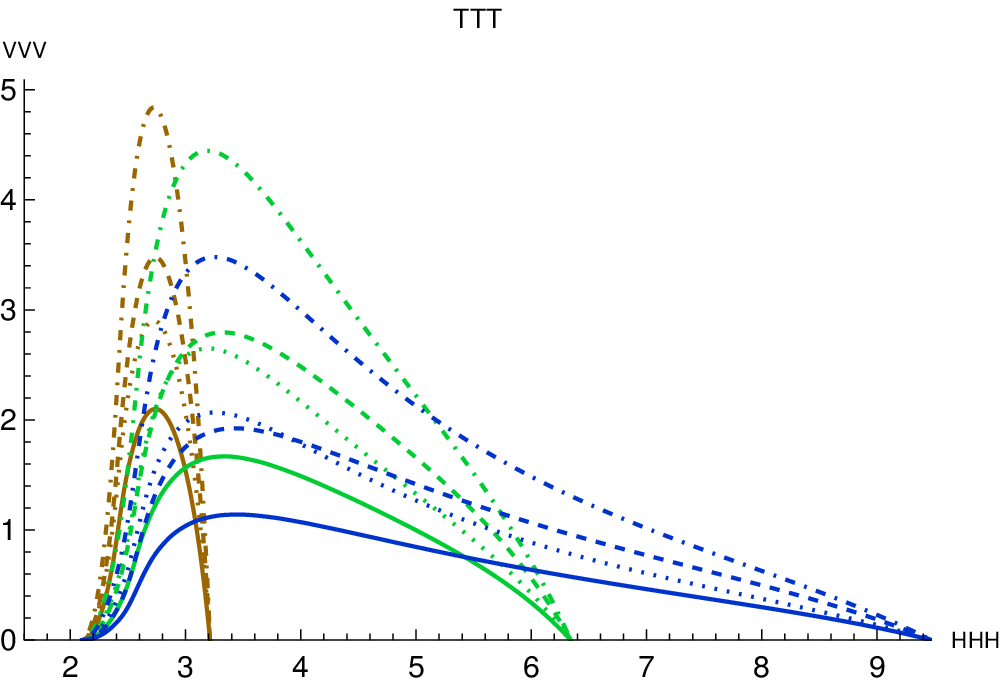}}}
	\vspace{0.2cm}
	\caption{\small The single differential cross section for {longitudinally polarised} $ \mesonzp,\,\mesonzn,\,\mesonpp,\,\mesonmn$ is shown as a function of $  M_{\gamma \meson}^{2}  $ on the top left, top right, bottom left and bottom right plots respectively for different values of $ S_{\gamma N} $. The brown, green and blue curves correspond to $ S_{\gamma N} = 8,\,14,\,20\,\GeV^{2} $. The dashed (non-dashed) lines correspond to holographic (asymptotic) DA, while the dotted (non-dotted) lines correspond to the standard (valence) scenario.}
	\label{fig:jlab-sing-diff}
\end{figure}

\begin{figure}[t!]
	\psfrag{HHH}{\hspace{-1.5cm}\raisebox{-.6cm}{\scalebox{.8}{$ M_{\gamma \meson}^2 ({\rm 
					GeV}^{2})$}}}
	\psfrag{VVV}{\raisebox{.3cm}{\scalebox{.9}{$\hspace{-.4cm}\displaystyle\frac{d 
					\sigma^{\mathrm{odd}}_{\gamma\mesonzp}}{d M^2_{\gamma\mesonzp}}({\rm pb} \cdot {\rm GeV}^{-2})$}}}
	\psfrag{TTT}{}
	{
		{\includegraphics[width=18pc]{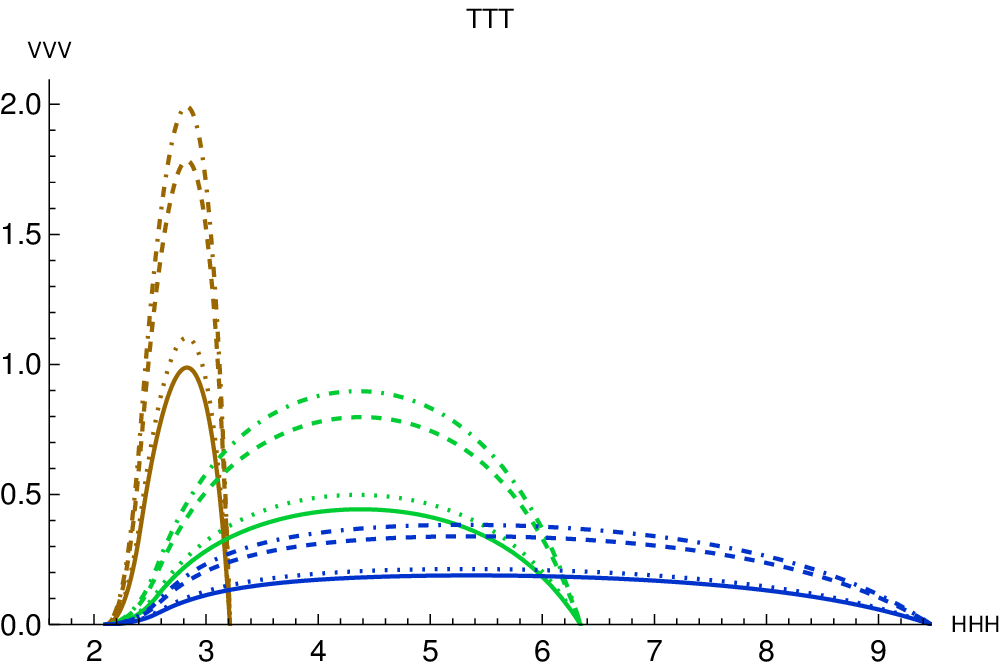}}
		\psfrag{VVV}{\raisebox{.3cm}{\scalebox{.9}{$\hspace{-.4cm}\displaystyle\frac{d 
						\sigma^{\mathrm{odd}}_{\gamma\mesonzn}}{d M^2_{\gamma \mesonzn}}({\rm pb} \cdot {\rm GeV}^{-2})$}}}
		{\includegraphics[width=18pc]{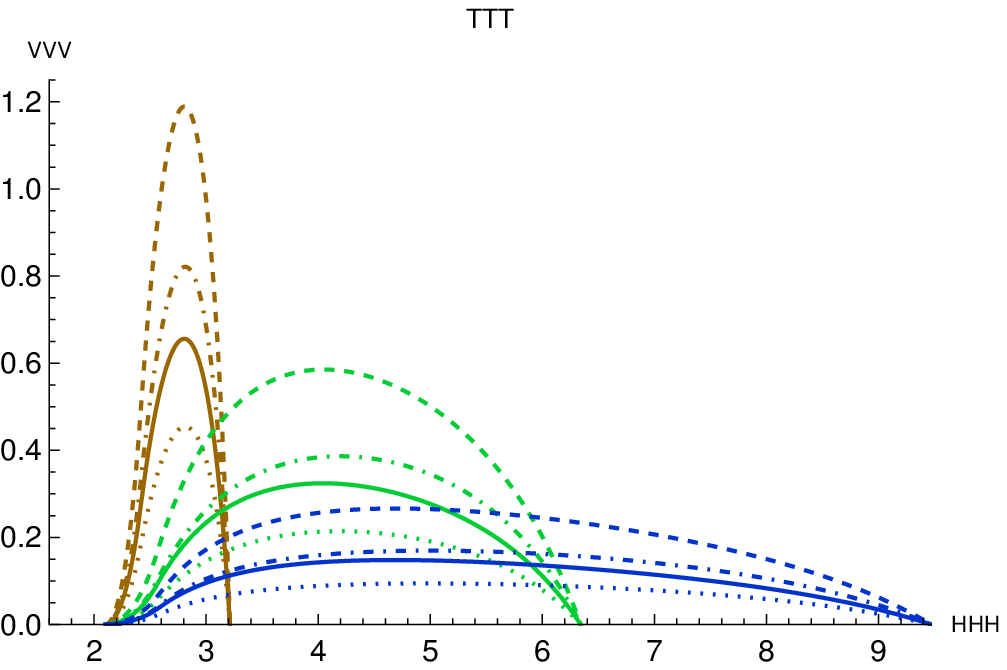}}}
	\\[25pt]
	\psfrag{VVV}{\raisebox{.3cm}{\scalebox{.9}{$\hspace{-.4cm}\displaystyle\frac{d 
					\sigma^{\mathrm{odd}}_{\gamma\mesonpp}}{d M^2_{\gamma\mesonpp}}({\rm pb} \cdot {\rm GeV}^{-2})$}}}
	\psfrag{TTT}{}
	{
		{\includegraphics[width=18pc]{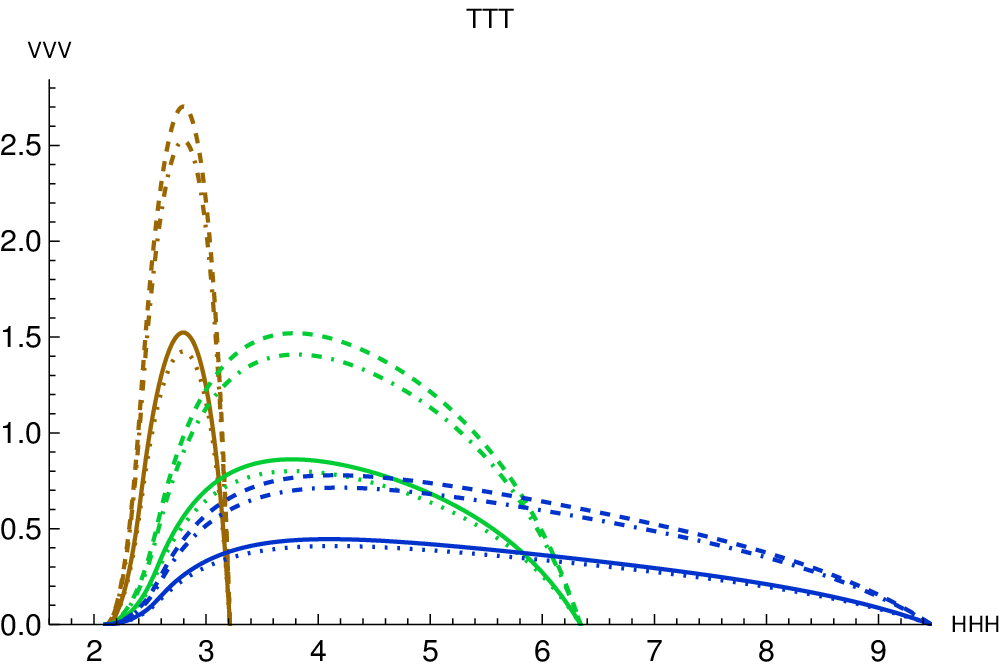}}
		\psfrag{VVV}{\raisebox{.3cm}{\scalebox{.9}{$\hspace{-.4cm}\displaystyle\frac{d 
						\sigma^{\mathrm{odd}}_{\gamma\mesonmn}}{d M^2_{\gamma\mesonmn}}({\rm pb} \cdot {\rm GeV}^{-2})$}}}
		{\includegraphics[width=18pc]{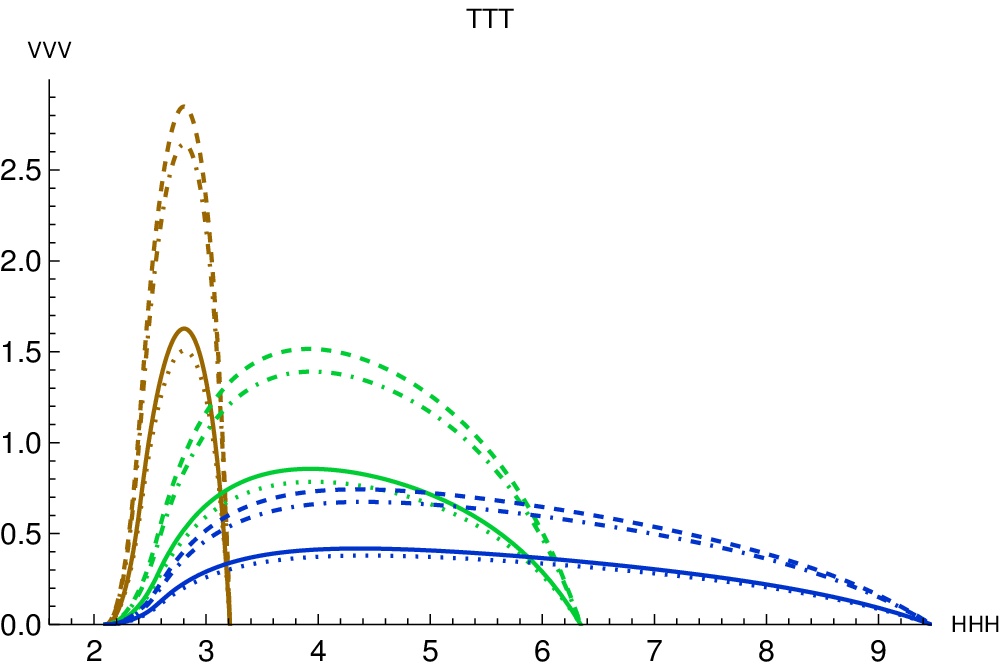}}}
	\vspace{0.2cm}
	\caption{\small The single differential cross section for {transversely polarised} $ \mesonzp,\,\mesonzn,\,\mesonpp,\,\mesonmn$ is shown as a function of $  M_{\gamma \meson}^{2}  $ on the top left, top right, bottom left and bottom right plots respectively  for different values of $ S_{\gamma N} $. The brown, green and blue curves correspond to $ S_{\gamma N} = 8,\,14,\,20\,\GeV^{2} $. The dashed (non-dashed) lines correspond to holographic (asymptotic) DA, while the dotted (non-dotted) lines correspond to the standard (valence) scenario.}
	\label{fig:co-jlab-sing-diff}
\end{figure}

We now integrate over the kinematical variables $ (-u') $ and $ (-t) $ and obtain the single differential cross section as a function of $ \Msq $. The details of this integration are given in \APP{}D of \cite{Boussarie:2016qop}, and in \APP{}E of \cite{Duplancic:2018bum}. The ansatz used for the $t$-dependence of the cross section has been modified in this work, see \eqref{dipole}, compared to the previous paper \cite{Boussarie:2016qop}, leading to slightly different values for the cross sections.  The effect of different values of $ S_{\gamma N} $ on the single differential cross section is shown in \FIG\ref{fig:jlab-sing-diff} for the chiral-even case and in \FIG\ref{fig:co-jlab-sing-diff} for the chiral-odd case. The different colours, brown, green and blue, correspond to $ \SgN $ values of 8, 14 and 20 GeV$ ^2 $ respectively. As $ \SgN $ increases, the maximum value of $ \Msq $ increases (simply due to the increase in the phase space), while the value of the cross section decreases.\footnote{A similar effect was observed in \FIG\ref{fig:jlab-fully-diff-diff-M2} with increasing $M_{\gamma \meson}^2$, instead of $\SgN$.}

As previously mentioned, the peaks in the plots in \FIGs\ref{fig:jlab-sing-diff} and \ref{fig:co-jlab-sing-diff} are the consequence of the competition between the decrease in the cross section and the increase in the volume of the phase space as $M_{\gamma \meson}^2$ increases. An interesting point to note is that the peak of the distribution is always found at \textit{low} $ \Msq $, around $ 3 \GeV^2 $. The reason for this is that the cross section grows rapidly as $M_{\gamma \meson}^2$ decreases, but at the same time, the kinematical cuts that we impose to use collinear QCD factorisation causes the volume of the phase space to vanish at a minimum value of $M_{\gamma \meson}^2$ of about 2.1 GeV$^2$. Furthermore, the height of the peak in the chiral-odd case decreases faster as $ \SgN $ increases. This can be traced back to the $  \xi ^2 $ prefactor in \eqref{all-rhoT}, since $ \xi $ decreases as $ \Msq $ decreases.

Like for the fully-differential cross section plots, we observe that the case of the photoproduction of $ \gamma \meson^{0}_{n} $ has the strongest dependence on the GPD model used for the chiral-odd case, while both $ \gamma \mesonzn $ and $ \gamma \mesonmn $ channels (\ie{}on neutron target) are very sensitive to the GPD model used for the chiral-even case. Finally, in both chiral-even and chiral-odd cases, using a holographic DA instead of an asymptotic DA gives a larger cross section, by roughly a factor of 2.

\FloatBarrier

\subsubsection{Integrated cross section}

\label{sec:int-X-section-JLab}

\begin{figure}[t!]
	\psfrag{HHH}{\hspace{-1.5cm}\raisebox{-.6cm}{\scalebox{.8}{$ S_{\gamma N} ({\rm 
					GeV}^{2})$}}}
	\psfrag{VVV}{\raisebox{.3cm}{\scalebox{.9}{$\hspace{-.4cm}\displaystyle
				\sigma^{\mathrm{even}}_{\gamma\mesonzp}({\rm pb})$}}}
	\psfrag{TTT}{}
	{
		{\includegraphics[width=18pc]{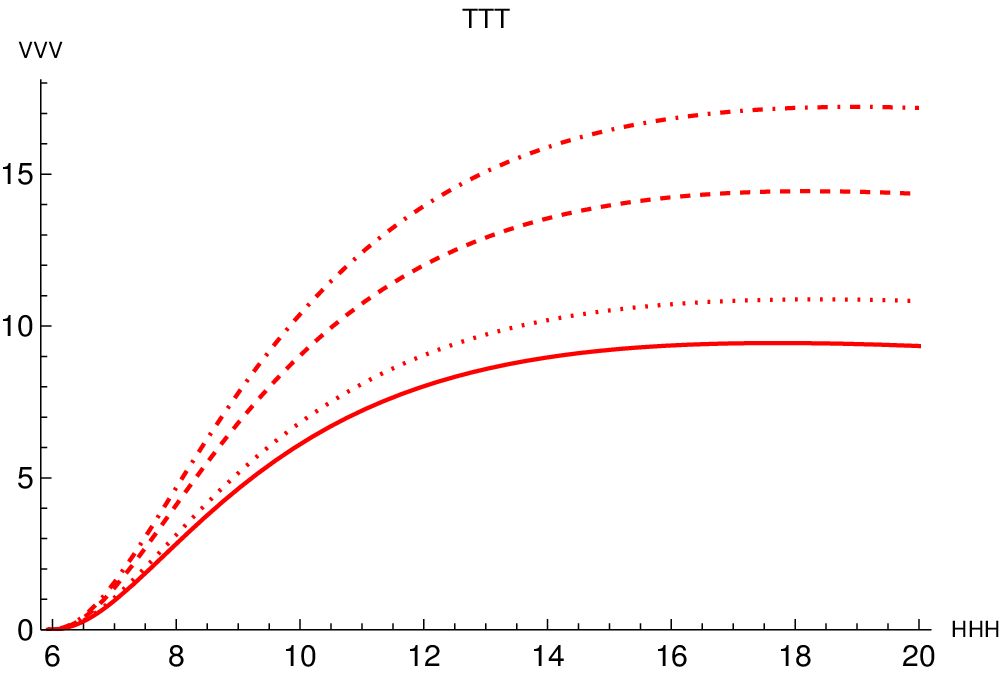}}
		\psfrag{VVV}{\raisebox{.3cm}{\scalebox{.9}{$\hspace{-.4cm}\displaystyle
					\sigma^{\mathrm{even}}_{\gamma\mesonzn}({\rm pb})$}}}
		{\includegraphics[width=18pc]{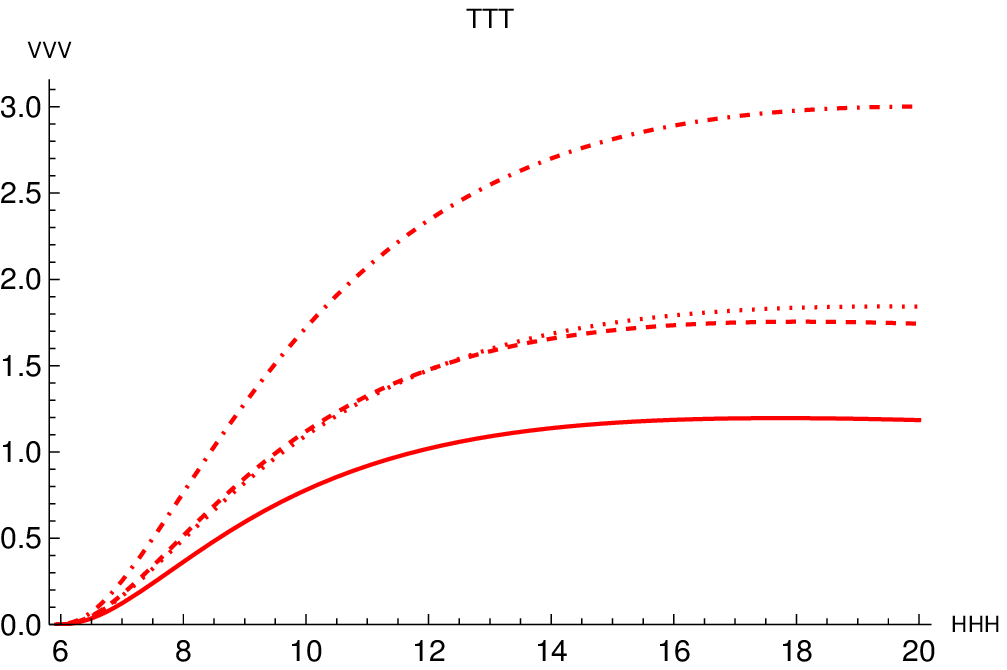}}}
	\\[25pt]
	\psfrag{VVV}{\raisebox{.3cm}{\scalebox{.9}{$\hspace{-.4cm}\displaystyle
				\sigma^{\mathrm{even}}_{\gamma\mesonpp}({\rm pb})$}}}
	\psfrag{TTT}{}
	{
		{\includegraphics[width=18pc]{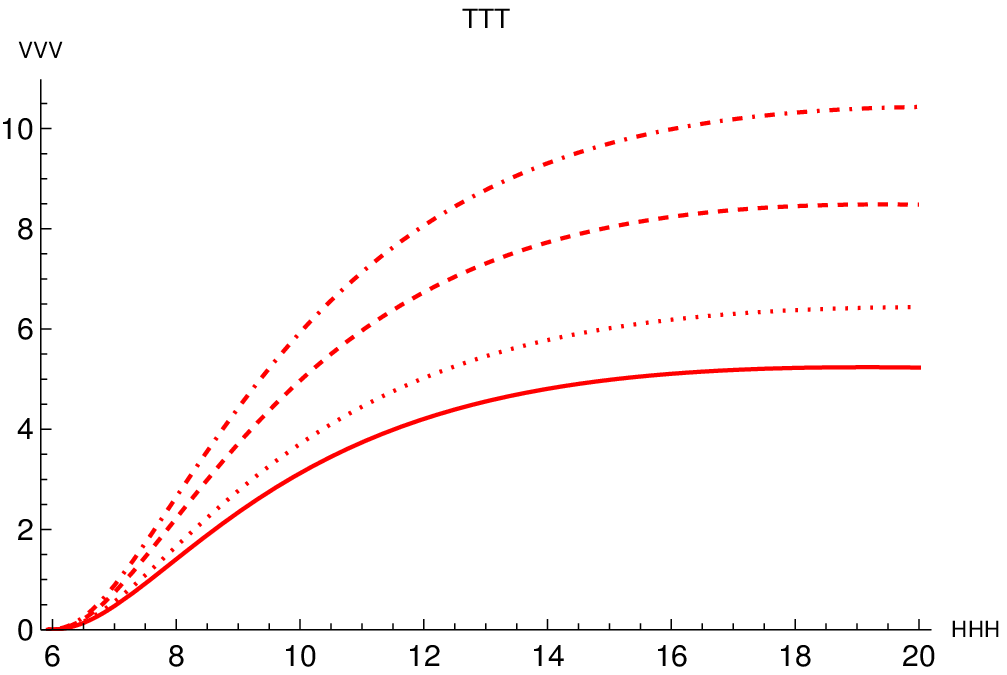}}
		\psfrag{VVV}{\raisebox{.3cm}{\scalebox{.9}{$\hspace{-.4cm}\displaystyle
					\sigma^{\mathrm{even}}_{\gamma\mesonmn}({\rm pb})$}}}
		{\includegraphics[width=18pc]{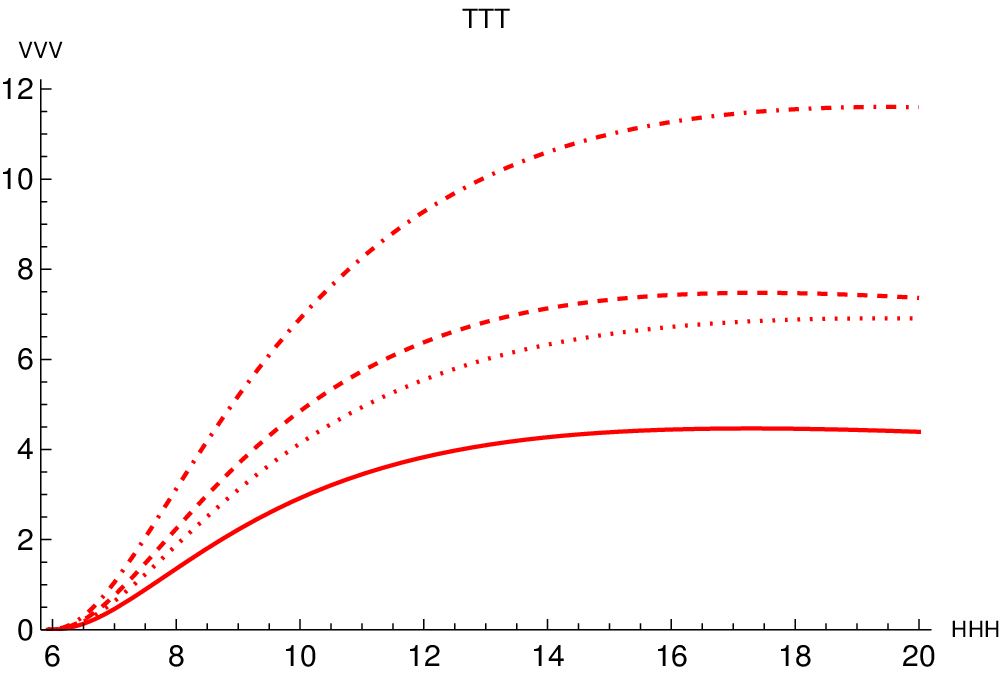}}}
	\vspace{0.2cm}
	\caption{\small The integrated cross section for {longitudinally polarised} $ \mesonzp,\,\mesonzn,\,\mesonpp,\,\mesonmn$ is shown as a function of $   S_{\gamma N}  $ on the top left, top right, bottom left and bottom right plots respectively. The dashed (non-dashed) lines correspond to holographic (asymptotic) DA, while the dotted (non-dotted) lines correspond to the standard (valence) scenario.}
	\label{fig:jlab-int-sigma}
\end{figure}

\begin{figure}[t!]
	\psfrag{HHH}{\hspace{-1.5cm}\raisebox{-.6cm}{\scalebox{.8}{$ S_{\gamma N} ({\rm 
					GeV}^{2})$}}}
	\psfrag{VVV}{\raisebox{.3cm}{\scalebox{.9}{$\hspace{-.4cm}\displaystyle
				\sigma^{\mathrm{odd}}_{\gamma\mesonzp}({\rm pb})$}}}
	\psfrag{TTT}{}
	{
		{\includegraphics[width=18pc]{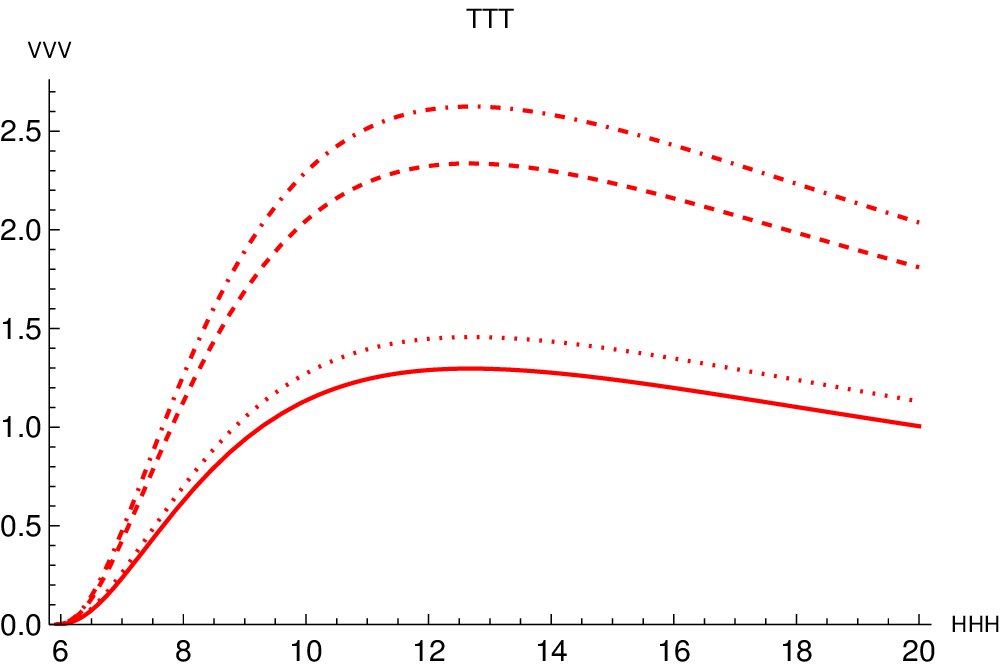}}
		\psfrag{VVV}{\raisebox{.3cm}{\scalebox{.9}{$\hspace{-.4cm}\displaystyle
					\sigma^{\mathrm{odd}}_{\gamma\mesonzn}({\rm pb})$}}}
		{\includegraphics[width=18pc]{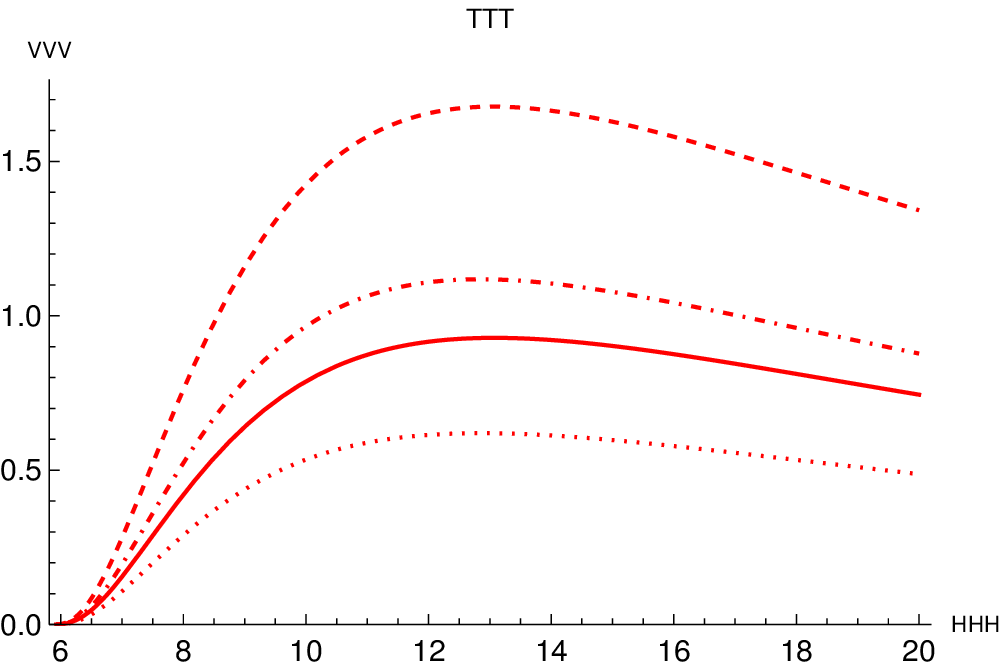}}}
	\\[25pt]
	\psfrag{VVV}{\raisebox{.3cm}{\scalebox{.9}{$\hspace{-.4cm}\displaystyle
				\sigma^{\mathrm{odd}}_{\gamma\mesonpp}({\rm pb})$}}}
	\psfrag{TTT}{}
	{
		{\includegraphics[width=18pc]{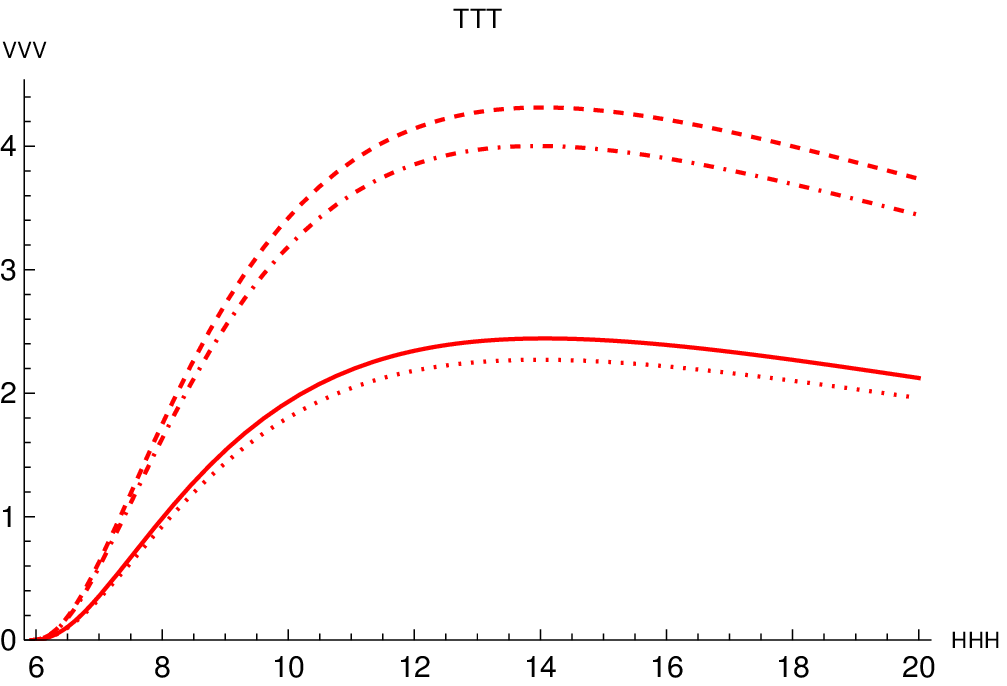}}
		\psfrag{VVV}{\raisebox{.3cm}{\scalebox{.9}{$\hspace{-.4cm}\displaystyle
					\sigma^{\mathrm{odd}}_{\gamma\mesonmn}({\rm pb})$}}}
		{\includegraphics[width=18pc]{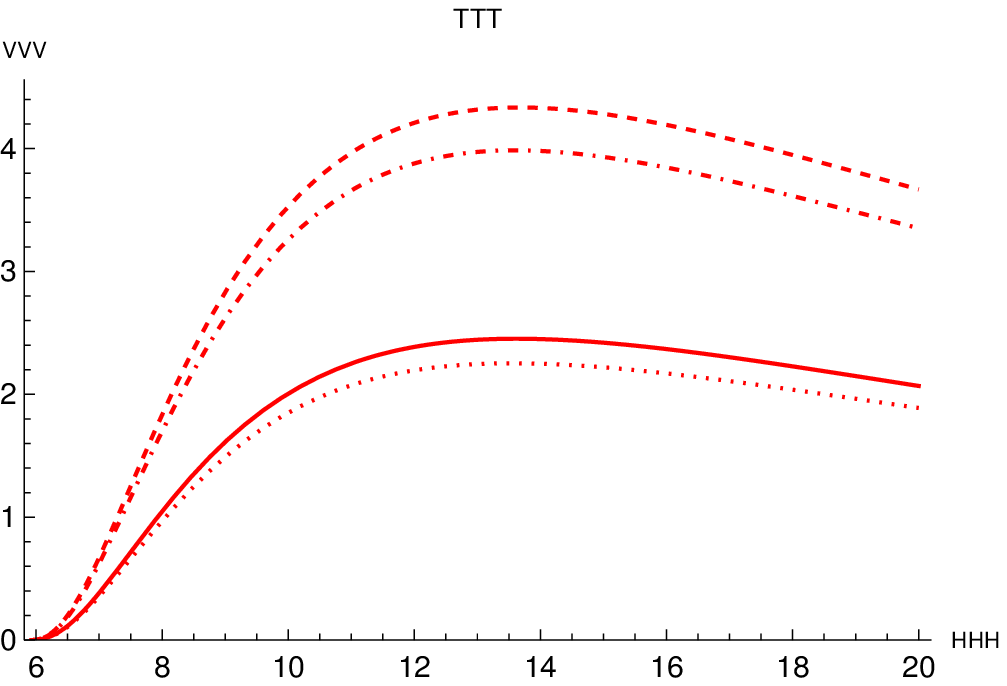}}}
	\vspace{0.2cm}
	\caption{\small The integrated cross section for {transversely polarised} $ \mesonzp,\,\mesonzn,\,\mesonpp,\,\mesonmn$ is shown as a function of $   S_{\gamma N}  $ on the top left, top right, bottom left and bottom right plots respectively. The dashed (non-dashed) lines correspond to holographic (asymptotic) DA, while the dotted (non-dotted) lines correspond to the standard (valence) scenario.}
	\label{fig:co-jlab-int-sigma}
\end{figure}

In this subsection, we discuss the variation of the cross section as a function of $ \SgN $, after integration over $ (-u') $, $ (-t) $ and $ \Msq $. The details of the integration are found in \APP{}D of \cite{Boussarie:2016qop} and \APP{}E of \cite{Duplancic:2018bum}. The variation of the cross section as a function of $ S_{\gamma N} $ is shown in \FIG\ref{fig:jlab-int-sigma} for the chiral-even case and \FIG\ref{fig:co-jlab-int-sigma} for the chiral-odd case. In both cases, the cross section has a peak, which occurs at around 20 $ \GeV^{2} $ for the chiral-even case, and around 12 $ \GeV^2 $ for the chiral-odd case (in the chiral-even case, the presence of the peak becomes evident in \FIG\ref{fig:compass-int-sigma} which corresponds to the same plot but extends to higher energies typical of COMPASS kinematics). In accordance with \FIGs\ref{fig:jlab-sing-diff} and \ref{fig:co-jlab-sing-diff}, where we observed that the peak of the single-differential cross section decreases more rapidly with increasing $ \SgN $ for the chiral-odd case, we observe here that the peak in the integrated cross section occurs at lower values of $ \SgN $  with chiral-odd GPDs.
Similar comments as in the previous subsection applies, \ie{}the case of the photoproduction of $ \gamma \meson^{0}_{n} $ has the strongest dependence on the GPD model used for the chiral-odd case, while both $ \gamma \mesonzn $ and $ \gamma \mesonmn $ channels (\ie{}on neutron target) are very sensitive to the GPD model used for the chiral-even case. Furthermore, in both chiral-even and chiral-odd cases, using a holographic DA instead of an asymptotic DA gives a larger cross section, by roughly a factor of 2.

\FloatBarrier

\subsubsection{Polarisation asymmetries}

\label{sec:pol-asym-jlab}

\begin{figure}[t!]
	\psfrag{HHH}{\hspace{-1.5cm}\raisebox{-.6cm}{\scalebox{.8}{$-u' ({\rm 
					GeV}^{2})$}}}
	\psfrag{VVV}{LPA$^{\gamma\mesonzp}_{\mathrm{max}} $}
	\psfrag{TTT}{}
	{
		{\includegraphics[width=18pc]{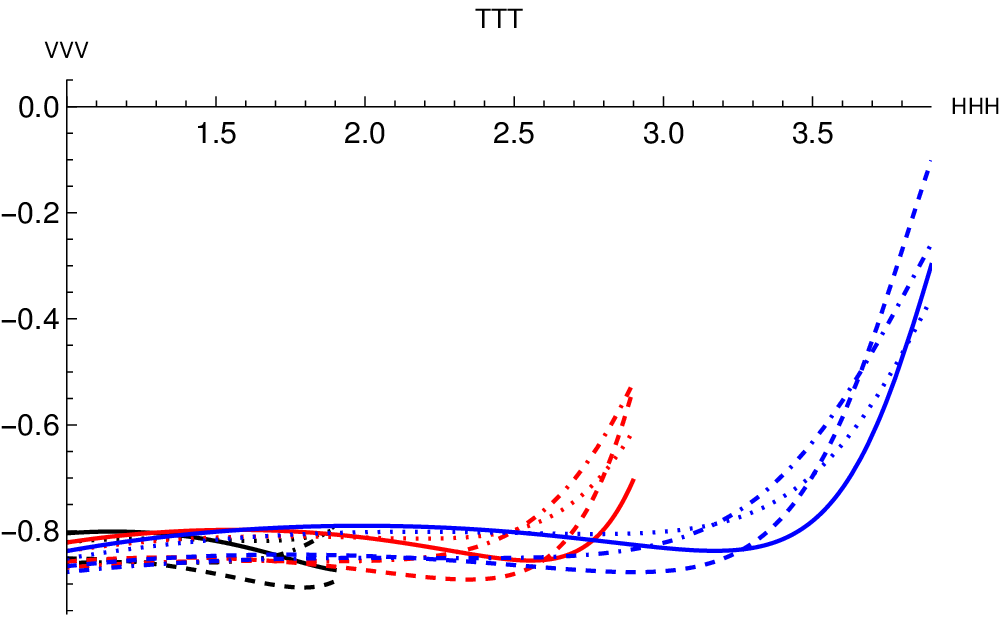}}
		\psfrag{VVV}{LPA$^{\gamma\mesonzn}_{\mathrm{max}} $}
		{\includegraphics[width=18pc]{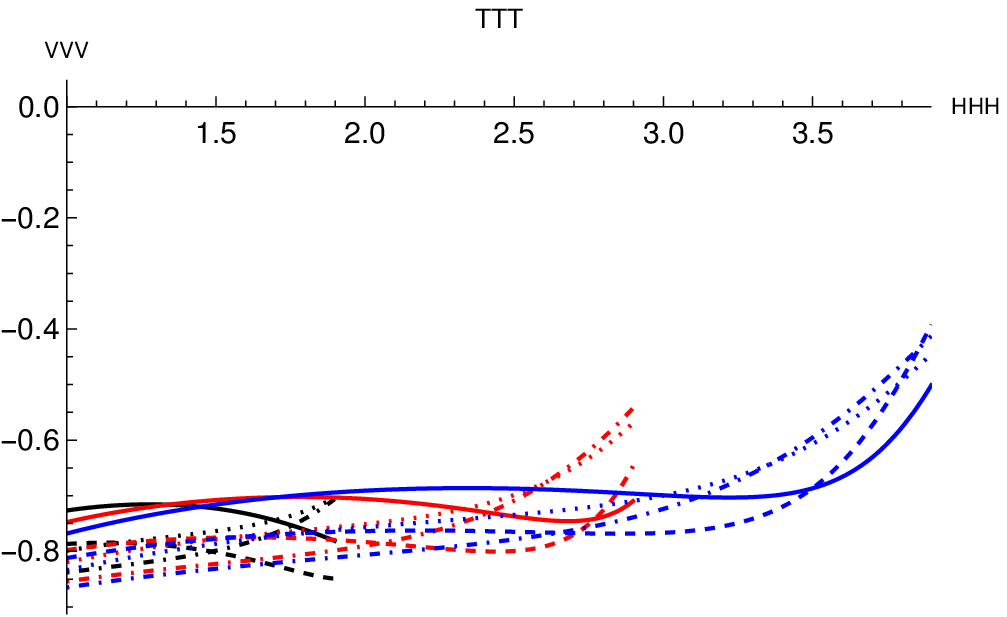}}}
	\\[25pt]
	{
		\psfrag{VVV}{LPA$^{\gamma\mesonpp}_{\mathrm{max}} $}
		{\includegraphics[width=18pc]{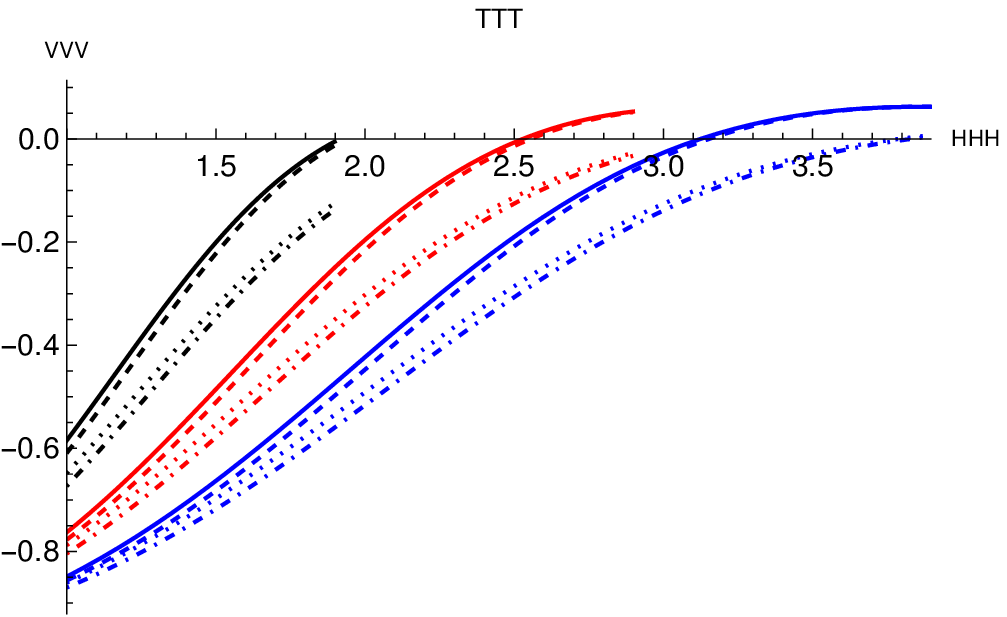}}
		\psfrag{VVV}{LPA$^{\gamma\mesonmn}_{\mathrm{max}} $}
		{\includegraphics[width=18pc]{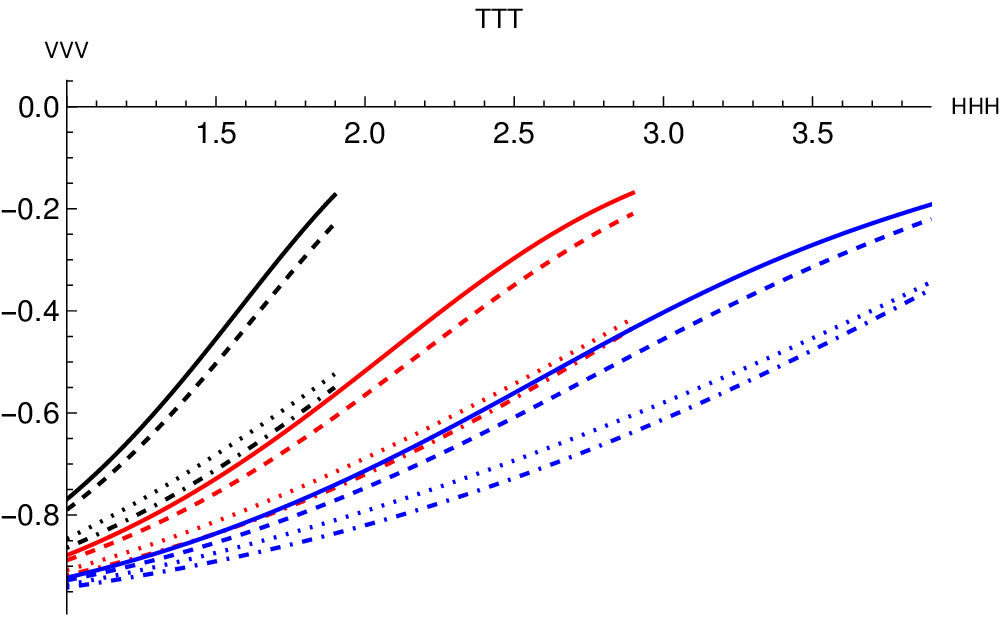}}}
	\vspace{0.2cm}
	\caption{\small The LPA at the fully-differential level for longitudinally polarised $ \mesonzp,\,\mesonzn,\,\mesonpp,\,\mesonmn$ is shown as a function of $  \left( -u' \right)  $ on the top left, top right, bottom left and bottom right plots respectively  for different values of $ M_{\gamma \meson}^2 $. The black, red and blue curves correspond to $ M_{\gamma \meson}^{2}=3,\,4,\,5\, $ GeV$ ^2 $ respectively, and $ \SgN = 20 \GeV^{2}$. The same conventions as in \FIG\ref{fig:jlab-fully-diff-diff-M2} are used here.}
	\label{fig:jlab-pol-asym-fully-diff-diff-M2}
\end{figure}

\begin{figure}[t!]
	\psfrag{HHH}{\hspace{-1.5cm}\raisebox{-.6cm}{\scalebox{.8}{$-u' ({\rm 
					GeV}^{2})$}}}
	\psfrag{VVV}{LPA$^{\gamma\mesonzp}_{\mathrm{max}} $}
	\psfrag{TTT}{}
	{
		{\includegraphics[width=18pc]{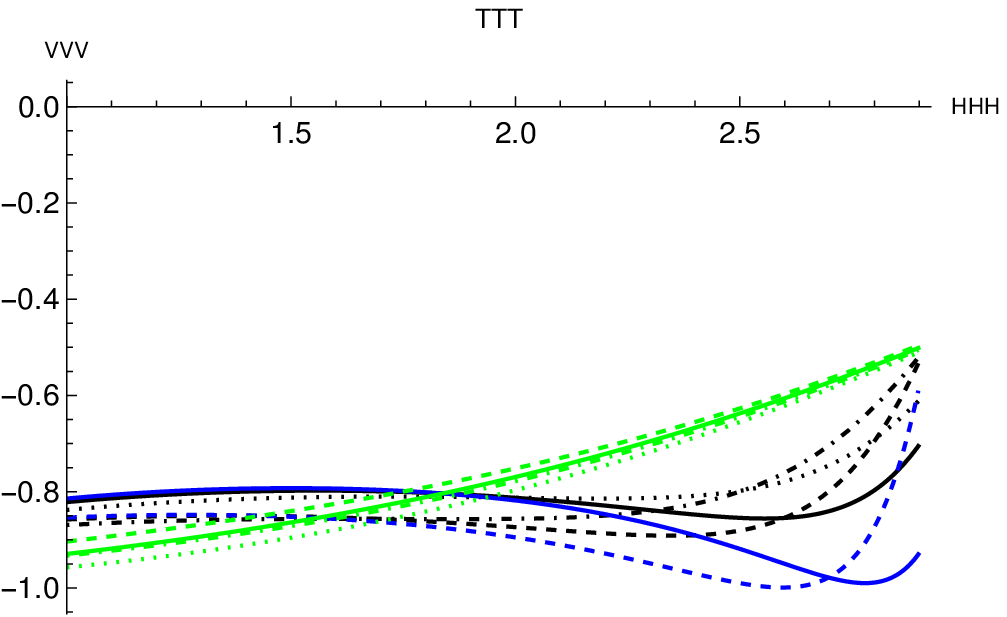}}
		\psfrag{VVV}{LPA$^{\gamma\mesonzn}_{\mathrm{max}} $}
		{\includegraphics[width=18pc]{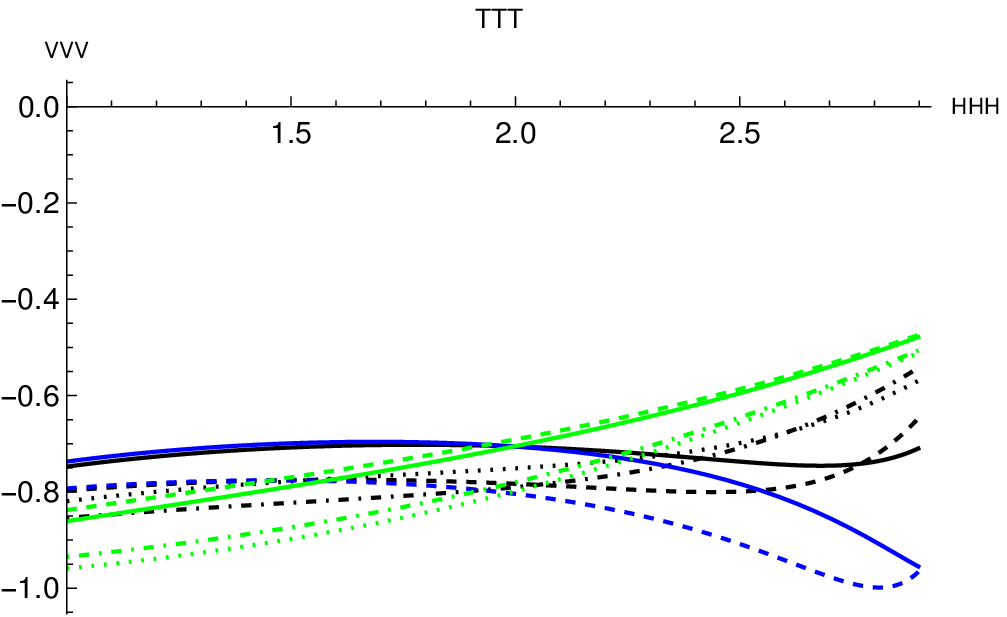}}}
	\\[25pt]
	{					\psfrag{VVV}{LPA$^{\gamma\mesonpp}_{\mathrm{max}} $}
		{\includegraphics[width=18pc]{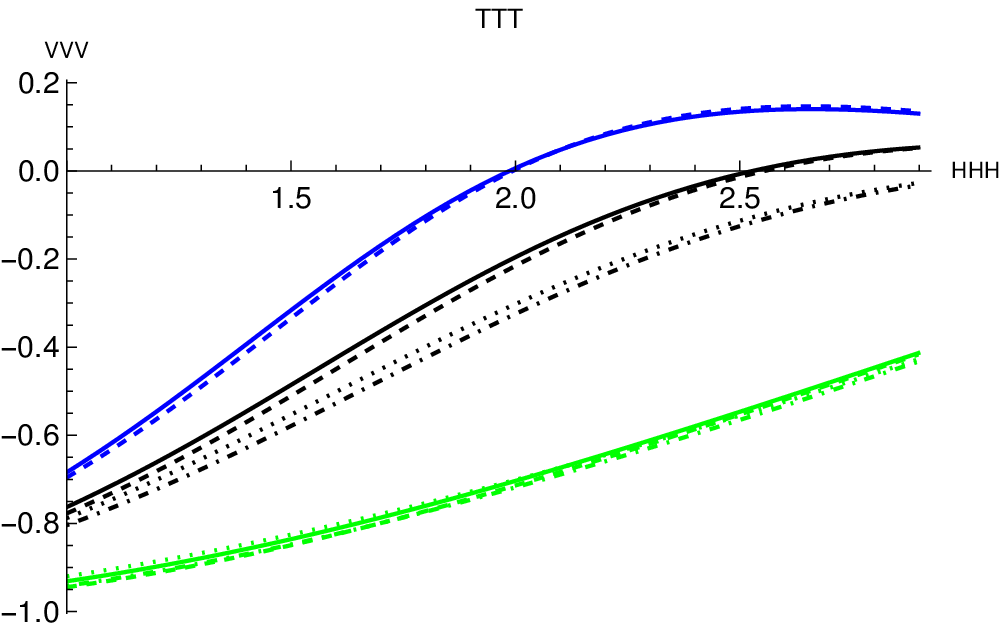}}
		\psfrag{VVV}{LPA$^{\gamma\mesonmn}_{\mathrm{max}} $}
		{\includegraphics[width=18pc]{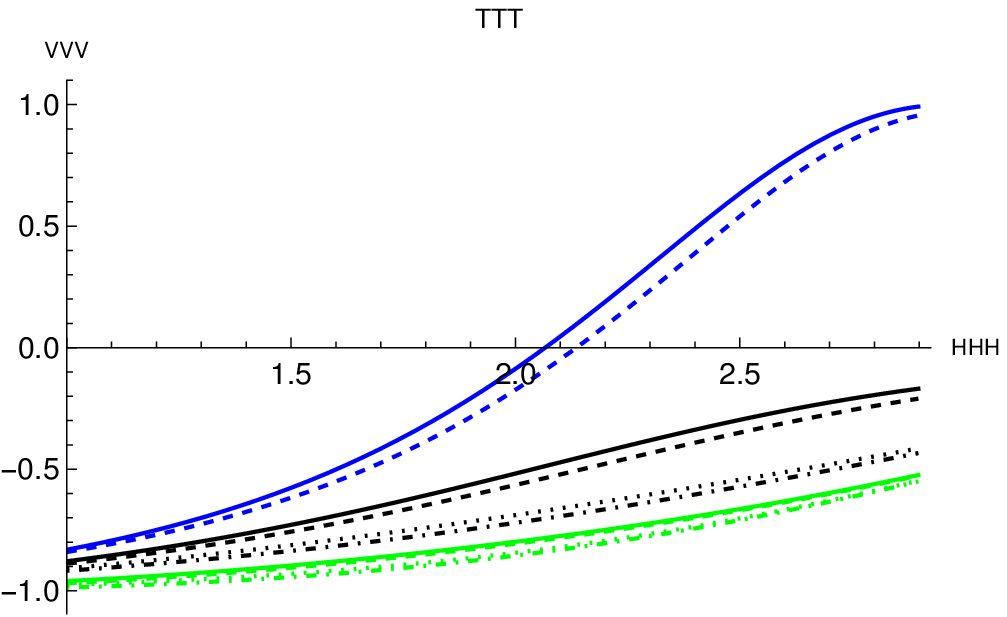}}}
	\vspace{0.2cm}
	\caption{\small The LPA at the fully-differential level for  longitudinally polarised $ \mesonzp,\,\mesonzn,\,\mesonpp,\,\mesonmn$ is shown as a function of $  \left( -u' \right)  $ on the top left, top right, bottom left and bottom right plots respectively, using $ \Msq = 4 \GeV^2$ and $ \SgN = 20 \GeV^2 $. The same conventions as in \FIG\ref{fig:jlab-fully-diff-VandA} are used here. Note that the vector contributions consist of only two curves in each case, since they are insensitive to either valence or standard scenarios.}
	\label{fig:jlab-pol-asym-fully-diff-VandA}
\end{figure}

\begin{figure}[t!]
	\psfrag{HHH}{\hspace{-1.5cm}\raisebox{-.6cm}{\scalebox{.8}{$-u' ({\rm 
					GeV}^{2})$}}}
	\psfrag{VVV}{LPA$^{\gamma\mesonzp}_{\mathrm{max}} $}
	\psfrag{TTT}{}
	{
		{\includegraphics[width=18pc]{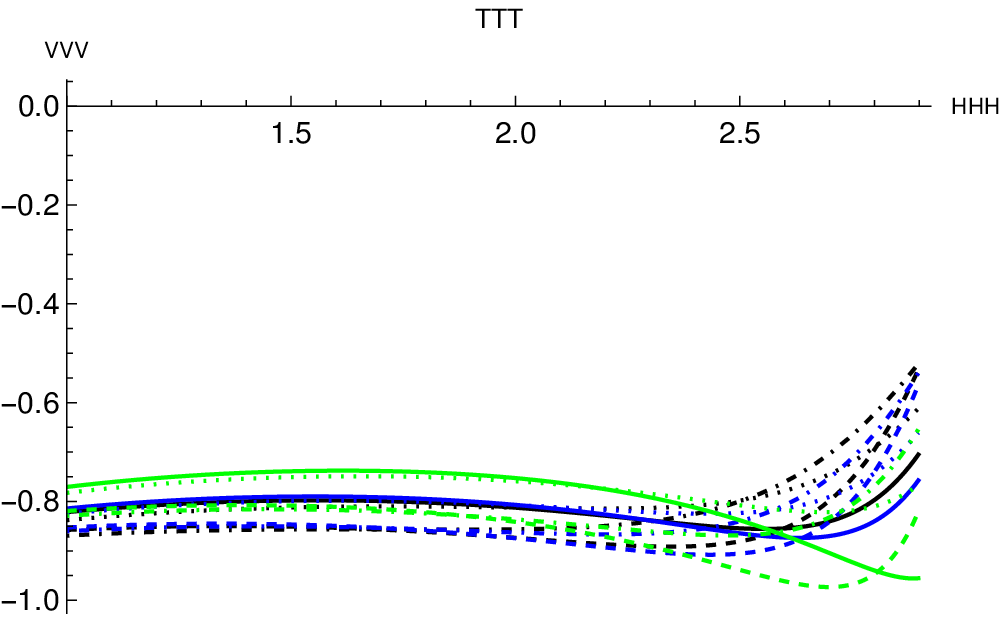}}
		\psfrag{VVV}{LPA$^{\gamma\mesonzn}_{\mathrm{max}} $}
		{\includegraphics[width=18pc]{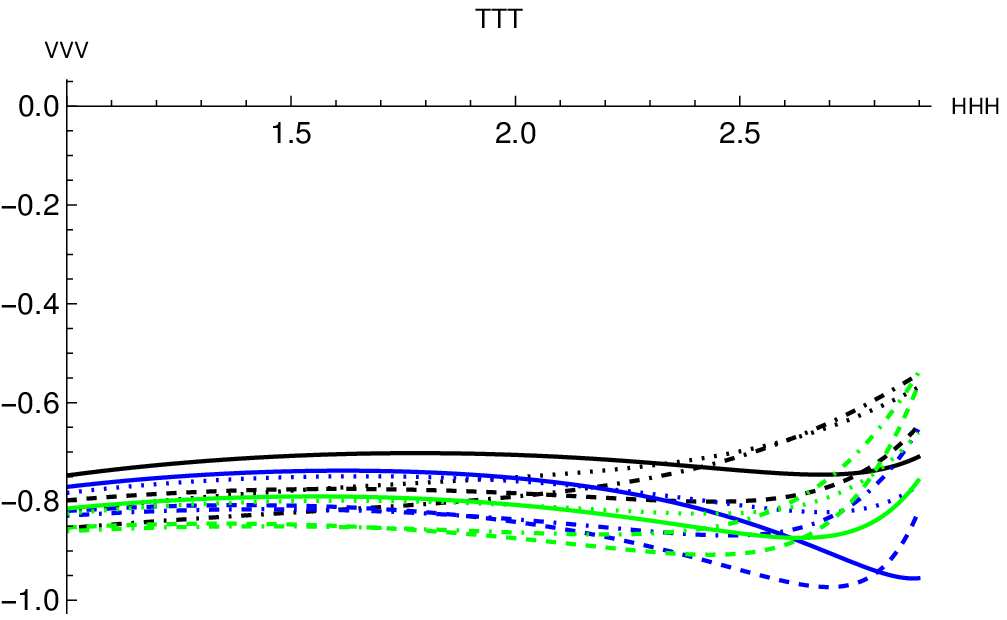}}}
	\\[25pt]
	{					\psfrag{VVV}{LPA$^{\gamma\mesonpp}_{\mathrm{max}} $}
		{\includegraphics[width=18pc]{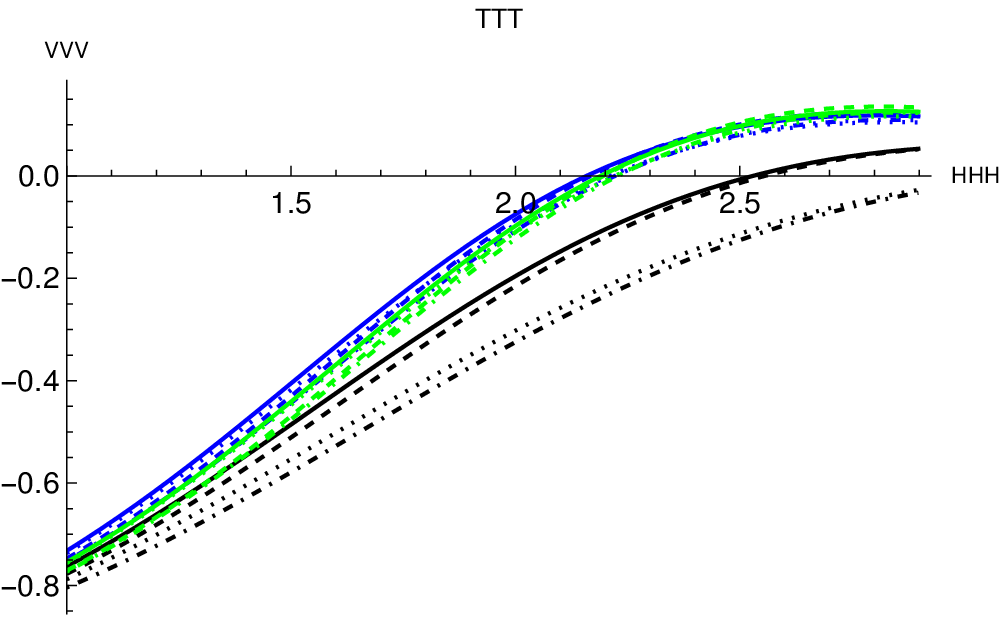}}
		\psfrag{VVV}{LPA$^{\gamma\mesonmn}_{\mathrm{max}} $}
		{\includegraphics[width=18pc]{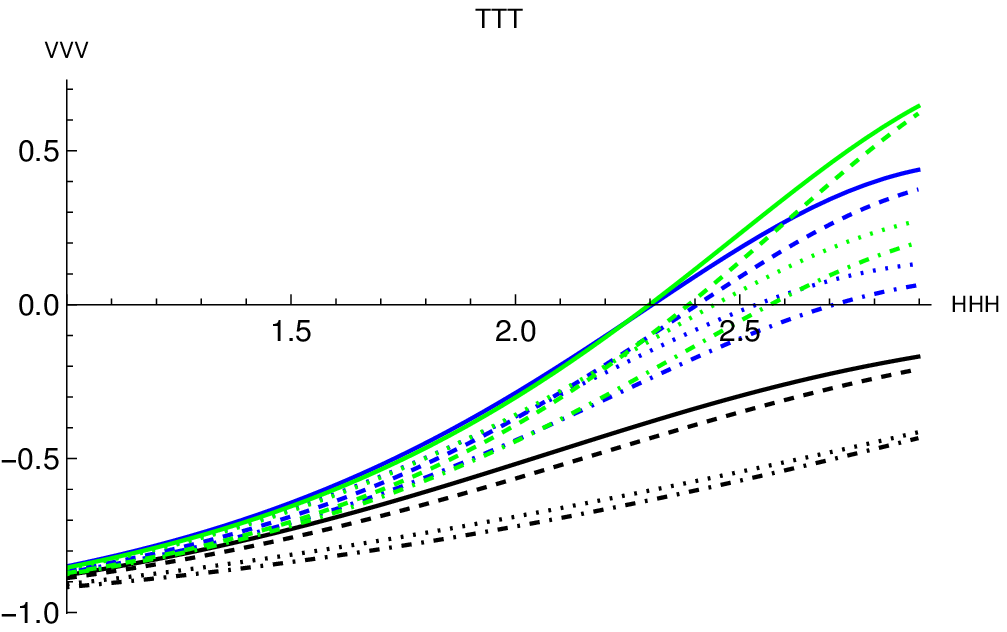}}}
	\vspace{0.2cm}
	\caption{\small The LPA at the fully-differential level for  longitudinally polarised $ \mesonzp,\,\mesonzn,\,\mesonpp,\,\mesonmn$ is shown as a function of $  \left( -u' \right)  $ on the top left, top right, bottom left and bottom right plots respectively, using $ \Msq = 4 \GeV^2$ and $ \SgN = 20 \GeV^2 $. The blue and green curves correspond to contributions from the $u$-quark ($ H_{u} $ and $  \tilde{H} _{u} $) and $d$-quark ($ H_{d} $ and $  \tilde{H} _{d} $) GPDs respectively. The black curves correspond to the total contribution. The same conventions as in \FIG\ref{fig:jlab-fully-diff-uandd} are used here.}
	\label{fig:jlab-pol-asym-fully-diff-uandd}
\end{figure}

We now discuss the plots for the LPA. First, we show the effect of different $ M_{\gamma \meson}^2 $ on the LPAs at the fully differential level (\ie{}differential in $ (-u') $, $ M_{\gamma \meson}^2 $ and $ (-t) $ as in \SEC\ref{sec:jlab-fully-diff-X-section}) in \FIG\ref{fig:jlab-pol-asym-fully-diff-diff-M2}. As in \FIG\ref{fig:jlab-fully-diff-diff-M2}, the values of $ \Msq $ used are 3, 4 and 5 GeV$ ^2 $. One thus finds that the process is dominated by incoming linearly polarised photons along the $ y $-direction, since the LPA is in general negative.

For the neutral $ \meson $-meson case, we observe that the LPA has a weak dependence on both the GPD and DA models used, especially at low $  \left( -u' \right)  $. Furthermore, the LPA remains quite flat and very sizeable except close to the maximum value of $  \left( -u' \right)  $. Finally, we also find that the LPA does not change significantly for different values of $ \Msq $. Both of the previous two observations make the LPA very promising for being measured at JLab.

For the charged $ \meson $-meson case, the LPA is very sizeable at low $  \left( -u' \right)  $ and its magnitude gradually decreases as $  \left( -u '\right)  $ increases. The shape is thus very different from the neutral  $ \meson $-meson one. On the other hand, the shape of the LPA is very similar to the one for the charged $  \pi ^{\pm} $ case (see \FIG{}8 in \cite{Duplancic:2022ffo}), except that the effect of the GPD model goes in the opposite way (\ie{}the more sizeable LPA comes from the standard scenario for the $ \meson $-meson case, but for the $ \pi^{\pm} $, this corresponds to the valence scenario). The LPA also becomes more sizeable when $ \Msq $ increases.

Next, we show how the relative contributions from the vector and axial GPDs affect the LPA at the fully differential level in \FIG\ref{fig:jlab-pol-asym-fully-diff-VandA}. To obtain the green (blue) curves, we set \textit{all} vector (axial) contributions to the polarised cross sections to zero, in both the numerator and denominator of \eqref{eq:LPA}. In the case of $ \meson^{0} $, the relative contribution of the axial GPD to the LPA is small at low $  \left( -u' \right)  $, but becomes important as $  \left( -u' \right)  $ increases, which can be implied from \FIG\ref{fig:jlab-fully-diff-VandA}. Interestingly, for $ \meson^{-} $, the contribution to the LPA from the vector GPD changes from -1 at low  $ \left( -u' \right)   $ to +1 at high $  \left( -u' \right)  $.
The LPA calculated from the axial GPD contribution has very little sensitivity on the GPD model used, in contrast with the $ \pi^{\pm} $ case, \cf\FIG 9 in \cite{Duplancic:2022ffo}.

At first sight, it may seem strange that the GPD model nevertheless has a sizeable effect on the total LPA for the $ \meson $-meson in \FIG\ref{fig:jlab-pol-asym-fully-diff-VandA}. This can be understood from the way the LPA is normalised. Taking the $ \mesonpp $ as a specific example, one observes that the cross section corresponding to the axial contribution, though small,  changes by a factor of 2 roughly between the two GPD models, see \FIG\ref{fig:jlab-fully-diff-VandA}. For the LPA, one finds the the axial GPD part contributes to a negative value of $ -0.5 $ (independent of the GPD model), and has a larger absolute size than the vector GPD part which is roughly $ 0.1 $. Therefore, the only difference between the two GPD models when computing the LPA for the total contribution occurs due to the factor of 2 coming from the cross section, which when coupled with the large negative value of $ -0.5 $, leads to a sizeable difference.

The relative contributions from the $ u $-quark GPDs ($ H_{u} $ and $  \tilde{H} _{u} $) and $ d $-quark GPDs ($ H_{d} $ and $  \tilde{H} _{d} $) to the LPA are shown in \FIG\ref{fig:jlab-pol-asym-fully-diff-uandd}.

\FloatBarrier

\begin{figure}[t!]
	\psfrag{HHH}{\hspace{-1.5cm}\raisebox{-.6cm}{\scalebox{.8}{ $ M_{\gamma \meson}^{2}({\rm 
					GeV}^{2}) $}}}
		\psfrag{VVV}{LPA$^{\gamma\mesonzp}_{\mathrm{max}} $}
	\psfrag{TTT}{}
	{
		{\includegraphics[width=18pc]{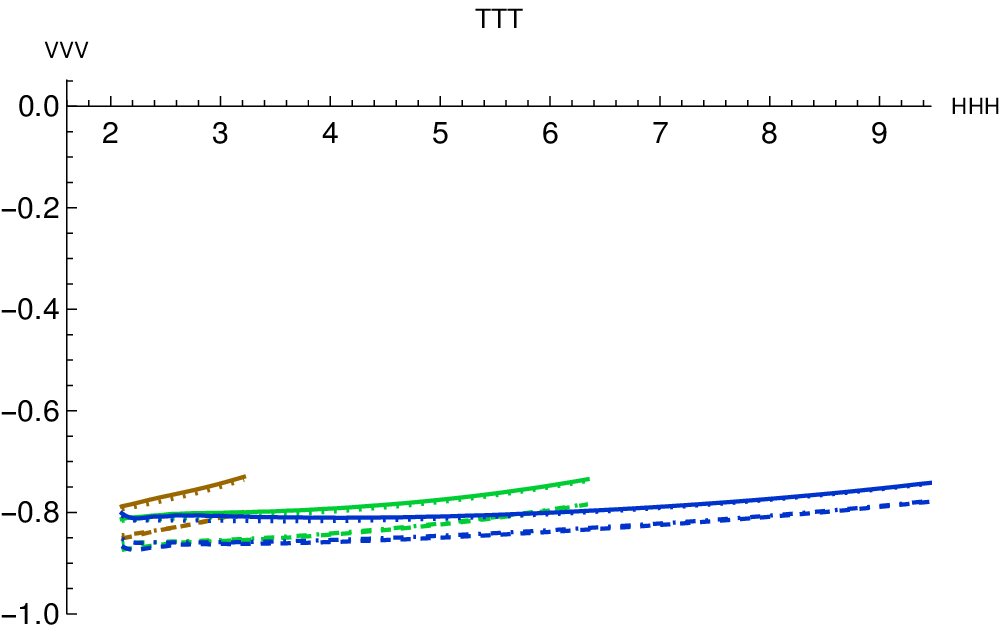}}
			\psfrag{VVV}{LPA$^{\gamma\mesonzn}_{\mathrm{max}} $}
		{\includegraphics[width=18pc]{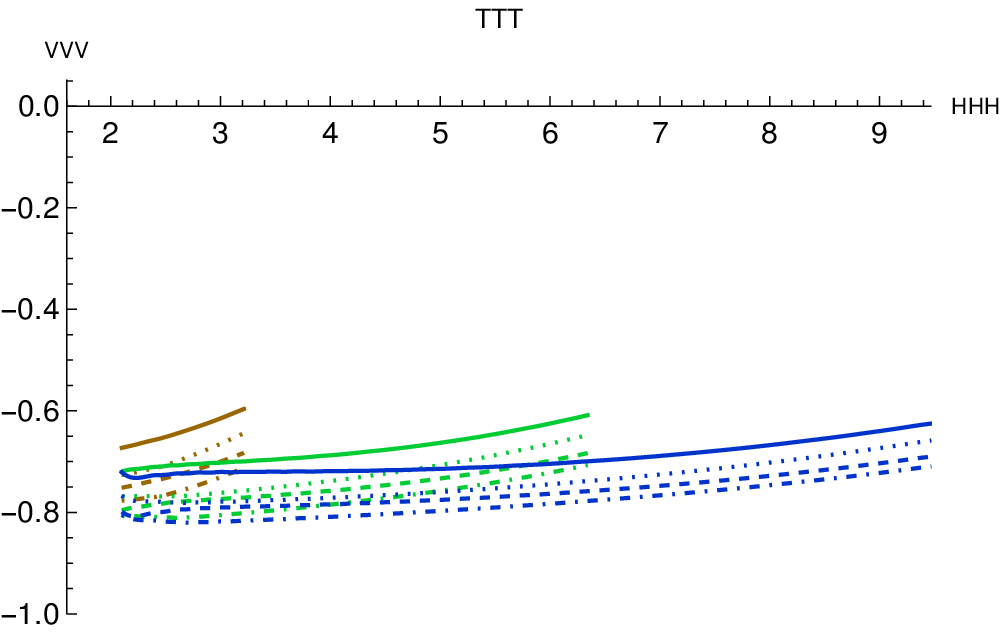}}}
	\\[25pt]
	{					\psfrag{VVV}{LPA$^{\gamma\mesonpp}_{\mathrm{max}} $}
		{\includegraphics[width=18pc]{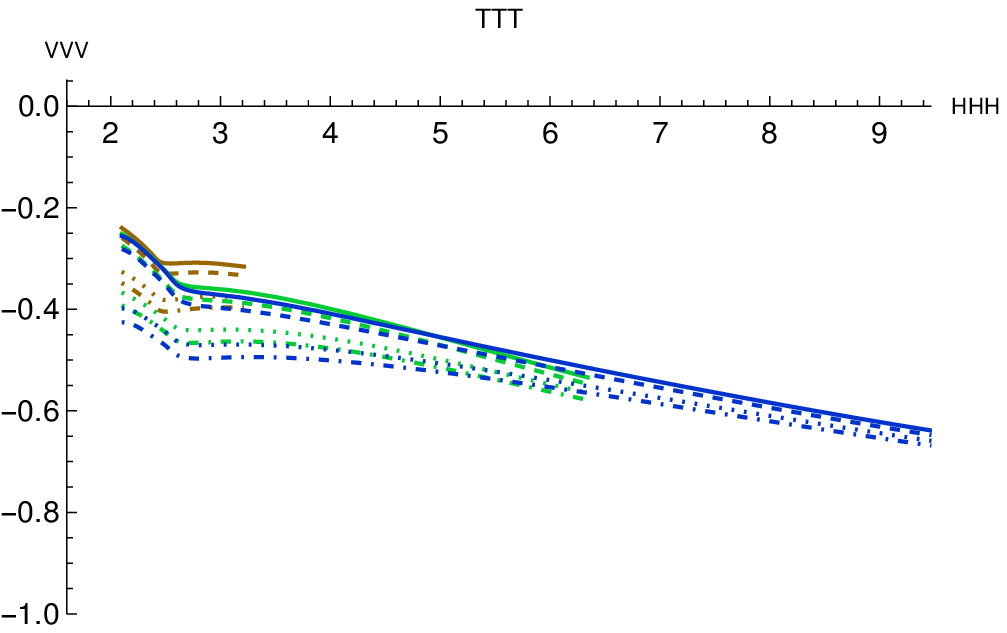}}
			\psfrag{VVV}{LPA$^{\gamma\mesonmn}_{\mathrm{max}} $}
		{\includegraphics[width=18pc]{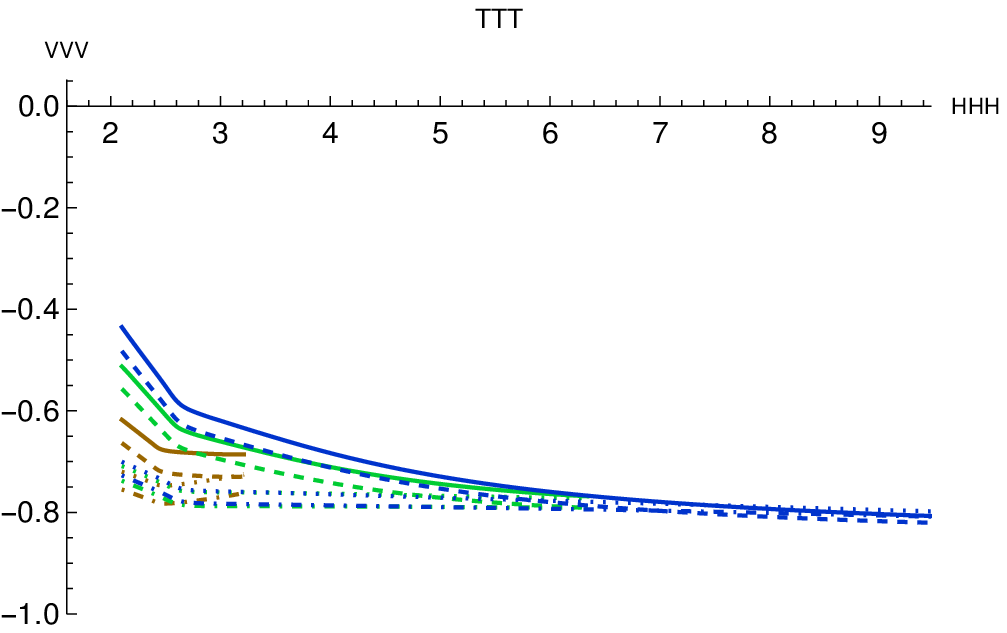}}}
	\vspace{0.2cm}
	\caption{\small The LPA at the single differential level for  longitudinally polarised $ \mesonzp,\,\mesonzn,\,\mesonpp,\,\mesonmn$ is shown as a function of $  M_{\gamma \meson}^{2}  $ on the top left, top right, bottom left and bottom right plots respectively. The brown, green and blue curves correspond to $ S_{\gamma N} = 8,\,14,\,20\,\GeV^{2} $. The same colour and line style conventions as in \FIG\ref{fig:jlab-sing-diff} are used here.}
	\label{fig:jlab-pol-asym-sing-diff}
\end{figure}

Next, we show the LPA, at the single differential level, for different values of $ S_{\gamma N} $ in \FIG\ref{fig:jlab-pol-asym-sing-diff}. As for the cross section plots in \SEC\ref{sec:sing-diff-X-section-JLab}, the values of $ \SgN $ used are 8, 14 and 20 GeV$ ^2 $. We note that neither the GPD nor the DA models have a significant effect on the LPA. This is contrast to the $ \pi ^{\pm} $ case, where the GPD model had an important effect, see \FIG 11 in \cite{Duplancic:2022ffo}. Moreoever, the magnitude of the LPA remains quite large throughout the range of $ \Msq $. This makes the LPA at the single differential level very promising to be measured experimentally.

\FloatBarrier

\begin{figure}[t!]
	\psfrag{HHH}{\hspace{-1.5cm}\raisebox{-.6cm}{\scalebox{.8}{ $ S_{\gamma N}({\rm 
					GeV}^{2}) $}}}
		\psfrag{VVV}{LPA$^{\gamma\mesonzp}_{\mathrm{max}} $}
	\psfrag{TTT}{}
	{
		{\includegraphics[width=18pc]{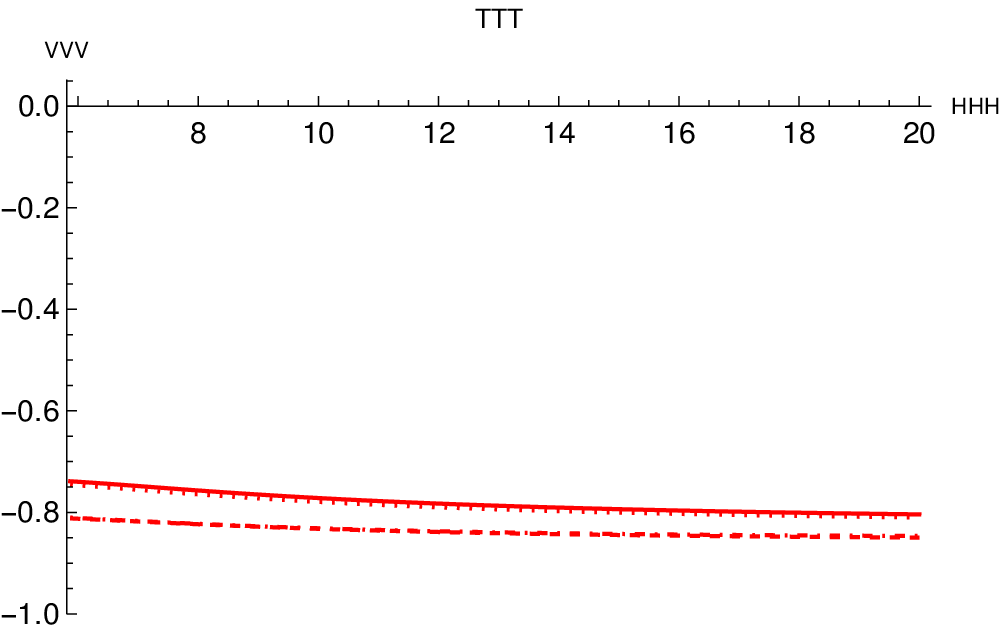}}
			\psfrag{VVV}{LPA$^{\gamma\mesonzn}_{\mathrm{max}} $}
		{\includegraphics[width=18pc]{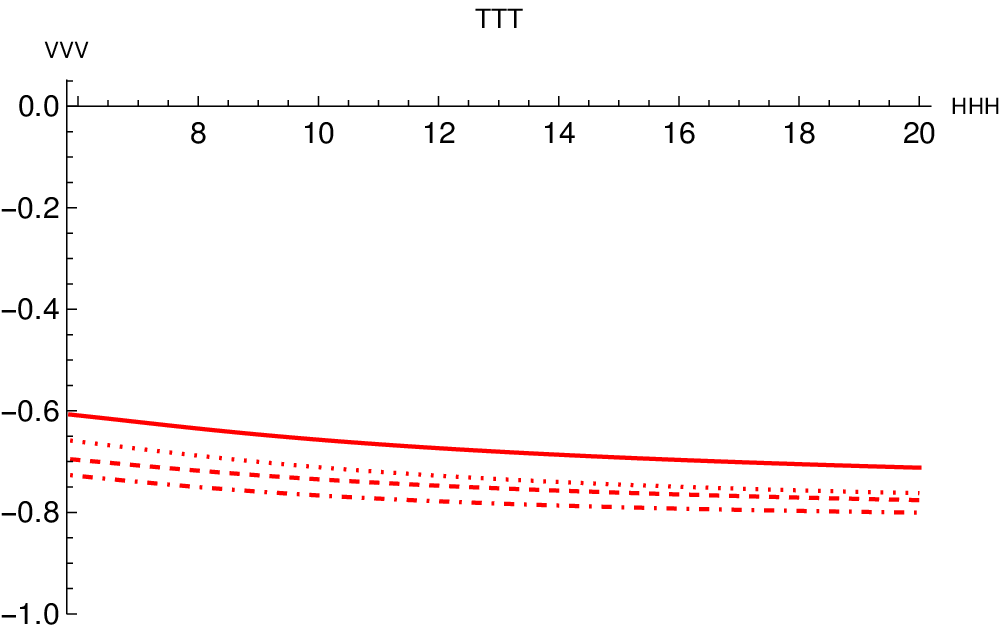}}}
	\\[25pt]
	{					\psfrag{VVV}{LPA$^{\gamma\mesonpp}_{\mathrm{max}} $}
		{\includegraphics[width=18pc]{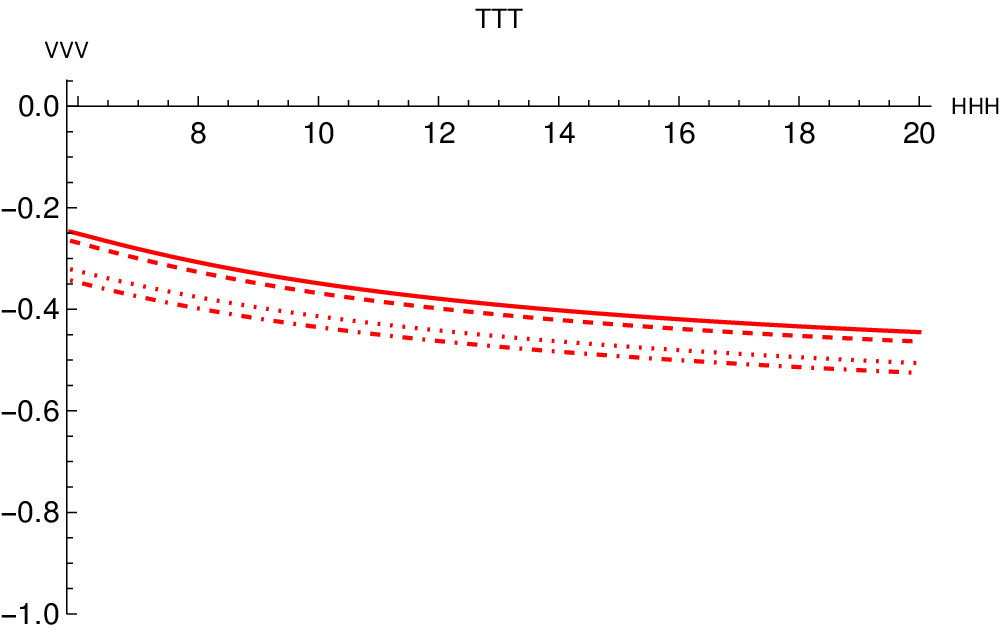}}
			\psfrag{VVV}{LPA$^{\gamma\mesonmn}_{\mathrm{max}} $}
		{\includegraphics[width=18pc]{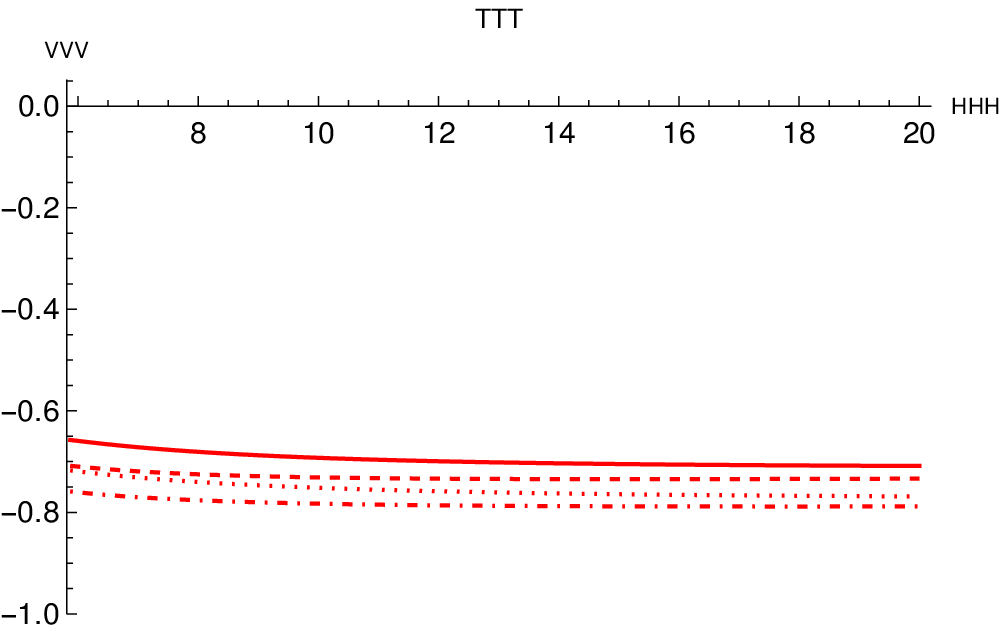}}}
	\vspace{0.2cm}
	\caption{\small The LPA integrated over all differential variables for  longitudinally polarised $ \mesonzp,\,\mesonzn,\,\mesonpp,\,\mesonmn$ is shown on the top left, top right, bottom left and bottom right plots respectively. The same colour and line style conventions as in \FIG\ref{fig:jlab-int-sigma} are used here.}
	\label{fig:jlab-pol-asym-int-sigma}
\end{figure}

Finally, we show the LPA, integrated over all differential variables, as a function of $ S_{\gamma N} $ in \FIG\ref{fig:jlab-pol-asym-int-sigma}. The LPA in all the four plots is rather flat, and is quite sizeable, with the $ \mesonpp $ having the smallest magnitude of roughly 40\%, while this goes up to about 80\% for the others. As at the single differential level, the LPA here has little sensitivity to the DA and GPD models used. Thus, the LPA is sizeable, and taking into account the fact that the expected counting rates found in \SEC\ref{sec:jlab-counting-rates} are large, the measurement of such an observable is very promising.

\FloatBarrier

\subsection{COMPASS kinematics}

\label{sec:COMPASS-kinematics}

Typically, COMPASS consists of colliding muons at an energy of 160 GeV onto a fixed target. This translates to a muon-nucleon centre-of-mass energy of roughly 301 GeV$ ^2 $. Since the skewness $  \xi  $ decreases with increasing $ S_{\gamma N} $ (see \EQ\eqref{skewness2}), COMPASS can in principle give us access to a kinematical region of small $  \xi  $ for GPDs ($0.0027\leq \xi \leq  0.35$), not accessible at JLab. The typical centre-of-mass energy $ S_{\gamma N} $ used for the plots that we show in this section is 200 GeV$ ^{2} $.

\subsubsection{Fully differential cross section}

\label{sec:compass-fully-diff}

\begin{figure}[t!]
	\psfrag{HHH}{\hspace{-1.5cm}\raisebox{-.6cm}{\scalebox{.8}{$-u' ({\rm 
					GeV}^{2})$}}}
	\psfrag{VVV}{\raisebox{.3cm}{\scalebox{.9}{$\hspace{-.4cm}\displaystyle\left.\frac{d 
					\sigma^{\mathrm{even}}_{\gamma\mesonzp}}{d M^2_{\gamma\mesonzp} d(-u') d(-t)}\right|_{(-t)_{\rm min}}({\rm pb} \cdot {\rm GeV}^{-6})$}}}
	\psfrag{TTT}{}
	{
		{\includegraphics[width=18pc]{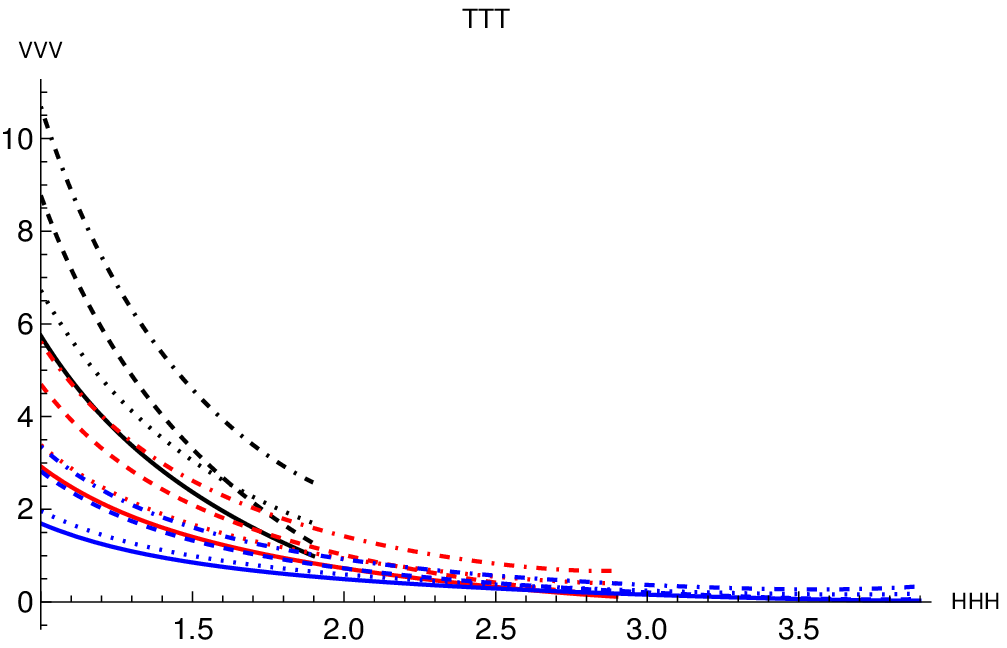}}
		\psfrag{VVV}{\raisebox{.3cm}{\scalebox{.9}{$\hspace{-.4cm}\displaystyle\left.\frac{d 
						\sigma^{\mathrm{even}}_{\gamma\mesonzn}}{d M^2_{\gamma \mesonzn} d(-u') d(-t)}\right|_{(-t)_{\rm min}}({\rm pb} \cdot {\rm GeV}^{-6})$}}}
		{\includegraphics[width=18pc]{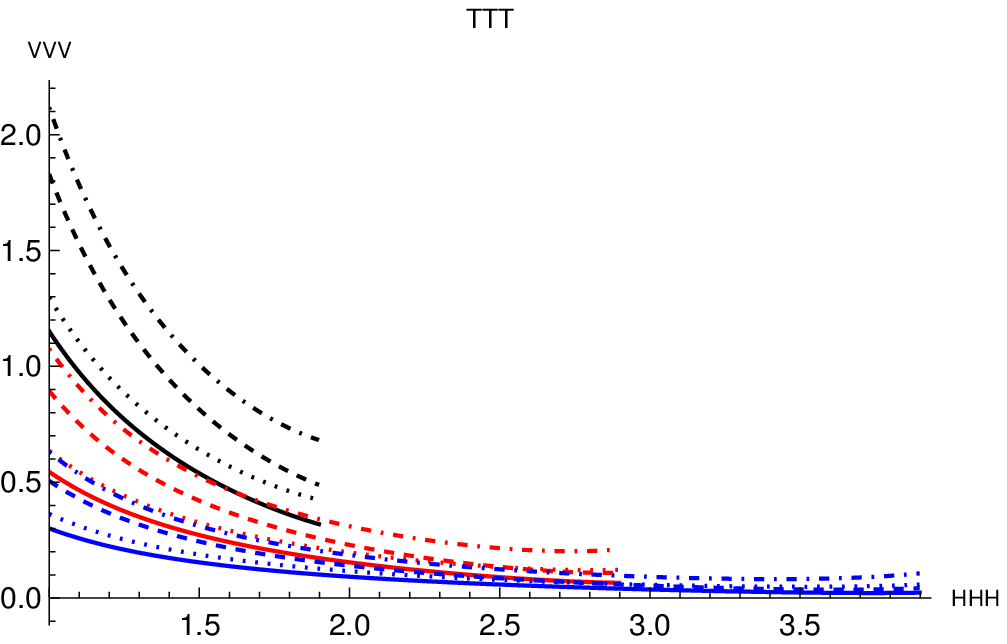}}}
	\\[25pt]
	\psfrag{VVV}{\raisebox{.3cm}{\scalebox{.9}{$\hspace{-.4cm}\displaystyle\left.\frac{d 
					\sigma^{\mathrm{even}}_{\gamma\mesonpp}}{d M^2_{\gamma\mesonpp} d(-u') d(-t)}\right|_{(-t)_{\rm min}}({\rm pb} \cdot {\rm GeV}^{-6})$}}}
	\psfrag{TTT}{}
	{
		{\includegraphics[width=18pc]{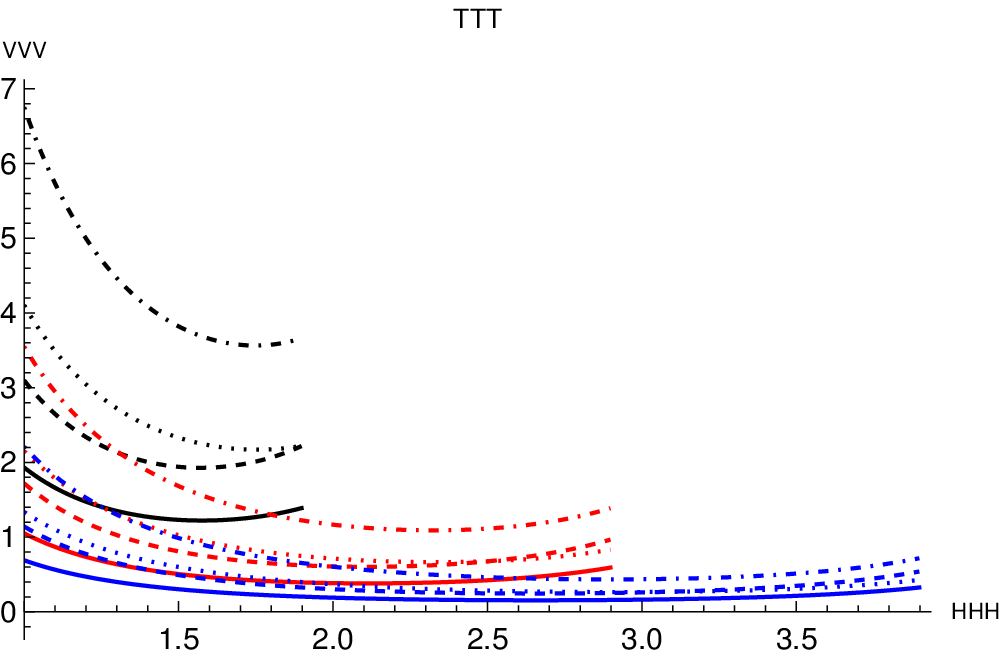}}
		\psfrag{VVV}{\raisebox{.3cm}{\scalebox{.9}{$\hspace{-.4cm}\displaystyle\left.\frac{d 
						\sigma^{\mathrm{even}}_{\gamma\mesonmn}}{d M^2_{\gamma\mesonmn} d(-u') d(-t)}\right|_{(-t)_{\rm min}}({\rm pb} \cdot {\rm GeV}^{-6})$}}}
		{\includegraphics[width=18pc]{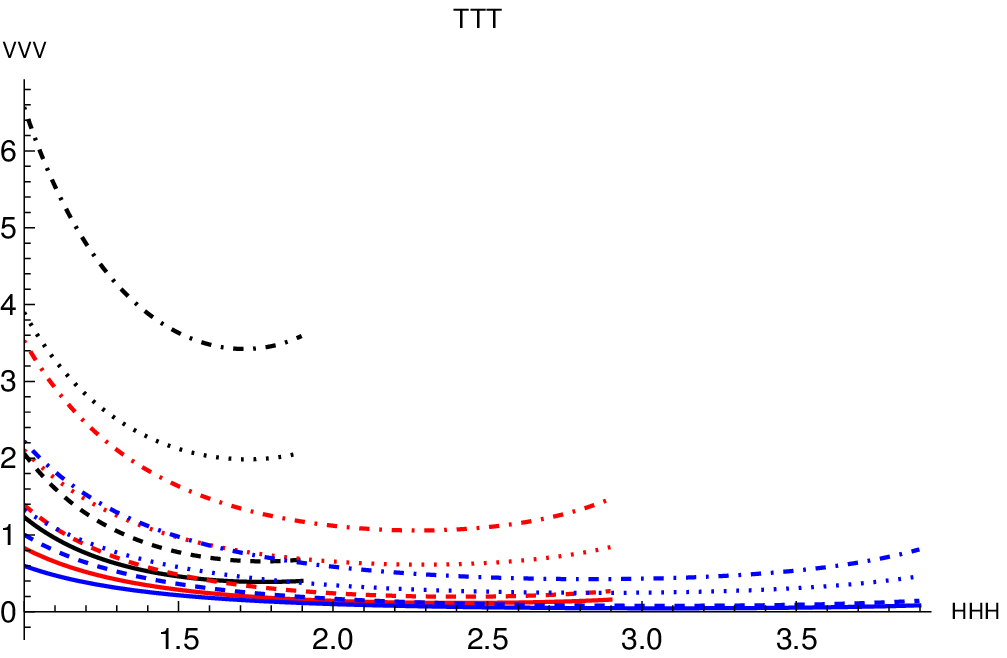}}}
	\vspace{0.2cm}
	\caption{\small The fully differential cross section for {longitudinally polarised} $ \mesonzp,\,\mesonzn,\,\mesonpp,\,\mesonmn$ is shown as a function of $  \left( -u' \right)  $ on the top left, top right, bottom left and bottom right plots respectively for different values of $ M_{\gamma \meson}^2 $. The black, red and blue curves correspond to $ M_{\gamma \meson}^{2}=3,\,4,\,5\, $ GeV$ ^2 $ respectively. The dashed (non-dashed) lines correspond to holographic (asymptotic) DA, while the dotted (non-dotted) lines correspond to the standard (valence) scenario. As mentioned in the main text, $ S_{\gamma N} $ is fixed at 200 GeV$ ^2 $ here.}
	\label{fig:compass-fully-diff-diff-M2}
\end{figure}

\begin{figure}[t!]
	\psfrag{HHH}{\hspace{-1.5cm}\raisebox{-.6cm}{\scalebox{.8}{$-u' ({\rm 
					GeV}^{2})$}}}
	\psfrag{VVV}{\raisebox{.3cm}{\scalebox{.9}{$\hspace{-.4cm}\displaystyle\left.\frac{d 
					\sigma^{\mathrm{odd}}_{\gamma\mesonzp}}{d M^2_{\gamma\mesonzp} d(-u') d(-t)}\right|_{(-t)_{\rm min}}({\rm pb} \cdot {\rm GeV}^{-6})$}}}
	\psfrag{TTT}{}
	{
		{\includegraphics[width=18pc]{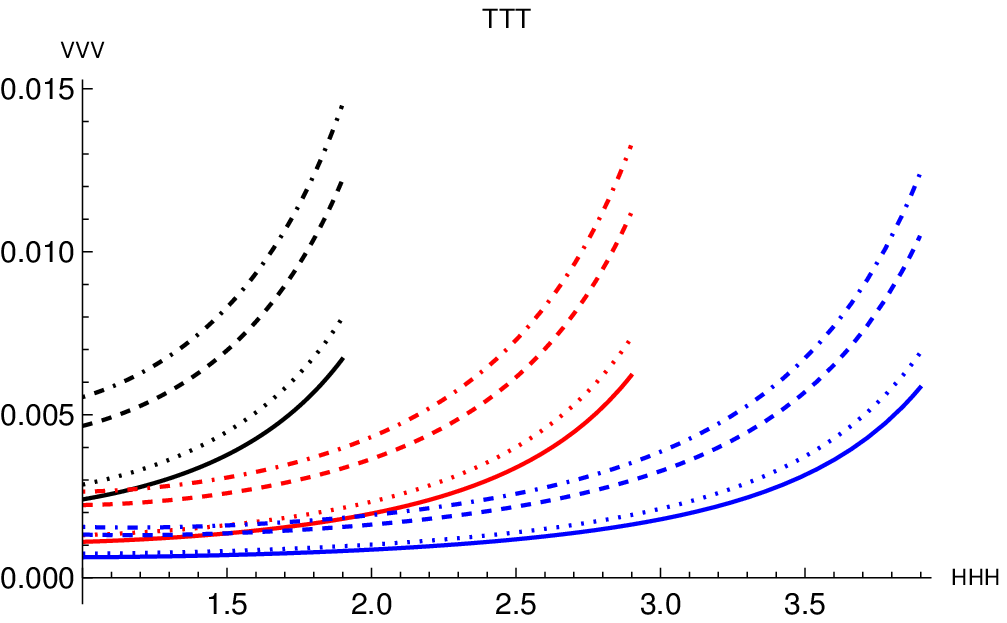}}
		\psfrag{VVV}{\raisebox{.3cm}{\scalebox{.9}{$\hspace{-.4cm}\displaystyle\left.\frac{d 
						\sigma^{\mathrm{odd}}_{\gamma\mesonzn}}{d M^2_{\gamma \mesonzn} d(-u') d(-t)}\right|_{(-t)_{\rm min}}({\rm pb} \cdot {\rm GeV}^{-6})$}}}
		{\includegraphics[width=18pc]{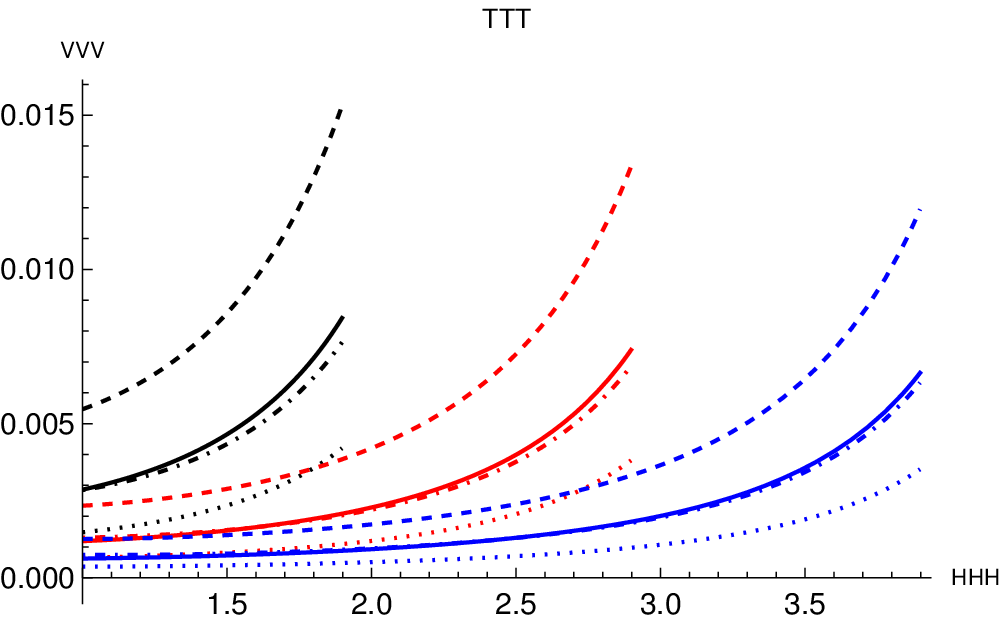}}}
	\\[25pt]
	\psfrag{VVV}{\raisebox{.3cm}{\scalebox{.9}{$\hspace{-.4cm}\displaystyle\left.\frac{d 
					\sigma^{\mathrm{odd}}_{\gamma\mesonpp}}{d M^2_{\gamma\mesonpp} d(-u') d(-t)}\right|_{(-t)_{\rm min}}({\rm pb} \cdot {\rm GeV}^{-6})$}}}
	\psfrag{TTT}{}
	{
		{\includegraphics[width=18pc]{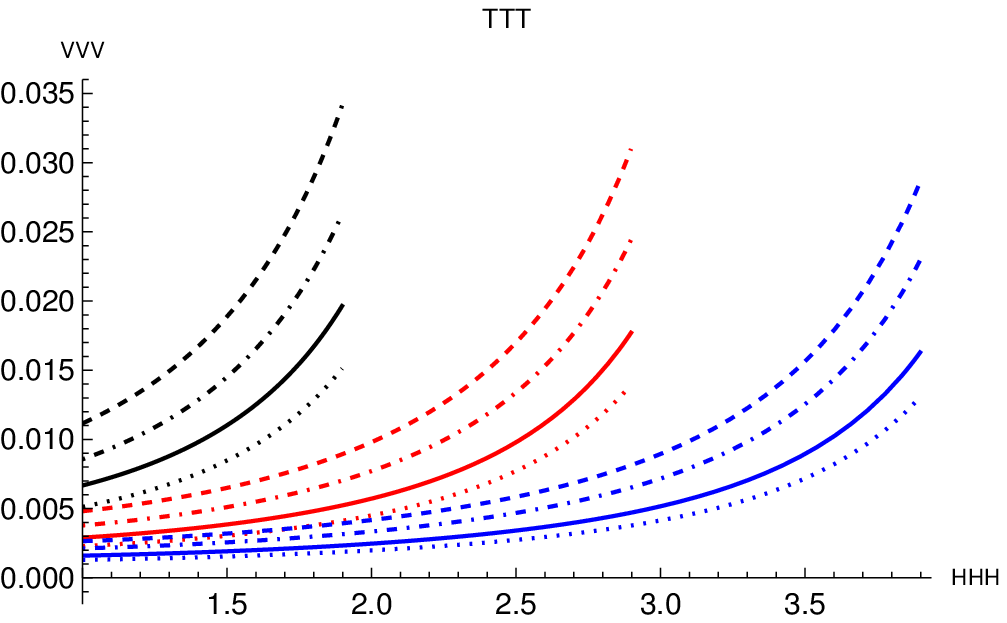}}
		\psfrag{VVV}{\raisebox{.3cm}{\scalebox{.9}{$\hspace{-.4cm}\displaystyle\left.\frac{d 
						\sigma^{\mathrm{odd}}_{\gamma\mesonmn}}{d M^2_{\gamma\mesonmn} d(-u') d(-t)}\right|_{(-t)_{\rm min}}({\rm pb} \cdot {\rm GeV}^{-6})$}}}
		{\includegraphics[width=18pc]{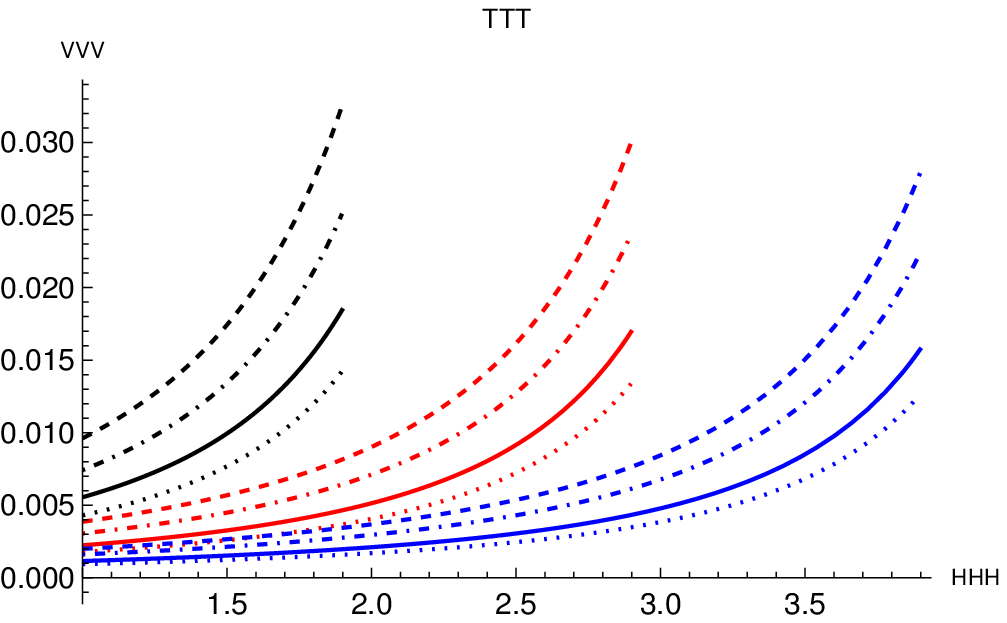}}}
	\vspace{0.2cm}
	\caption{\small The fully differential cross section for {transversely polarised} $ \mesonzp,\,\mesonzn,\,\mesonpp,\,\mesonmn$ is shown as a function of $  \left( -u' \right)  $ on the top left, top right, bottom left and bottom right plots respectively for different values of $ M_{\gamma \meson}^2 $. The black, red and blue curves correspond to $ M_{\gamma \meson}^{2}=3,\,4,\,5\, $ GeV$ ^2 $ respectively. The dashed (non-dashed) lines correspond to holographic (asymptotic) DA, while the dotted (non-dotted) lines correspond to the standard (valence) scenario. As mentioned in the main text, $ S_{\gamma N} $ is fixed at 200 GeV$ ^2 $ here.}
	\label{fig:co-compass-fully-diff-diff-M2}
\end{figure}

\FIG\ref{fig:compass-fully-diff-diff-M2} shows the effect of different value of $ \Msq $ on the fully-differential cross section for the chiral-even case. We choose 3 different values for $ \Msq $, namely $ \Msq = 3,\,4,\,5 \GeV^2 $. Compared to the corresponding plots at $ \SgN = 20 \GeV^2 $ in \FIG\ref{fig:jlab-fully-diff-diff-M2}, the cross sections here are smaller by a factor of 8 roughly. We note that the uncertainty due to the model used is significant for the charged $ \meson $-meson case, and is particularly driven by the GPD model. This allows in principle to discriminate between the 2 GPD models that are investigated.

The variation of the differential cross section with $  \left( -u' \right)  $ for the chiral-odd case is shown in \FIG\ref{fig:co-compass-fully-diff-diff-M2}. Here, we note that the cross section is much smaller than the chiral-even case, a feature which is more pronounced than in the JLab kinematics case. This can be attributed to the fact that the amplitude squared is proportional to $ \xi^2 $ (see \eqref{all-rhoT}), and $ \xi $ becomes smaller at higher centre-of-mass energies $ \SgN $. On the other hand, this $ \xi^2  $ factor is absent for the chiral-even case, see \eqref{squareCEresultsmallxi}.

\begin{figure}[t!]
	\psfrag{HHH}{\hspace{-1.5cm}\raisebox{-.6cm}{\scalebox{.8}{$-u' ({\rm 
					GeV}^{2})$}}}
	\psfrag{VVV}{\raisebox{.3cm}{\scalebox{.9}{$\hspace{-.4cm}\displaystyle\left.\frac{d 
					\sigma^{\mathrm{even}}_{\gamma\mesonzp}}{d M^2_{\gamma\mesonzp} d(-u') d(-t)}\right|_{(-t)_{\rm min}}({\rm pb} \cdot {\rm GeV}^{-6})$}}}
	\psfrag{TTT}{}
	{
		{\includegraphics[width=18pc]{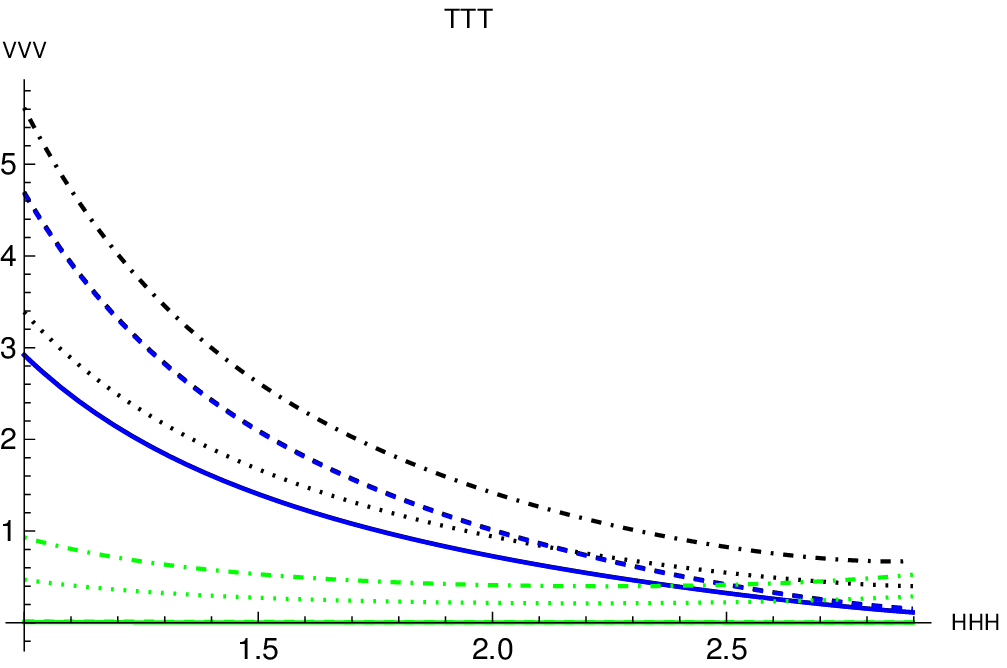}}
		\psfrag{VVV}{\raisebox{.3cm}{\scalebox{.9}{$\hspace{-.4cm}\displaystyle\left.\frac{d 
						\sigma^{\mathrm{even}}_{\gamma\mesonzn}}{d M^2_{\gamma \mesonzn} d(-u') d(-t)}\right|_{(-t)_{\rm min}}({\rm pb} \cdot {\rm GeV}^{-6})$}}}
		{\includegraphics[width=18pc]{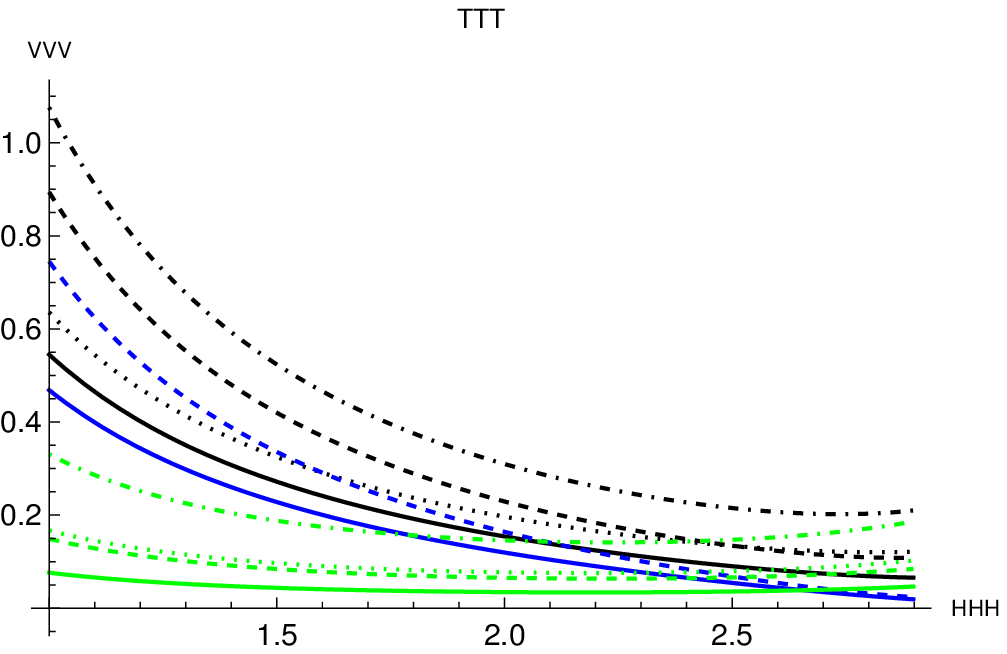}}}
	\\[25pt]
	\psfrag{VVV}{\raisebox{.3cm}{\scalebox{.9}{$\hspace{-.4cm}\displaystyle\left.\frac{d 
					\sigma^{\mathrm{even}}_{\gamma\mesonpp}}{d M^2_{\gamma\mesonpp} d(-u') d(-t)}\right|_{(-t)_{\rm min}}({\rm pb} \cdot {\rm GeV}^{-6})$}}}
	\psfrag{TTT}{}
	{
		{\includegraphics[width=18pc]{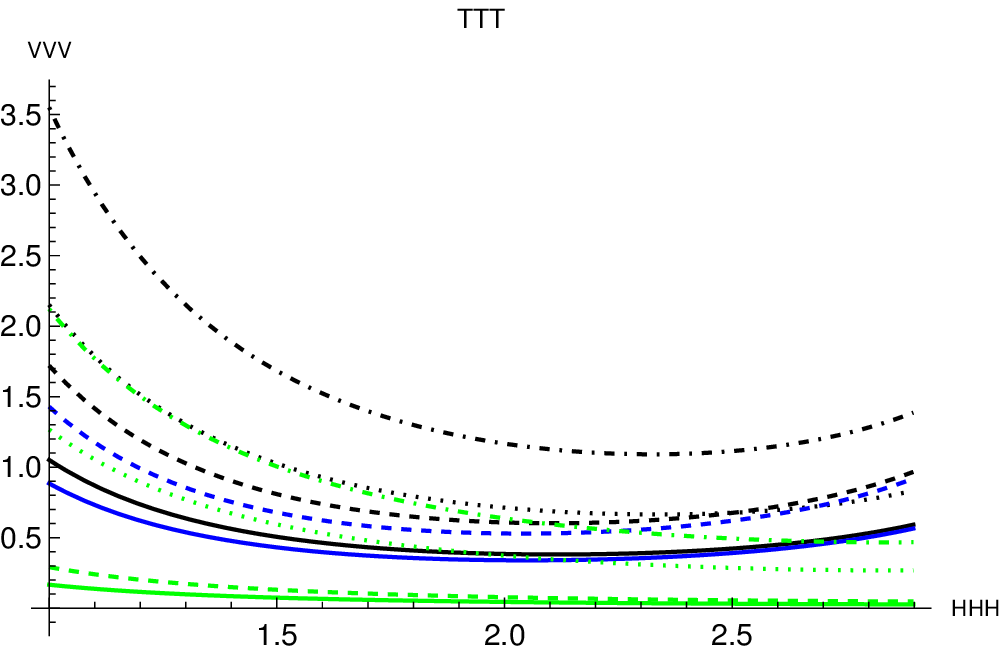}}
		\psfrag{VVV}{\raisebox{.3cm}{\scalebox{.9}{$\hspace{-.4cm}\displaystyle\left.\frac{d 
						\sigma^{\mathrm{even}}_{\gamma\mesonmn}}{d M^2_{\gamma\mesonmn} d(-u') d(-t)}\right|_{(-t)_{\rm min}}({\rm pb} \cdot {\rm GeV}^{-6})$}}}
		{\includegraphics[width=18pc]{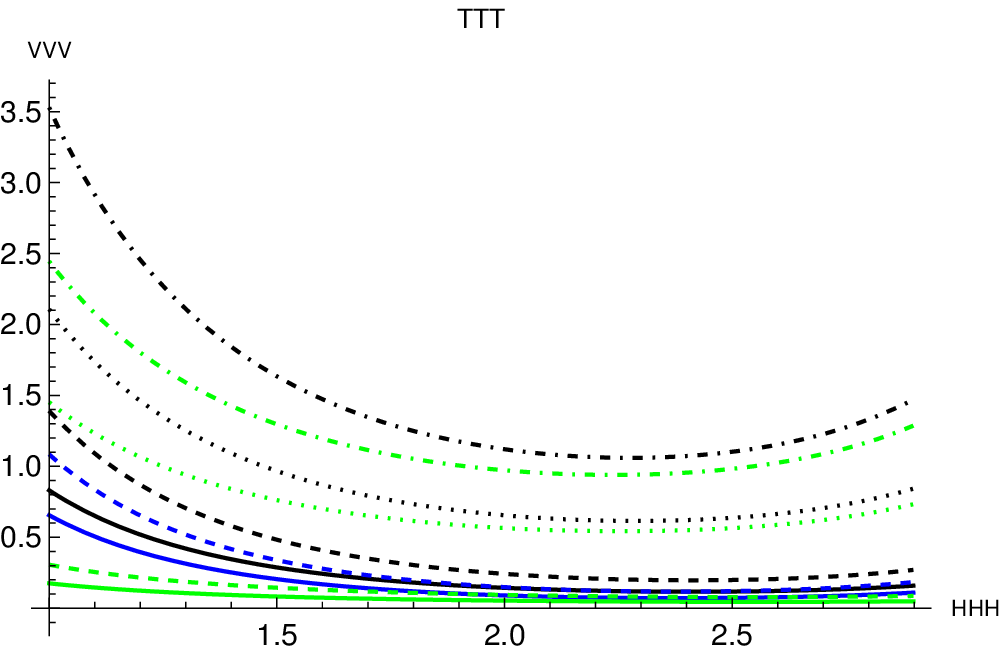}}}
	\vspace{0.2cm}
	\caption{\small The fully differential cross section for {longitudinally polarised} $ \mesonzp,\,\mesonzn,\,\mesonpp,\,\mesonmn$ is shown as a function of $  \left( -u' \right)  $ on the top left, top right, bottom left and bottom right plots respectively. The blue and green curves correspond to contributions from the vector and axial GPDs respectively. The black curves correspond to the total contribution, \ie{}vector and axial GPD contributions combined. As before, the dashed (non-dashed) lines correspond to holographic (asymptotic) DA, while the dotted (non-dotted) lines correspond to the standard (valence) scenario. We fix $ S_{\gamma N}= 200\,  \mathrm{GeV}^{2}  $ and $ M_{\gamma \meson}^{2}= 4\,  \mathrm{GeV}^{2}  $. Note that the vector contributions consist of only two curves in each case, since they are insensitive to either valence or standard scenarios.}
	\label{fig:compass-fully-diff-VandA}
\end{figure}

The relative contributions of the vector and axial GPDs to the cross section for the longitudinally-polarised $ \meson $-meson case are shown in \FIG\ref{fig:compass-fully-diff-VandA}. Similar comments as in \SEC\ref{sec:jlab-fully-diff-X-section} apply.

\begin{figure}[t!]
	\psfrag{HHH}{\hspace{-1.5cm}\raisebox{-.6cm}{\scalebox{.8}{$-u' ({\rm 
					GeV}^{2})$}}}
	\psfrag{VVV}{\raisebox{.3cm}{\scalebox{.9}{$\hspace{-.4cm}\displaystyle\left.\frac{d 
					\sigma^{\mathrm{even}}_{\gamma\mesonzp}}{d M^2_{\gamma\mesonzp} d(-u') d(-t)}\right|_{(-t)_{\rm min}}({\rm pb} \cdot {\rm GeV}^{-6})$}}}
	\psfrag{TTT}{}
	{
		{\includegraphics[width=18pc]{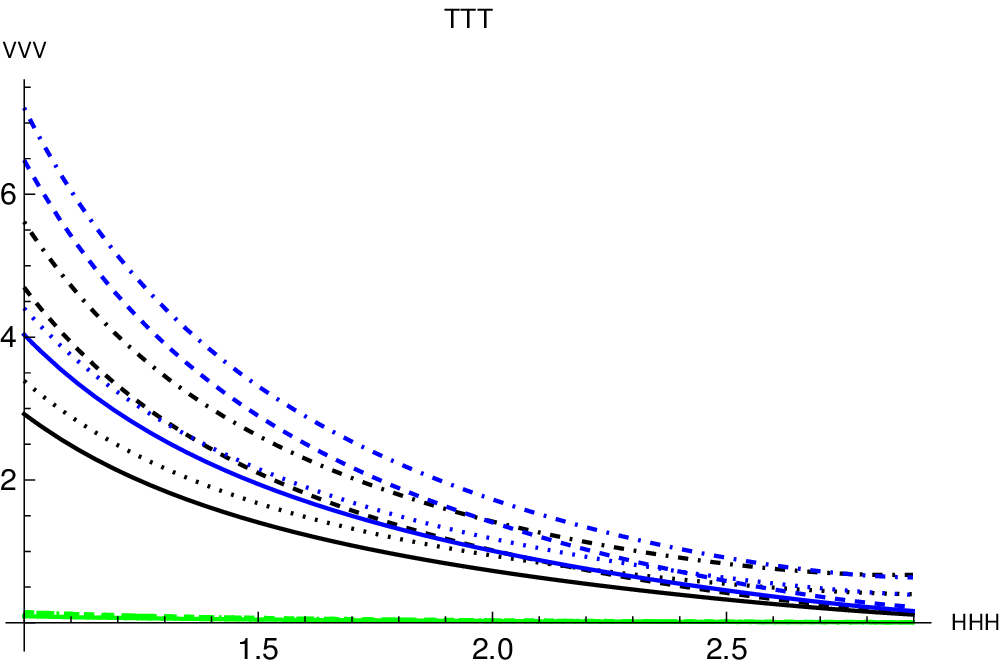}}
		\psfrag{VVV}{\raisebox{.3cm}{\scalebox{.9}{$\hspace{-.4cm}\displaystyle\left.\frac{d 
						\sigma^{\mathrm{even}}_{\gamma\mesonzn}}{d M^2_{\gamma \mesonzn} d(-u') d(-t)}\right|_{(-t)_{\rm min}}({\rm pb} \cdot {\rm GeV}^{-6})$}}}
		{\includegraphics[width=18pc]{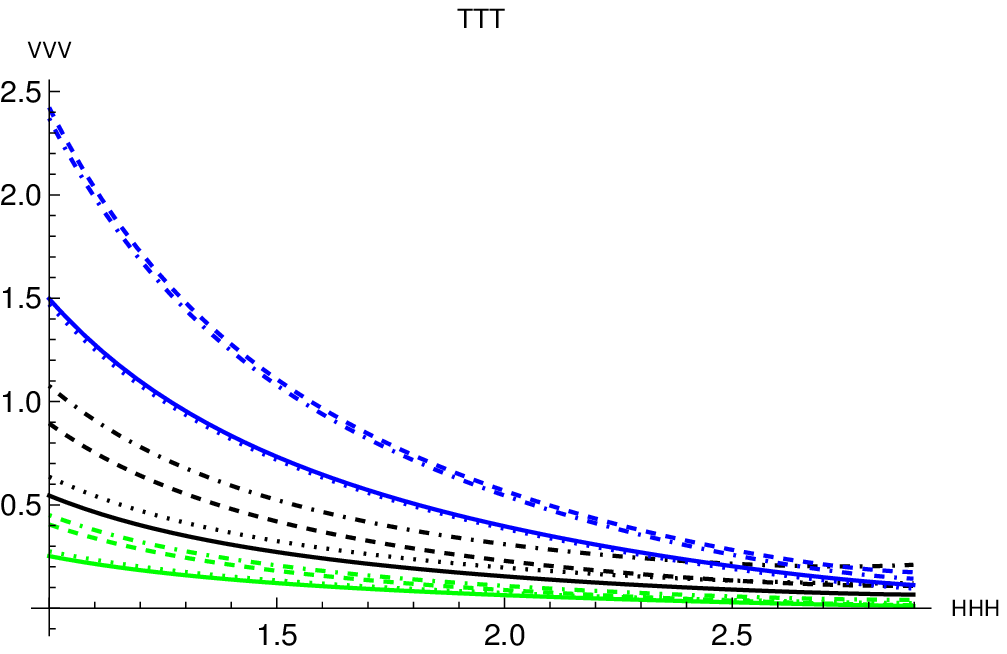}}}
	\\[25pt]
	\psfrag{VVV}{\raisebox{.3cm}{\scalebox{.9}{$\hspace{-.4cm}\displaystyle\left.\frac{d 
					\sigma^{\mathrm{even}}_{\gamma\mesonpp}}{d M^2_{\gamma\mesonpp} d(-u') d(-t)}\right|_{(-t)_{\rm min}}({\rm pb} \cdot {\rm GeV}^{-6})$}}}
	\psfrag{TTT}{}
	{
		{\includegraphics[width=18pc]{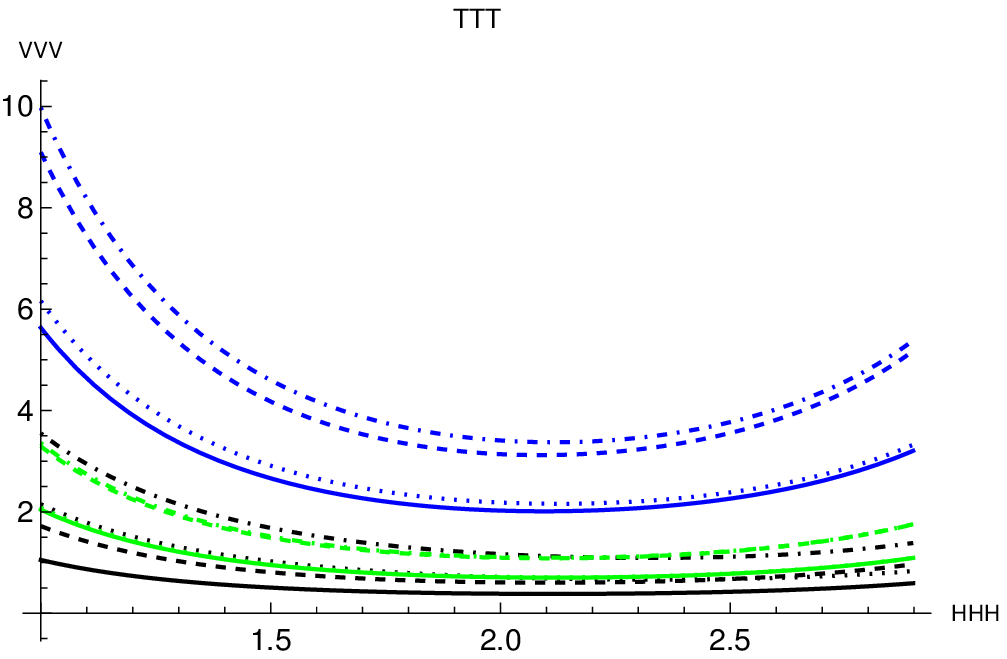}}
		\psfrag{VVV}{\raisebox{.3cm}{\scalebox{.9}{$\hspace{-.4cm}\displaystyle\left.\frac{d 
						\sigma^{\mathrm{even}}_{\gamma\mesonmn}}{d M^2_{\gamma\mesonmn} d(-u') d(-t)}\right|_{(-t)_{\rm min}}({\rm pb} \cdot {\rm GeV}^{-6})$}}}
		{\includegraphics[width=18pc]{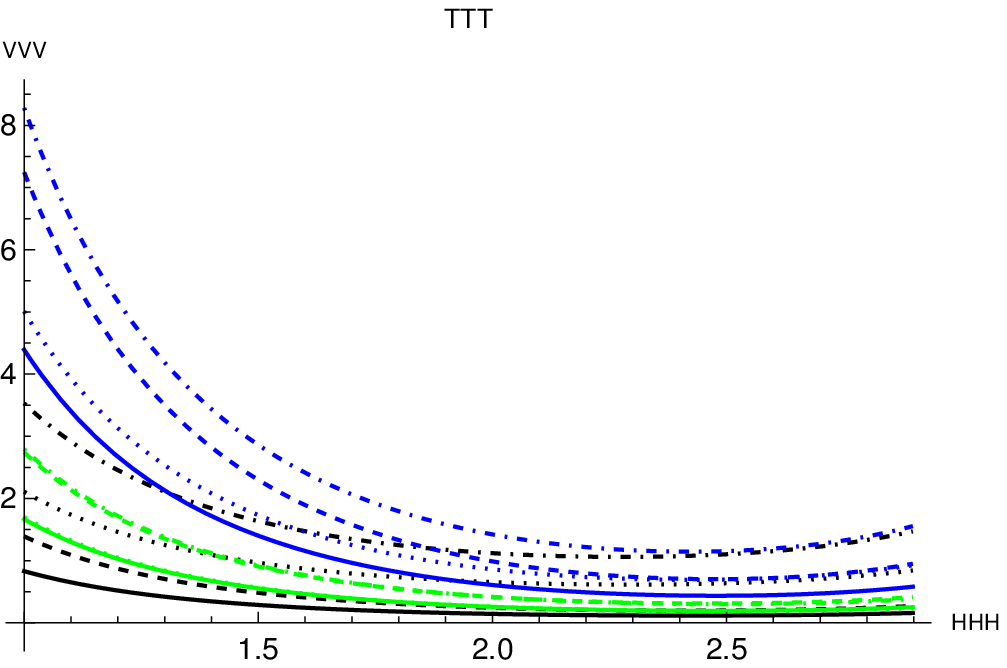}}}
	\vspace{0.2cm}
	\caption{\small The fully differential cross section for {longitudinally polarised} $ \mesonzp,\,\mesonzn,\,\mesonpp,\,\mesonmn$ is shown as a function of $  \left( -u' \right)  $ on the top left, top right, bottom left and bottom right plots respectively. The blue and green curves correspond to contributions from the $u$-quark ($ H_{u} $ and $  \tilde{H} _{u} $) and $d$-quark ($ H_{d} $ and $  \tilde{H} _{d} $) GPDs respectively. The black curves correspond to the total contribution. Otherwise, conventions are the same as in previous plots. We fix $ S_{\gamma N}= 200\,  \mathrm{GeV}^{2}  $ and $ M_{\gamma \meson}^{2}= 4\,  \mathrm{GeV}^{2}  $.}
	\label{fig:compass-fully-diff-uandd}
\end{figure}

\begin{figure}[t!]
	\psfrag{HHH}{\hspace{-1.5cm}\raisebox{-.6cm}{\scalebox{.8}{$-u' ({\rm 
					GeV}^{2})$}}}
	\psfrag{VVV}{\raisebox{.3cm}{\scalebox{.9}{$\hspace{-.4cm}\displaystyle\left.\frac{d 
					\sigma^{\mathrm{odd}}_{\gamma\mesonzp}}{d M^2_{\gamma\mesonzp} d(-u') d(-t)}\right|_{(-t)_{\rm min}}({\rm pb} \cdot {\rm GeV}^{-6})$}}}
	\psfrag{TTT}{}
	{
		{\includegraphics[width=18pc]{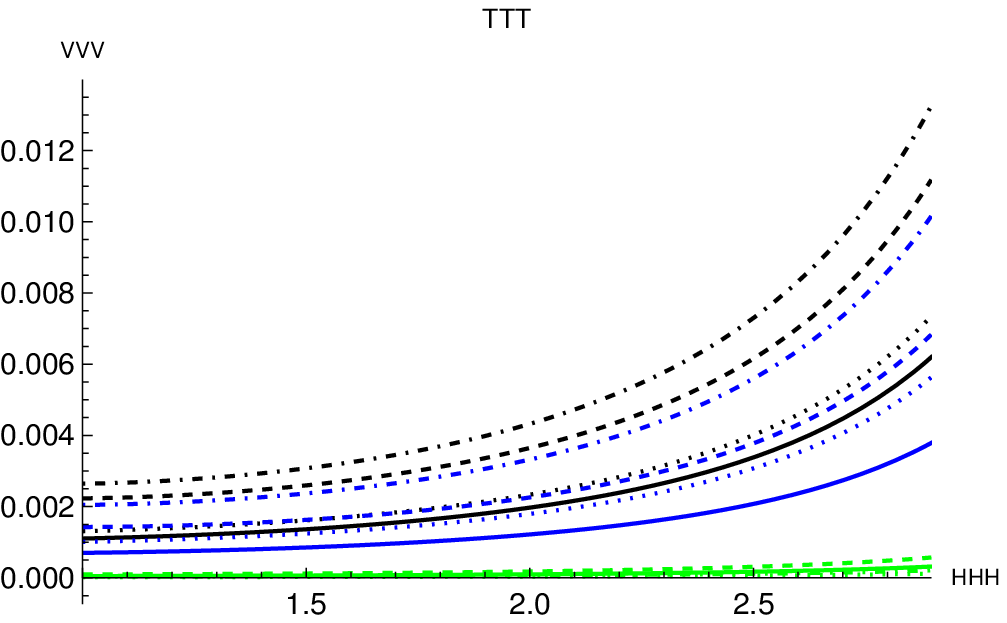}}
		\psfrag{VVV}{\raisebox{.3cm}{\scalebox{.9}{$\hspace{-.4cm}\displaystyle\left.\frac{d 
						\sigma^{\mathrm{odd}}_{\gamma\mesonzn}}{d M^2_{\gamma \mesonzn} d(-u') d(-t)}\right|_{(-t)_{\rm min}}({\rm pb} \cdot {\rm GeV}^{-6})$}}}
		{\includegraphics[width=18pc]{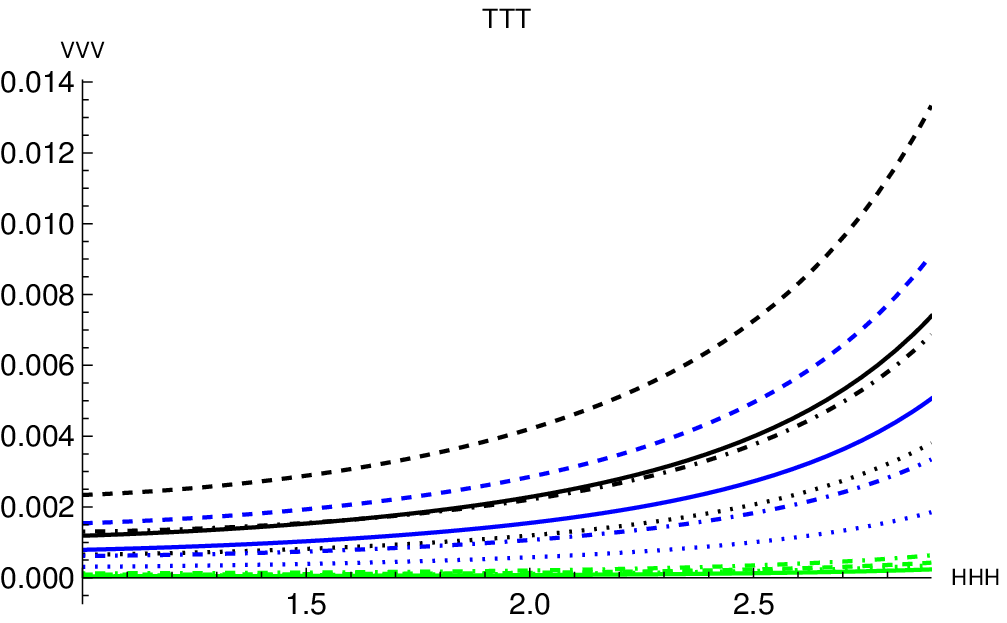}}}
	\\[25pt]
	\psfrag{VVV}{\raisebox{.3cm}{\scalebox{.9}{$\hspace{-.4cm}\displaystyle\left.\frac{d 
					\sigma^{\mathrm{odd}}_{\gamma\mesonpp}}{d M^2_{\gamma\mesonpp} d(-u') d(-t)}\right|_{(-t)_{\rm min}}({\rm pb} \cdot {\rm GeV}^{-6})$}}}
	\psfrag{TTT}{}
	{
		{\includegraphics[width=18pc]{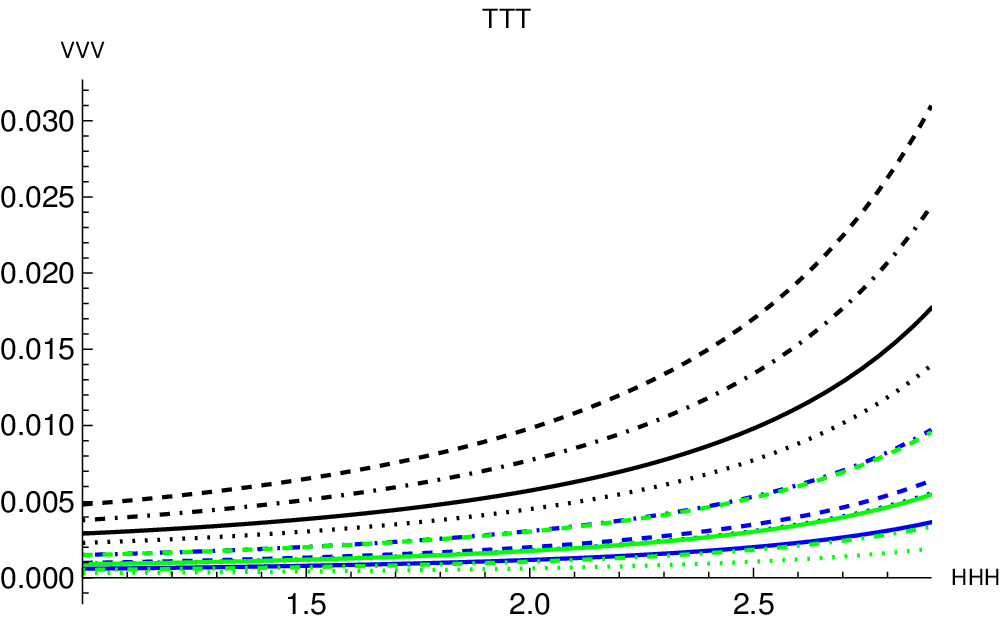}}
		\psfrag{VVV}{\raisebox{.3cm}{\scalebox{.9}{$\hspace{-.4cm}\displaystyle\left.\frac{d 
						\sigma^{\mathrm{odd}}_{\gamma\mesonmn}}{d M^2_{\gamma\mesonmn} d(-u') d(-t)}\right|_{(-t)_{\rm min}}({\rm pb} \cdot {\rm GeV}^{-6})$}}}
		{\includegraphics[width=18pc]{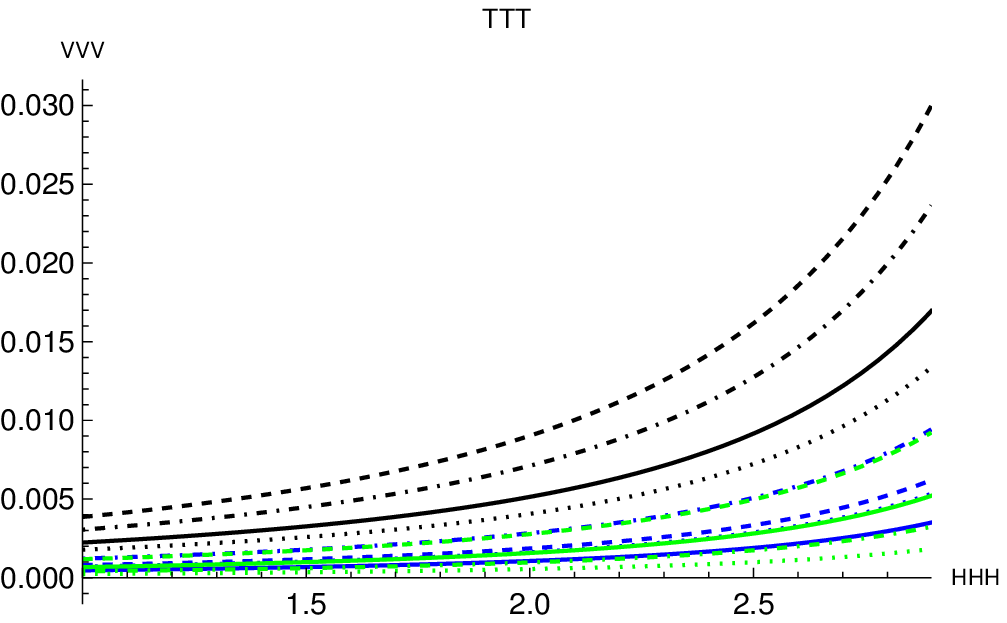}}}
	\vspace{0.2cm}
	\caption{\small The fully differential cross section for {transversely polarised} $ \mesonzp,\,\mesonzn,\,\mesonpp,\,\mesonmn$ is shown as a function of $  \left( -u' \right)  $ on the top left, top right, bottom left and bottom right plots respectively. The blue and green curves correspond to contributions from the $u$-quark ($ H_{u} $ and $  \tilde{H} _{u} $) and $d$-quark ($ H_{d} $ and $  \tilde{H} _{d} $) GPDs respectively. The black curves correspond to the total contribution. Otherwise, conventions are the same as in previous plots. We fix $ S_{\gamma N}= 200\,  \mathrm{GeV}^{2}  $ and $ M_{\gamma \meson}^{2}= 4\,  \mathrm{GeV}^{2}  $.}
	\label{fig:co-compass-fully-diff-uandd}
\end{figure}

Finally, to conclude this subsection, the relative contributions of the $ u $-quark and $ d $-quark GPDs to the cross section are shown in \FIG\ref{fig:compass-fully-diff-uandd} for the chiral-even case, and in \FIG\ref{fig:co-compass-fully-diff-uandd} for the chiral-odd case.

\FloatBarrier

\subsubsection{Single differential cross section}

\begin{figure}[t!]
	\psfrag{HHH}{\hspace{-1.5cm}\raisebox{-.6cm}{\scalebox{.8}{$\Msq ({\rm 
					GeV}^{2})$}}}
	\psfrag{VVV}{\raisebox{.3cm}{\scalebox{.9}{$\hspace{-.4cm}\displaystyle\frac{d 
					\sigma^{\mathrm{even}}_{\gamma\mesonzp}}{d M^2_{\gamma\mesonzp}}({\rm pb} \cdot {\rm GeV}^{-2})$}}}
	\psfrag{TTT}{}
	{
		{\includegraphics[width=18pc]{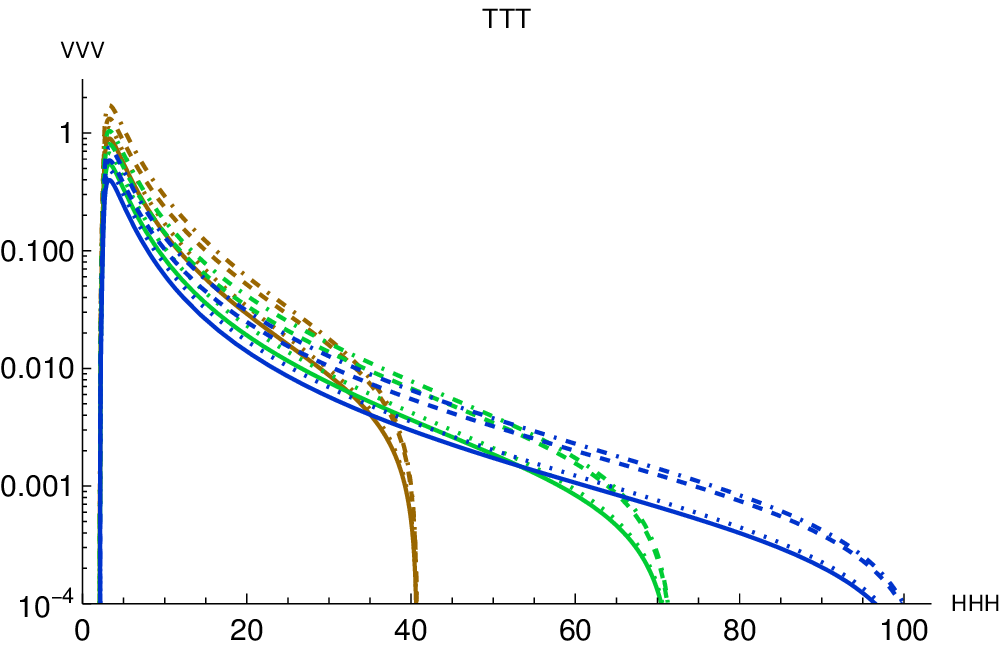}}
		\psfrag{VVV}{\raisebox{.3cm}{\scalebox{.9}{$\hspace{-.4cm}\displaystyle\frac{d 
						\sigma^{\mathrm{even}}_{\gamma\mesonzn}}{d M^2_{\gamma \mesonzn}}({\rm pb} \cdot {\rm GeV}^{-2})$}}}
		{\includegraphics[width=18pc]{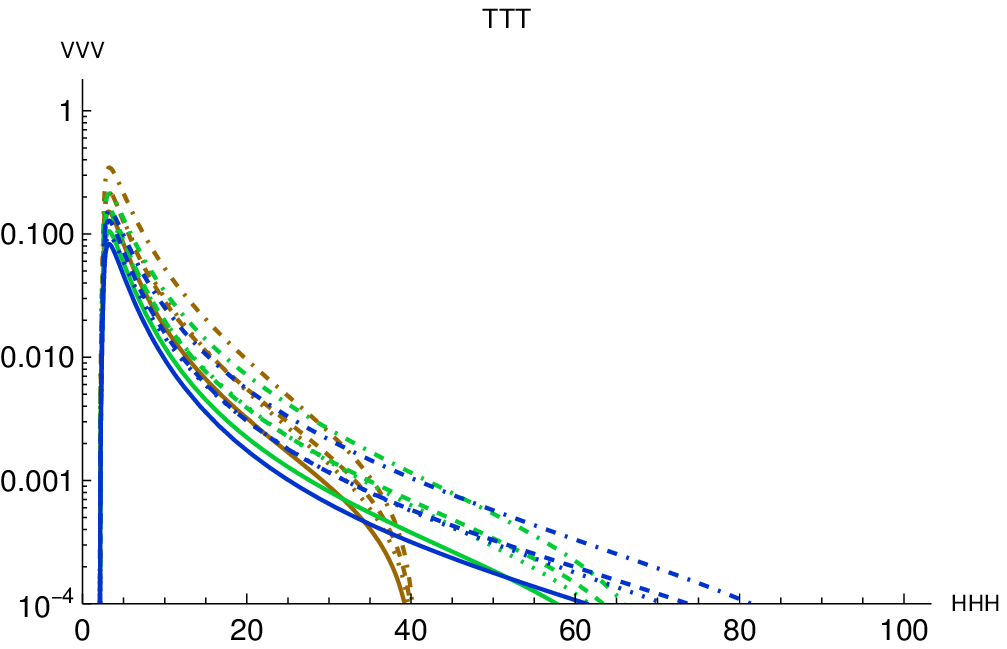}}}
	\\[25pt]
	\psfrag{VVV}{\raisebox{.3cm}{\scalebox{.9}{$\hspace{-.4cm}\displaystyle\frac{d 
					\sigma^{\mathrm{even}}_{\gamma\mesonpp}}{d M^2_{\gamma\mesonpp}}({\rm pb} \cdot {\rm GeV}^{-2})$}}}
	\psfrag{TTT}{}
	{
		{\includegraphics[width=18pc]{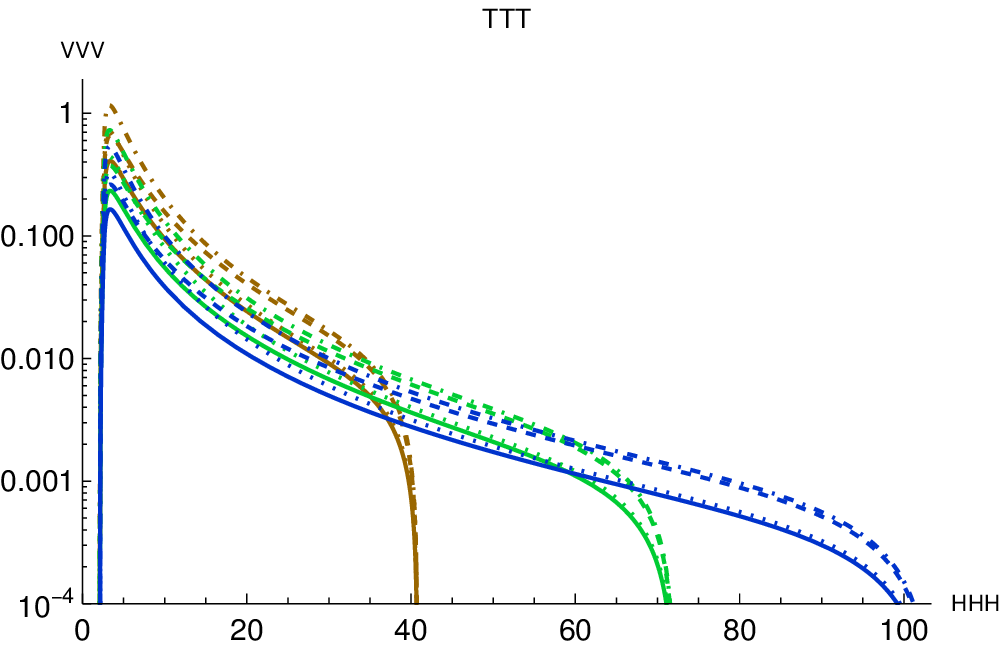}}
		\psfrag{VVV}{\raisebox{.3cm}{\scalebox{.9}{$\hspace{-.4cm}\displaystyle\frac{d 
						\sigma^{\mathrm{even}}_{\gamma\mesonmn}}{d M^2_{\gamma\mesonmn}}({\rm pb} \cdot {\rm GeV}^{-2})$}}}
		{\includegraphics[width=18pc]{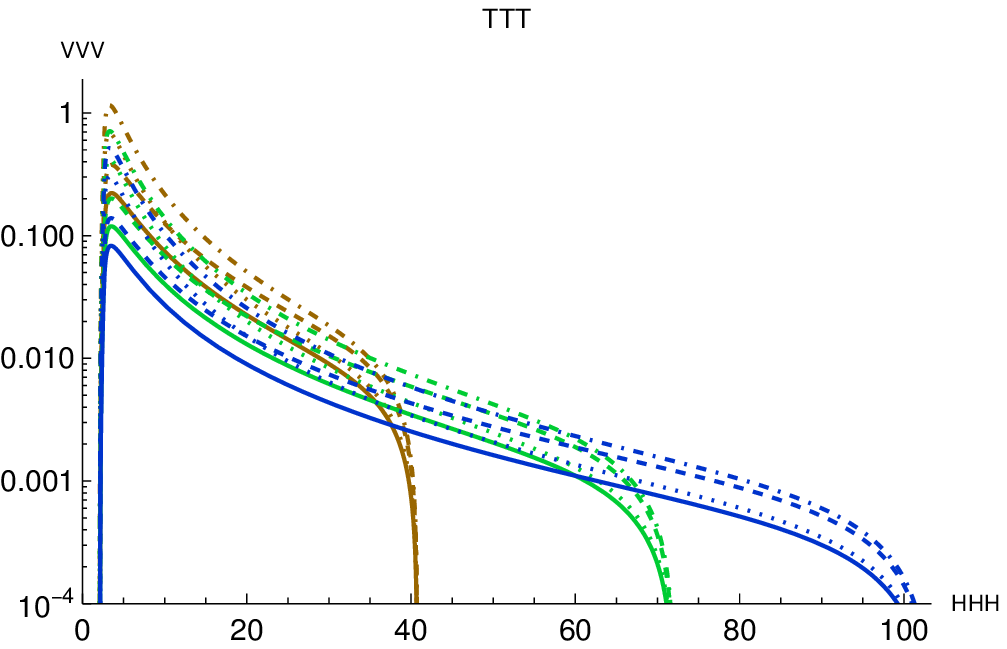}}}
	\vspace{0.2cm}
	\caption{\small The single differential cross section for {longitudinally polarised} $ \mesonzp,\,\mesonzn,\,\mesonpp,\,\mesonmn$ is shown as a function of $  M_{\gamma \meson}^{2}  $ on the top left, top right, bottom left and bottom right plots respectively for different values of $ S_{\gamma N} $. The brown, green and blue curves correspond to $ S_{\gamma N} = 80,\,140,\,200\,\GeV^{2} $. The dashed (non-dashed) lines correspond to holographic (asymptotic) DA, while the dotted (non-dotted) lines correspond to the standard (valence) scenario. The holographic DA with the standard scenario has the largest contribution for every $ \SgN $.}
	\label{fig:compass-sing-diff}
\end{figure}

\begin{figure}[t!]
	\psfrag{HHH}{\hspace{-1.5cm}\raisebox{-.6cm}{\scalebox{.8}{$\Msq ({\rm 
					GeV}^{2})$}}}
	\psfrag{VVV}{\raisebox{.3cm}{\scalebox{.9}{$\hspace{-.4cm}\displaystyle\frac{d 
					\sigma^{\mathrm{odd}}_{\gamma\mesonzp}}{d M^2_{\gamma\mesonzp}}({\rm pb} \cdot {\rm GeV}^{-2})$}}}
	\psfrag{TTT}{}
	{
		{\includegraphics[width=18pc]{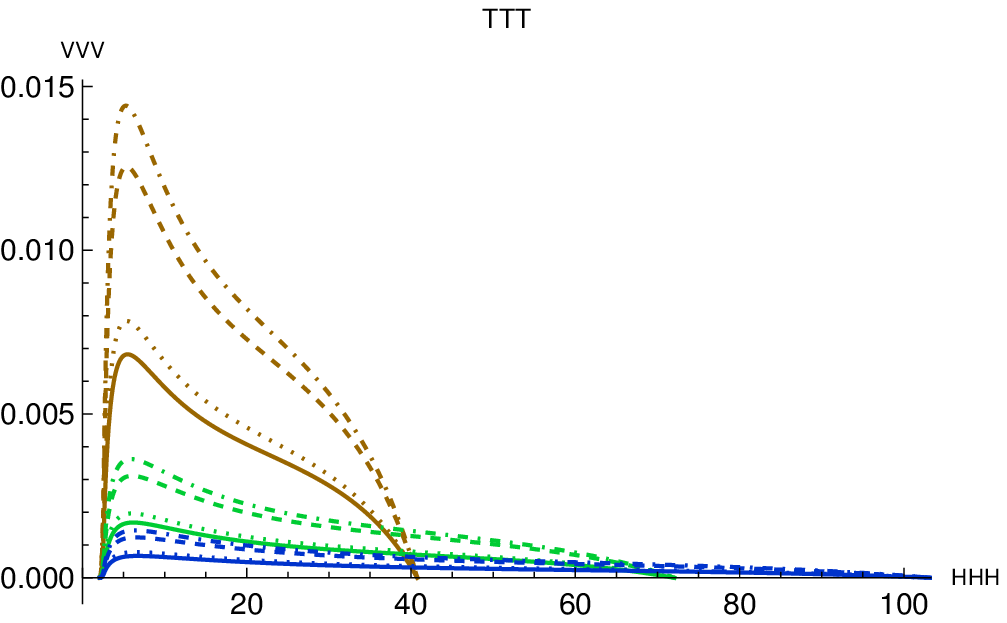}}
		\psfrag{VVV}{\raisebox{.3cm}{\scalebox{.9}{$\hspace{-.4cm}\displaystyle\frac{d 
						\sigma^{\mathrm{odd}}_{\gamma\mesonzn}}{d M^2_{\gamma \mesonzn}}({\rm pb} \cdot {\rm GeV}^{-2})$}}}
		{\includegraphics[width=18pc]{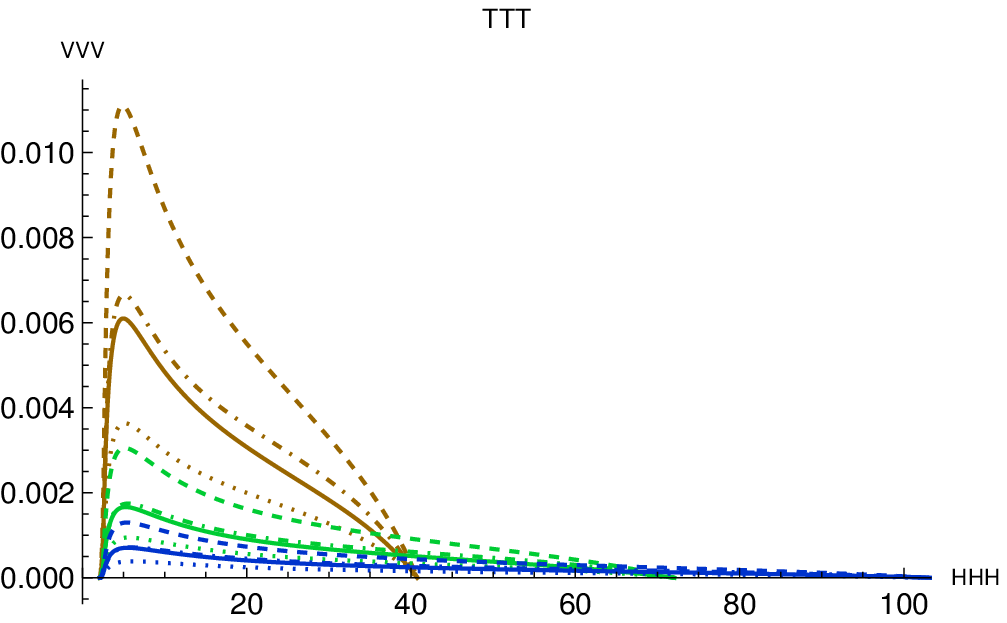}}}
	\\[25pt]
	\psfrag{VVV}{\raisebox{.3cm}{\scalebox{.9}{$\hspace{-.4cm}\displaystyle\frac{d 
					\sigma^{\mathrm{odd}}_{\gamma\mesonpp}}{d M^2_{\gamma\mesonpp}}({\rm pb} \cdot {\rm GeV}^{-2})$}}}
	\psfrag{TTT}{}
	{
		{\includegraphics[width=18pc]{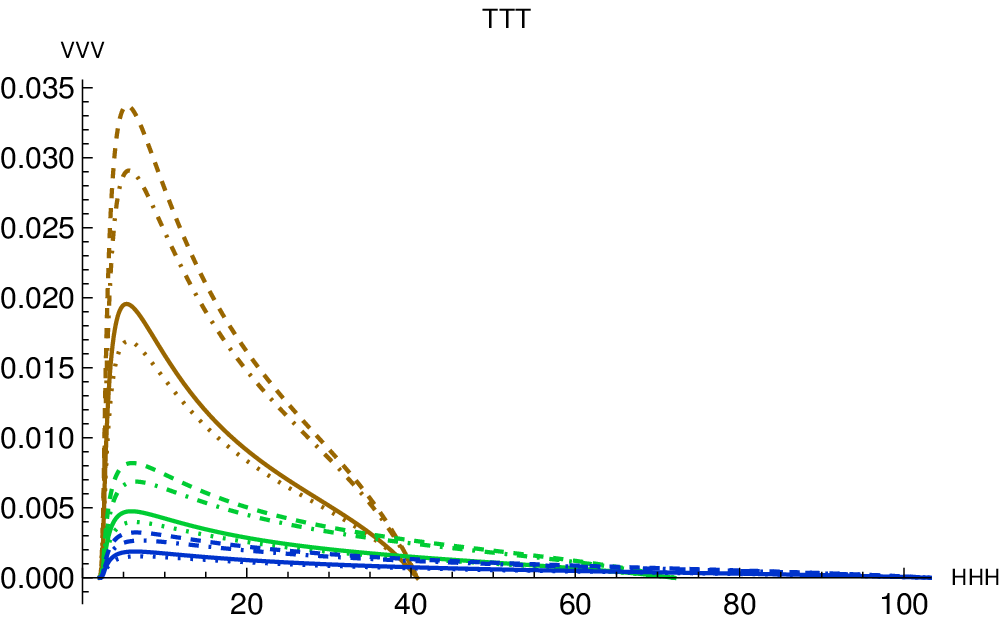}}
		\psfrag{VVV}{\raisebox{.3cm}{\scalebox{.9}{$\hspace{-.4cm}\displaystyle\frac{d 
						\sigma^{\mathrm{odd}}_{\gamma\mesonmn}}{d M^2_{\gamma\mesonmn}}({\rm pb} \cdot {\rm GeV}^{-2})$}}}
		{\includegraphics[width=18pc]{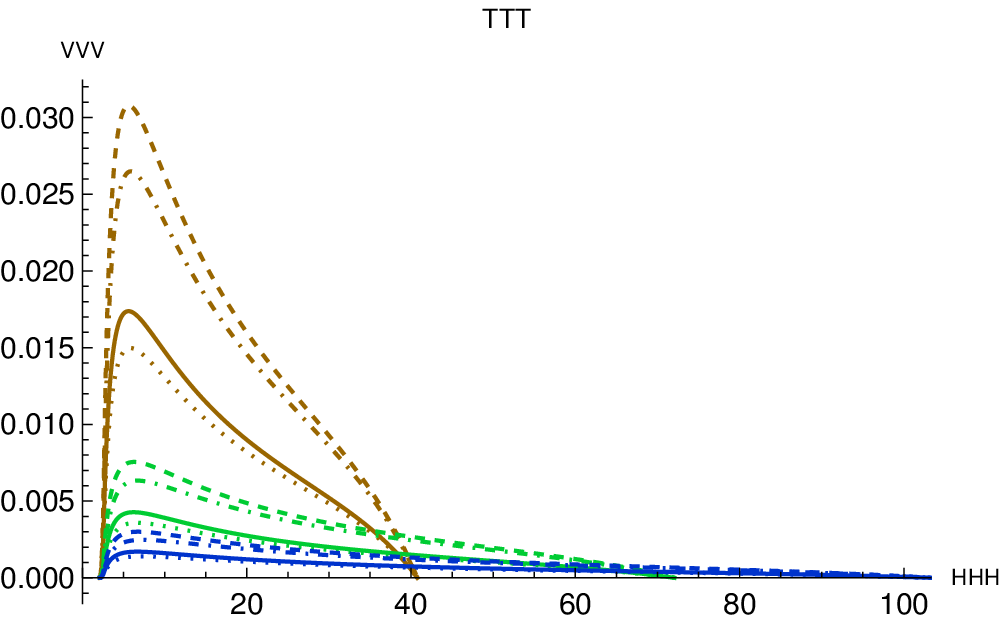}}}
	\vspace{0.2cm}
	\caption{\small The single differential cross section for {transversely polarised} $ \mesonzp,\,\mesonzn,\,\mesonpp,\,\mesonmn$ is shown as a function of $  M_{\gamma \meson}^{2}  $ on the top left, top right, bottom left and bottom right plots respectively for different values of $ S_{\gamma N} $. The brown, green and blue curves correspond to $ S_{\gamma N} = 80,\,140,\,200\,\GeV^{2} $. The dashed (non-dashed) lines correspond to holographic (asymptotic) DA, while the dotted (non-dotted) lines correspond to the standard (valence) scenario.}
	\label{fig:co-compass-sing-diff}
\end{figure}

\FIG\ref{fig:compass-sing-diff} shows the variation of the single differential cross section with $ M_{\gamma \meson}^2 $ for different values of $ S_{\gamma N} $ for the chiral-even case. We choose 3 different values for $ \SgN $, namely 80, 140 and 200 GeV$ ^2 $. Due to large variations in the cross section over the full range of $ \Msq $, a log scale is used for the vertical axis.  We observe that the cross section is dominated by the region of very small $ \Msq $.

For the chiral-odd case, shown in \FIG\ref{fig:co-compass-sing-diff}, we first note that the cross section is smaller wrt the chiral-even case, by a factor of 100 roughly. In particular, the height of the peak for the chiral-odd case is much lower than the chiral-even case. Again, this is related to the $ \xi^2 $ suppression factor that comes from the square of the chiral-odd amplitude.

\FloatBarrier

\subsubsection{Integrated cross section}

\begin{figure}[t!]
	\psfrag{HHH}{\hspace{-1.5cm}\raisebox{-.6cm}{\scalebox{.8}{$ S_{\gamma N} ({\rm 
					GeV}^{2})$}}}
	\psfrag{VVV}{\raisebox{.3cm}{\scalebox{.9}{$\hspace{-.4cm}\displaystyle
				\sigma^{\mathrm{even}}_{\gamma\mesonzp}({\rm pb})$}}}
	\psfrag{TTT}{}
	{
		{\includegraphics[width=18pc]{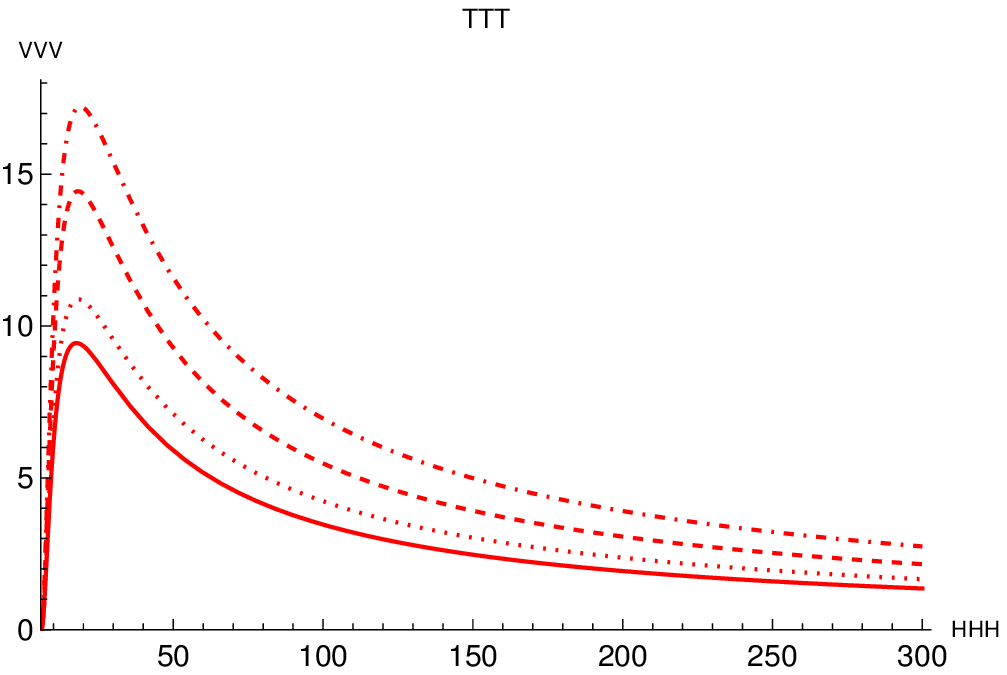}}
		\psfrag{VVV}{\raisebox{.3cm}{\scalebox{.9}{$\hspace{-.4cm}\displaystyle
					\sigma^{\mathrm{even}}_{\gamma\mesonzn}({\rm pb})$}}}
		{\includegraphics[width=18pc]{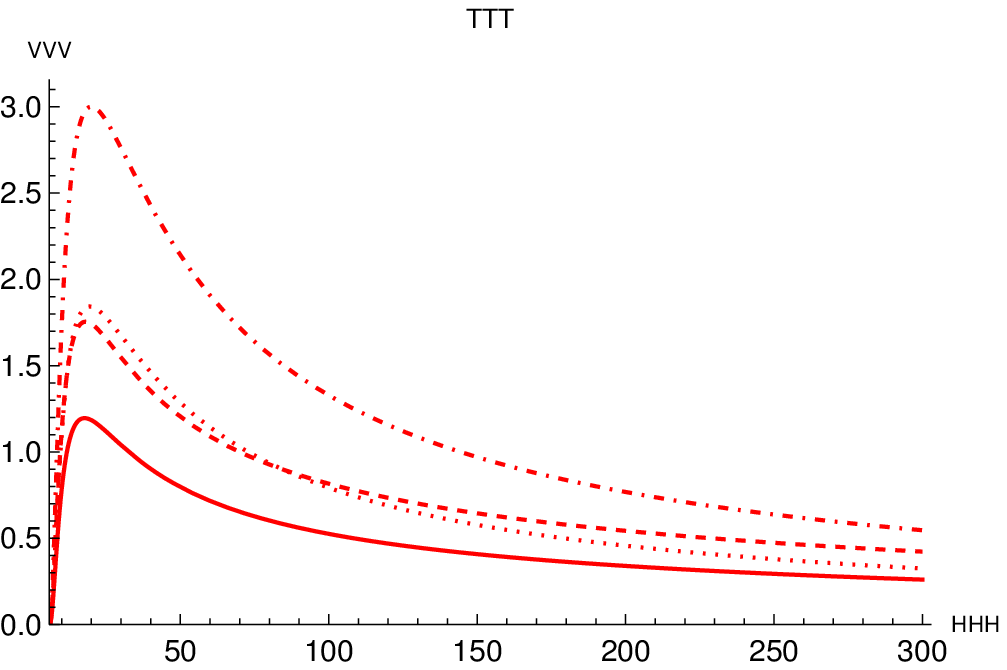}}}
	\\[25pt]
	\psfrag{VVV}{\raisebox{.3cm}{\scalebox{.9}{$\hspace{-.4cm}\displaystyle
				\sigma^{\mathrm{even}}_{\gamma\mesonpp}({\rm pb})$}}}
	\psfrag{TTT}{}
	{
		{\includegraphics[width=18pc]{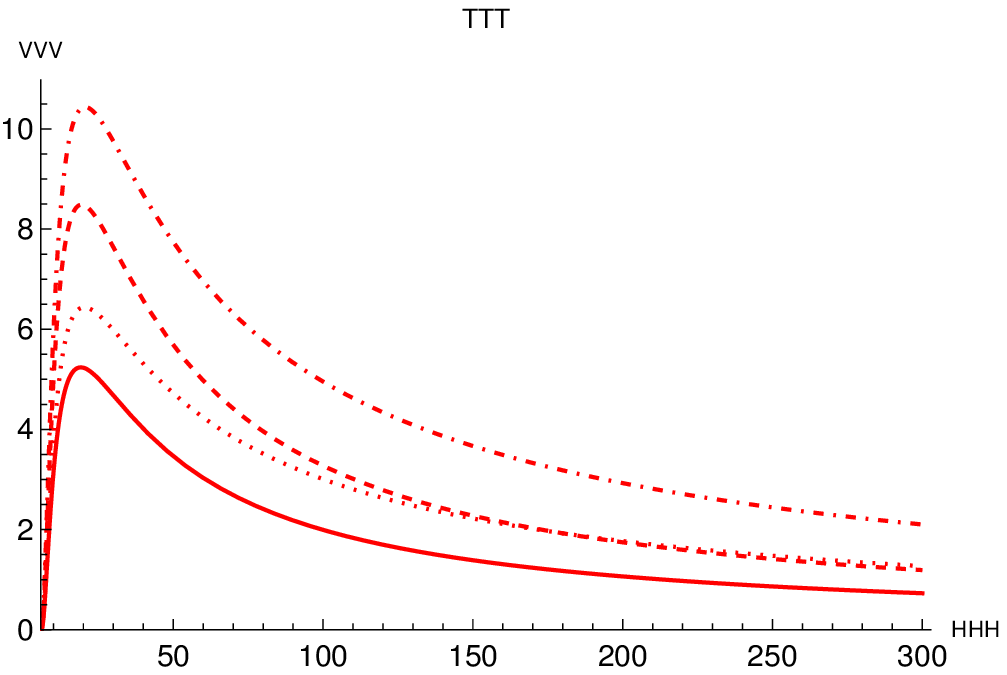}}
		\psfrag{VVV}{\raisebox{.3cm}{\scalebox{.9}{$\hspace{-.4cm}\displaystyle
					\sigma^{\mathrm{even}}_{\gamma\mesonmn}({\rm pb})$}}}
		{\includegraphics[width=18pc]{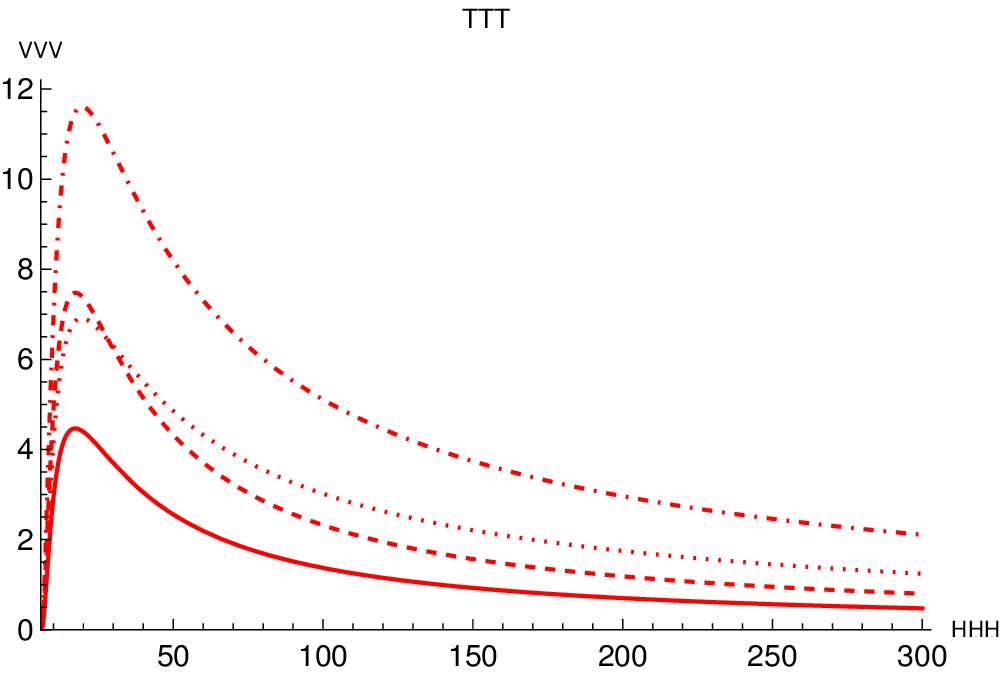}}}
	\vspace{0.2cm}
	\caption{\small The integrated cross section for {longitudinally polarised} $ \mesonzp,\,\mesonzn,\,\mesonpp,\,\mesonmn$ is shown as a function of $   S_{\gamma N}  $ on the top left, top right, bottom left and bottom right plots respectively. The dashed (non-dashed) lines correspond to holographic (asymptotic) DA, while the dotted (non-dotted) lines correspond to the standard (valence) scenario. We thus find that the maximum cross section appears at around 20 GeV$ ^2 $, a feature which was not clear in \FIG\ref{fig:jlab-int-sigma}.}
	\label{fig:compass-int-sigma}
\end{figure}

\begin{figure}[t!]
	\psfrag{HHH}{\hspace{-1.5cm}\raisebox{-.6cm}{\scalebox{.8}{$ S_{\gamma N} ({\rm 
					GeV}^{2})$}}}
	\psfrag{VVV}{\raisebox{.3cm}{\scalebox{.9}{$\hspace{-.4cm}\displaystyle
				\sigma^{\mathrm{odd}}_{\gamma\mesonzp}({\rm pb})$}}}
	\psfrag{TTT}{}
	{
		{\includegraphics[width=18pc]{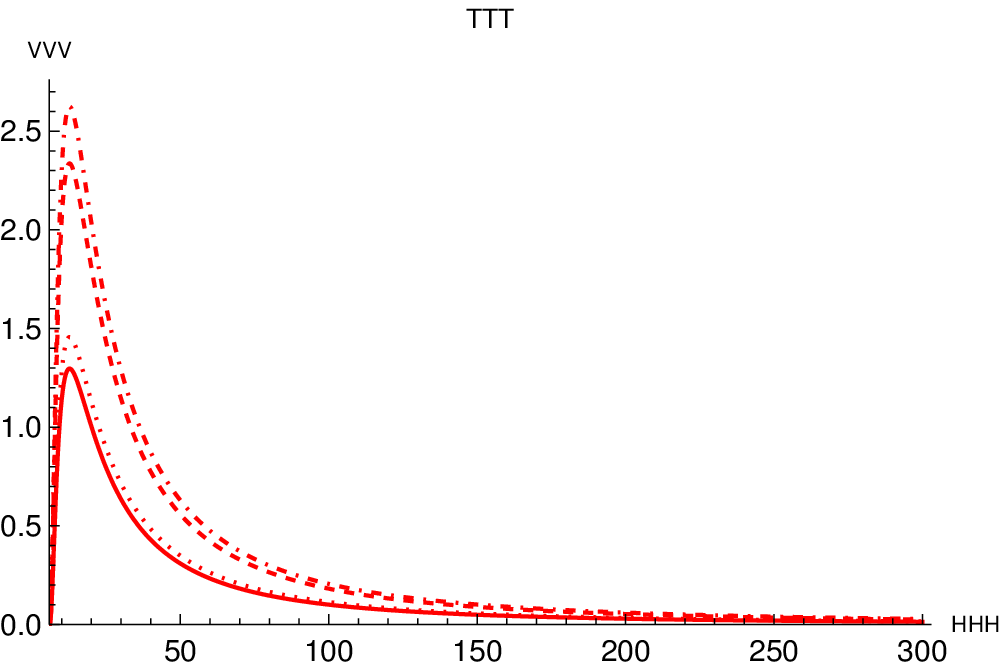}}
		\psfrag{VVV}{\raisebox{.3cm}{\scalebox{.9}{$\hspace{-.4cm}\displaystyle
					\sigma^{\mathrm{odd}}_{\gamma\mesonzn}({\rm pb})$}}}
		{\includegraphics[width=18pc]{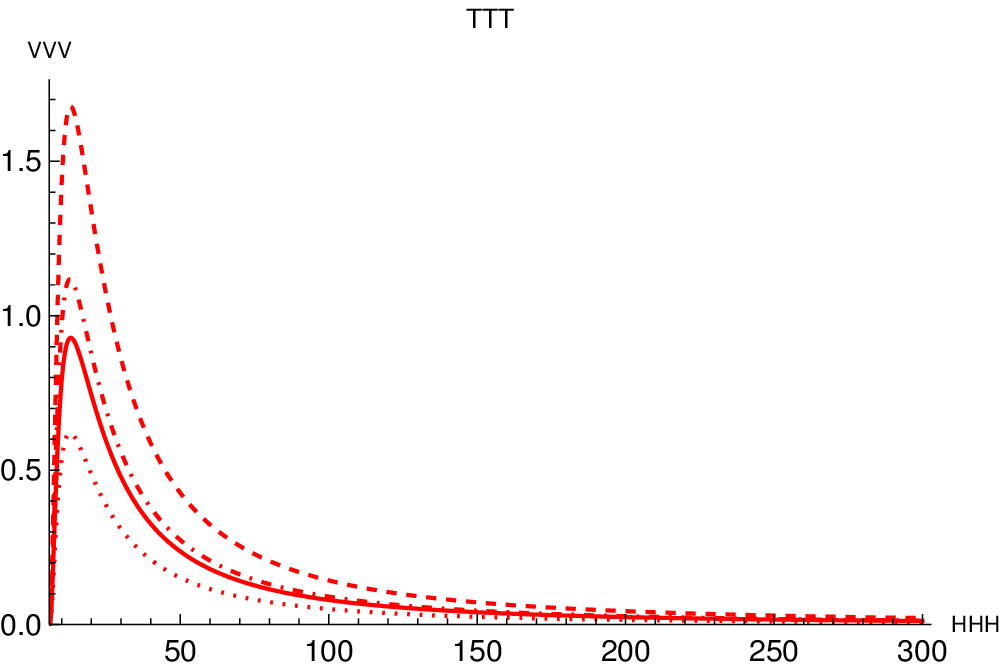}}}
	\\[25pt]
	\psfrag{VVV}{\raisebox{.3cm}{\scalebox{.9}{$\hspace{-.4cm}\displaystyle
				\sigma^{\mathrm{odd}}_{\gamma\mesonpp}({\rm pb})$}}}
	\psfrag{TTT}{}
	{
		{\includegraphics[width=18pc]{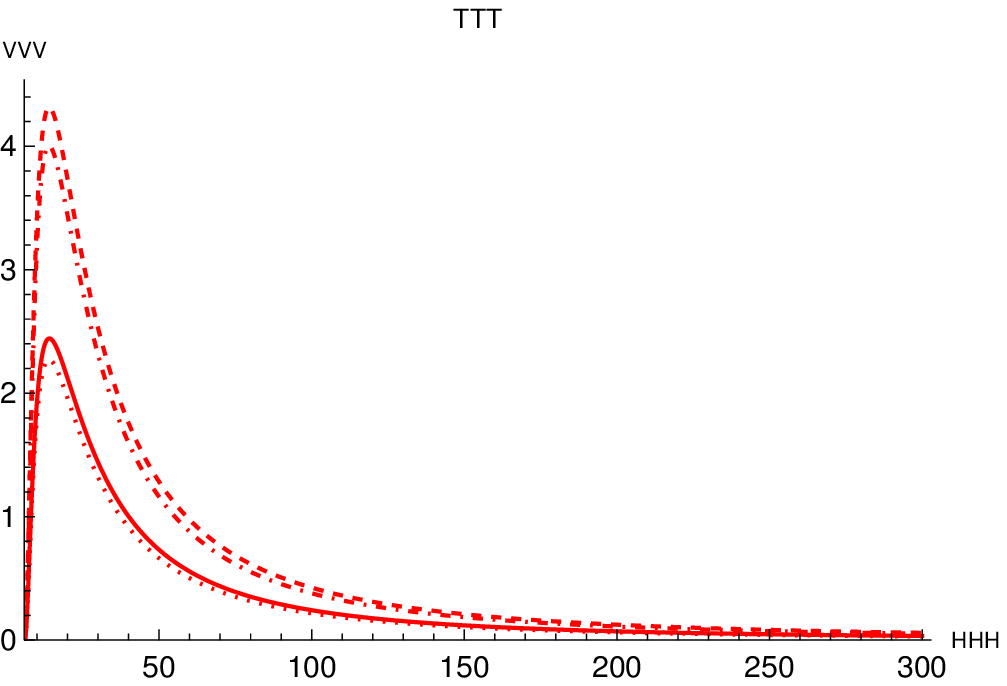}}
		\psfrag{VVV}{\raisebox{.3cm}{\scalebox{.9}{$\hspace{-.4cm}\displaystyle
					\sigma^{\mathrm{odd}}_{\gamma\mesonmn}({\rm pb})$}}}
		{\includegraphics[width=18pc]{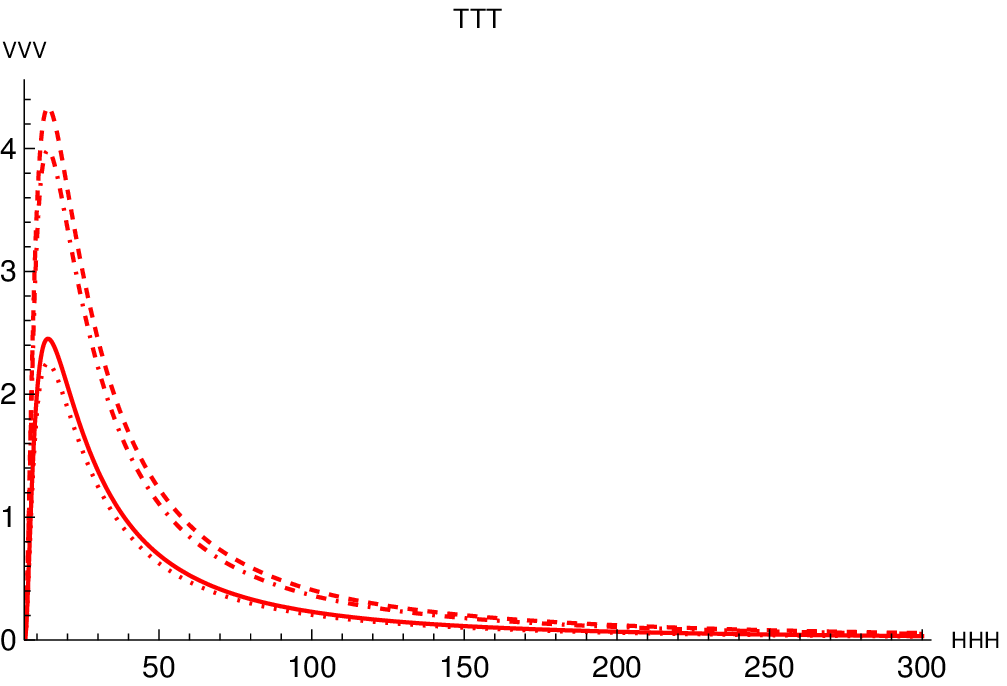}}}
	\vspace{0.2cm}
	\caption{\small The integrated cross section for {transversely polarised} $ \mesonzp,\,\mesonzn,\,\mesonpp,\,\mesonmn$ is shown as a function of $   S_{\gamma N}  $ on the top left, top right, bottom left and bottom right plots respectively. The dashed (non-dashed) lines correspond to holographic (asymptotic) DA, while the dotted (non-dotted) lines correspond to the standard (valence) scenario.}
	\label{fig:co-compass-int-sigma}
\end{figure}

In \FIG\ref{fig:compass-int-sigma}, we show the variation of the integrated cross section as a function of $ \SgN $ for the chiral-even case. We cover all the kinematical range of COMPASS by going to $ \SgN = 300 \GeV^2 $. The peak of the cross section occurs at around $ 20 \GeV^2 $.

For the chiral-odd case, the variation of the integrated cross section as a function of $ \SgN $ is shown in \FIG\ref{fig:co-compass-int-sigma}. The cross section here also has a peak at around $ 20 \GeV^2 $, but falls more rapidly with increasing $ \SgN $ thereafter. Consequently, only the region of $ \SgN $ close to the peak is relevant for the chiral-odd case.

\FloatBarrier

\subsubsection{Polarisation asymmetries}

\label{sec:pol-asym-COMPASS}

\begin{figure}[t!]
	\psfrag{HHH}{\hspace{-1.5cm}\raisebox{-.6cm}{\scalebox{.8}{$-u' ({\rm 
					GeV}^{2})$}}}
	\psfrag{VVV}{LPA$^{\gamma\mesonzp}_{\mathrm{max}} $}
	\psfrag{TTT}{}
	{
		{\includegraphics[width=18pc]{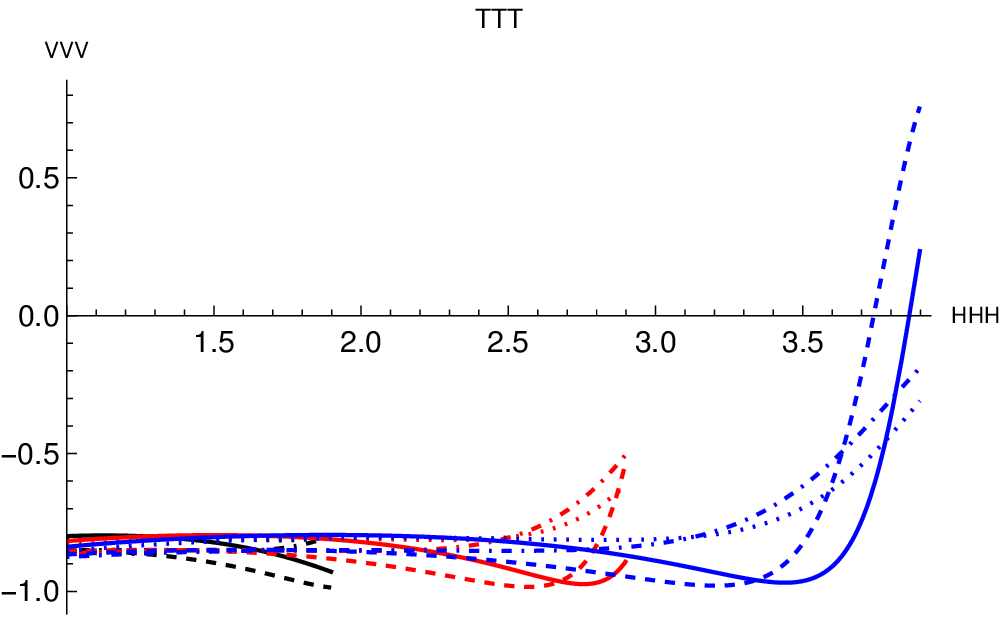}}
		\psfrag{VVV}{LPA$^{\gamma\mesonzn}_{\mathrm{max}} $}
		{\includegraphics[width=18pc]{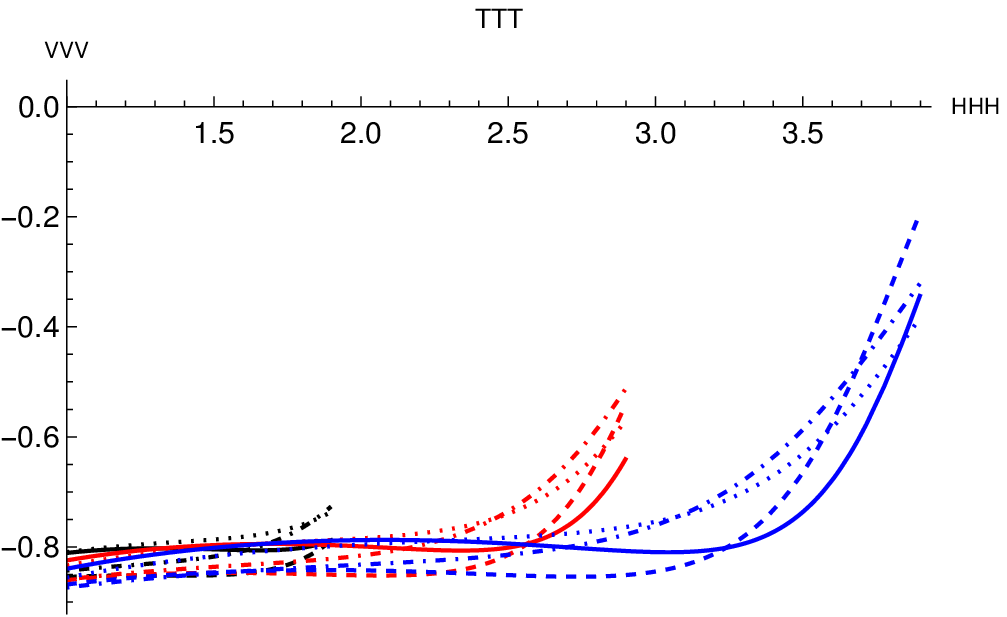}}}
	\\[25pt]
	{					\psfrag{VVV}{LPA$^{\gamma\mesonpp}_{\mathrm{max}} $}
		{\includegraphics[width=18pc]{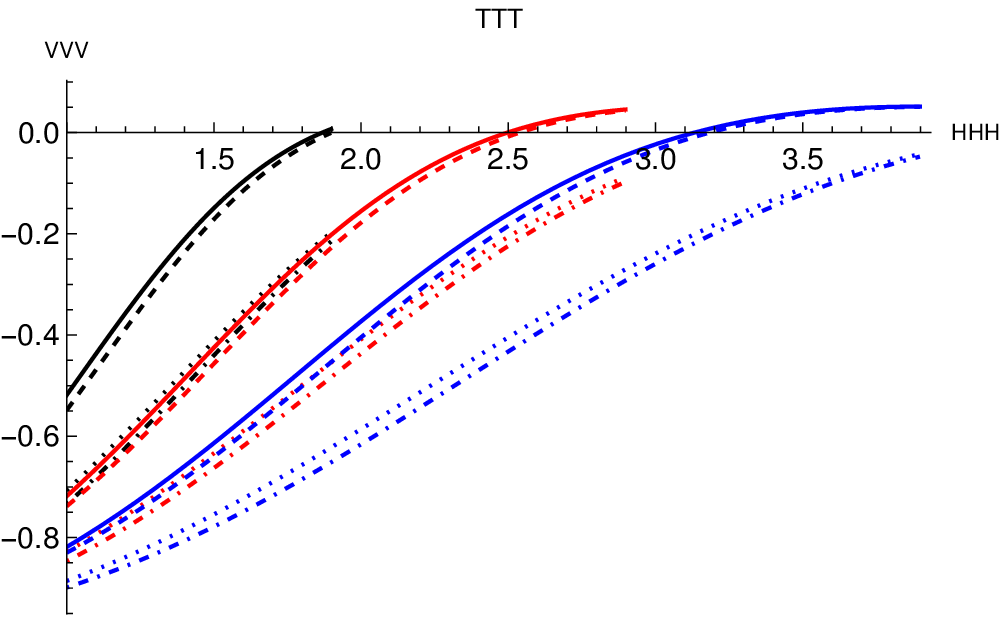}}
		\psfrag{VVV}{LPA$^{\gamma\mesonmn}_{\mathrm{max}} $}
		{\includegraphics[width=18pc]{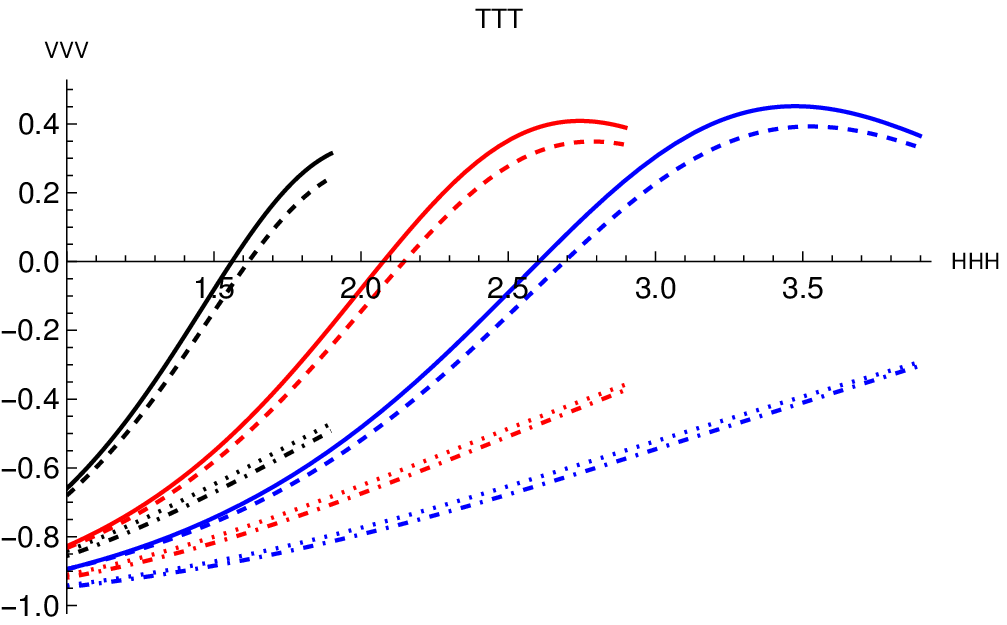}}}
	\vspace{0.2cm}
	\caption{\small The LPA at the fully-differential level for  longitudinally polarised $ \mesonzp,\,\mesonzn,\,\mesonpp,\,\mesonmn$ is shown as a function of $  \left( -u' \right)  $ on the top left, top right, bottom left and bottom right plots respectively for different values of $ M_{\gamma \meson}^2 $. The black, red and blue curves correspond to $ M_{\gamma \meson}^{2}=3,\,4,\,5\, $ GeV$ ^2 $ respectively, and $ \SgN = 200 \GeV^{2}$. The same conventions as in \FIG\ref{fig:compass-fully-diff-diff-M2} are used here.}
	\label{fig:compass-pol-asym-fully-diff-diff-M2}
\end{figure}

\begin{figure}[t!]
	\psfrag{HHH}{\hspace{-1.5cm}\raisebox{-.6cm}{\scalebox{.8}{$-u' ({\rm 
					GeV}^{2})$}}}
	\psfrag{VVV}{LPA$^{\gamma\mesonzp}_{\mathrm{max}} $}
	\psfrag{TTT}{}
	{
		{\includegraphics[width=18pc]{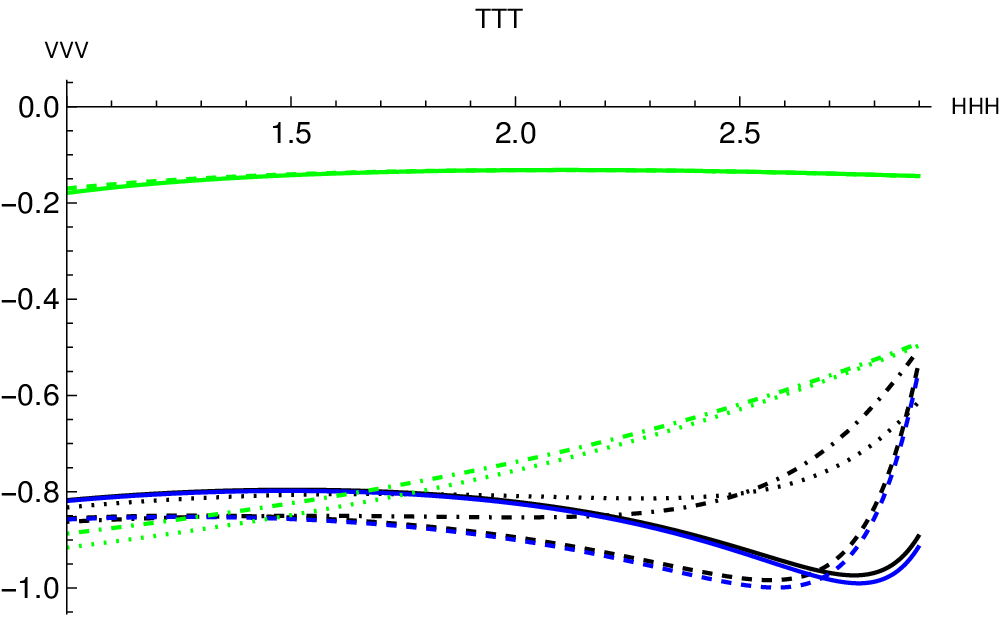}}
		\psfrag{VVV}{LPA$^{\gamma\mesonzn}_{\mathrm{max}} $}
		{\includegraphics[width=18pc]{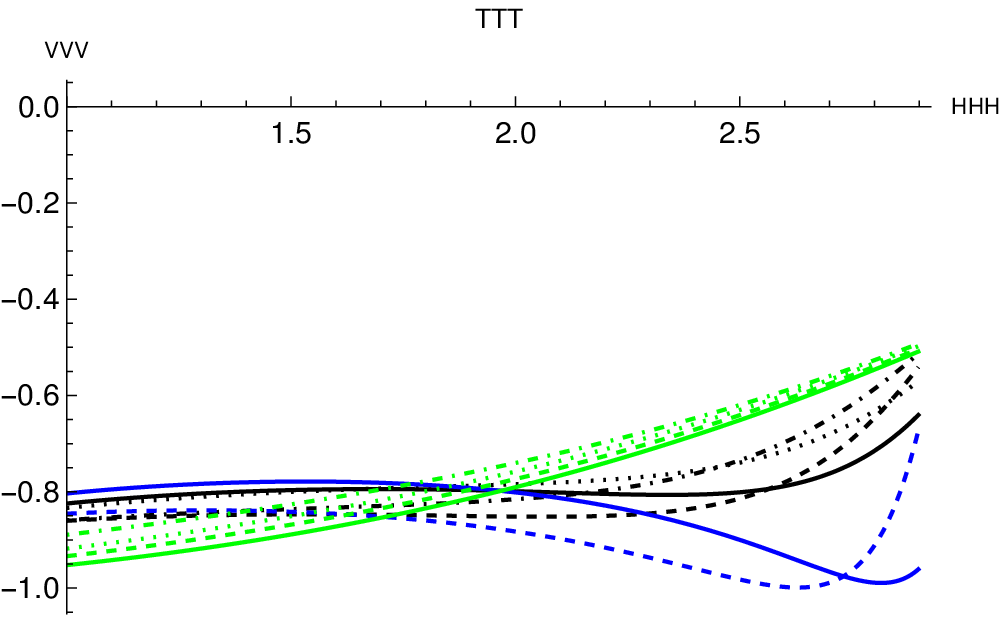}}}
	\\[25pt]
	{					\psfrag{VVV}{LPA$^{\gamma\mesonpp}_{\mathrm{max}} $}
		{\includegraphics[width=18pc]{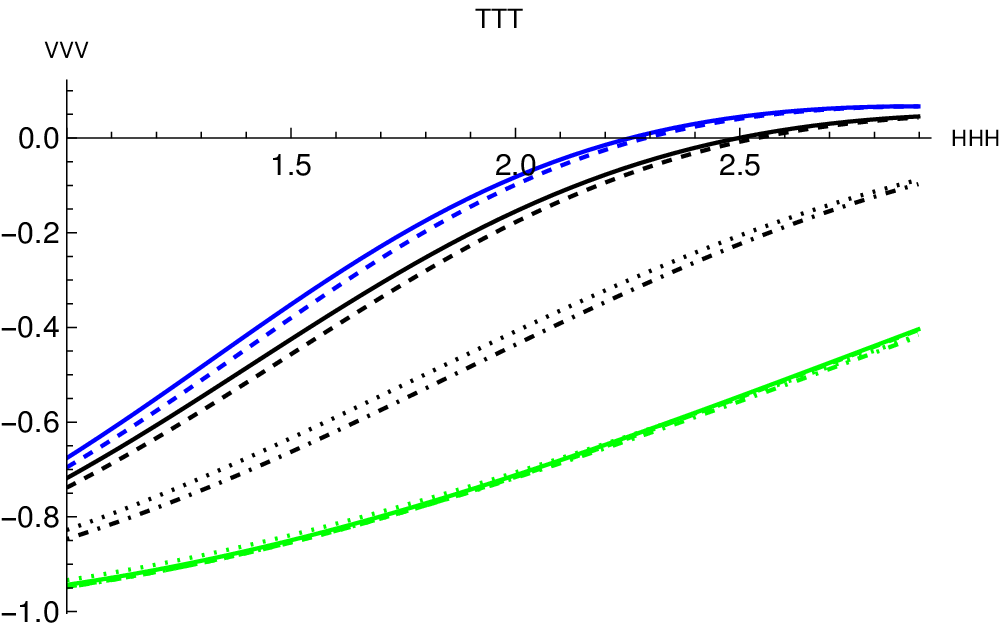}}
		\psfrag{VVV}{LPA$^{\gamma\mesonmn}_{\mathrm{max}} $}
		{\includegraphics[width=18pc]{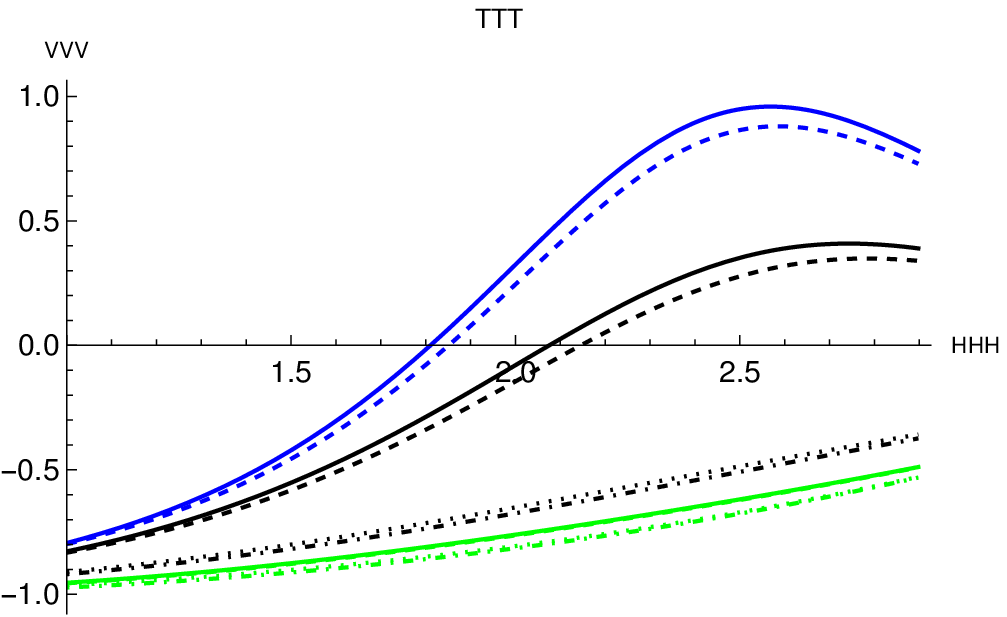}}}
	\vspace{0.2cm}
	\caption{\small The LPA at the fully-differential level for  longitudinally polarised $ \mesonzp,\,\mesonzn,\,\mesonpp,\,\mesonmn$ is shown as a function of $  \left( -u' \right)  $ on the top left, top right, bottom left and bottom right plots respectively, using $ \Msq = 4 \GeV^2$ and $ \SgN = 200 \GeV^2 $. The same conventions as in \FIG\ref{fig:compass-fully-diff-VandA} are used here. Note that the vector contributions consist of only two curves in each case, since they are insensitive to either valence or standard scenarios.}
	\label{fig:compass-pol-asym-fully-diff-VandA}
\end{figure}

In this section, we show the results for the linear polarisation asymmetries (LPAs) for COMPASS kinematics. As for the cross section plots in the previous section, we choose the reference value of 200 GeV$ ^2 $ for $ \SgN $ for the fully differential and single differential plots. As discussed in \SEC\ref{sec:pol-asym-jlab}, the plots that we show here correspond to $  \mathrm{LPA}_{ \mathrm{max} }  $. Furthermore, we note that only plots of the LPAs for the chiral-even case are shown here, since the LPAs for the chiral-odd case vanish.

In \FIG\ref{fig:compass-pol-asym-fully-diff-diff-M2}, the LPAs at the fully-differential level are shown as a function of $  \left( -u' \right)  $, for different values of $ \Msq $. The three values of $ \Msq $ that we use are $ \Msq = 3,\,4,\,5 \GeV^2 $, which correspond to the black, red and blue curves respectively. As in the JLab kinematics case, we observe that for the charged $ \meson $-meson case, the LPAs can be used to discriminate between the GPD models used. Furthermore, for the $ \mesonmn $ case, the sign of the LPA changes flips from negative to positive as $  \left( -u' \right)  $ increases for the valence scenario only.

In \FIG\ref{fig:compass-pol-asym-fully-diff-VandA}, we show the relative contributions of the vector and axial GPDs to the LPA at the fully differential level. The values $ \SgN =200 \GeV^2$ and $ \Msq=4 \GeV^2 $ are used to generate the plots. For the $ \mesonzp $ case, the LPA is remains very negative and relatively flat, except for the axial contribution in the standard GPD scenario. For $ \mesonzn $, we observe that the LPA remains rather flat at very negative values throughout the range of $  \left( -u' \right)  $. Finally, for the charged $ \meson $-meson case, the LPA covers a wider range, starting at a sizeable value at low $  \left( -u' \right)  $.

\begin{figure}[t!]
	\psfrag{HHH}{\hspace{-1.5cm}\raisebox{-.6cm}{\scalebox{.8}{$-u' ({\rm 
					GeV}^{2})$}}}
		\psfrag{VVV}{LPA$^{\gamma\mesonzp}_{\mathrm{max}} $}
	\psfrag{TTT}{}
	{
		{\includegraphics[width=18pc]{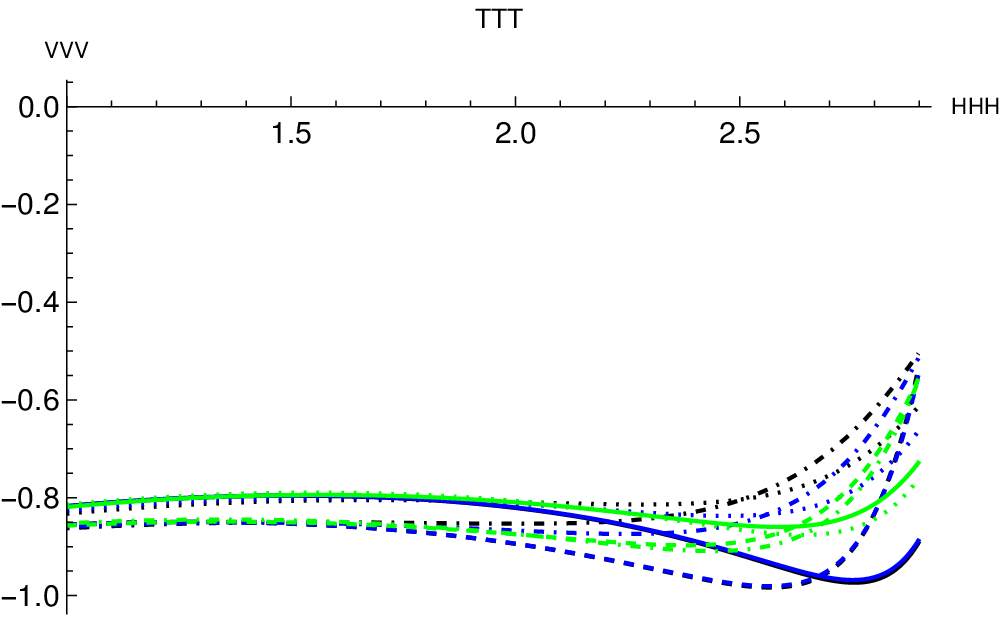}}
			\psfrag{VVV}{LPA$^{\gamma\mesonzn}_{\mathrm{max}} $}
		{\includegraphics[width=18pc]{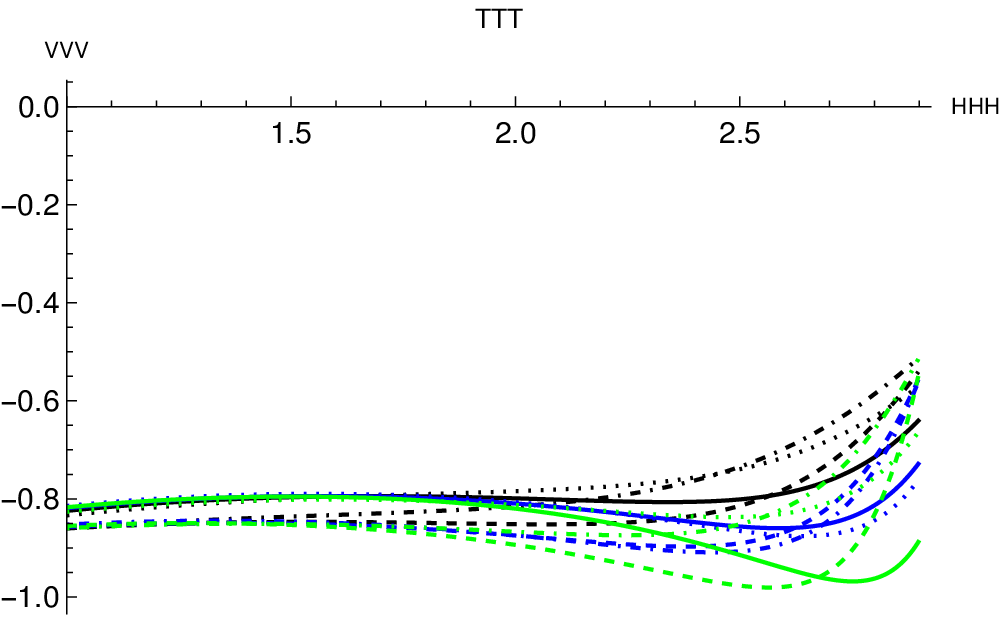}}}
	\\[25pt]
	{					\psfrag{VVV}{LPA$^{\gamma\mesonpp}_{\mathrm{max}} $}
		{\includegraphics[width=18pc]{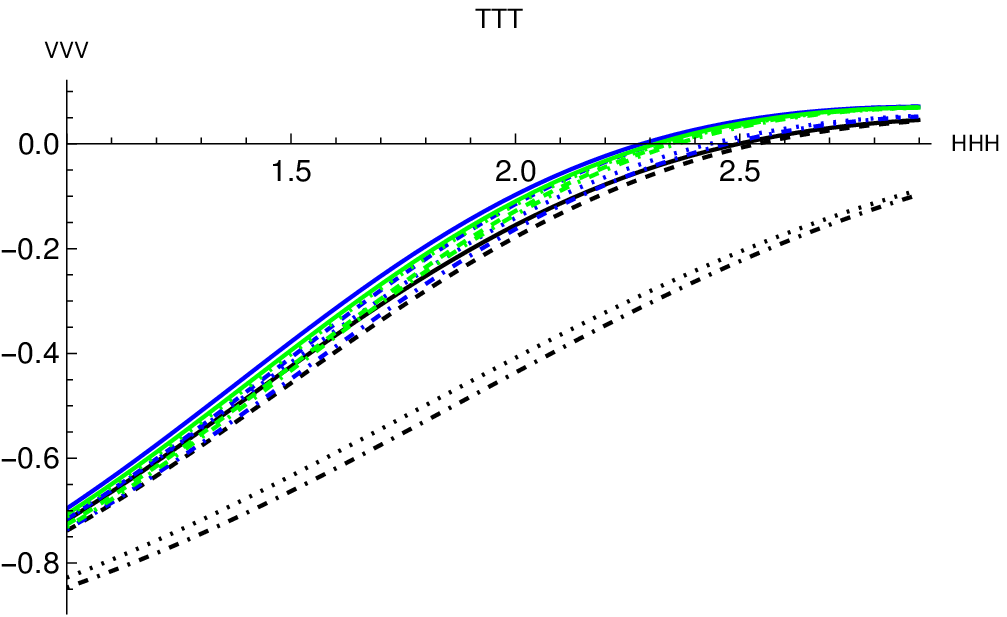}}
			\psfrag{VVV}{LPA$^{\gamma\mesonmn}_{\mathrm{max}} $}
		{\includegraphics[width=18pc]{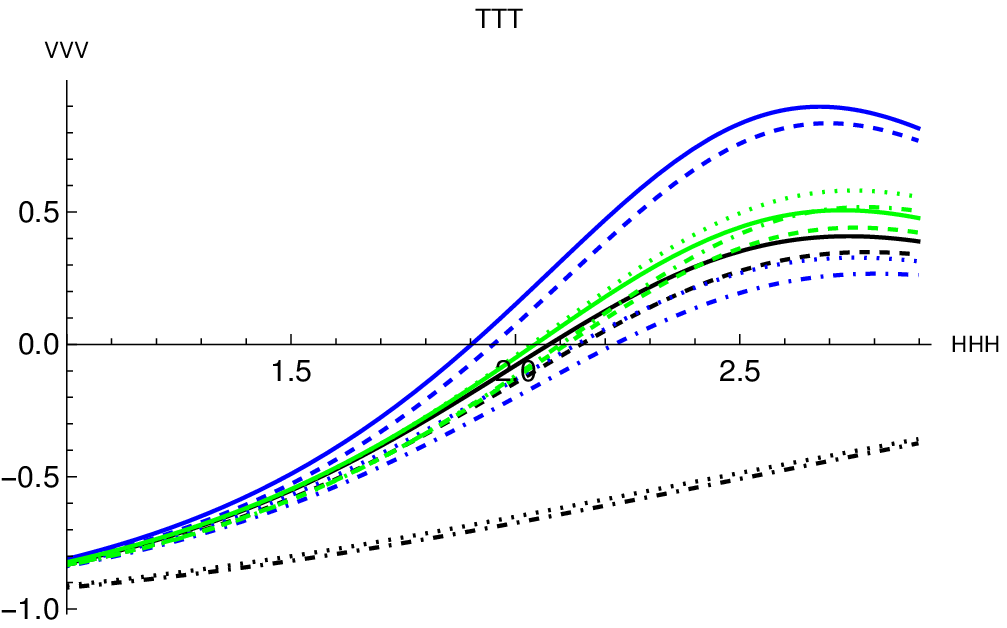}}}
	\vspace{0.2cm}
	\caption{\small The LPA at the fully-differential level for  longitudinally polarised $ \mesonzp,\,\mesonzn,\,\mesonpp,\,\mesonmn$ is shown as a function of $  \left( -u' \right)  $ on the top left, top right, bottom left and bottom right plots respectively, using $ \Msq = 4 \GeV^2$ and $ \SgN = 200 \GeV^2 $. The blue and green curves correspond to contributions from the $u$-quark ($ H_{u} $ and $  \tilde{H} _{u} $) and $d$-quark ($ H_{d} $ and $  \tilde{H} _{d} $) GPDs respectively. The black curves correspond to the total contribution. The same conventions as in \FIG\ref{fig:compass-fully-diff-uandd} are used here.}
	\label{fig:compass-pol-asym-fully-diff-uandd}
\end{figure}

The relative contributions of the $ u $-quark and $ d $-quark GPDs to the LPA at the fully differential level are shown in \FIG\ref{fig:compass-pol-asym-fully-diff-uandd}. We choose $ \SgN =200 \GeV^2$ and $ \Msq=4 \GeV^2 $ to generate the plots.

\FloatBarrier

\begin{figure}[t!]
	\psfrag{HHH}{\hspace{-1.5cm}\raisebox{-.6cm}{\scalebox{.8}{ $ M_{\gamma \meson}^{2}({\rm 
					GeV}^{2}) $}}}
		\psfrag{VVV}{LPA$^{\gamma\mesonzp}_{\mathrm{max}} $}
	\psfrag{TTT}{}
	{
		{\includegraphics[width=18pc]{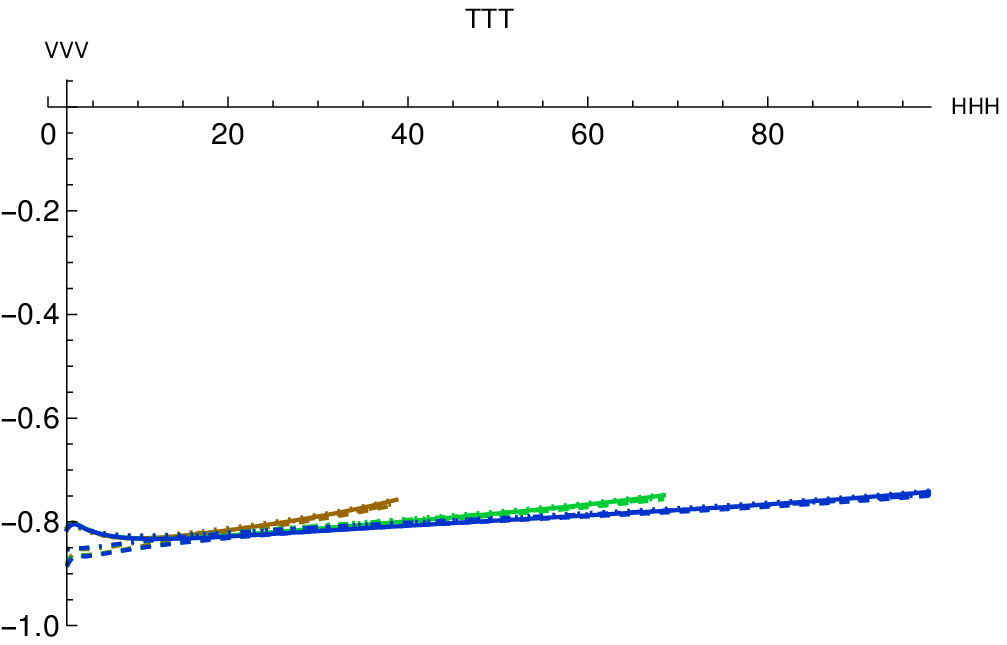}}
			\psfrag{VVV}{LPA$^{\gamma\mesonzn}_{\mathrm{max}} $}
		{\includegraphics[width=18pc]{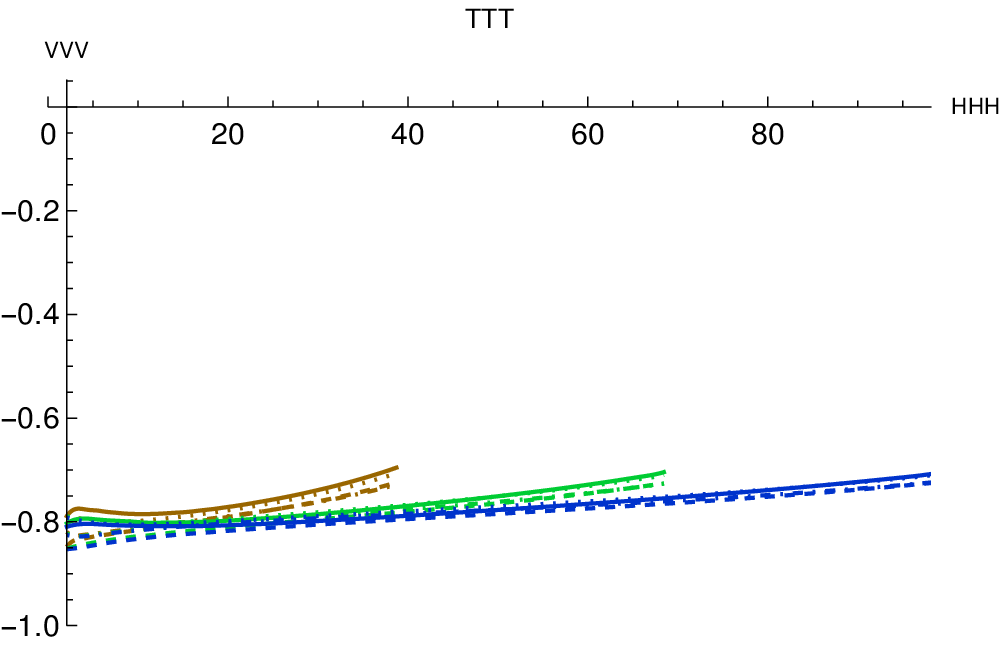}}}
	\\[25pt]
	{					\psfrag{VVV}{LPA$^{\gamma\mesonpp}_{\mathrm{max}} $}
		{\includegraphics[width=18pc]{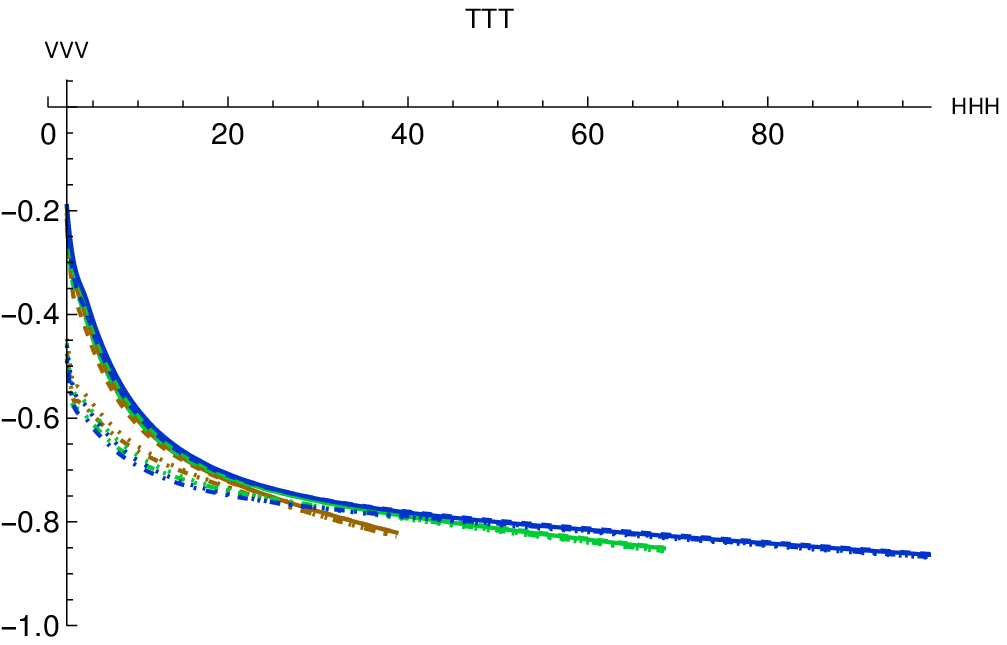}}
			\psfrag{VVV}{LPA$^{\gamma\mesonmn}_{\mathrm{max}} $}
		{\includegraphics[width=18pc]{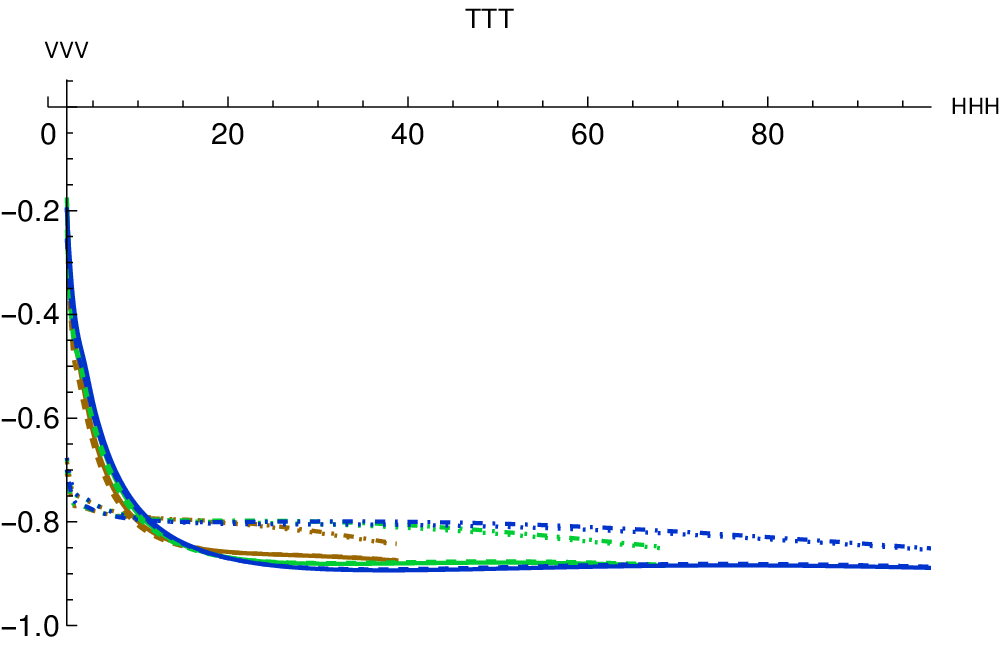}}}
	\vspace{0.2cm}
	\caption{\small The LPA at the single differential level for  longitudinally polarised $ \mesonzp,\,\mesonzn,\,\mesonpp,\,\mesonmn$ is shown as a function of $  M_{\gamma \meson}^{2}  $ on the top left, top right, bottom left and bottom right plots respectively. The brown, green and blue curves correspond to $ S_{\gamma N} = 80,\,140,\,200\,\GeV^{2} $. The same colour and line style conventions as in \FIG\ref{fig:compass-sing-diff} are used here.}
	\label{fig:compass-pol-asym-sing-diff}
\end{figure}

In \FIG\ref{fig:compass-pol-asym-sing-diff}, we show the variation of the LPA at the single-differential level as a function of $ \Msq $ for different values of $ \SgN $. We observe that the LPA is rather flat at a value of about $ -0.8 $, with the exception of the charged $ \meson $-meson case at low $ \Msq $. Furthermore, we note that the GPD or DA model used has little effect on the LPA, with the exception of the $ \mesonmn $ case, where the GPD model nevertheless has non-negligible effect. This is in contrast to the charged $ \pi^{\pm} $, where we found that the LPA is very sensitive to the GPD model, see \FIG 21 in \cite{Duplancic:2022ffo}.

\FloatBarrier

\begin{figure}[t!]
	\psfrag{HHH}{\hspace{-1.5cm}\raisebox{-.6cm}{\scalebox{.8}{ $ S_{\gamma N}({\rm 
					GeV}^{2}) $}}}
		\psfrag{VVV}{LPA$^{\gamma\mesonzp}_{\mathrm{max}} $}
	\psfrag{TTT}{}
	{
		{\includegraphics[width=18pc]{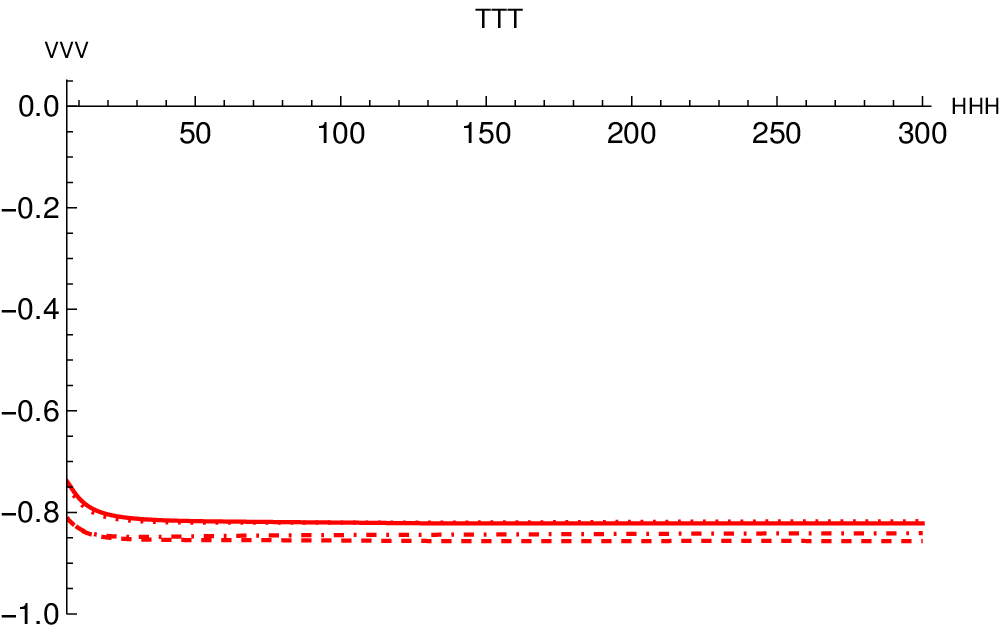}}
			\psfrag{VVV}{LPA$^{\gamma\mesonzn}_{\mathrm{max}} $}
		{\includegraphics[width=18pc]{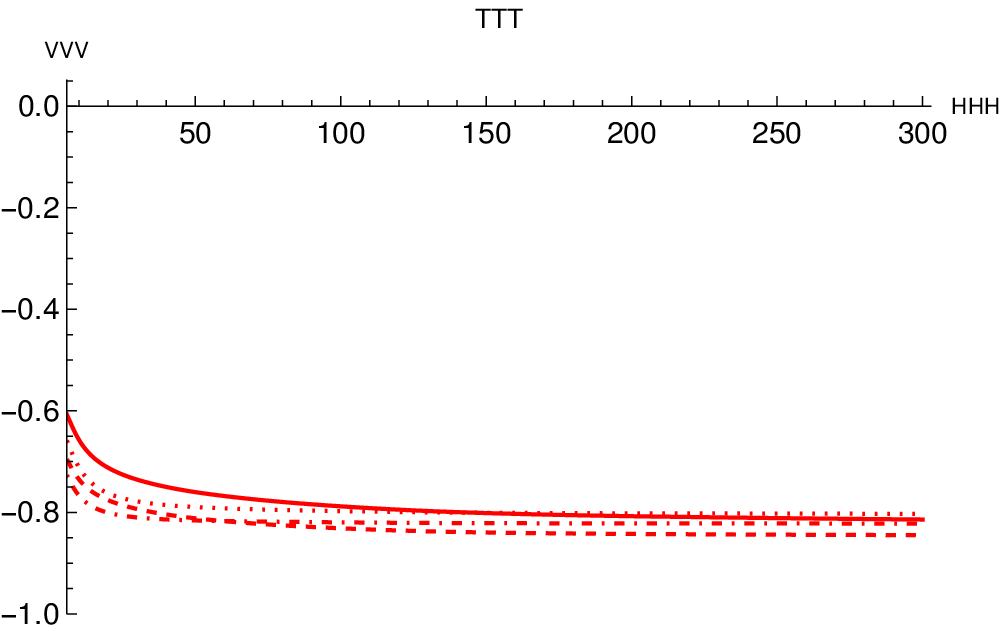}}}
	\\[25pt]
	{					\psfrag{VVV}{LPA$^{\gamma\mesonpp}_{\mathrm{max}} $}
		{\includegraphics[width=18pc]{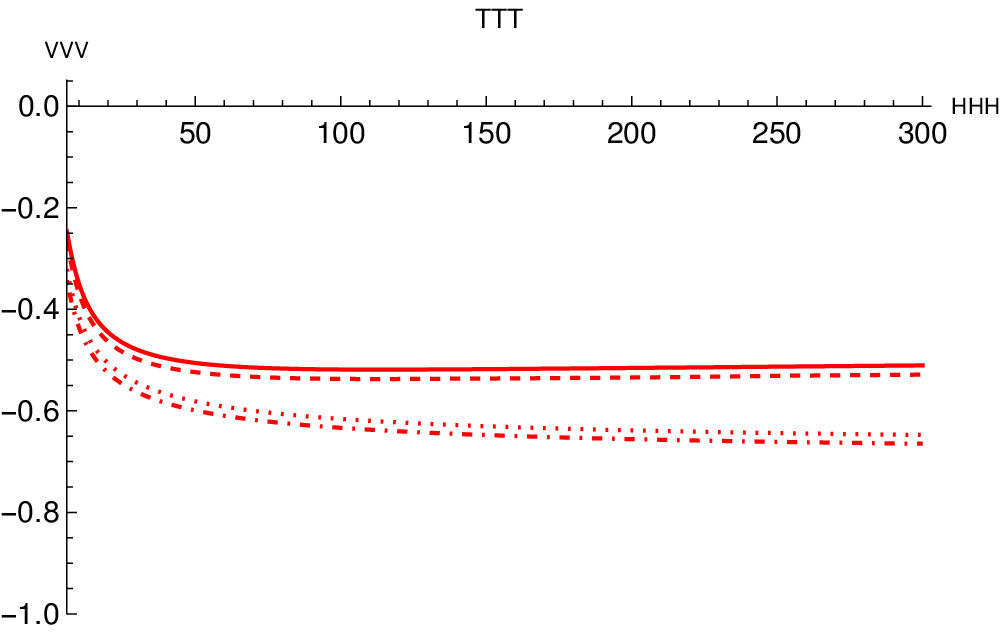}}
			\psfrag{VVV}{LPA$^{\gamma\mesonmn}_{\mathrm{max}} $}
		{\includegraphics[width=18pc]{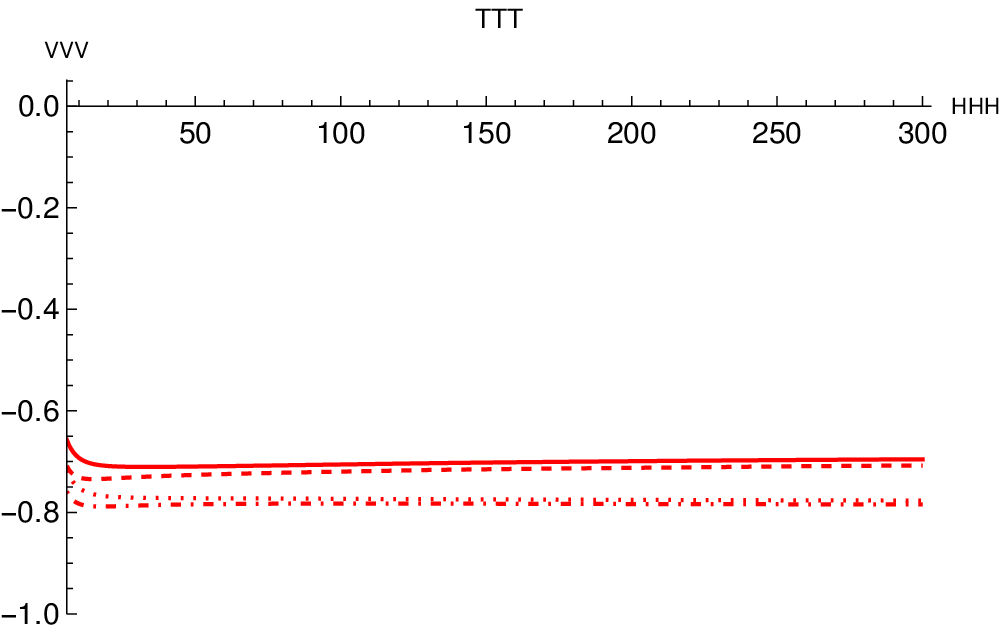}}}
	\vspace{0.2cm}
	\caption{\small The LPA integrated over all differential variables for  longitudinally polarised $ \mesonzp,\,\mesonzn,\,\mesonpp,\,\mesonmn$ is shown on the top left, top right, bottom left and bottom right plots respectively as a function of $ \SgN $. The same colour and line style conventions as in \FIG\ref{fig:compass-int-sigma} are used here.}
	\label{fig:compass-pol-asym-int-sigma}
\end{figure}

To conclude this section on COMPASS kinematics, the variation of the LPA, integrated over all differential variables, is shown as a function of $ \SgN $ in \FIG\ref{fig:compass-pol-asym-int-sigma}. Here, the LPA is again rather flat at roughly $ -0.8 $ with the exception of the $ \mesonpp $ case, for which the magnitude of the LPA is smaller. We also observe that LPA is rather insensitive to the GPD or DA model used, except for the charged $ \meson $-meson cases, for which the GPD model has an effect.

\FloatBarrier

\subsection{EIC and UPC at LHC kinematics}

\label{sec:EIC-LHC-UPC-kinematics}

We consider photon-nucleon centre-of-mass energies $ \SgN $ of up to 20000 GeV$ ^2 $. Such a choice covers the whole range of the expected EIC kinematics (with maximum centre of mass energy of the electron-proton system roughly 19600 GeV$ ^2 $ \cite{AbdulKhalek:2021gbh}), and the most relevant part of UPCs at LHC kinematics (which in principle centre of mass energies of the order of the TeV scale).

For UPCs at LHC kinematics, we note that both the cross section and the photon flux drop very rapidly as $ \SgN $ increases. Therefore, only a tiny contribution is lost by neglecting contributions which are beyond the kinematics of EIC, \ie{}above $ \SgN=20000 \GeV^2 $.

\subsubsection{Fully differential cross section}

\label{sec:EIC-fully-diff}

\begin{figure}[t!]
	\psfrag{HHH}{\hspace{-1.5cm}\raisebox{-.6cm}{\scalebox{.8}{$-u' ({\rm 
					GeV}^{2})$}}}
	\psfrag{VVV}{\raisebox{.3cm}{\scalebox{.9}{$\hspace{-.4cm}\displaystyle\left.\frac{d 
					\sigma^{\mathrm{even}}_{\gamma\mesonzp}}{d M^2_{\gamma\mesonzp} d(-u') d(-t)}\right|_{(-t)_{\rm min}}({\rm pb} \cdot {\rm GeV}^{-6})$}}}
	\psfrag{TTT}{}
	{
		{\includegraphics[width=18pc]{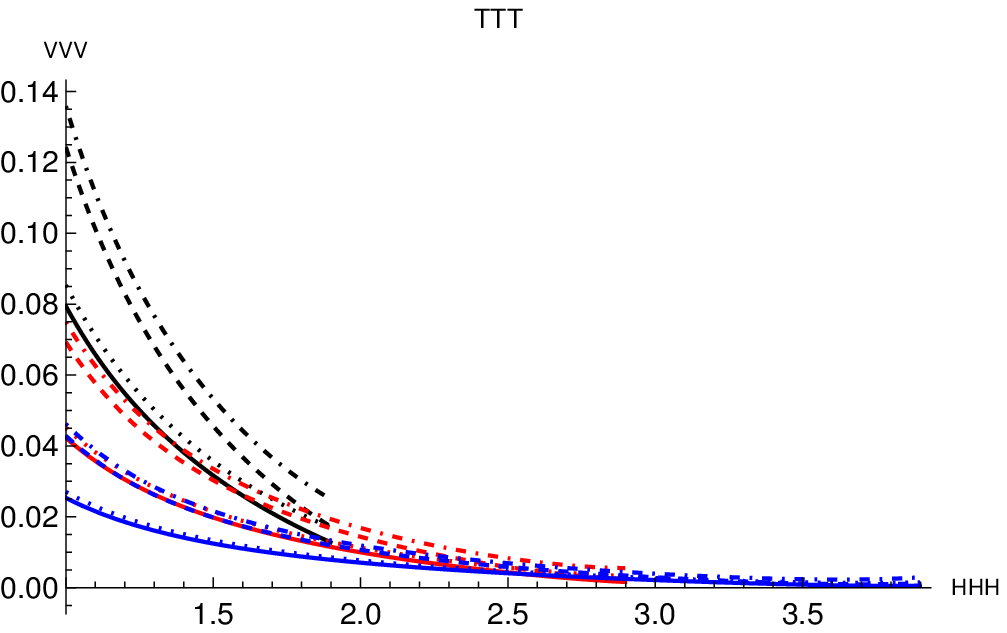}}
		\psfrag{VVV}{\raisebox{.3cm}{\scalebox{.9}{$\hspace{-.4cm}\displaystyle\left.\frac{d 
						\sigma^{\mathrm{even}}_{\gamma\mesonzn}}{d M^2_{\gamma \mesonzn} d(-u') d(-t)}\right|_{(-t)_{\rm min}}({\rm pb} \cdot {\rm GeV}^{-6})$}}}
		{\includegraphics[width=18pc]{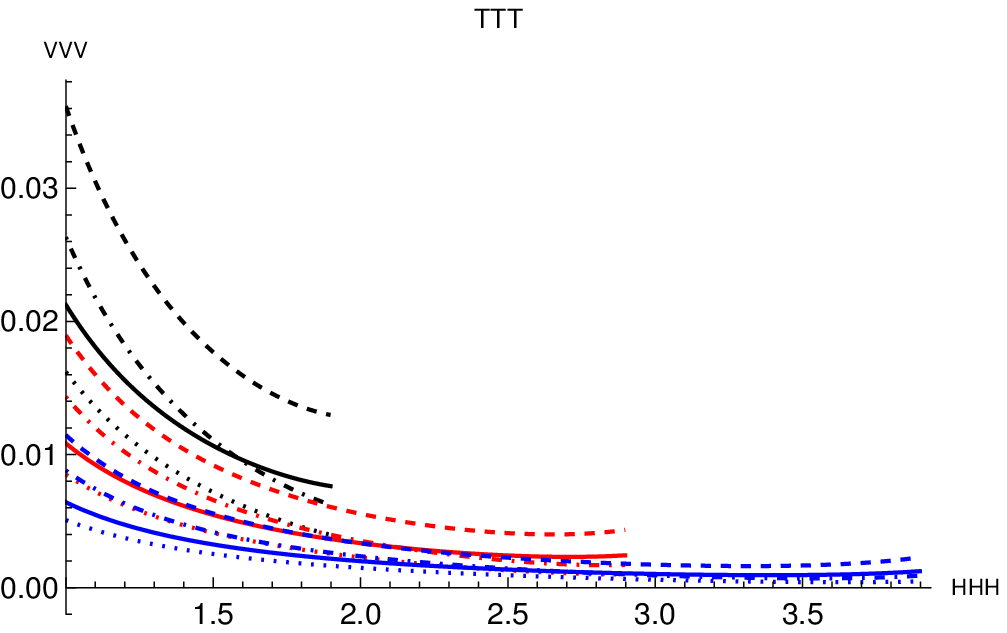}}}
	\\[25pt]
	{
		\psfrag{VVV}{\raisebox{.3cm}{\scalebox{.9}{$\hspace{-.4cm}\displaystyle\left.\frac{d 
						\sigma^{\mathrm{even}}_{\gamma\mesonpp}}{d M^2_{\gamma\mesonpp} d(-u') d(-t)}\right|_{(-t)_{\rm min}}({\rm pb} \cdot {\rm GeV}^{-6})$}}}
		{\includegraphics[width=18pc]{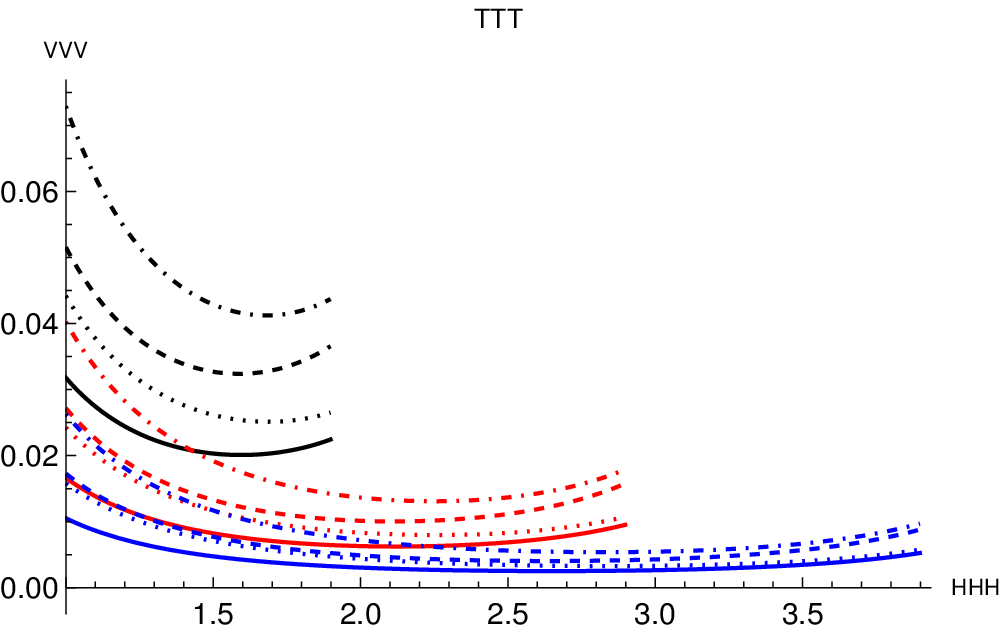}}
		\psfrag{VVV}{\raisebox{.3cm}{\scalebox{.9}{$\hspace{-.4cm}\displaystyle\left.\frac{d 
						\sigma^{\mathrm{even}}_{\gamma\mesonmn}}{d M^2_{\gamma\mesonmn} d(-u') d(-t)}\right|_{(-t)_{\rm min}}({\rm pb} \cdot {\rm GeV}^{-6})$}}}
		{\includegraphics[width=18pc]{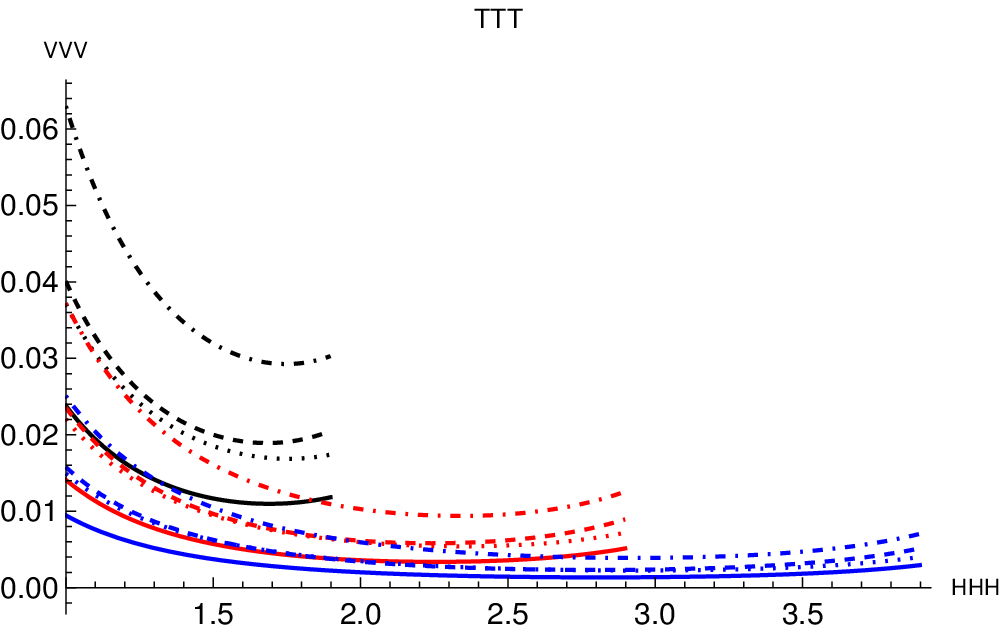}}}
	\vspace{0.2cm}
	\caption{\small The fully differential cross section for {longitudinally polarised} $ \mesonzp,\,\mesonzn,\,\mesonpp,\,\mesonmn$ is shown as a function of $  \left( -u' \right)  $ on the top left, top right, bottom left and bottom right plots respectively for different values of $ M_{\gamma \meson}^2 $. The black, red and blue curves correspond to $ M_{\gamma \meson}^{2}=3,\,4,\,5\, $ GeV$ ^2 $ respectively. The dashed (non-dashed) lines correspond to holographic (asymptotic) DA, while the dotted (non-dotted) lines correspond to the standard (valence) scenario. As mentioned in the main text, $ S_{\gamma N} $ is fixed at 20000 GeV$ ^2 $ here.}
	\label{fig:EIC-LHC-UPC-fully-diff-diff-M2}
\end{figure}

At large $ \SgN $, typical of EIC and UPCs at LHC kinematics, the cross section for the chiral-odd case is heavily suppressed compared to the chiral-even case. As mentioned before, this is due to the $ \xi^2 $ factor that appears in front of the squared amplitude for the chiral-odd case. Consequently, in this section, we only show the plots for the chiral-even case.

In \FIG\ref{fig:EIC-LHC-UPC-fully-diff-diff-M2}, the fully differential cross section as a function of $  \left( -u' \right)  $ is shown for different values of $ \Msq $. We choose $ \SgN = 20000 \GeV^2 $. For $ \Msq $, we take $ \Msq =3,\,4,\,5 \GeV^2 $, since the cross section become much smaller at higher values of $ \Msq $. We observe a decrease of the cross section by a factor of 100 roughly compared to the COMPASS kinematics case in \SEC\ref{sec:compass-fully-diff}.

\begin{figure}[t!]
	\psfrag{HHH}{\hspace{-1.5cm}\raisebox{-.6cm}{\scalebox{.8}{$-u' ({\rm 
					GeV}^{2})$}}}
	\psfrag{VVV}{\raisebox{.3cm}{\scalebox{.9}{$\hspace{-.4cm}\displaystyle\left.\frac{d 
					\sigma^{\mathrm{even}}_{\gamma\mesonzp}}{d M^2_{\gamma\mesonzp} d(-u') d(-t)}\right|_{(-t)_{\rm min}}({\rm pb} \cdot {\rm GeV}^{-6})$}}}
	\psfrag{TTT}{}
	{
		{\includegraphics[width=18pc]{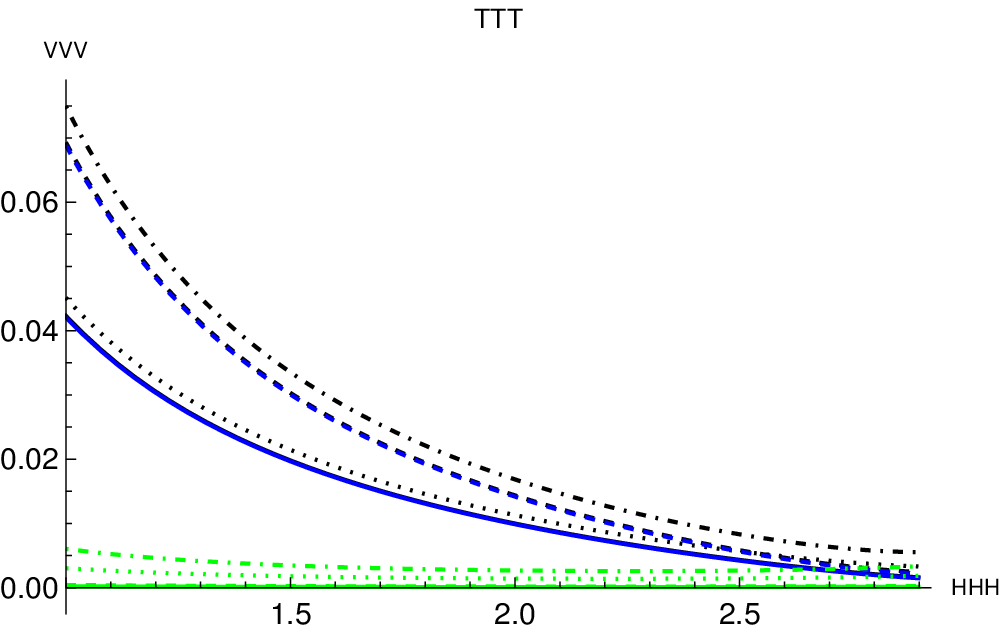}}
		\psfrag{VVV}{\raisebox{.3cm}{\scalebox{.9}{$\hspace{-.4cm}\displaystyle\left.\frac{d 
						\sigma^{\mathrm{even}}_{\gamma\mesonzn}}{d M^2_{\gamma \mesonzn} d(-u') d(-t)}\right|_{(-t)_{\rm min}}({\rm pb} \cdot {\rm GeV}^{-6})$}}}
		{\includegraphics[width=18pc]{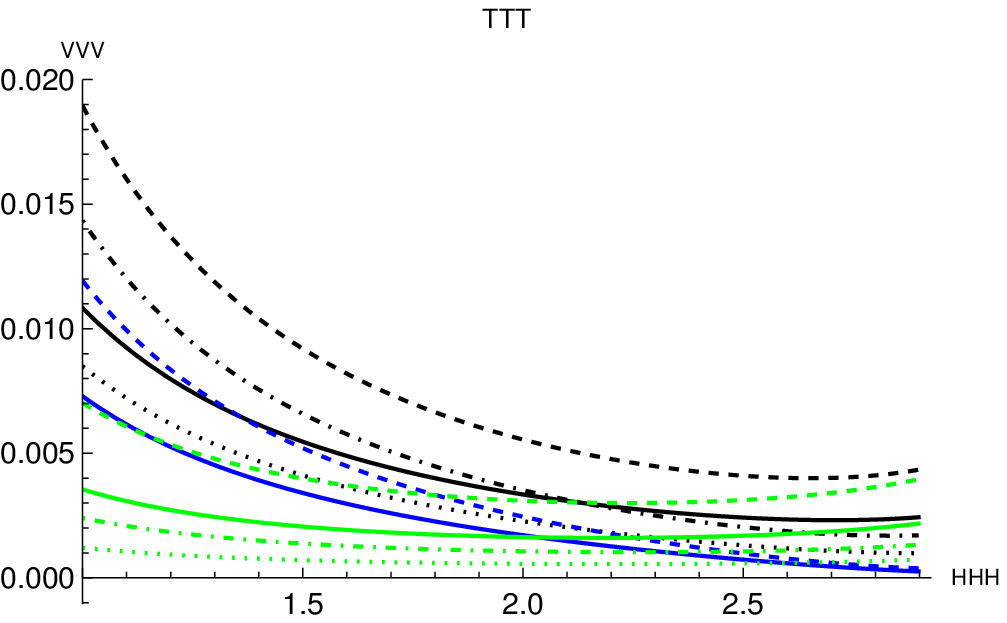}}}
	\\[25pt]
	\psfrag{VVV}{\raisebox{.3cm}{\scalebox{.9}{$\hspace{-.4cm}\displaystyle\left.\frac{d 
					\sigma^{\mathrm{even}}_{\gamma\mesonpp}}{d M^2_{\gamma\mesonpp} d(-u') d(-t)}\right|_{(-t)_{\rm min}}({\rm pb} \cdot {\rm GeV}^{-6})$}}}
	\psfrag{TTT}{}
	{
		{\includegraphics[width=18pc]{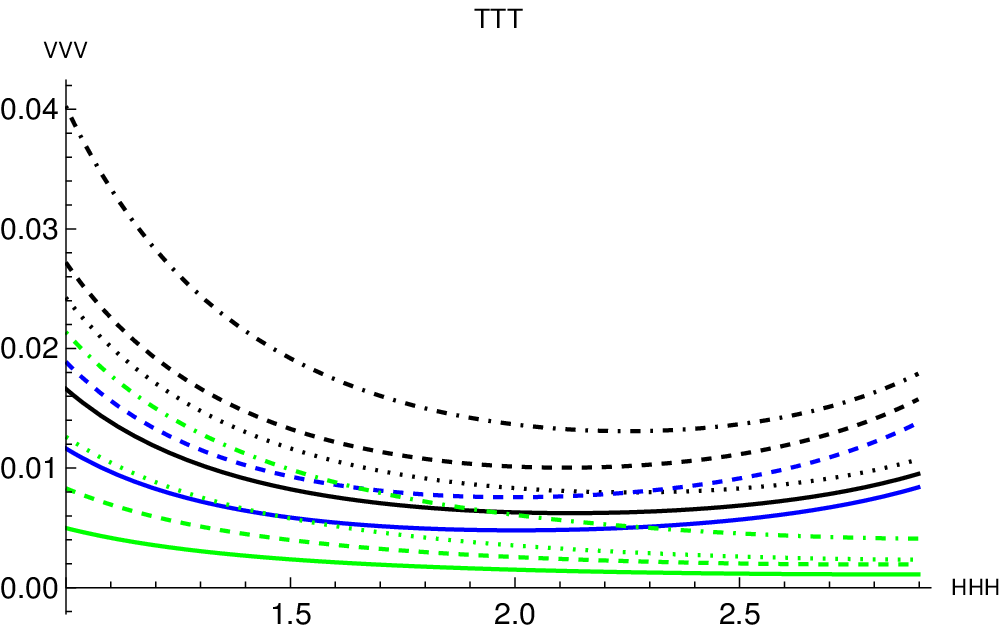}}
		\psfrag{VVV}{\raisebox{.3cm}{\scalebox{.9}{$\hspace{-.4cm}\displaystyle\left.\frac{d 
						\sigma^{\mathrm{even}}_{\gamma\mesonmn}}{d M^2_{\gamma\mesonmn} d(-u') d(-t)}\right|_{(-t)_{\rm min}}({\rm pb} \cdot {\rm GeV}^{-6})$}}}
		{\includegraphics[width=18pc]{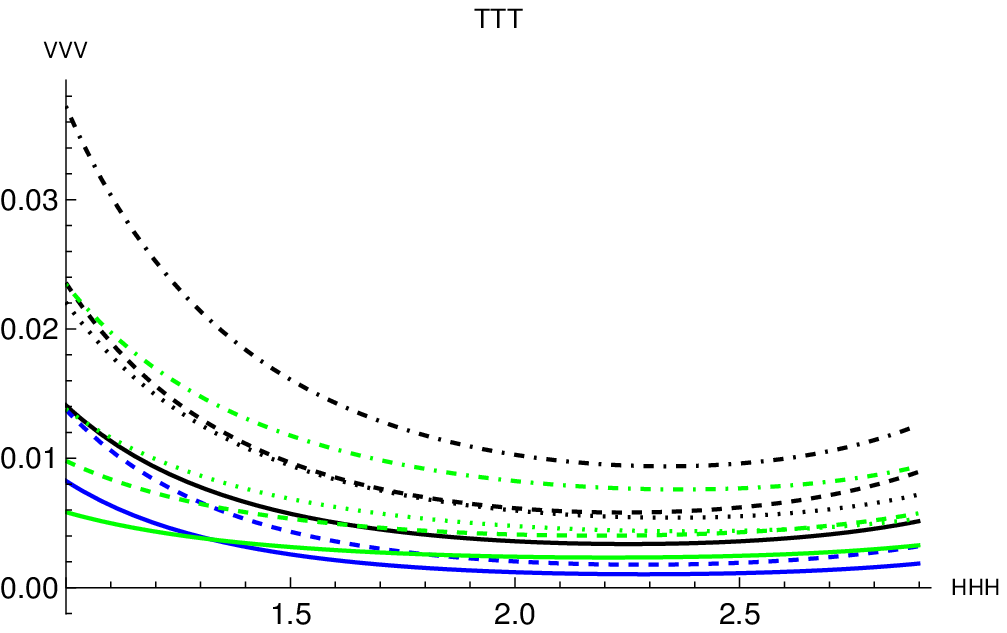}}}
	\vspace{0.2cm}
	\caption{\small The fully differential cross section for {longitudinally polarised} $ \mesonzp,\,\mesonzn,\,\mesonpp,\,\mesonmn$ is shown as a function of $  \left( -u' \right)  $ on the top left, top right, bottom left and bottom right plots respectively. The blue and green curves correspond to contributions from the vector and axial GPDs respectively. The black curves correspond to the total contribution, \ie{}vector and axial GPD contributions combined. As before, the dashed (non-dashed) lines correspond to holographic (asymptotic) DA, while the dotted (non-dotted) lines correspond to the standard (valence) scenario. We fix $ S_{\gamma N}= 20000\,  \mathrm{GeV}^{2}  $ and $ M_{\gamma \meson}^{2}= 4\,  \mathrm{GeV}^{2}  $. Note that the vector contributions consist of only two curves in each case, since they are insensitive to either valence or standard scenarios.}
	\label{fig:EIC-LHC-UPC-fully-diff-VandA}
\end{figure}

The relative contributions of the vector and axial GPDs to the fully differential cross section is shown in \FIG\ref{fig:EIC-LHC-UPC-fully-diff-VandA} as a function of $  \left( -u' \right)  $. The plots are generated using $ \Msq= 4\GeV^2 $ and $ \SgN = 20000 \GeV^2 $.

\begin{figure}[t!]
	\psfrag{HHH}{\hspace{-1.5cm}\raisebox{-.6cm}{\scalebox{.8}{$-u' ({\rm 
					GeV}^{2})$}}}
	\psfrag{VVV}{\raisebox{.3cm}{\scalebox{.9}{$\hspace{-.4cm}\displaystyle\left.\frac{d 
					\sigma^{\mathrm{even}}_{\gamma\mesonzp}}{d M^2_{\gamma\mesonzp} d(-u') d(-t)}\right|_{(-t)_{\rm min}}({\rm pb} \cdot {\rm GeV}^{-6})$}}}
	\psfrag{TTT}{}
	{
		{\includegraphics[width=18pc]{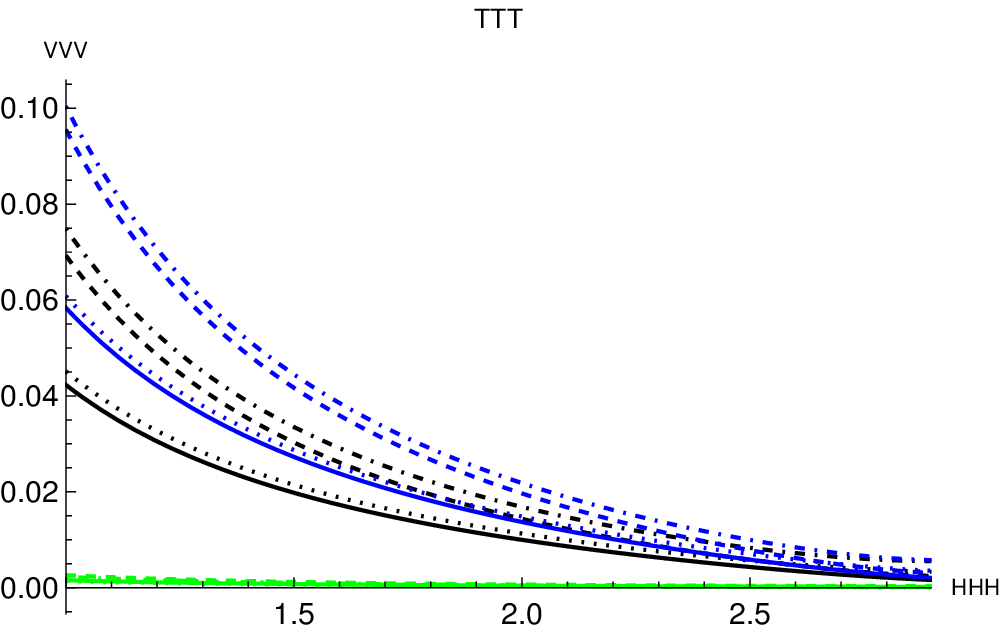}}
		\psfrag{VVV}{\raisebox{.3cm}{\scalebox{.9}{$\hspace{-.4cm}\displaystyle\left.\frac{d 
						\sigma^{\mathrm{even}}_{\gamma\mesonzn}}{d M^2_{\gamma \mesonzn} d(-u') d(-t)}\right|_{(-t)_{\rm min}}({\rm pb} \cdot {\rm GeV}^{-6})$}}}
		{\includegraphics[width=18pc]{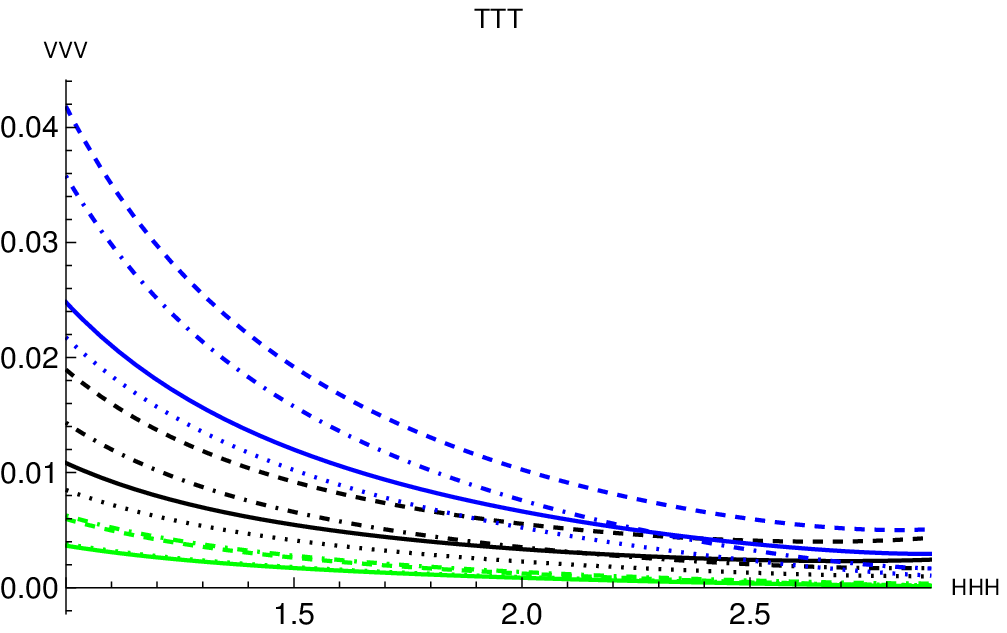}}}
	\\[25pt]
	\psfrag{VVV}{\raisebox{.3cm}{\scalebox{.9}{$\hspace{-.4cm}\displaystyle\left.\frac{d 
					\sigma^{\mathrm{even}}_{\gamma\mesonpp}}{d M^2_{\gamma\mesonpp} d(-u') d(-t)}\right|_{(-t)_{\rm min}}({\rm pb} \cdot {\rm GeV}^{-6})$}}}
	\psfrag{TTT}{}
	{
		{\includegraphics[width=18pc]{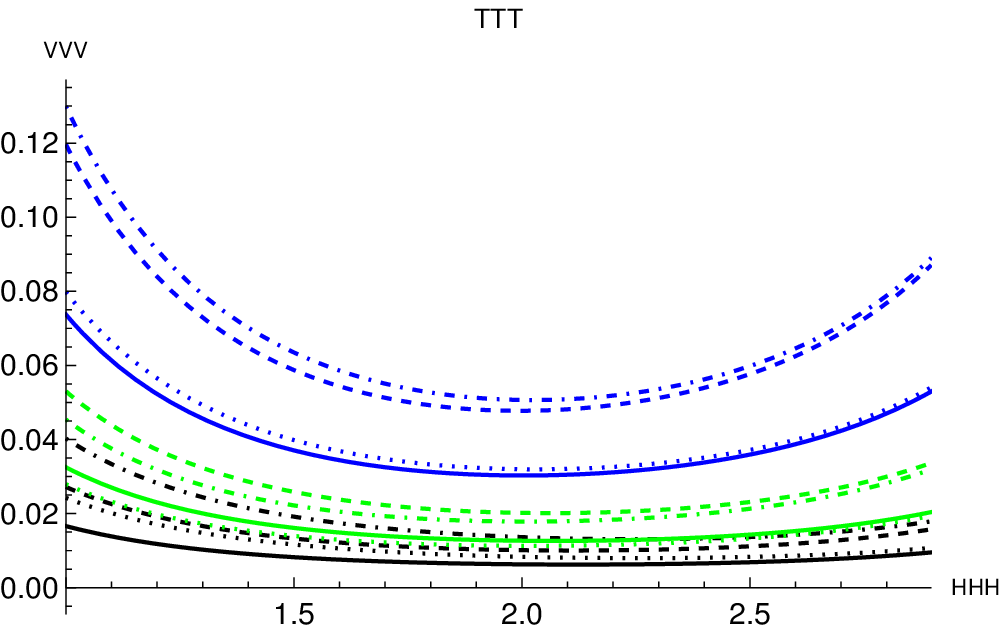}}
		\psfrag{VVV}{\raisebox{.3cm}{\scalebox{.9}{$\hspace{-.4cm}\displaystyle\left.\frac{d 
						\sigma^{\mathrm{even}}_{\gamma\mesonmn}}{d M^2_{\gamma\mesonmn} d(-u') d(-t)}\right|_{(-t)_{\rm min}}({\rm pb} \cdot {\rm GeV}^{-6})$}}}
		{\includegraphics[width=18pc]{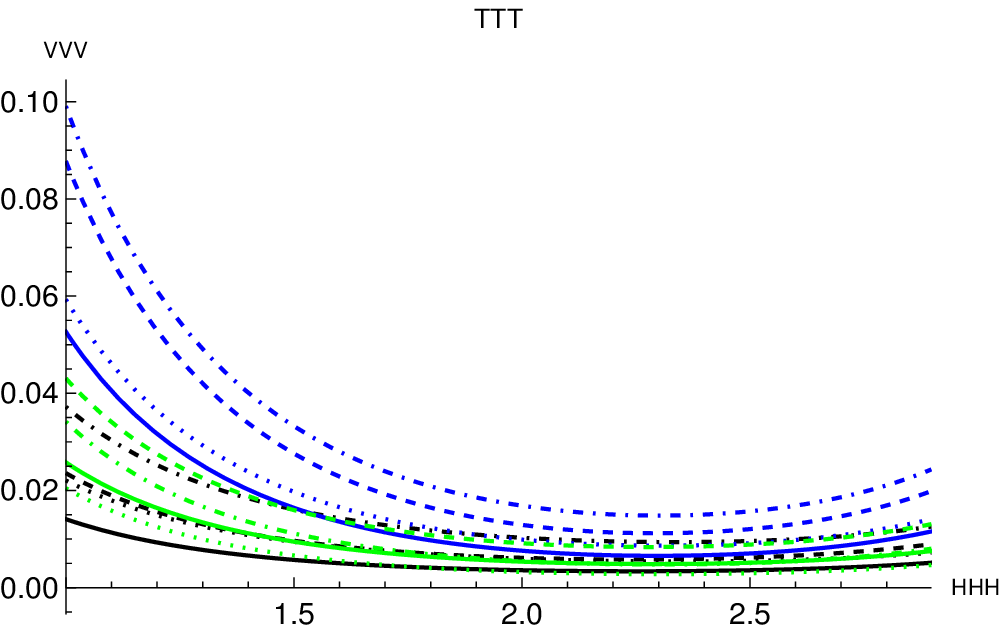}}}
	\vspace{0.2cm}
	\caption{\small The fully differential cross section for {longitudinally polarised} $ \mesonzp,\,\mesonzn,\,\mesonpp,\,\mesonmn$ is shown as a function of $  \left( -u' \right)  $ on the top left, top right, bottom left and bottom right plots respectively. The blue and green curves correspond to contributions from the $u$-quark ($ H_{u} $ and $  \tilde{H} _{u} $) and $d$-quark ($ H_{d} $ and $  \tilde{H} _{d} $) GPDs respectively. The black curves correspond to the total contribution. Otherwise, conventions are the same as in previous plots. We fix $ S_{\gamma N}= 20000\,  \mathrm{GeV}^{2}  $ and $ M_{\gamma \meson}^{2}= 4\,  \mathrm{GeV}^{2}  $.}
	\label{fig:EIC-LHC-UPC-fully-diff-uandd}
\end{figure}

To conclude this subsection, we show the relative contributions of the $ u $-quark and $ d $-quark GPDs to the fully differential cross section in \FIG\ref{fig:EIC-LHC-UPC-fully-diff-uandd}, as a function of $ (-u') $. The value of $ \SgN $ is fixed at $ 20000 \GeV^2 $ and $ \Msq $ at  $ 4 \GeV^2 $.

\FloatBarrier

\subsubsection{Single differential cross section}

\label{sec:EIC-LHC-UPC-single-diff-X-section}

\begin{figure}[t!]
	\psfrag{HHH}{\hspace{-1.5cm}\raisebox{-.6cm}{\scalebox{.8}{$\Msq ({\rm 
					GeV}^{2})$}}}
	\psfrag{VVV}{\raisebox{.3cm}{\scalebox{.9}{$\hspace{-.4cm}\displaystyle\frac{d 
					\sigma^{\mathrm{even}}_{\gamma\mesonzp}}{d M^2_{\gamma\mesonzp}}({\rm pb} \cdot {\rm GeV}^{-2})$}}}
	\psfrag{TTT}{}
	{
		{\includegraphics[width=18pc]{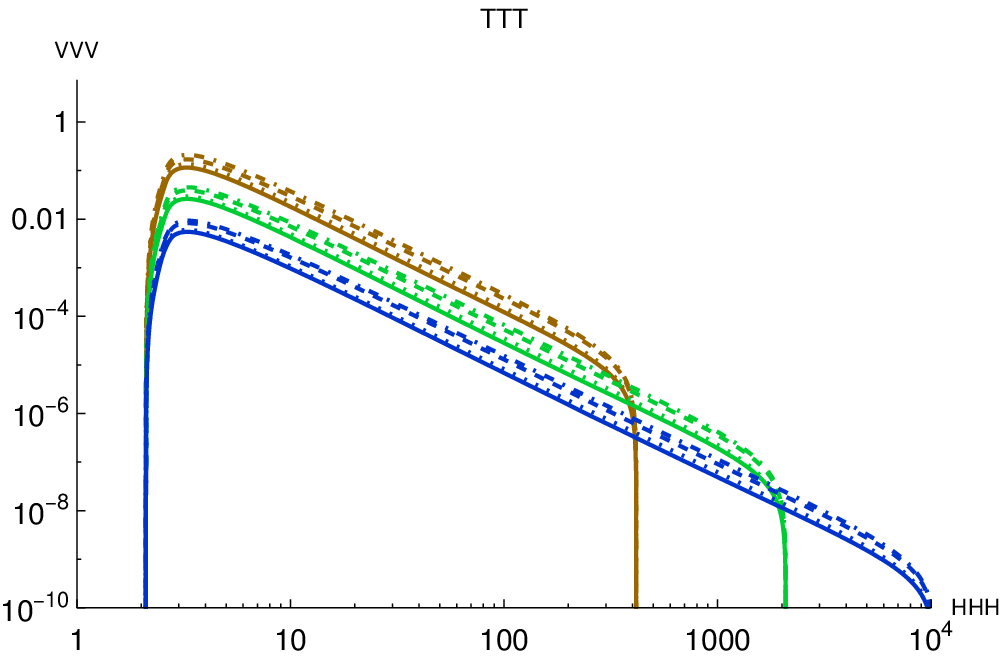}}
		\psfrag{VVV}{\raisebox{.3cm}{\scalebox{.9}{$\hspace{-.4cm}\displaystyle\frac{d 
						\sigma^{\mathrm{even}}_{\gamma\mesonzn}}{d M^2_{\gamma \mesonzn}}({\rm pb} \cdot {\rm GeV}^{-2})$}}}
		{\includegraphics[width=18pc]{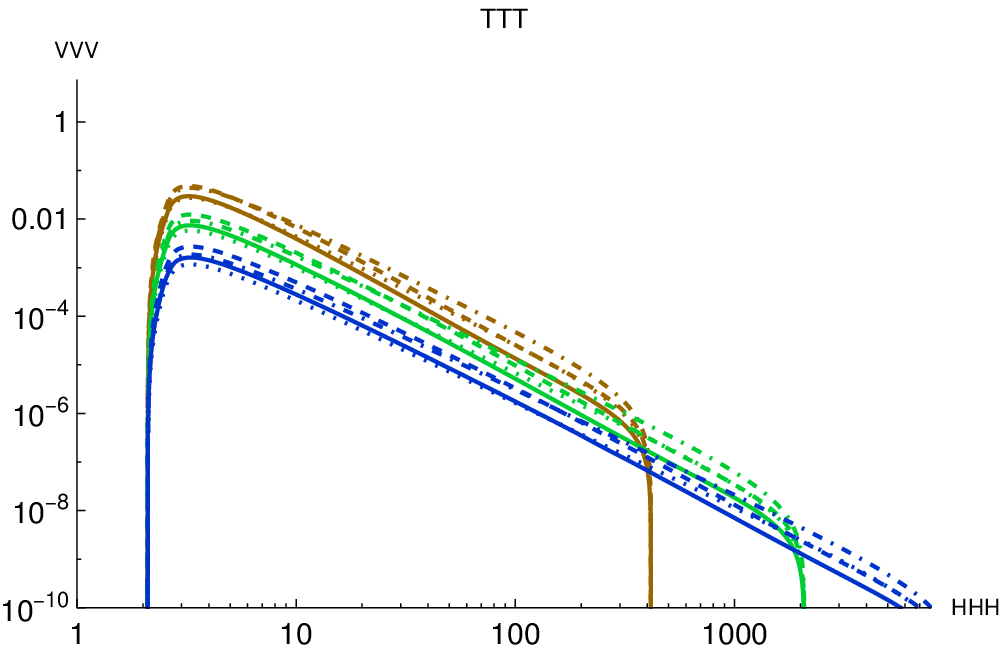}}}
	\\[25pt]
	\psfrag{VVV}{\raisebox{.3cm}{\scalebox{.9}{$\hspace{-.4cm}\displaystyle\frac{d 
					\sigma^{\mathrm{even}}_{\gamma\mesonpp}}{d M^2_{\gamma\mesonpp}}({\rm pb} \cdot {\rm GeV}^{-2})$}}}
	\psfrag{TTT}{}
	{
		{\includegraphics[width=18pc]{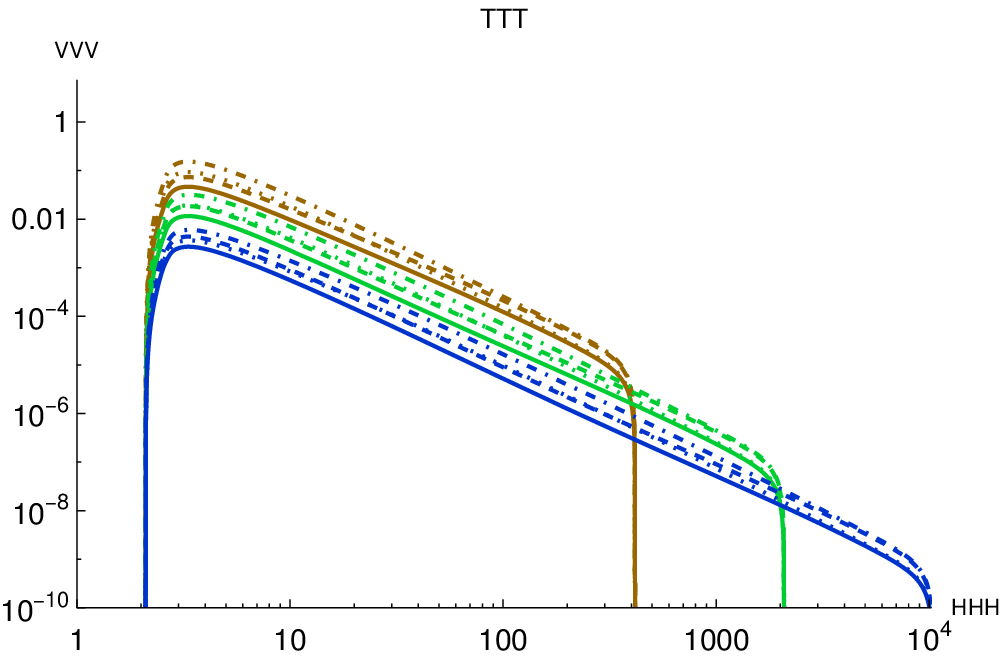}}
		\psfrag{VVV}{\raisebox{.3cm}{\scalebox{.9}{$\hspace{-.4cm}\displaystyle\frac{d 
						\sigma^{\mathrm{even}}_{\gamma\mesonmn}}{d M^2_{\gamma\mesonmn}}({\rm pb} \cdot {\rm GeV}^{-2})$}}}
		{\includegraphics[width=18pc]{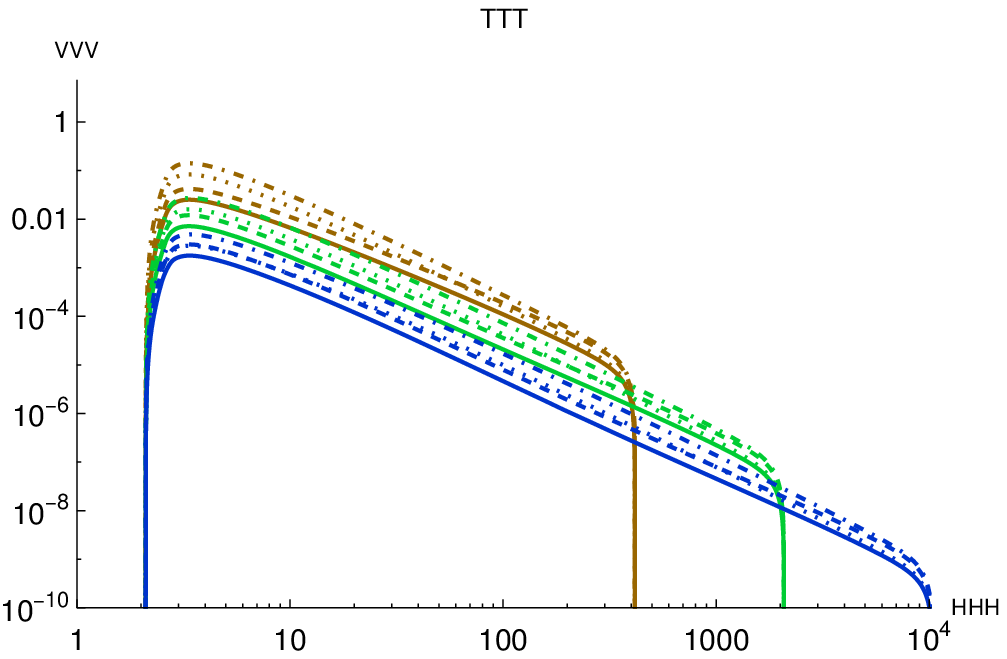}}}
	\vspace{0.2cm}
	\caption{\small The single differential cross section for {longitudinally polarised} $ \mesonzp,\,\mesonzn,\,\mesonpp,\,\mesonmn$ is shown as a function of $  M_{\gamma \meson}^{2}  $ on the top left, top right, bottom left and bottom right plots respectively for different values of $ S_{\gamma N} $. The brown, green and blue curves correspond to $ S_{\gamma N} = 800,\,4000,\,20000\,\GeV^{2} $. The dashed (non-dashed) lines correspond to holographic (asymptotic) DA, while the dotted (non-dotted) lines correspond to the standard (valence) scenario. Note that both axes are log scales.}
	\label{fig:EIC-LHC-UPC-sing-diff}
\end{figure}

We proceed as in \SEC\ref{sec:EIC-fully-diff}, and show only plots for the dominant chiral-even case.

Here, we show the variation of the cross section at the single-differential level as a function of $ \Msq $ for different values of $ \SgN $ in \FIG\ref{fig:EIC-LHC-UPC-sing-diff}. We choose 3 different values for $ \SgN $, namely 800, 4000 and 20000 GeV$ ^{2} $ for the brown, green and blue curves respectively. We observe that the peak of the cross section lies at low values of $ \Msq $ (roughly  3-4 $ \GeV^2 $).

\FloatBarrier

\subsubsection{Integrated cross section}

\begin{figure}[t!]
	\psfrag{HHH}{\hspace{-1.5cm}\raisebox{-.6cm}{\scalebox{.8}{$ S_{\gamma N} ({\rm 
					GeV}^{2})$}}}
	\psfrag{VVV}{\raisebox{.3cm}{\scalebox{.9}{$\hspace{-.4cm}\displaystyle
				\sigma^{\mathrm{even}}_{\gamma\mesonzp}({\rm pb})$}}}
	\psfrag{TTT}{}
	{
		{\includegraphics[width=18pc]{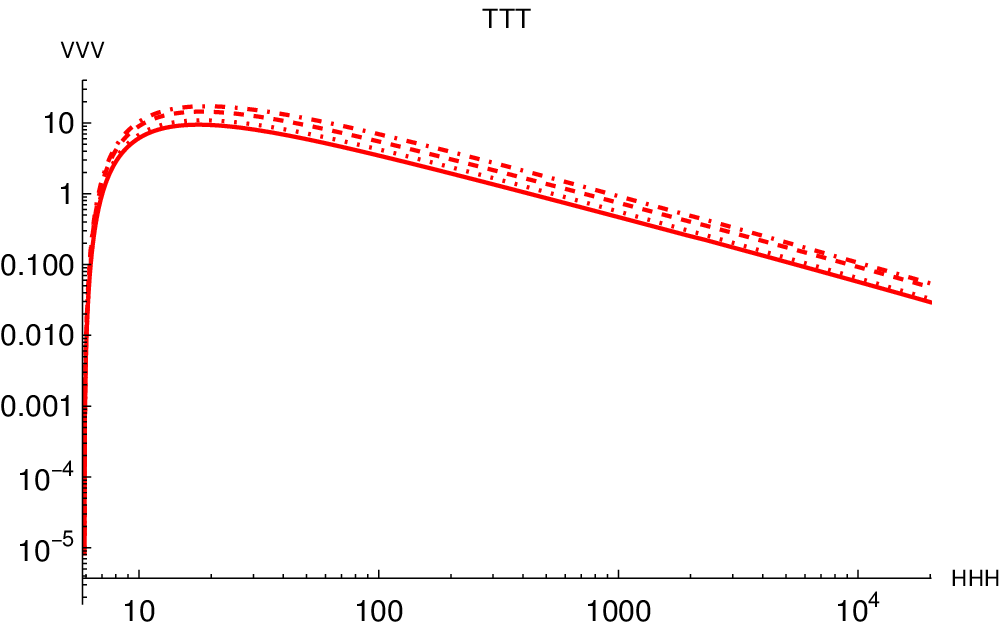}}
		\psfrag{VVV}{\raisebox{.3cm}{\scalebox{.9}{$\hspace{-.4cm}\displaystyle
					\sigma^{\mathrm{even}}_{\gamma\mesonzn}({\rm pb})$}}}
		{\includegraphics[width=18pc]{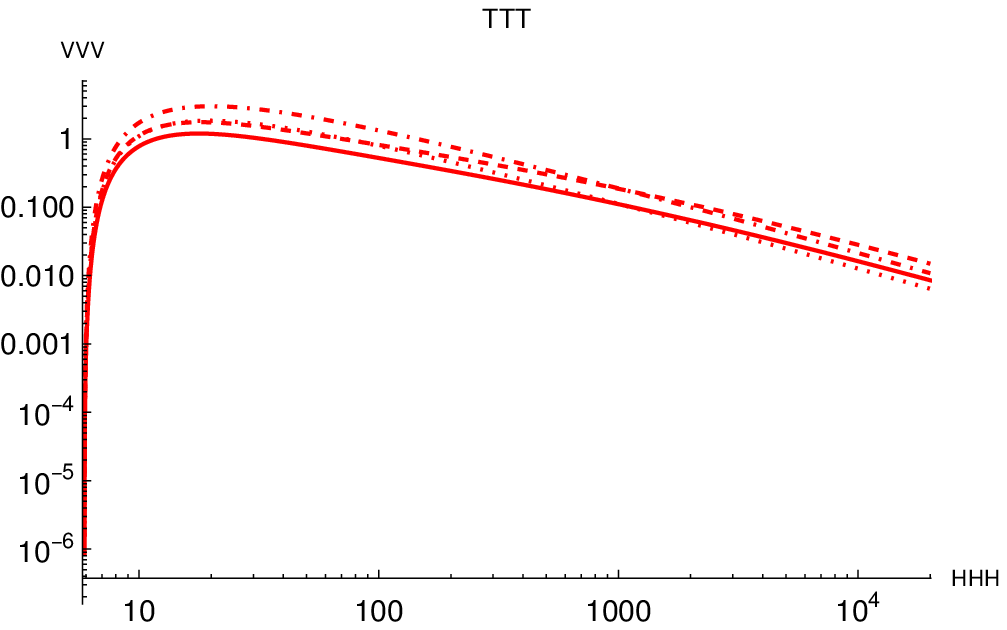}}}
	\\[25pt]
	\psfrag{VVV}{\raisebox{.3cm}{\scalebox{.9}{$\hspace{-.4cm}\displaystyle
				\sigma^{\mathrm{even}}_{\gamma\mesonpp}({\rm pb})$}}}
	\psfrag{TTT}{}
	{
		{\includegraphics[width=18pc]{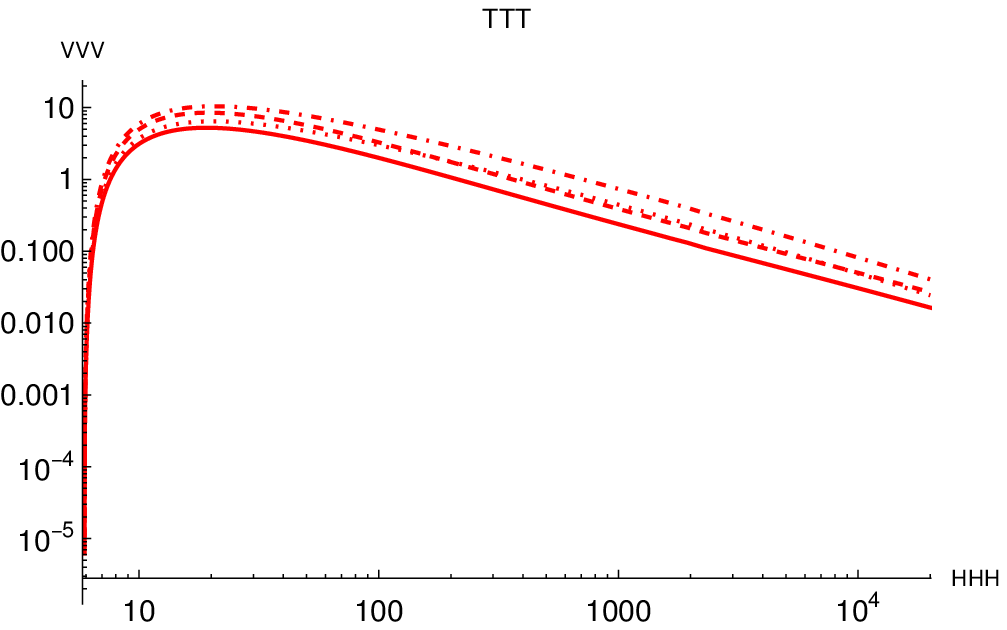}}
		\psfrag{VVV}{\raisebox{.3cm}{\scalebox{.9}{$\hspace{-.4cm}\displaystyle
					\sigma^{\mathrm{even}}_{\gamma\mesonmn}({\rm pb})$}}}
		{\includegraphics[width=18pc]{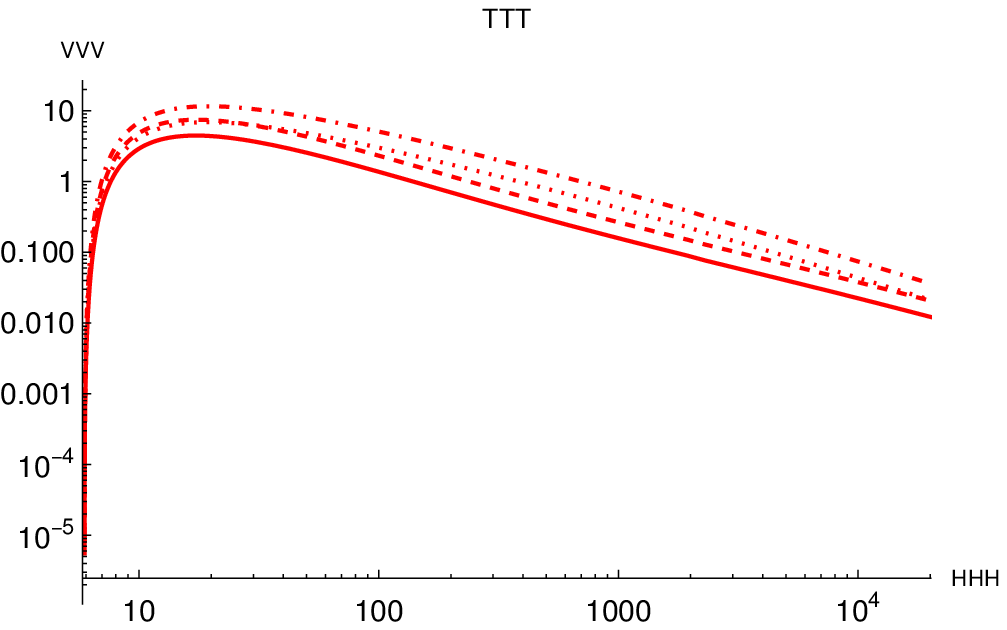}}}
	\vspace{0.2cm}
	\caption{\small The integrated cross section for {longitudinally polarised} $ \mesonzp,\,\mesonzn,\,\mesonpp,\,\mesonmn$ is shown as a function of $   S_{\gamma N}  $ on the top left, top right, bottom left and bottom right plots respectively. The dashed (non-dashed) lines correspond to holographic (asymptotic) DA, while the dotted (non-dotted) lines correspond to the standard (valence) scenario.}
	\label{fig:EIC-LHC-UPC-int-sigma}
\end{figure}

\begin{figure}[t!]
	\psfrag{HHH}{\hspace{-1.5cm}\raisebox{-.6cm}{\scalebox{.8}{$ S_{\gamma N} ({\rm 
					GeV}^{2})$}}}
	\psfrag{VVV}{\raisebox{.3cm}{\scalebox{.9}{$\hspace{-.4cm}\displaystyle
				\sigma^{\mathrm{odd}}_{\gamma\mesonzp}({\rm pb})$}}}
	\psfrag{TTT}{}
	{
		{\includegraphics[width=18pc]{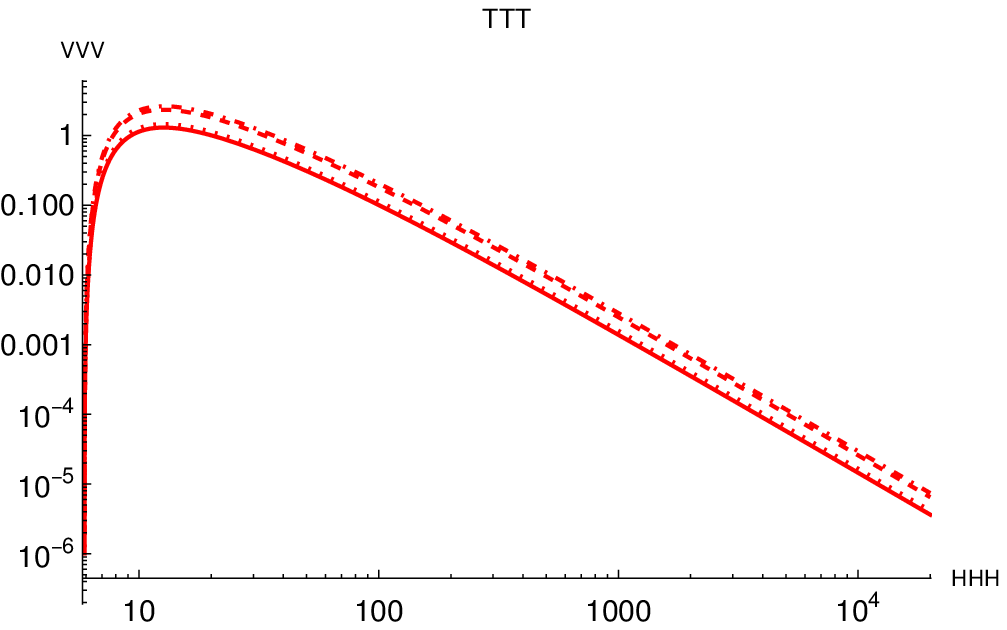}}
		\psfrag{VVV}{\raisebox{.3cm}{\scalebox{.9}{$\hspace{-.4cm}\displaystyle
					\sigma^{\mathrm{odd}}_{\gamma\mesonzn}({\rm pb})$}}}
		{\includegraphics[width=18pc]{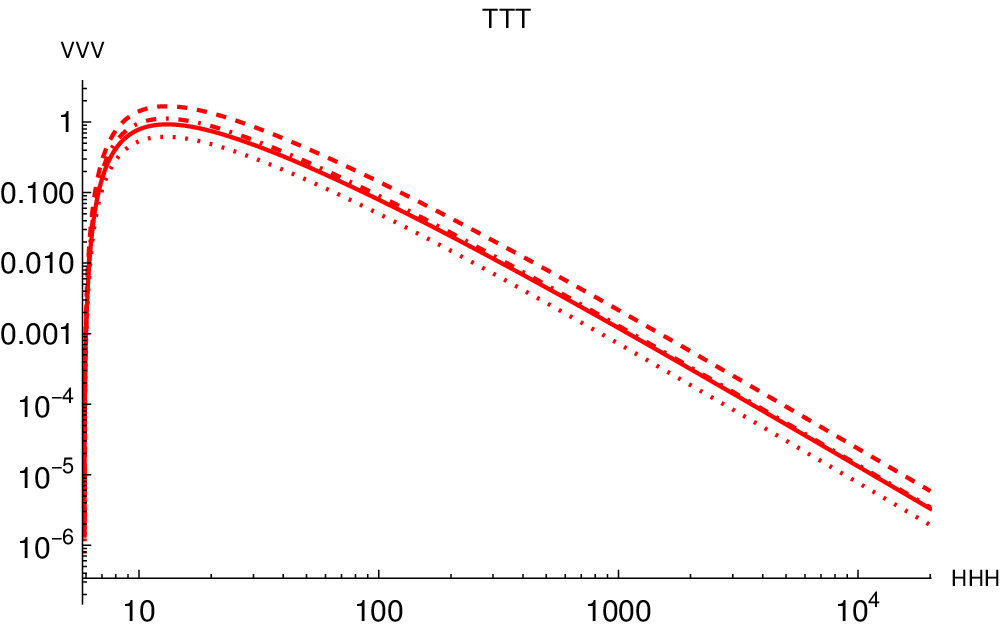}}}
	\\[25pt]
	\psfrag{VVV}{\raisebox{.3cm}{\scalebox{.9}{$\hspace{-.4cm}\displaystyle
				\sigma^{\mathrm{odd}}_{\gamma\mesonpp}({\rm pb})$}}}
	\psfrag{TTT}{}
	{
		{\includegraphics[width=18pc]{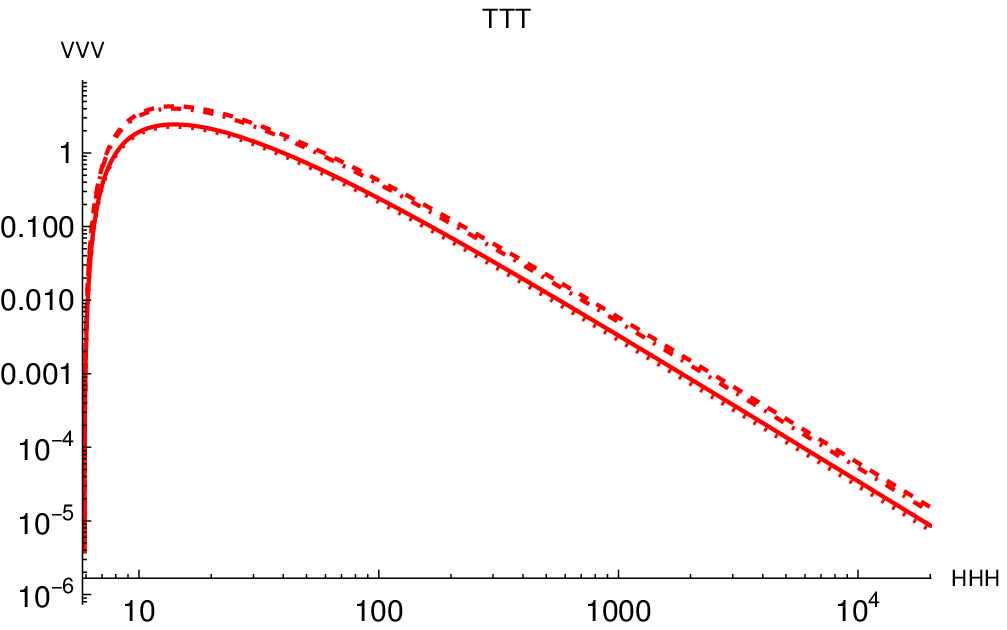}}
		\psfrag{VVV}{\raisebox{.3cm}{\scalebox{.9}{$\hspace{-.4cm}\displaystyle
					\sigma^{\mathrm{odd}}_{\gamma\mesonmn}({\rm pb})$}}}
		{\includegraphics[width=18pc]{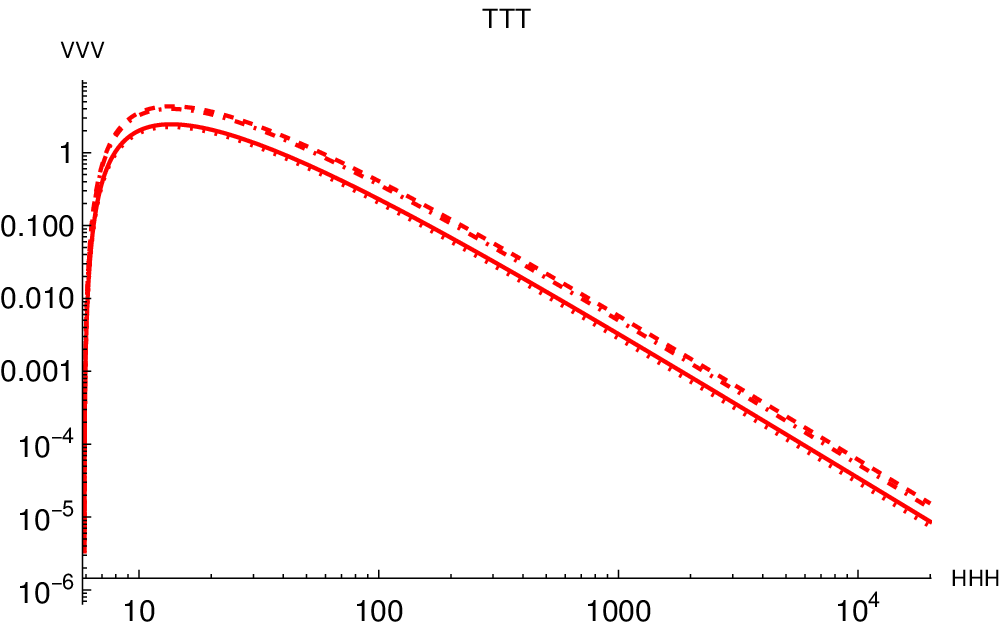}}}
	\vspace{0.2cm}
	\caption{\small The integrated cross section for {transversely polarised} $ \mesonzp,\,\mesonzn,\,\mesonpp,\,\mesonmn$ is shown as a function of $   S_{\gamma N}  $ on the top left, top right, bottom left and bottom right plots respectively. The dashed (non-dashed) lines correspond to holographic (asymptotic) DA, while the dotted (non-dotted) lines correspond to the standard (valence) scenario.}
	\label{fig:co-EIC-LHC-UPC-int-sigma}
\end{figure}

In \FIG\ref{fig:EIC-LHC-UPC-int-sigma}, the variation of the integrated  cross section as a function of $ \SgN $ is shown for the chiral-even case. We observe that the largest cross section is obtained by using a holographic DA model and the GPD model  corresponding to the standard scenario (dashed-dotted line). We note that the cross section falls to very low values at $ \SgN=20000 \GeV^2 $, roughly 200 times less than its value at the peak, which occurs at around $ 20 \GeV^2 $. This, coupled with the fact that the photon flux in UPCs also decreases with $ \SgN $, justifies the truncation at $ \SgN=20000 \GeV^2 $ when considering UPCs at LHC kinematics, which involves TeV energies.

The corresponding plots for the chiral-odd case is shown in \FIG\ref{fig:co-EIC-LHC-UPC-int-sigma}. Here, we observe that the cross section, after the peak, drops at a much faster rate compared to the chiral-even case. In fact, the cross section at $ \SgN = 20000 \GeV^2 $ drops by a factor of $ 10^6 $ roughly compared to its value at the peak.

\FloatBarrier

\subsubsection{Polarisation asymmetries}

\begin{figure}[t!]
	\psfrag{HHH}{\hspace{-1.5cm}\raisebox{-.6cm}{\scalebox{.8}{$-u' ({\rm 
					GeV}^{2})$}}}
	\psfrag{VVV}{LPA$^{\gamma\mesonzp}_{\mathrm{max}} $}
	\psfrag{TTT}{}
	{
		{\includegraphics[width=18pc]{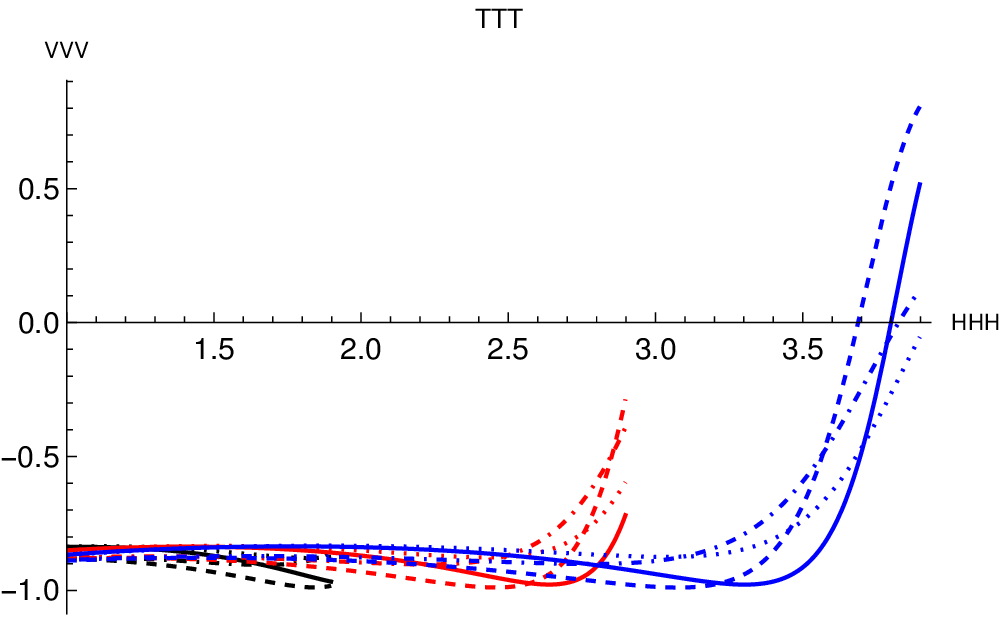}}
		\psfrag{VVV}{LPA$^{\gamma\mesonzn}_{\mathrm{max}} $}
		{\includegraphics[width=18pc]{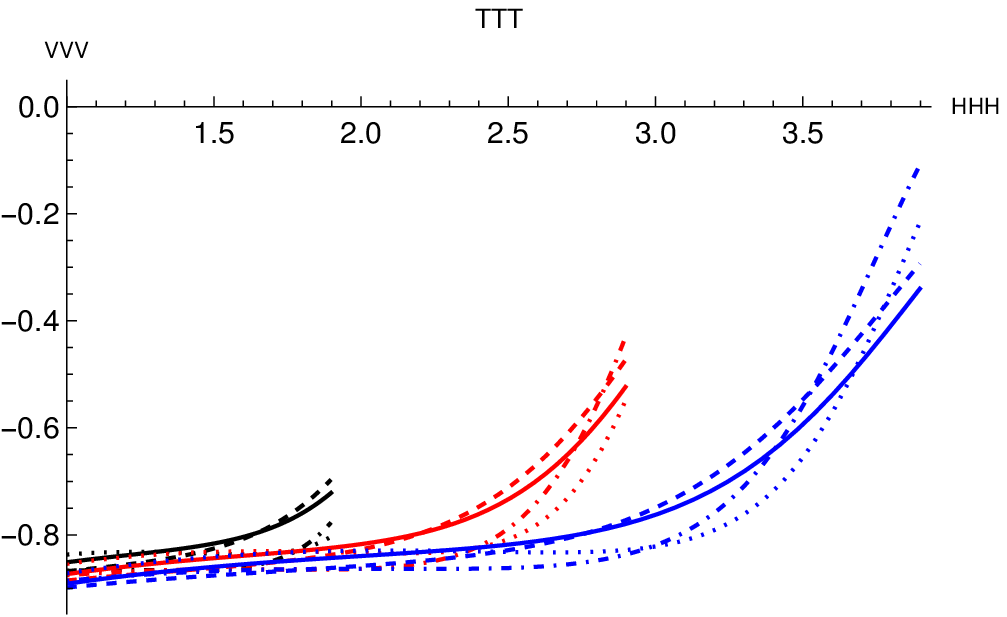}}}
	\\[25pt]
	{
		{\includegraphics[width=18pc]{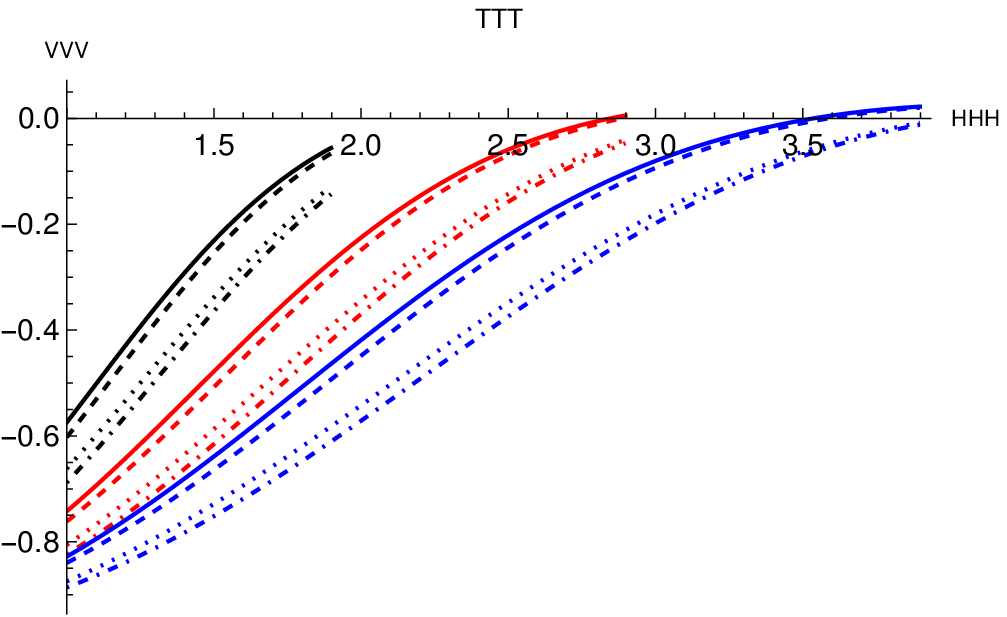}}
		\psfrag{VVV}{LPA$^{\gamma\mesonmn}_{\mathrm{max}} $}
		{\includegraphics[width=18pc]{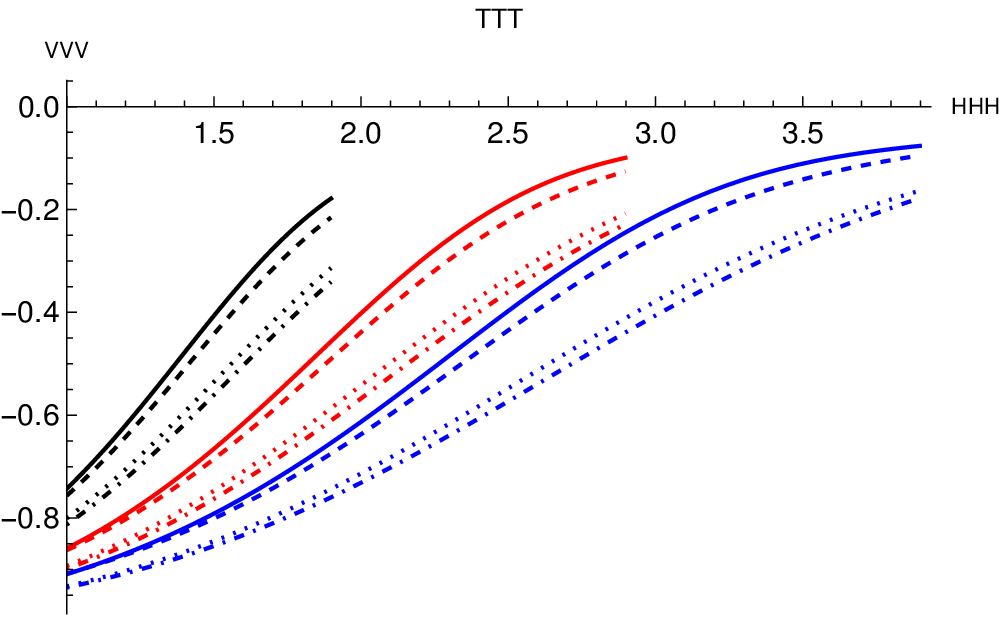}}}
	\vspace{0.2cm}
	\caption{\small The LPA at the fully-differential level for  longitudinally polarised $ \mesonzp,\,\mesonzn,\,\mesonpp,\,\mesonmn$ is shown as a function of $  \left( -u' \right)  $ on the top left, top right, bottom left and bottom right plots respectively for different values of $ M_{\gamma \meson}^2 $. The black, red and blue curves correspond to $ M_{\gamma \meson}^{2}=3,\,4,\,5\, $ GeV$ ^2 $ respectively, and $ \SgN = 20000 \GeV^{2}$. The same conventions as in \FIG\ref{fig:EIC-LHC-UPC-fully-diff-diff-M2} are used here.}
	\label{fig:EIC-LHC-UPC-pol-asym-fully-diff-diff-M2}
\end{figure}

\begin{figure}[t!]
	\psfrag{HHH}{\hspace{-1.5cm}\raisebox{-.6cm}{\scalebox{.8}{$-u' ({\rm 
					GeV}^{2})$}}}
	\psfrag{VVV}{LPA$^{\gamma\mesonzp}_{\mathrm{max}} $}
	\psfrag{TTT}{}
	{
		{\includegraphics[width=18pc]{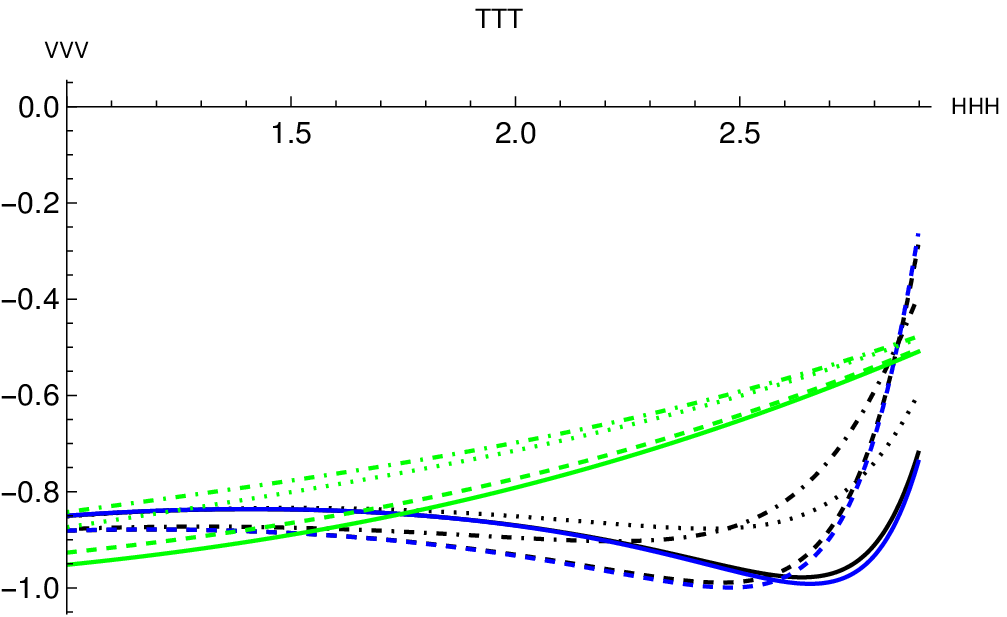}}
		\psfrag{VVV}{LPA$^{\gamma\mesonzn}_{\mathrm{max}} $}
		{\includegraphics[width=18pc]{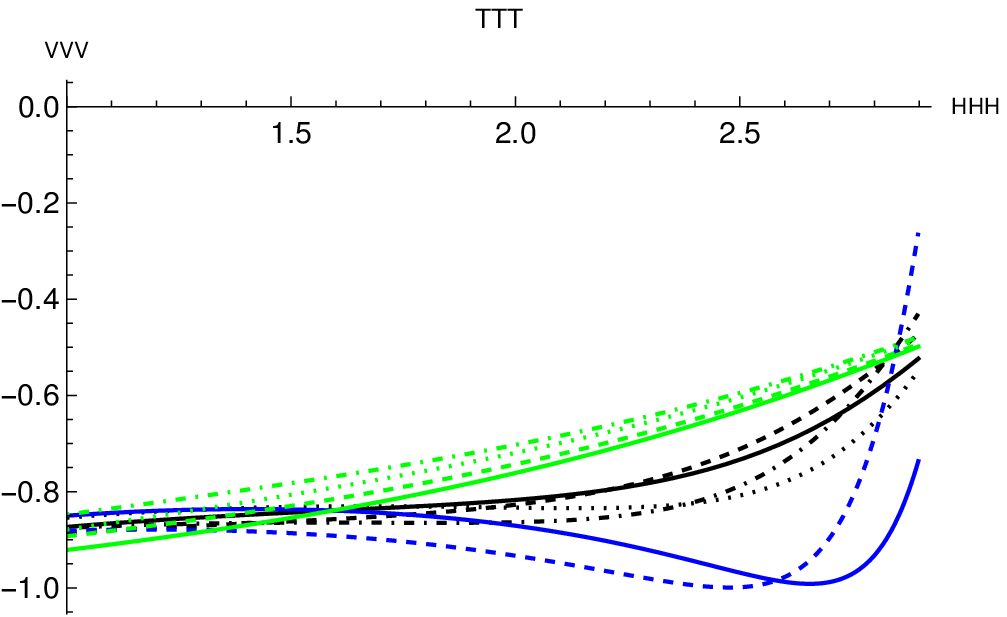}}}
	\\[25pt]
	{					\psfrag{VVV}{LPA$^{\gamma\mesonpp}_{\mathrm{max}} $}
		{\includegraphics[width=18pc]{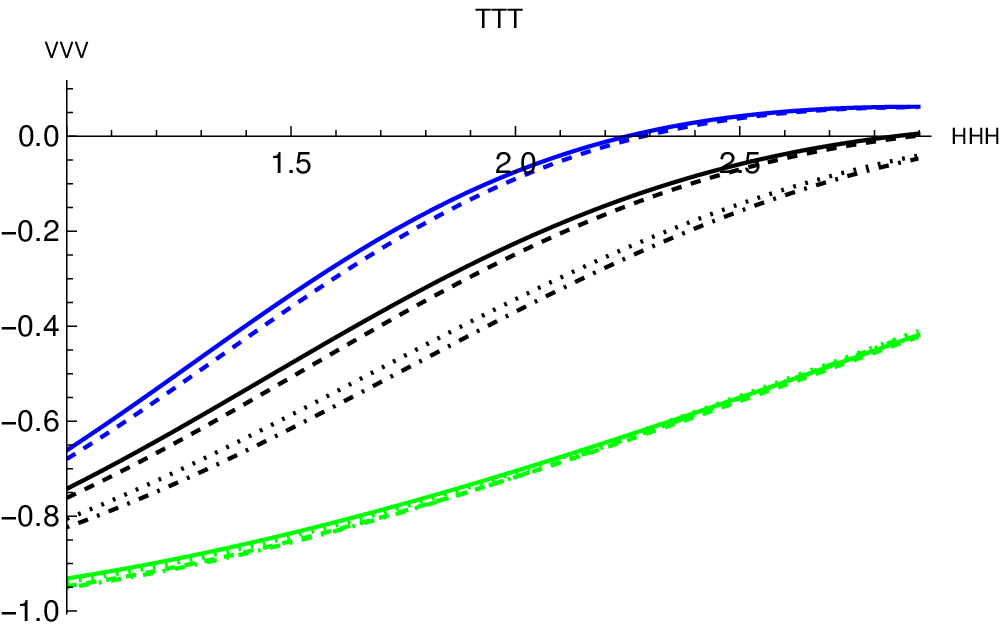}}
		\psfrag{VVV}{LPA$^{\gamma\mesonmn}_{\mathrm{max}} $}
		{\includegraphics[width=18pc]{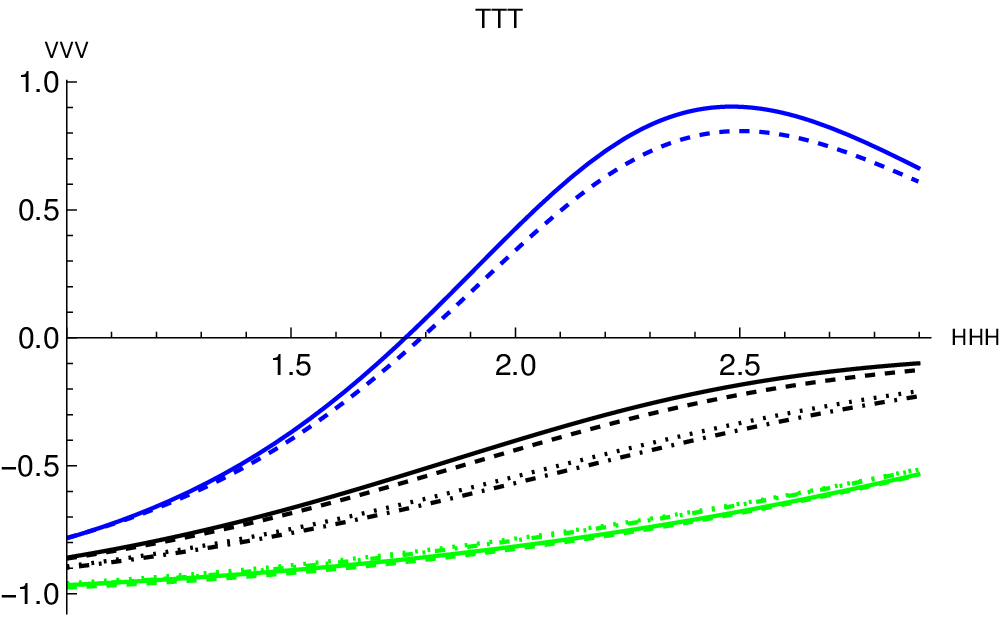}}}
	\vspace{0.2cm}
	\caption{\small The LPA at the fully-differential level for  longitudinally polarised $ \mesonzp,\,\mesonzn,\,\mesonpp,\,\mesonmn$ is shown as a function of $  \left( -u' \right)  $ on the top left, top right, bottom left and bottom right plots respectively, using $ \Msq = 4 \GeV^2$ and $ \SgN = 20000 \GeV^2 $. The same conventions as in \FIG\ref{fig:EIC-LHC-UPC-fully-diff-VandA} are used here. Note that the vector contributions consist of only two curves in each case, since they are insensitive to either valence or standard scenarios.}
	\label{fig:EIC-LHC-UPC-pol-asym-fully-diff-VandA}
\end{figure}

We recall that the LPA for the chiral-odd case vanishes, and therefore, we only show plots for the chiral-even case in this section.

In \FIG\ref{fig:EIC-LHC-UPC-pol-asym-fully-diff-diff-M2}, the LPA at the fully-differential level is shown as a function of $  \left( -u' \right)  $, for different value of $ \Msq $. The kinematical values chosen are $ \SgN = 20000 \GeV^2 $, and $ \Msq = 3,\,4,\,5 \GeV^2 $. The behaviour of the LPA is similar to the ones described in previous sections.

We show the relative contributions from the vector and axial GPDs to the LPA at the fully-differential level in \FIG\ref{fig:EIC-LHC-UPC-pol-asym-fully-diff-VandA}, as a function of $  \left( -u' \right)  $. We choose $ \SgN =20000 \GeV^2 $ and 
$ \Msq=4 \GeV^2 $ to generate the plots.

Next, the relative contributions from the vector and axial GPDs to the LPA are shown in \FIG\ref{fig:EIC-LHC-UPC-pol-asym-fully-diff-VandA}. $ \SgN =20000 \GeV^2$ and $ \Msq=4 \GeV^2 $ were used to generate the plots. As before, the axial GPD contributions using the standard and valence scenarios are significantly different, while the DA model has little effect on the LPA. Similar comments as before apply.

\begin{figure}[t!]
	\psfrag{HHH}{\hspace{-1.5cm}\raisebox{-.6cm}{\scalebox{.8}{$-u' ({\rm 
					GeV}^{2})$}}}
		\psfrag{VVV}{LPA$^{\gamma\mesonzp}_{\mathrm{max}} $}
	\psfrag{TTT}{}
	{
		{\includegraphics[width=18pc]{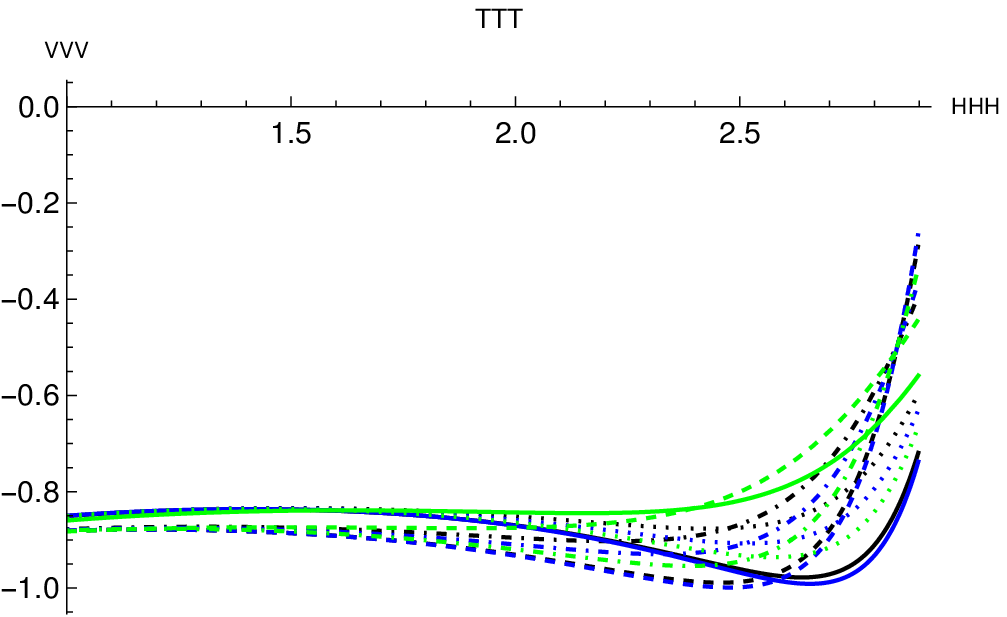}}
			\psfrag{VVV}{LPA$^{\gamma\mesonzn}_{\mathrm{max}} $}
		{\includegraphics[width=18pc]{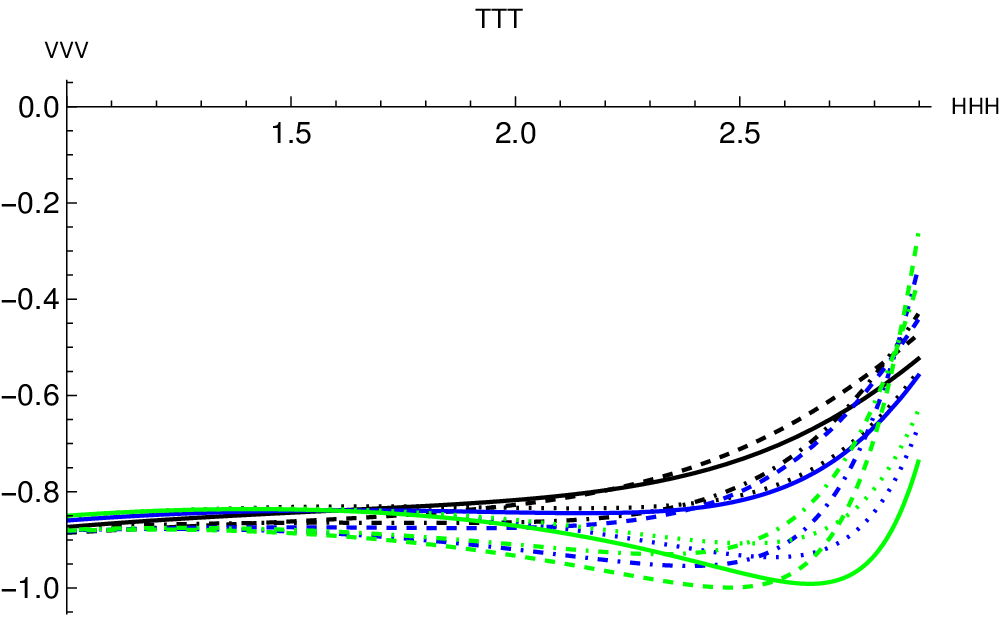}}}
	\\[25pt]
	{					\psfrag{VVV}{LPA$^{\gamma\mesonpp}_{\mathrm{max}} $}
		{\includegraphics[width=18pc]{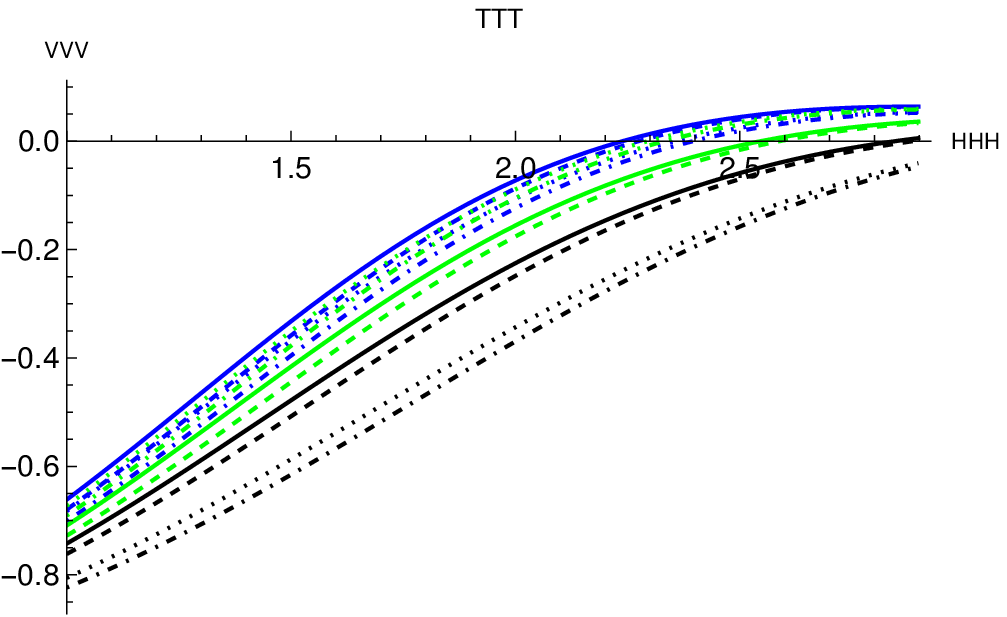}}
			\psfrag{VVV}{LPA$^{\gamma\mesonmn}_{\mathrm{max}} $}
		{\includegraphics[width=18pc]{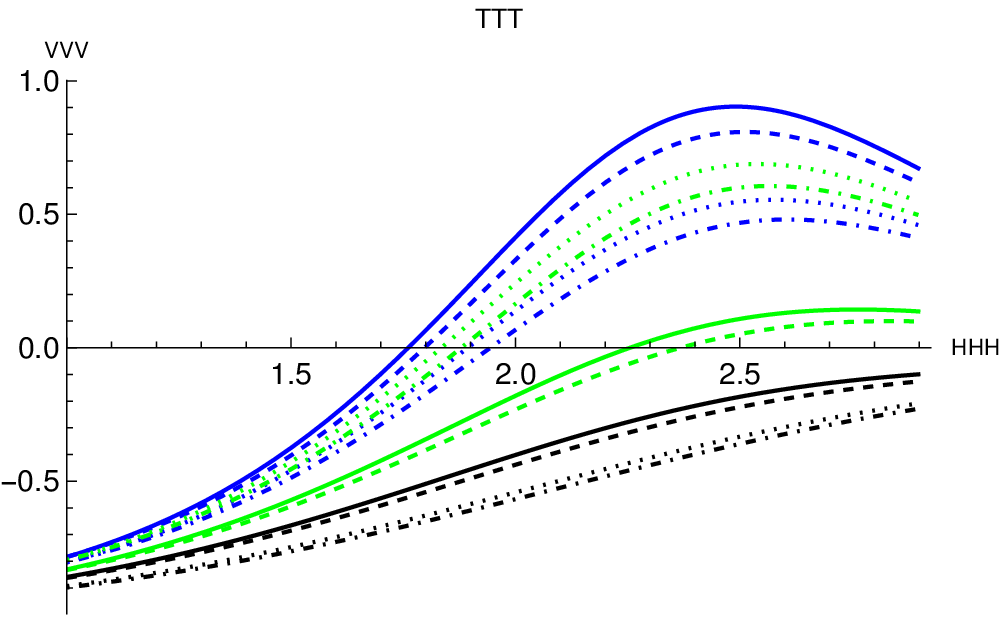}}}
	\vspace{0.2cm}
	\caption{\small The LPA at the fully-differential level for longitudinally polarised  $ \mesonzp,\,\mesonzn,\,\mesonpp,\,\mesonmn$ is shown as a function of $  \left( -u' \right)  $ on the top left, top right, bottom left and bottom right plots respectively, using $ \Msq = 4 \GeV^2$ and $ \SgN = 20000 \GeV^2 $. The blue and green curves correspond to contributions from the $u$-quark ($ H_{u} $ and $  \tilde{H} _{u} $) and $d$-quark ($ H_{d} $ and $  \tilde{H} _{d} $) GPDs respectively. The black curves correspond to the total contribution. The same conventions as in \FIG\ref{fig:EIC-LHC-UPC-fully-diff-uandd} are used here.}
	\label{fig:EIC-LHC-UPC-pol-asym-fully-diff-uandd}
\end{figure}

Finally, the relative contributions to the LPA from the $ u $-quark and $ d $-quark GPDs are shown in \FIG\ref{fig:EIC-LHC-UPC-pol-asym-fully-diff-uandd} as a function of $  \left( -u' \right)  $. The kinematical values used to generate the plots are $ \SgN = 20000 \GeV^2 $ and $ \Msq = 4 \GeV^2 $.

\FloatBarrier

\begin{figure}[t!]
	\psfrag{HHH}{\hspace{-1.5cm}\raisebox{-.6cm}{\scalebox{.8}{ $ M_{\gamma \meson}^{2}({\rm 
					GeV}^{2}) $}}}
		\psfrag{VVV}{LPA$^{\gamma\mesonzp}_{\mathrm{max}} $}
	\psfrag{TTT}{}
	{
		{\includegraphics[width=18pc]{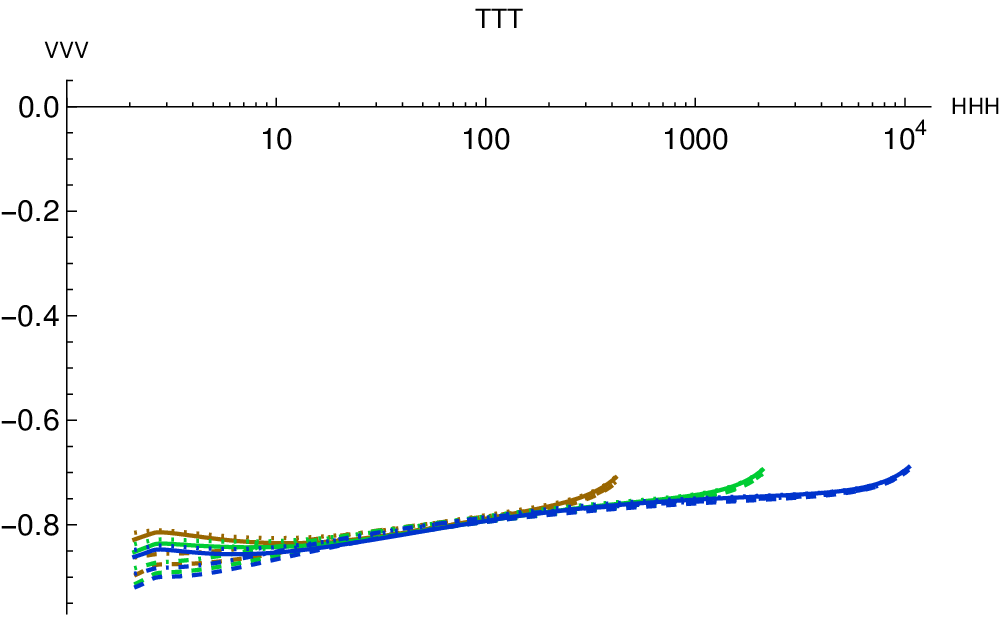}}
			\psfrag{VVV}{LPA$^{\gamma\mesonzn}_{\mathrm{max}} $}
		{\includegraphics[width=18pc]{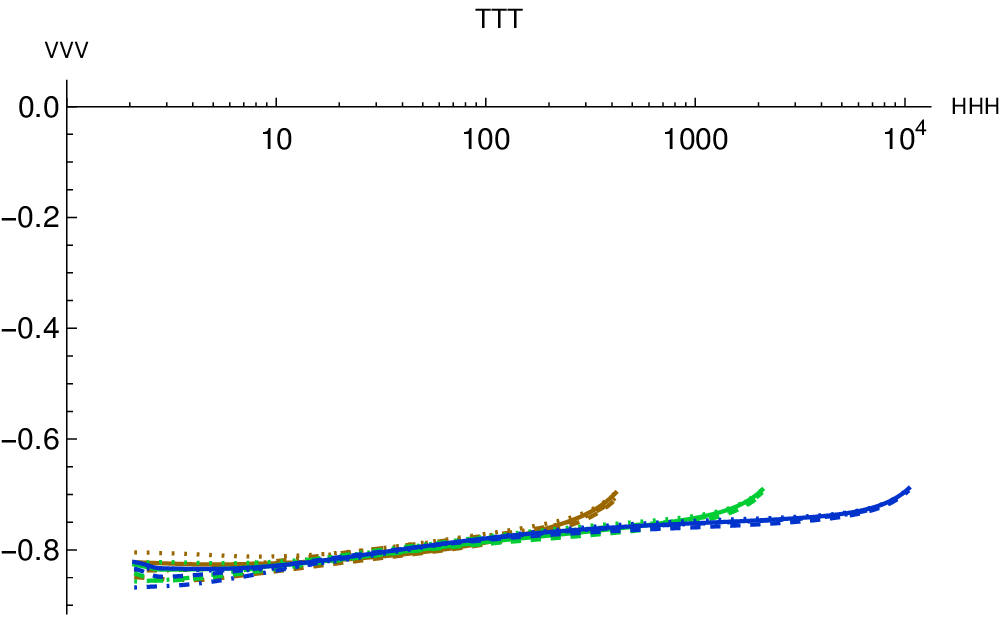}}}
	\\[25pt]
	{					\psfrag{VVV}{LPA$^{\gamma\mesonpp}_{\mathrm{max}} $}
		{\includegraphics[width=18pc]{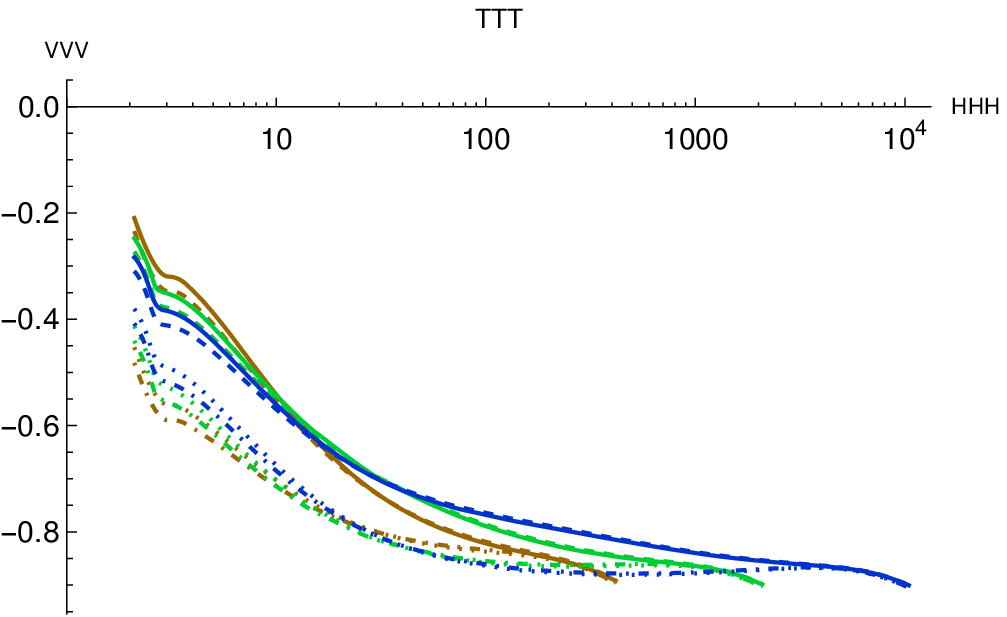}}
			\psfrag{VVV}{LPA$^{\gamma\mesonmn}_{\mathrm{max}} $}
		{\includegraphics[width=18pc]{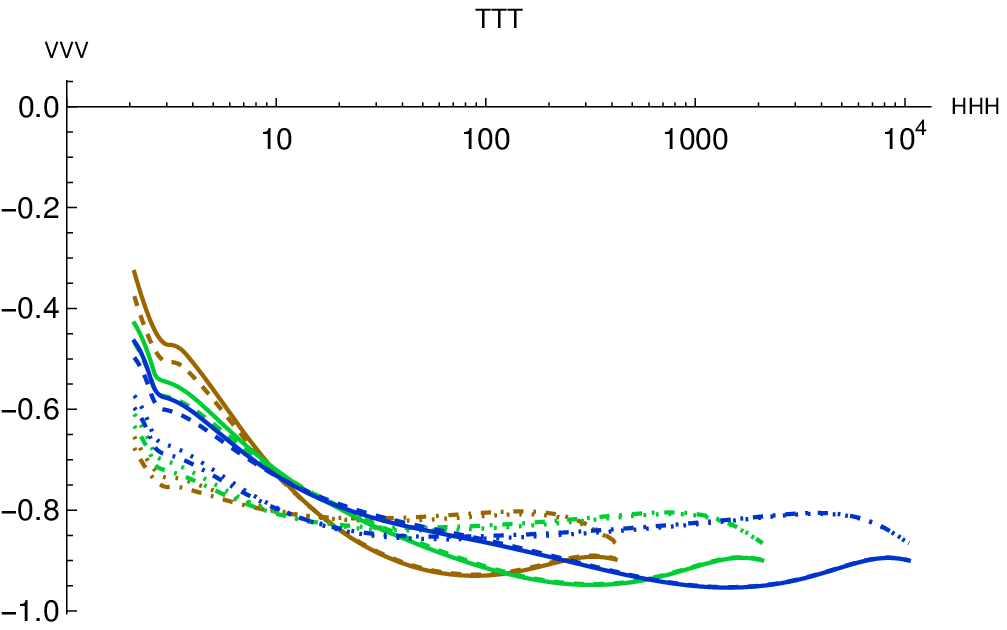}}}
	\vspace{0.2cm}
	\caption{\small The LPA at the single differential level for longitudinally polarised  $ \mesonzp,\,\mesonzn,\,\mesonpp,\,\mesonmn$ is shown as a function of $  M_{\gamma \meson}^{2}  $ on the top left, top right, bottom left and bottom right plots respectively. The brown, green and blue curves correspond to $ S_{\gamma N} = 800,\,4000$ and 20000 GeV$ ^2 $. The same colour and line style conventions as in \FIG\ref{fig:EIC-LHC-UPC-sing-diff} are used here. Note that a log scale is used for the horizontal axis.}
	\label{fig:EIC-LHC-UPC-pol-asym-sing-diff}
\end{figure}

The LPA at the single differential level is shown in \FIG\ref{fig:EIC-LHC-UPC-pol-asym-sing-diff} as a function of $ \Msq $ for different values of $ \SgN $. The 3 values of $ \SgN $ chosen are 800, 4000 and 20000 $ \GeV^2 $ corresponding to the brown, green and blue curves respectively. We observe that the behaviour of the LPA is very similar to the one for COMPASS kinematics in \FIG\ref{fig:compass-pol-asym-sing-diff}.

\FloatBarrier

\begin{figure}[t!]
	\psfrag{HHH}{\hspace{-1.5cm}\raisebox{-.6cm}{\scalebox{.8}{ $ S_{\gamma N}({\rm 
					GeV}^{2}) $}}}
		\psfrag{VVV}{LPA$^{\gamma\mesonzp}_{\mathrm{max}} $}
	\psfrag{TTT}{}
	{
		{\includegraphics[width=18pc]{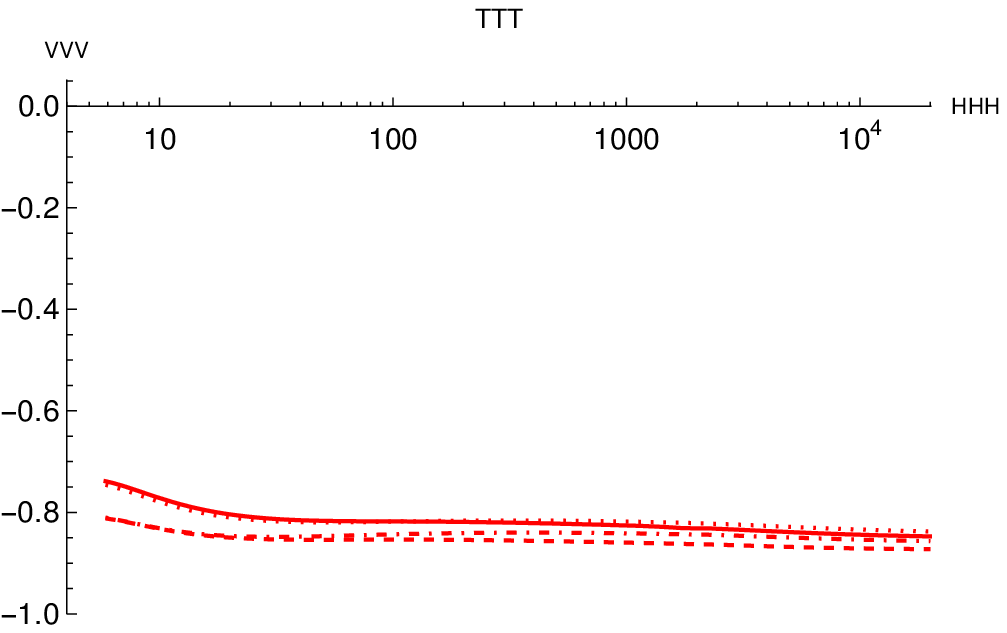}}
			\psfrag{VVV}{LPA$^{\gamma\mesonzn}_{\mathrm{max}} $}
		{\includegraphics[width=18pc]{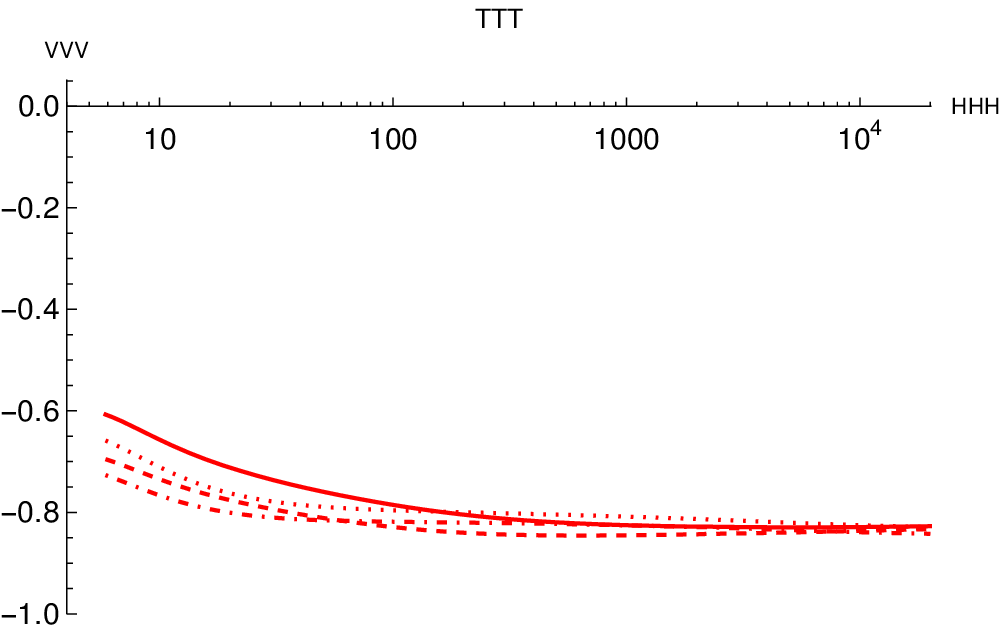}}}
	\\[25pt]
	{					\psfrag{VVV}{LPA$^{\gamma\mesonpp}_{\mathrm{max}} $}
		{\includegraphics[width=18pc]{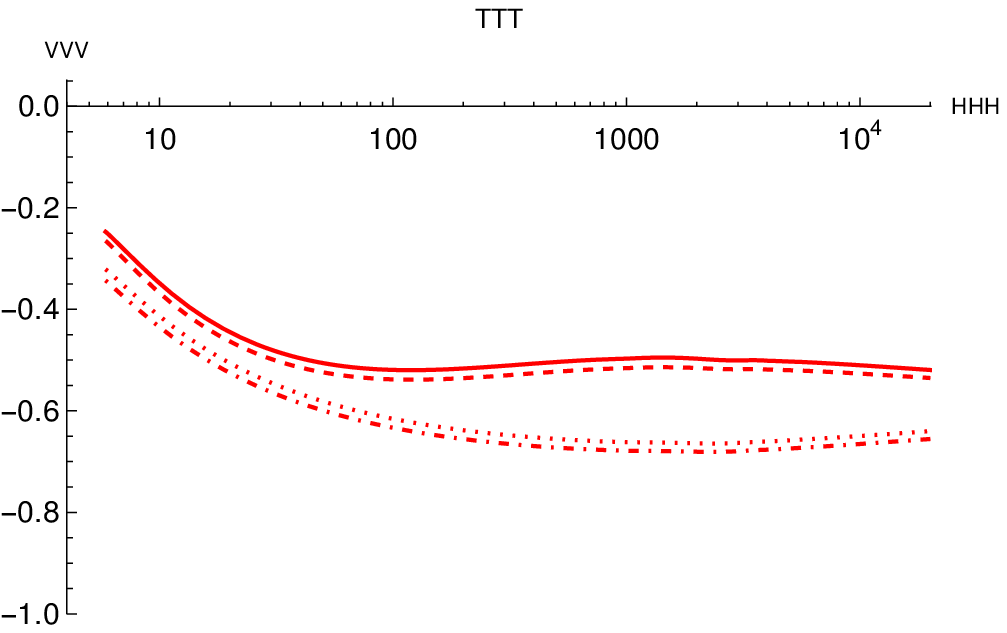}}
			\psfrag{VVV}{LPA$^{\gamma\mesonmn}_{\mathrm{max}} $}
		{\includegraphics[width=18pc]{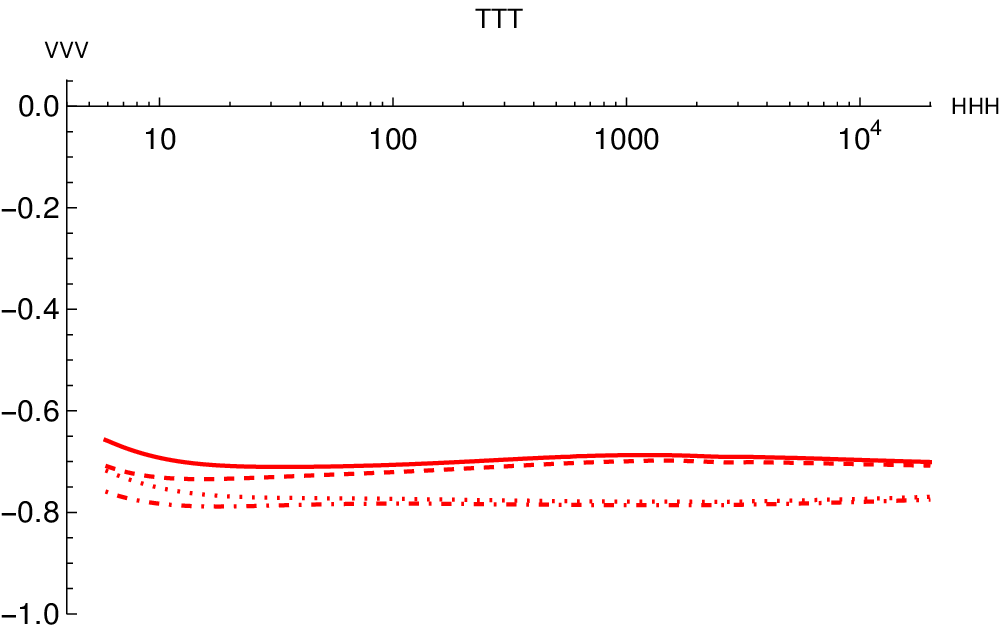}}}
	\vspace{0.2cm}
	\caption{\small The LPA integrated over all differential variables for  longitudinally polarised $ \mesonzp,\,\mesonzn,\,\mesonpp,\,\mesonmn$ is shown on the top left, top right, bottom left and bottom right plots respectively as a function of $ \SgN $. The same colour and line style conventions as in \FIG\ref{fig:EIC-LHC-UPC-int-sigma} are used here.}
	\label{fig:EIC-LHC-UPC-pol-asym-int-sigma}
\end{figure}

To conclude this section, the LPA, computed after integration over the differential variables, is shown as a function of $ \SgN $ in \FIG\ref{fig:EIC-LHC-UPC-pol-asym-int-sigma}. Again, we note that the behaviour of the LPA is very similar to the one corresponding to COMPASS kinematics, see \FIG\ref{fig:compass-pol-asym-int-sigma}.

\FloatBarrier

\subsection{Counting rates}

\label{sec:counting-rates}

\subsubsection{JLab}

\label{sec:jlab-counting-rates}

At JLab, to calculate the photon flux, we use the Weizs\"acker-Williams distribution. The details of the formulae used are found in \APP D.1 of \cite{Duplancic:2022ffo}.

The lepton beam at JLab forces one to also consider Bethe-Heitler-type 
processes.
However, such contributions are 
suppressed with respect to the photoproduction mechanism studied here, see \cite{Boussarie:2016qop}.

{The angular coverage of the final state particles is in principle a potential 
experimental issue. It can be shown that the angular distribution of the outgoing photon at JLab Hall B, which 
might evade detection, does not affect our predictions. The discussion on this subject is presented in \APP\ref{app:angle-outgoing-photon}.
}

The counting rates expected at JLab for our process are shown in Table \ref{tab:jlab-counting-rates}, assuming a luminosity of 100~nb$ ^{-1} $s$ ^{-1} $, and 100 days of run. The minimum and maximum values of the counting rates correspond to the boundaries obtained by considering all the 4 different possibilities, \ie{}the 2 models for the GPDs (standard and valence scenarios) and the 2 models for the DAs (asymptotic and holographic DA). The smallest value is in general obtained for an asymptotic DA with valence scenario GPD model, while the largest value is obtained for a holographic DA with standard scenario GPD model. The values obtained for the JLab experiment are very promising.

\begin{table}[h!]
	\centering
\begin{tabular}{|c|l|c|}
\hline
GPD &Meson & Counting rates \\
\hline
\hline
\multirow{4}{*}{Chiral-even}
& $  {\mesonzp} $ & 1.3-2.4 $  \times 10^5  $\\
\cline{2-3}
&$  {\mesonzn} $ & 1.7-4.0 $  \times 10^4$ \\
\cline{2-3}
&$  {\mesonpp} $ & 0.9-1.4 $  \times 10^5  $\\
\cline{2-3}
&$  {\mesonmn} $ & 0.3-1.8 $  \times 10^5$ \\
\hline
\hline
\multirow{4}{*}{Chiral-odd}
& $  {\mesonzp} $ & 2.1-4.2 $  \times 10^4  $\\
\cline{2-3}
&$  {\mesonzn} $ & 1.0-2.6 $  \times 10^4$ \\
\cline{2-3}
&$  {\mesonpp} $ & 3.5-6.7 $  \times 10^4  $\\
\cline{2-3}
&$  {\mesonmn} $ & 3.5-6.8 $  \times 10^4$ \\
\hline
\end{tabular}
\caption{Estimated counting rates at JLab for $  \gamma  \meson$ photoproduction.}
\label{tab:jlab-counting-rates}
\end{table}

\subsubsection{COMPASS}

At COMPASS,  one again uses the Weizs\"acker-Williams distribution to obtain the photon flux from the muon beam, Here, we also fix $ Q^{2}_{ \mathrm{max} } = 0.1 \GeV ^2 $.

The counting rates expected at COMPASS for our process is shown in Table \ref{tab:compass-counting-rates}. Like before, the minimum and maximum values of the counting rates correspond to the boundaries obtained by considering all the different possibilities, \ie{}the 2 models for the GPDs (standard and valence scenarios) and the 2 models for the DAs (asymptotic and holographic DA). In general, the lowest value is obtained for an asymptotic DA with valence scenario, while the largest value is obtained for a holographic DA with standard scenario. We have assumed a luminosity of 0.1 nb$ ^{-1} $s$ ^{-1} $, and 300 days of run.

\begin{table}[h!]
	\centering
	\begin{tabular}{|c|l|c|}
		\hline
		GPD &Meson & Counting rates \\
		\hline
		\hline
		\multirow{4}{*}{Chiral-even}
		& $  {\mesonzp} $ & 0.7-1.2 $  \times 10^3  $\\
		\cline{2-3}
		&$  {\mesonzn} $ & 0.8-2.1 $  \times 10^2$ \\
		\cline{2-3}
		&$  {\mesonpp} $ & 3.6-7.4 $  \times 10^2  $\\
		\cline{2-3}
		&$  {\mesonmn} $ & 3.0-8.1 $  \times 10^2$ \\
		\hline
		\hline
		\multirow{4}{*}{Chiral-odd}
		& $  {\mesonzp} $ & 75-152 \\
		\cline{2-3}
		&$  {\mesonzn} $ & 36-98 \\
		\cline{2-3}
		&$  {\mesonpp} $ & 135-257\\
		\cline{2-3}
		&$  {\mesonmn} $ & 133-257 \\
		\hline
	\end{tabular}
	\caption{Estimated counting rates at COMPASS for $ \gamma\meson  $ photoproduction.}
	\label{tab:compass-counting-rates}
\end{table}

\subsubsection{EIC}

\label{sec:EIC-counting-rates}

The counting rates for EIC, assuming a total integrated luminosity of $ 10^7 $ nb$ ^{-1} $, is shown in Table \ref{tab:EIC-counting-rates}. In particular, we use the highest expected electron-nucleon centre of mass energy, corresponding to  $ S_{eN}= 19600 \GeV^2 $ \cite{AbdulKhalek:2021gbh}. Since the centre-of-mass energies available at EIC are high enough, one can study the kinematic region where the skewness $  \xi  $ is small. Therefore, we also show the counting rates with the constraint that $ \SgN > 300 \GeV^2 $, which corresponds roughly to $  \xi \lesssim 5  \cdot 10^{-3} $.\footnote{Note that the relation between $ \SgN $ and $  \xi  $ involves $ \Msq $, which is why a cut in $\SgN$ does not directly correspond to a cut in $\xi$. However, as can be seen in \SEC\ref{sec:EIC-LHC-UPC-single-diff-X-section}, the cross section is dominated by small $ \Msq $, so the region of small $  \xi  $ is actually the one where most of the contribution comes from.} In fact, values of the skewness $  \xi  $ as small as $ 7.5 \cdot 10^{-6} $ can be probed. By imposing the cut in $ \SgN $, the counting rates decrease by roughly a factor of 20 in the chiral-even case. This significant decrease is due to the fact that the peak of the cross section is located at low $ \SgN $, roughly $ 20 \GeV^2 $, as can be seen in \FIGs\ref{fig:compass-int-sigma} and \ref{fig:EIC-LHC-UPC-int-sigma}. In the chiral-odd case, the region of small $ \xi $ is heavily suppressed, since the cross section is multiplied by a factor of $ \xi^2 $, compared to the chiral-even case, see \EQs \eqref{all-rho} and \eqref{all-rhoT}. As a result, we do not show the counting rates for the chiral-odd case when imposing the $ \SgN > 300 \GeV^2 $ cut. The minimum and maximum values for the counting rates in Table \ref{tab:EIC-counting-rates} are obtained as described in previous sections.

\begin{table}[h!]
	\centering
	\begin{tabular}{|c|l|c|c|}
		\hline
		GPD &Meson & Total Counting rates & Counting rates with $ \SgN > 300 \GeV^2 $ \\
		\hline
		\hline
		\multirow{4}{*}{Chiral-even}
		& $  {\mesonzp} $ & 1.3-2.4 $  \times 10^4  $& 0.6-1.2 $ \times 10^3 $\\
		\cline{2-4}
		&$  {\mesonzn} $ & 1.7-4.3 $  \times 10^3$ & 1.3-2.4 $ \times 10^2 $\\
		\cline{2-4}
		&$  {\mesonpp} $ & 0.7-1.5 $  \times 10^4  $& 3.1-9.3 $ \times 10^2 $\\
		\cline{2-4}
		&$  {\mesonmn} $ & 0.6-1.6 $  \times 10^4$ & 2.0-9.1 $ \times 10^2 $\\
		\hline
		\hline
		\multirow{4}{*}{Chiral-odd}
		& $  {\mesonzp} $ & 1.2-2.4 $ \times 10^3 $ & -\\
		\cline{2-4}
		&$  {\mesonzn} $ & 0.6-1.5 $ \times 10^3 $ & - \\
		\cline{2-4}
		&$  {\mesonpp} $ & 2.1-4.2 $ \times 10^3 $& -\\
		\cline{2-4}
		&$  {\mesonmn} $ & 2.1-4.1 $ \times 10^3 $ &-\\
		\hline
	\end{tabular}
	\caption{Estimated counting rates at EIC kinematics for $ \gamma \meson $ photoproduction.}
	\label{tab:EIC-counting-rates}
\end{table}

\subsubsection{Ultraperipheral collisions at LHC}

\label{sec:UPC-LHC}

In ultraperipheral collisions (UPCs), the beam and target are far enough apart such that there are no hadronic interactions between them, such that the  nucleus/proton interacts by the exchange of photons. In particular, heavy nuclei, such as lead, can act as a good source of photons, since the photon flux scales as $ Z^2 $, where $ Z $ is the charge of the nucleus. The details on how the photon flux is obtained can be found in \APP D.2 of \cite{Duplancic:2022ffo}.

The counting rates corresponding to p-Pb UPCs at LHC, assuming an integrated luminosity of 1200 nb$ ^{-1} $, is shown in Table \ref{tab:LHC-UPC-counting-rates}. This corresponds to the expected data taking for runs 3 and 4 \cite{Citron:2018lsq}. As in \SEC\ref{sec:EIC-counting-rates}, there is an order of magnitude drop in the counting rates for the chiral-even case when a cut of $ \SgN > 300 \GeV^2 $ is imposed. The chiral-odd counting rates are also very small when the $ \SgN > 300 \GeV^2 $ cut is applied, and they are thus omitted from Table \ref{tab:LHC-UPC-counting-rates}.

\begin{table}[h!]
	\centering
	\begin{tabular}{|c|l|c|c|}
		\hline
		GPD &Meson & Total Counting rates & Counting rates with $ \SgN > 300 \GeV^2 $ \\
		\hline
		\hline
		\multirow{2}{*}{Chiral-even}
		& $  {\mesonzp} $ & 0.9-1.6 $  \times 10^4  $& 4.1-8.1 $ \times 10^2 $\\
		\cline{2-4}
		&$  {\mesonpp} $ & 0.5-1.1 $  \times 10^4  $& 2.1-6.4 $ \times 10^2 $\\
		\hline
		\hline
		\multirow{2}{*}{Chiral-odd}
		& $  {\mesonzp} $ & 0.8-1.7 $ \times 10^3 $ & -\\
		\cline{2-4}
		&$  {\mesonpp} $ & 1.5-2.9 $ \times 10^3 $& -\\
		\hline
	\end{tabular}
	\caption{Estimated counting rates at p-Pb UPCs at LHC for $ \gamma \meson $ photoproduction.}
	\label{tab:LHC-UPC-counting-rates}
\end{table}

\section{Conclusion}

\label{sec:conclusion}


In this work, we extend the analysis 
of $\gamma N \to \gamma \meson^{0} N'$ process
introduced in \cite{Boussarie:2016qop} 
by including the linear polarisation asymmetries, 
extending the kinematics to selected future experiments (COMPASS, EIC and UPCs at LHC),  computing predictions for an alternative `holographic' DA
(\ref{DA-hol}) and considering charged $ \meson $-mesons. 
Since we consider the  large angle scattering kinematics, 
which amounts to large $ (-u') $ and $ \Msq $, and small $ (-t) $, we are able to employ the collinear factorisation.
In fact, QCD factorisation has been recently proven to hold for a family of $2 \to 3 $ exclusive processes \cite{Qiu:2022bpq, Qiu:2022pla}, which includes our process,
for large $   |\vec{p}_{t}|  $. We find that imposing kinematical cuts on $  \left( -u' \right),\, \left( -t' \right)   $ and $  \left( -t \right)  $ in \eqref{eq:kinematical-cuts} is sufficient to push the $ \meson N' $ invariant mass above the resonance region.

Our results show that the exclusive photoproduction of a $ \gamma {\meson} $ pair  provides another interesting channel to study GPDs, besides the extensively studied channels such as DVCS, DVMP and TCS. We have estimated the counting rates at various experiments in \SEC\ref{sec:counting-rates}, and the values obtained are promising, especially at JLab where they were found to be of the order of $ 10^5 $, assuming a luminosity of 100~nb$ ^{-1} $s$ ^{-1} $, and 100 days of run. In fact, the GPD model corresponding to the standard scenario, which is favoured by lattice results \cite{Alexandrou:2017oeh}, as well as its recent update in \cite{ Alexandrou:2020sml}, gives larger cross sections in general. Furthermore, we found that the linear polarisation asymmetries wrt the incoming photon are sizeable. Moreover, by exploiting the high energies available at EIC and UPCs at LHC, one is able to probe GPDs in the region of small skewness $ \xi $, a region where very little is known about GPDs. 
We found that by restricting the kinematics to the region of $ \xi  \leq 5\times 10^{-3} $, 
the counting rates drop by roughly a factor of 10,
which still leaves sufficient statistics. 

We intend to extend the present computation by including NLO corrections in $ \alpha_{s} $. While QCD collinear factorisation was proved for our process, the knowledge of such corrections, which are often significant for phenomenology, will increase the precision of our predictions and will give us the opportunity to estimate the uncertainties related to our process based on the collinear factorisation approach.

\section*{Acknowledgements}

We thank Nicole D'Hose,  Aude Glaenzer, Cedric Lorc\'e, Ronan McNulty, Kenneth Osterberg, Marco Santimaria, Daria Sokhan, Pawel Sznajder, Daniel Tapia Takaki, Charlotte van Hulse and Michael Winn for useful discussions. SN is supported by the GLUODYNAMICS project funded by the ``P2IO LabEx (ANR-10-LABEX-0038)" in the framework ``Investissements d’Avenir" (ANR-11-IDEX-0003-01) managed by the Agence Nationale de la Recherche (ANR), France. SN also acknowledges the hospitality of NCBJ where part of this work was done. The work of L. S. is supported by the grant 2019/33/B/ST2/02588 of the National Science Center in Poland. L. S. thanks the P2IO Laboratory of Excellence (Programme Investissements d’Avenir ANR-10-LABEX-0038) and the P2I - Graduate School of Physics of Paris-Saclay University for support. 
This publication is supported by the Croatian Science Foundation project IP-2019-04-9709,
and by the EU Horizon 2020 research and innovation programme, STRONG-2020
project, under grant agreement No 824093.

\appendix

\section{Chiral-odd amplitudes}

\label{app:CO-amplitudes}

For the chiral-odd case, diagrams $A_3$ and $A_4$ contribute
to the structure $T^i_{A \perp}$ while 
diagrams $B_1$ and $B_5$ contribute
to the structure $T^i_{B \perp}.$
Thus,
\begin{eqnarray}
	{N}^q_{T\,A}[(AB)_{123}]&\equiv& N_{T\, A}^q [A_3]\,,\\
	{N}^q_{T\,A}[(AB)_{45}]&\equiv& N_{T\, A}^q[A_4]\,,\\
	{N}^q_{T\,B}[(AB)_{123}]&\equiv& N_{T\, B}^q[B_1] \,,\\
	{N}^q_{T\,B}[(AB)_{45}]&\equiv& 	N_{T\, B}^q[B_5] \,.
\end{eqnarray}

For convenience, we define the coefficients $ T^{\perp}_{A} [A_3],\,T^{\perp}_{A} [A_4],\,T^{\perp}_{B} [B_1]$ and $T^{\perp}_{B} [B_5] $, given by
\begin{eqnarray}
	\label{eq:NqTA}
	N_{T\, A}^q [A_3]&\equiv&  \int_{-1}^1 \int_{0}^1 T^{\perp}_{A} [A_3]\phi(z) \, dz \, H_T(x,\xi) \,dx \,,\\
	N_{T\, A}^q[A_4] &\equiv&  \int_{-1}^1 \int_{0}^1 T^{\perp}_{A}[A_4] \phi(z) \, dz \, H_T(x,\xi) \,dx \,,\\
	N_{T\, B}^q[B_1] &\equiv& \int_{-1}^1 \int_{0}^1 T^{\perp}_{B}[B_1] \phi(z) \, dz \, H_T(x,\xi) \, dx
	\,,\\
		\label{eq:NqTB}
	N_{T\, B}^q[B_5] &\equiv& \int_{-1}^1 \int_{0}^1 T^{\perp}_{B}[B_5] \phi(z) \, dz \, H_T(x,\xi) \, dx\,.
\end{eqnarray}

\subsection{Asymptotic DA case}

\label{app:as-DA-case}

For the case of the asymptotic DA in \eqref{DA-asymp}, we get\footnote{We note that some typos have been corrected here wrt results in \APP{}B.2 of our previous publication \cite{Boussarie:2016qop}.}
\beqa
T^{\perp}_{A}[A_3] \phi^{\mathrm{as}}(z) &=& -\frac{3}{\alpha^2 \bar{\alpha}
	\xi  (\xi -x-i \epsilon ) (\xi +x-i
	\epsilon )}\;, \\[5pt]
T^{\perp}_{A}[A_4] \phi^{\mathrm{as}}(z) &=& \frac{3 (1-z)}{\alpha ^2 \xi  (\xi -x-i
	\epsilon ) (\alpha  (-\xi +x+i \epsilon
	)+(1-z) (2 \xi +(1-\alpha ) (-\xi +x+i
	\epsilon )))}\,,\quad\;\; \\[5pt]
T^{\perp}_{B}[B_1] \phi^{\mathrm{as}}(z) &=&
-\frac{3}{(1-\alpha ) \xi  (\xi -x-i
	\epsilon ) (\xi +x+i \epsilon )}
\;,\\[5pt]
T^{\perp}_{B}[B_5] \phi^{\mathrm{as}}(z) &=&
 \frac{3 z}{\xi  (\xi +x+i \epsilon ) (\alpha
	(-\xi +x+i \epsilon )+(1-z) (2 \xi
	+(1-\alpha ) (-\xi +x+i \epsilon )))}
.
\eqa
The integral with respect to $z$ is trivially performed in this case. Thus, one gets
\begin{eqnarray}
	\label{int_z-diagA-chiral-odd}
	&&\hspace{-1cm} \int_{0}^1 T^{\perp}_{A}[A_3] \phi^{\mathrm{as}}(z) \, dz =  -\frac{3}{\alpha^2 \bar{\alpha}
		\xi  (\xi -x-i \epsilon ) (\xi +x-i
		\epsilon )} \;, \\[5pt]
	&&\hspace{-1cm} \int_{0}^1 T^{\perp}_{A}[A_4] \phi^{\mathrm{as}}(z) \, dz =  \frac{3}{\alpha^2 \xi  (\xi
		-x-i \epsilon ) (2 \xi +(1-\alpha ) (-\xi
		+x+i \epsilon ))} +
	\frac{3 \ln \left(\frac{\xi +x+i \epsilon
		}{\alpha  (-\xi +x+i \epsilon
			)}\right)}{\alpha \xi  (2 \xi +(1-\alpha ) (-\xi
		+x+i \epsilon ))^2}\,, \nonumber\\[5pt]
	\label{int_z-diagB-chiral-odd}
	&&\hspace{-1cm} \int_{0}^1 T^{\perp}_{B}[B_1] \phi^{\mathrm{as}}(z) \, dz =  -\frac{3}{(1-\alpha ) \xi  (\xi -x-i
		\epsilon ) (\xi +x+i \epsilon )} \;, \\[5pt]
	&&\hspace{-1cm} \int_{0}^1 T^{\perp}_{B}[B_5] \phi^{\mathrm{as}}(z) \, dz =  -\frac{3}{\xi  (\xi
		+x+i \epsilon ) (2 \xi +(1-\alpha ) (-\xi
		+x+i \epsilon ))}  
	+   
	\frac{3 \ln \left(\frac{\xi +x+i \epsilon
		}{\alpha  (-\xi +x+i \epsilon
			)}\right)}{\xi  (2 \xi +(1-\alpha ) (-\xi
		+x+i \epsilon ))^2}. \nonumber
\end{eqnarray}
Let us note that the last term in the previous expressions (\ref{int_z-diagA-chiral-odd}) and 
(\ref{int_z-diagB-chiral-odd})
might seem to have a double pole when $x = -\frac{1+\alpha}{\bar{\alpha}}\xi -i \epsilon$. However, the logarithm cancels under such conditions, so this pole is actually a simple pole.

Using \eqref{eq:NqTA} to \eqref{eq:NqTB}, we can write the integrals with respect to $x$ in terms of building block integrals, given in \APP{}D of \cite{Duplancic:2018bum}. Thus, we have
\begin{eqnarray}
	\label{Res-int-T_CO_A}
	N_{T\, A}^q [A_3]&=&\frac{3}{2 \alpha ^2 \bar{\alpha} \xi ^2}(I_e-I_g)\;,\\
	N_{T\, A}^q[A_4] &=&
	-\frac{3}{\alpha ^2 \xi }I_a+\frac{3}{\alpha  \xi
	}I_d\;,\\
	\label{Res-int-T_CO_B}
	N_{T\, B}^q[B_1] &=&
	\frac{3 }{2 \bar{\alpha} \xi ^2}(I_e-I_f)\;,\\
	N_{T\, B}^q[B_5] &	=&
	-\frac{3 }{\xi }I_l+\frac{3 }{\xi }I_d\;.
\end{eqnarray}

For symmetric GPDs, we have
\begin{eqnarray}
	N_{T\, A}^q [A_3]^{s} &=&\frac{3}{2 \alpha ^2 \bar{\alpha} \xi ^2}(2I_e)\;,\\[5pt]
	N_{T\, A}^q[A_4]^{s} & =&
	-\frac{3}{\alpha ^2 \xi }(\frac{1}{2 \xi }I_e-\frac{ \bar{\alpha }}{2 \xi }I_i)+\frac{3}{\alpha  \xi
	}I_d\;,\\[5pt]
	N_{T\, B}^q[B_1]^{s}&
	=&
	\frac{3 }{2 \bar{\alpha} \xi ^2}(I_e+\bar{I}_{e})\;,\\[5pt]
	N_{T\, B}^q[B_5]^{s} &	=&
	-\frac{3 }{\xi }(-\frac{1}{2 \alpha  \xi }\bar{I}_{e}-\frac{\bar{ \alpha }}{2 \alpha  \xi }I_i)+\frac{3 }{\xi }I_d\;.
\end{eqnarray}

For anti-symmetric GPDs, we have
\begin{eqnarray}
	\label{eq:NTA-a-A3}
N_{T\, A}^q [A_3]^{a}&=&0\;,\\[5pt]
N_{T\, A}^q[A_4]^{a}  &=&
	-\frac{3}{\alpha ^2 \xi }(\frac{1}{2 \xi }I_e-\frac{ \bar{\alpha }}{2 \xi }I_i)+\frac{3}{\alpha  \xi
	}I_d\;,\\[5pt]
		\label{eq:NTB-a-B1}
N_{T\, B}^q[B_1]^{a} &
	=&
	\frac{3 }{2 \bar{\alpha} \xi ^2}(I_e-\bar{I}_e)\;,\\[5pt]
N_{T\, B}^q[B_5]^{a} 	&=&
	-\frac{3 }{\xi }(\frac{1}{2 \alpha  \xi }\bar{I}_{e}-\frac{\bar{ \alpha }}{2 \alpha  \xi }I_i)+\frac{3 }{\xi }I_d\;.
\end{eqnarray}
So, only the building block integrals $ I_e,\,I_{i} $ and $ I_{d} $ are needed in the asymptotical DA case.

\subsection{Holographic DA case}

\label{app:hol-DA-case}

Here, we essentially repeat the above steps, but with a holographic DA whose form is given in \eqref{DA-hol}, instead of an asymptotic DA. For the contributions to diagrams $ A_3 $ and $ B_1 $, the same results as in the asymptotic DA case can be used, with a change of overall prefactor from 6 to 8, see \EQs{}C.3 and C.4 in \cite{Duplancic:2022ffo}. Therefore, we only focus on the results for the $ A_4 $ and $ B_5 $ diagrams here. The results, in terms of building block integrals given in \APP{}D of \cite{Duplancic:2018bum} and \APP{}C of \cite{Duplancic:2022ffo}, read
\begin{eqnarray}
	\label{eq:hol-Res-int-T_CO_A}
	N_{T\, A}^q[A_4] &\equiv&  {s^3} \int_{-1}^1 \int_{0}^1 T^{\perp}_{A}[A_4] \phi^{\mathrm{hol}}(z) \, dz \, H_T(x,\xi) \,dx =-\frac{4}{ \alpha ^2  \xi } \left[ \frac{1}{2 \xi }I_e-\sqrt{ \alpha }  \chi _{a} \right] \;,\\[5pt]
	\label{eq:hol-Res-int-T_CO_B}
	N_{T\, B}^q[B_5] &\equiv& {s^3} \int_{-1}^1 \int_{0}^1 T^{\perp}_{B}[B_5] \phi^{\mathrm{hol}}(z) \, dz \, H_T(x,\xi) \, dx
	=\frac{4}{ \xi } \left[ -\frac{1}{2 \alpha  \xi }I_f +\frac{1}{\sqrt{ \alpha }} \chi _{a}\right] \;.
\end{eqnarray}

For symmetric GPDs, we have
\begin{eqnarray}
	N_{T\, A}^q[A_4]^{s} &\equiv&  {s^3} \int_{-1}^1 \int_{0}^1 T^{\perp}_{A}[A_4]^{s} \phi^{\mathrm{hol}}(z) \, dz \, H_T(x,\xi) \,dx =-\frac{4}{ \alpha ^2  \xi } \left[ \frac{1}{2 \xi }I_e-\sqrt{ \alpha }  \chi _{a} \right] \;,\\[5pt]
	N_{T\, B}^q[B_5]^{s} &\equiv& {s^3} \int_{-1}^1 \int_{0}^1 T^{\perp}_{B}[B_5]^{s} \phi^{\mathrm{hol}}(z) \, dz \, H_T(x,\xi) \, dx
	=\frac{4}{ \xi } \left[ \frac{1}{2 \alpha  \xi }\bar{I}_e +\frac{1}{\sqrt{ \alpha }} \chi _{a}\right] \;,
\end{eqnarray}

and for anti-symmetric GPDs, we have
\begin{eqnarray}
	N_{T\, A}^q[A_4]^{a} &\equiv&  {s^3} \int_{-1}^1 \int_{0}^1 T^{\perp}_{A}[A_4]^{a} \phi^{\mathrm{hol}}(z) \, dz \, H_T(x,\xi) \,dx =-\frac{4}{ \alpha ^2  \xi } \left[ \frac{1}{2 \xi }I_e-\sqrt{ \alpha }  \chi _{a} \right] \;,\\[5pt]
	N_{T\, B}^q[B_5]^{a} &\equiv& {s^3} \int_{-1}^1 \int_{0}^1 T^{\perp}_{B}[B_5]^{a} \phi^{\mathrm{hol}}(z) \, dz \, H_T(x,\xi) \, dx
	=\frac{4}{ \xi } \left[- \frac{1}{2 \alpha  \xi }\bar{I}_e +\frac{1}{\sqrt{ \alpha }} \chi _{a}\right] \;.
\end{eqnarray}
So, only the extra building block integral $  \chi _{a} $ is needed.

\section{Effect of angular cuts on the outgoing photon at JLab}

\label{app:angle-outgoing-photon}

In this appendix, we show the influence of angular cuts on the outgoing photon at JLab on the cross section.

\subsection{Angular distribution}

\begin{figure}[h!]
	\begin{center}
		\psfrag{T}{}
		\psfrag{H}{\hspace{-1.5cm}\raisebox{-.6cm}{\scalebox{.7}{$\theta$}}}
		\psfrag{V}{\raisebox{.3cm}{\scalebox{.7}{$\hspace{-.4cm}\displaystyle\frac{1}{\sigma^{ \mathrm{even} }_{\gamma \mesonzp}}\frac{d \sigma^{ \mathrm{even} }_{\gamma \mesonzp}}{d \theta}$}}}
		\includegraphics[width=7cm]{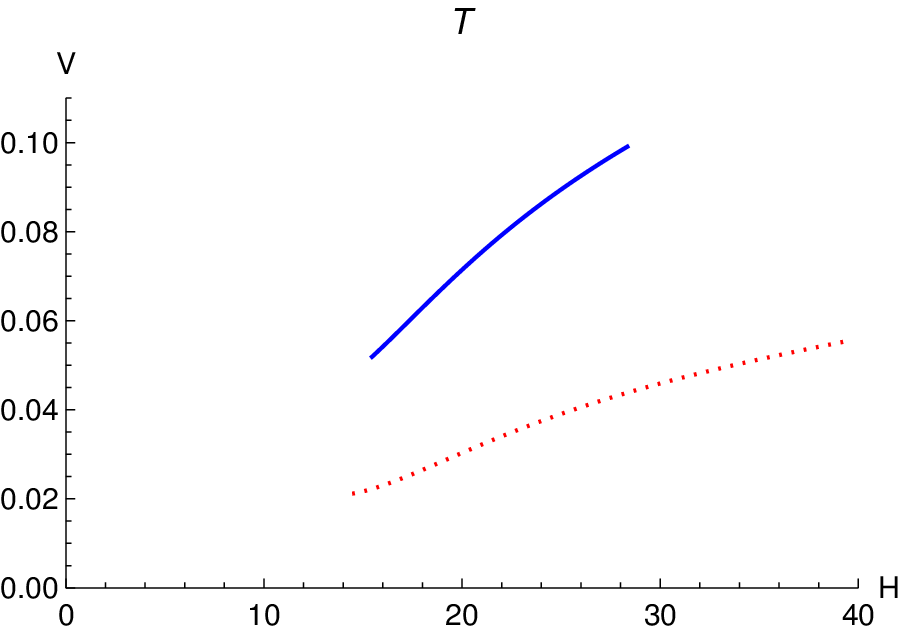}
		\includegraphics[width=7cm]{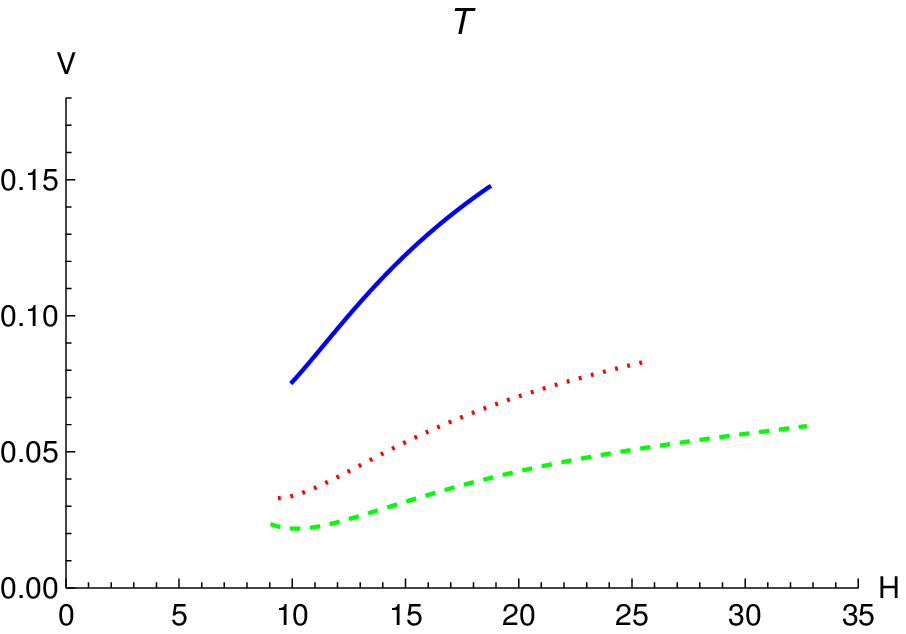}
		
		\vspace{.8cm}
		\includegraphics[width=7cm]{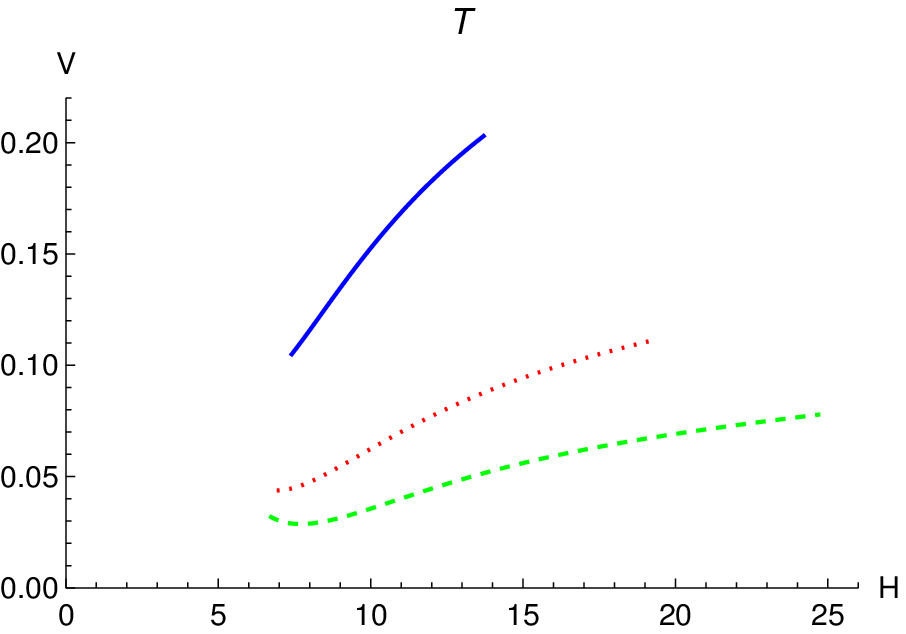}
		\vspace{.4cm}
		\caption{Angular distribution in the chiral-even case for $\gamma  \mesonzp $ photoproduction.
			Up, left: $S_{\gamma N}=10~{\rm GeV}^2$, for $M^2_{\gamma \mesonzp}=
			3~{\rm GeV}^2$ (solid blue) and  $M^2_{\gamma \mesonzp}=
			4~{\rm GeV}^2$ (dotted red).  Up, right:
			$S_{\gamma N}=15~{\rm GeV}^2$, for $M^2_{\gamma \mesonzp}=
			3.5~{\rm GeV}^2$ (solid blue), $M^2_{\gamma \mesonzp}=
			5~{\rm GeV}^2$ (dotted red) and 
			$M^2_{\gamma \mesonzp}=
			6.5~{\rm GeV}^2$ (dashed green).
			Down: 
			$S_{\gamma N}=20~{\rm GeV}^2$, for $M^2_{\gamma \mesonzp}=
			4~{\rm GeV}^2$ (solid blue), $M^2_{\gamma \mesonzp}=
			6~{\rm GeV}^2$ (dotted red) and 
			$M^2_{\gamma \mesonzp}=
			8~{\rm GeV}^2$ (dashed green).}
		\label{fig:thetacut-rho0-even}
	\end{center}
\end{figure}

\begin{figure}[h!]
	\begin{center}
		\psfrag{T}{}
		\psfrag{H}{\hspace{-1.5cm}\raisebox{-.6cm}{\scalebox{.7}{$\theta$}}}
		\psfrag{V}{\raisebox{.3cm}{\scalebox{.7}{$\hspace{-.4cm}\displaystyle\frac{1}{\sigma^{ \mathrm{even} }_{\gamma \mesonzn}}\frac{d \sigma^{ \mathrm{even} }_{\gamma \mesonzn}}{d \theta}$}}}
		\includegraphics[width=7cm]{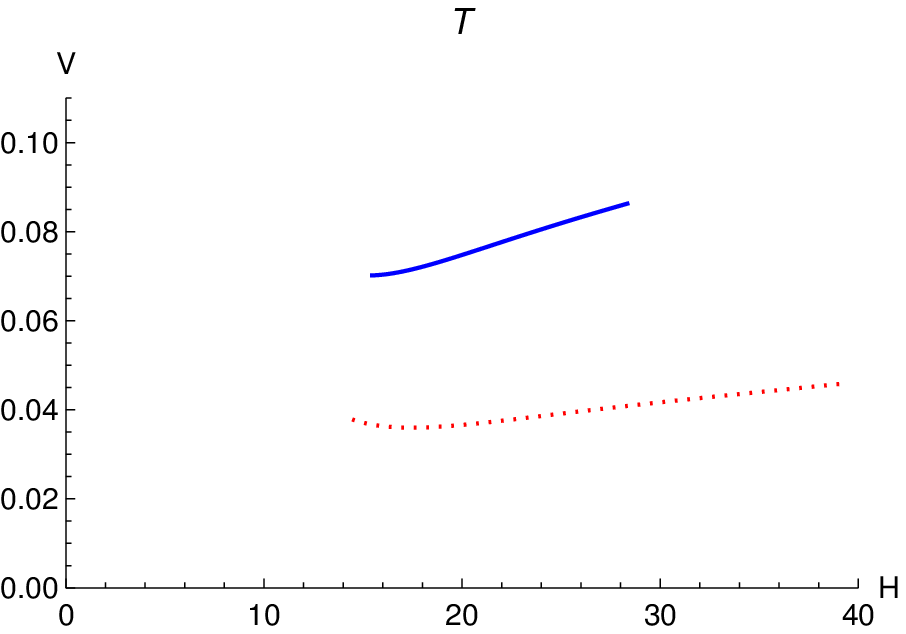}
		\includegraphics[width=7cm]{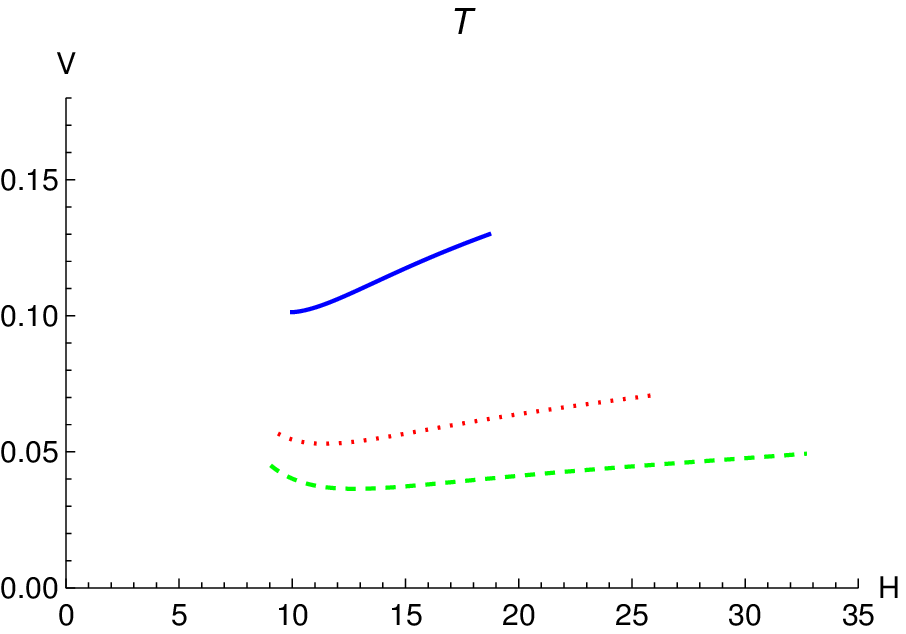}
		
		\vspace{.8cm}
		\includegraphics[width=7cm]{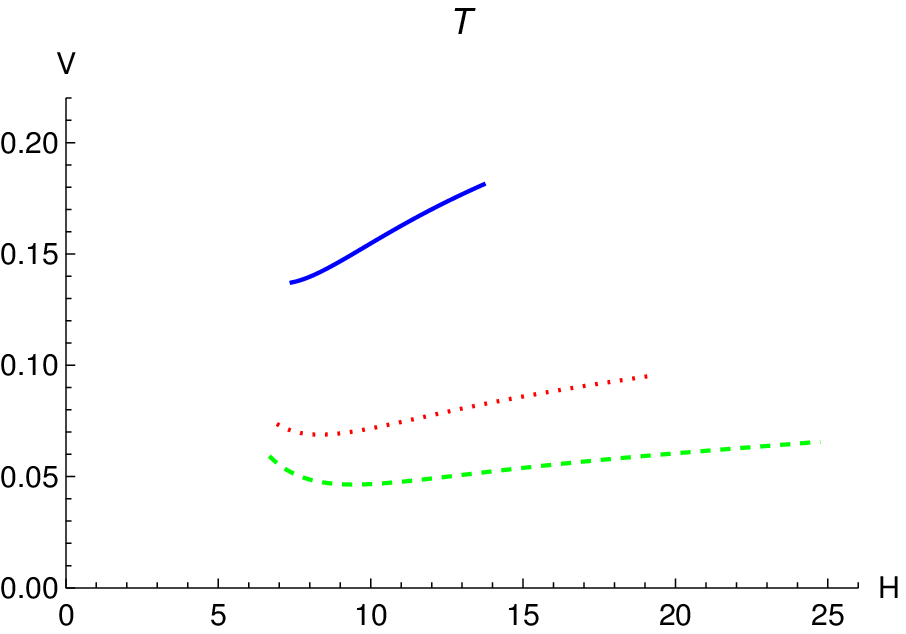}
		\vspace{.4cm}
		\caption{Angular distribution in the chiral-even case for $\gamma  \mesonzn $ photoproduction.
			Up, left: $S_{\gamma N}=10~{\rm GeV}^2$, for $M^2_{\gamma \mesonzn}=
			3~{\rm GeV}^2$ (solid blue) and  $M^2_{\gamma \mesonzn}=
			4~{\rm GeV}^2$ (dotted red).  Up, right:
			$S_{\gamma N}=15~{\rm GeV}^2$, for $M^2_{\gamma \mesonzn}=
			3.5~{\rm GeV}^2$ (solid blue), $M^2_{\gamma \mesonzn}=
			5~{\rm GeV}^2$ (dotted red) and 
			$M^2_{\gamma \mesonzn}=
			6.5~{\rm GeV}^2$ (dashed green).
			Down: 
			$S_{\gamma N}=20~{\rm GeV}^2$, for $M^2_{\gamma \mesonzn}=
			4~{\rm GeV}^2$ (solid blue), $M^2_{\gamma \mesonzn}=
			6~{\rm GeV}^2$ (dotted red) and 
			$M^2_{\gamma \mesonzn}=
			8~{\rm GeV}^2$ (dashed green).}
		\label{fig:thetacut-rho0neutron-even}
	\end{center}
\end{figure}

\begin{figure}[h!]
	\begin{center}
		\psfrag{T}{}
		\psfrag{H}{\hspace{-1.5cm}\raisebox{-.6cm}{\scalebox{.7}{$\theta$}}}
		\psfrag{V}{\raisebox{.3cm}{\scalebox{.7}{$\hspace{-.4cm}\displaystyle\frac{1}{\sigma^{ \mathrm{even} }_{\gamma \mesonpp}}\frac{d \sigma^{ \mathrm{even} }_{\gamma \mesonpp}}{d \theta}$}}}
		\includegraphics[width=7cm]{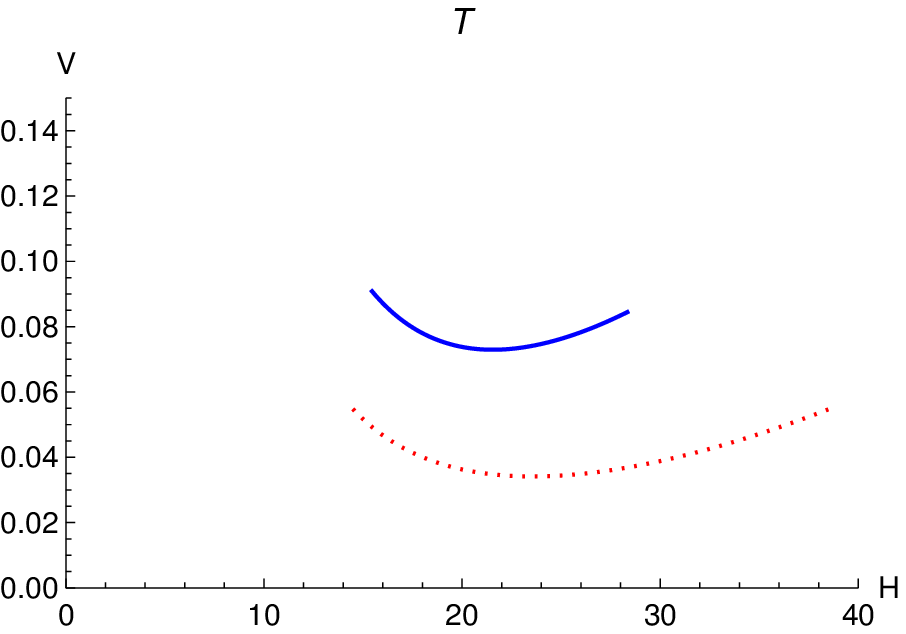}
		\includegraphics[width=7cm]{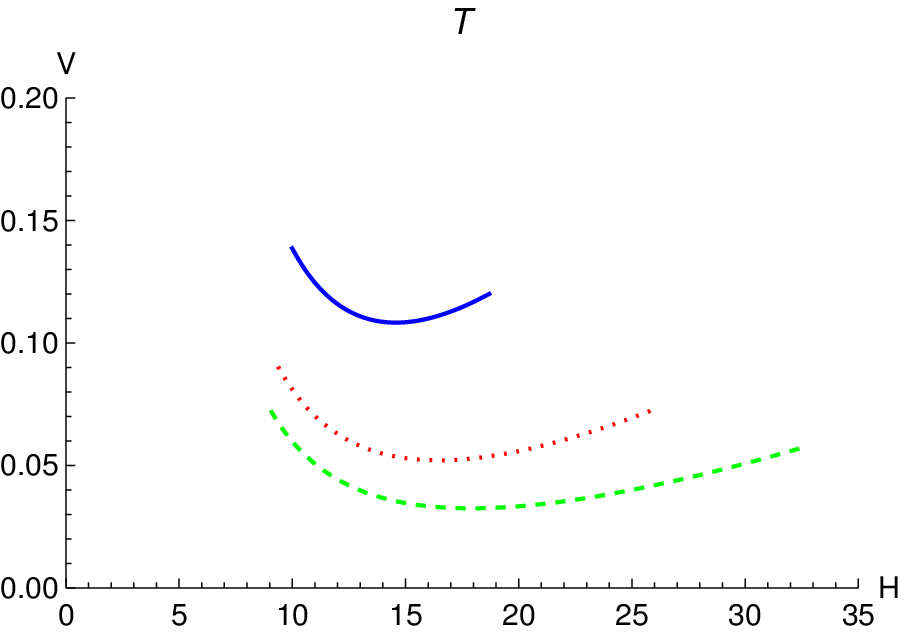}
		
		\vspace{.8cm}
		\includegraphics[width=7cm]{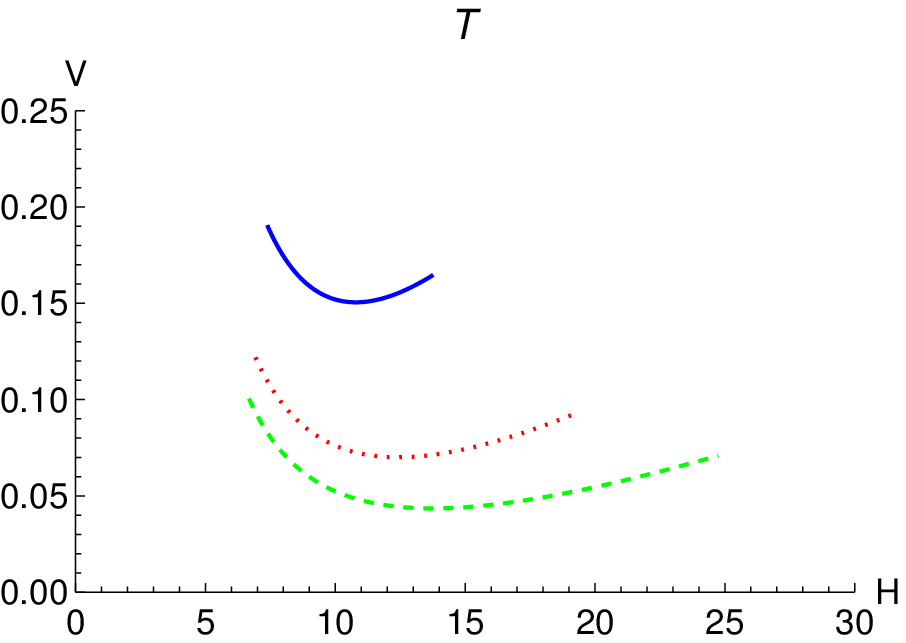}
		\vspace{.4cm}
		\caption{Angular distribution in the chiral-even case for $\gamma  \mesonpp $ photoproduction.
			Up, left: $S_{\gamma N}=10~{\rm GeV}^2$, for $M^2_{\gamma \mesonpp}=
			3~{\rm GeV}^2$ (solid blue) and  $M^2_{\gamma \mesonpp}=
			4~{\rm GeV}^2$ (dotted red).  Up, right:
			$S_{\gamma N}=15~{\rm GeV}^2$, for $M^2_{\gamma \mesonpp}=
			3.5~{\rm GeV}^2$ (solid blue), $M^2_{\gamma \mesonpp}=
			5~{\rm GeV}^2$ (dotted red) and 
			$M^2_{\gamma \mesonpp}=
			6.5~{\rm GeV}^2$ (dashed green).
			Down: 
			$S_{\gamma N}=20~{\rm GeV}^2$, for $M^2_{\gamma \mesonpp}=
			4~{\rm GeV}^2$ (solid blue), $M^2_{\gamma \mesonpp}=
			6~{\rm GeV}^2$ (dotted red) and 
			$M^2_{\gamma \mesonpp}=
			8~{\rm GeV}^2$ (dashed green).}
		\label{fig:thetacut-rhoplus-even}
	\end{center}
\end{figure}

\begin{figure}[h!]
	\begin{center}
		\psfrag{T}{}
		\psfrag{H}{\hspace{-1.5cm}\raisebox{-.6cm}{\scalebox{.7}{$\theta$}}}
		\psfrag{V}{\raisebox{.3cm}{\scalebox{.7}{$\hspace{-.4cm}\displaystyle\frac{1}{\sigma^{ \mathrm{even} }_{\gamma \mesonmn}}\frac{d \sigma^{ \mathrm{even} }_{\gamma \mesonmn}}{d \theta}$}}}
		\includegraphics[width=7cm]{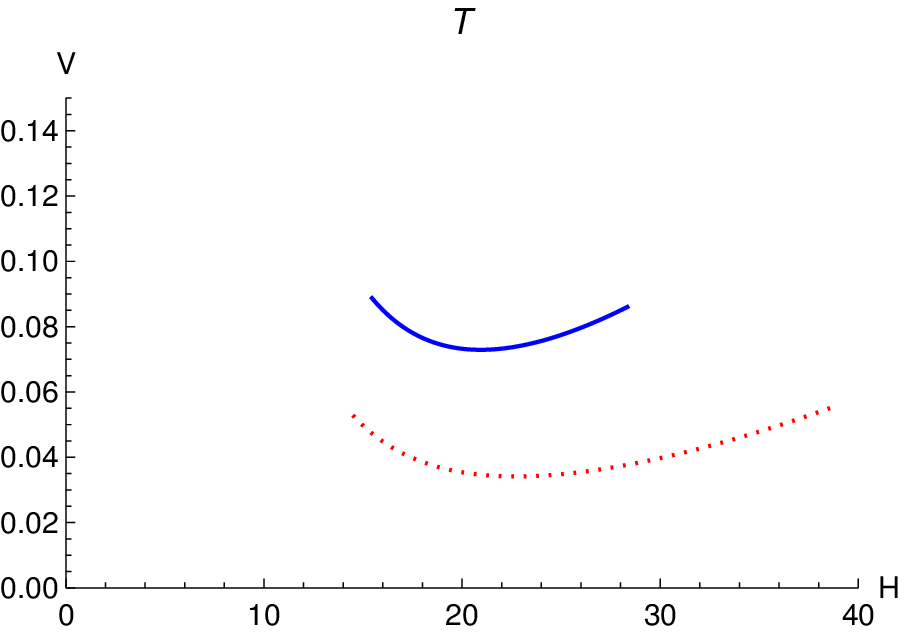}
		\includegraphics[width=7cm]{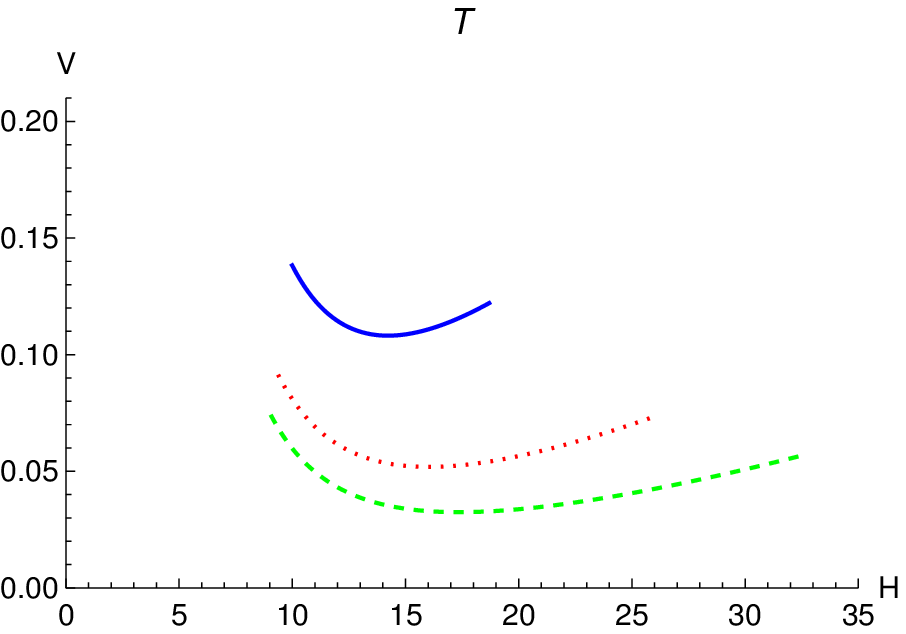}
		
		\vspace{.8cm}
		\includegraphics[width=7cm]{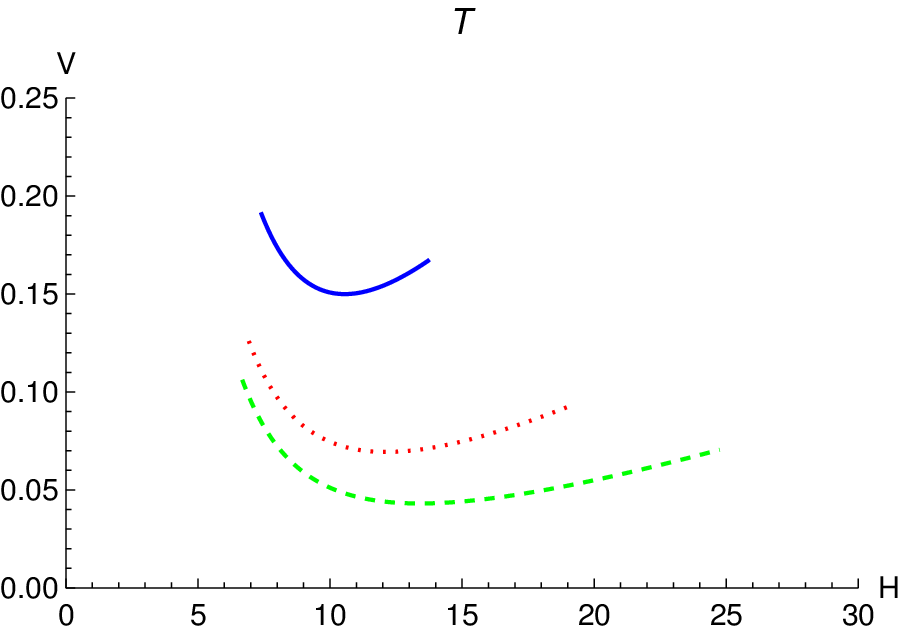}
		\vspace{.4cm}
		\caption{Angular distribution in the chiral-even case for $ \gamma \mesonmn $ photoproduction.
			Up, left: $S_{\gamma N}=10~{\rm GeV}^2$, for $M^2_{\gamma \mesonmn}=
			3~{\rm GeV}^2$ (solid blue) and  $M^2_{\gamma \mesonmn}=
			4~{\rm GeV}^2$ (dotted red).  Up, right:
			$S_{\gamma N}=15~{\rm GeV}^2$, for $M^2_{\gamma \mesonmn}=
			3.5~{\rm GeV}^2$ (solid blue), $M^2_{\gamma \mesonmn}=
			5~{\rm GeV}^2$ (dotted red) and 
			$M^2_{\gamma \mesonmn}=
			6.5~{\rm GeV}^2$ (dashed green).
			Down: 
			$S_{\gamma N}=20~{\rm GeV}^2$, for $M^2_{\gamma \mesonmn}=
			4~{\rm GeV}^2$ (solid blue), $M^2_{\gamma \mesonmn}=
			6~{\rm GeV}^2$ (dotted red) and 
			$M^2_{\gamma \mesonmn}=
			8~{\rm GeV}^2$ (dashed green).}
		\label{fig:thetacut-rhominus-even}
	\end{center}
\end{figure}

\begin{figure}[h!]
	\begin{center}
		\psfrag{T}{}
		\psfrag{H}{\hspace{-1.5cm}\raisebox{-.6cm}{\scalebox{.7}{$\theta$}}}
		\psfrag{V}{\raisebox{.3cm}{\scalebox{.7}{$\hspace{-.4cm}\displaystyle\frac{1}{\sigma^{ \mathrm{odd} }_{\gamma \mesonzp}}\frac{d \sigma^{ \mathrm{odd} }_{\gamma \mesonzp}}{d \theta}$}}}
		\includegraphics[width=7cm]{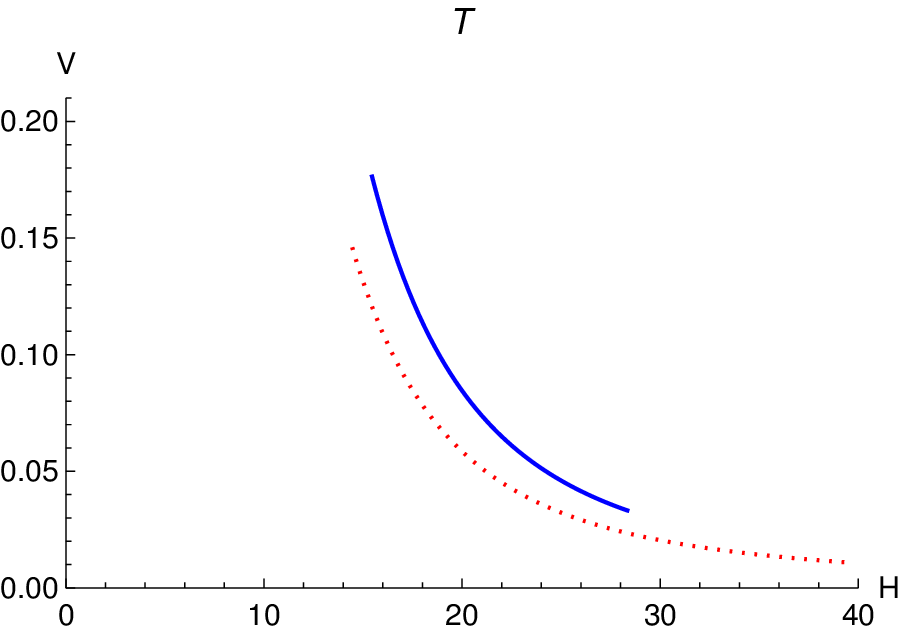}
		\includegraphics[width=7cm]{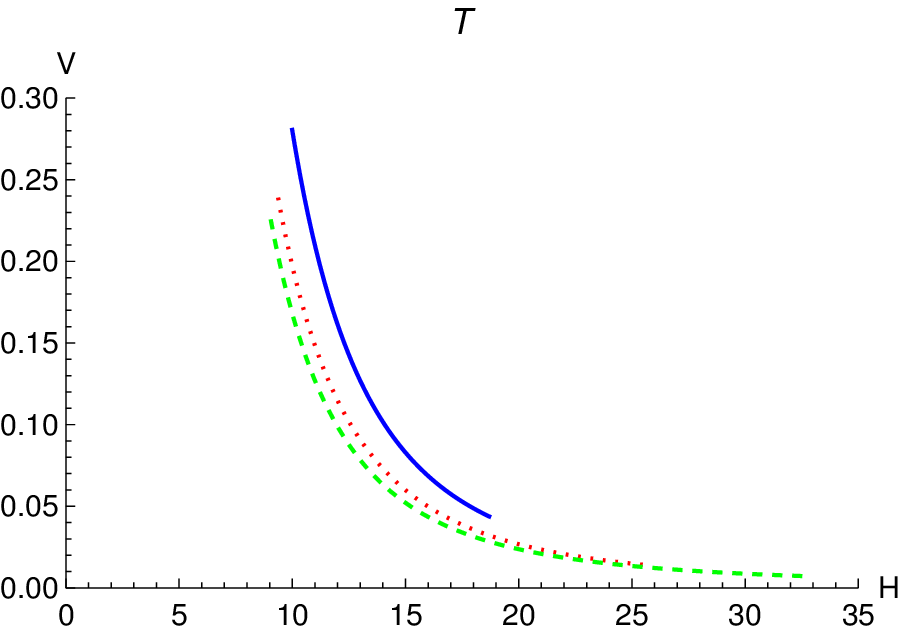}
		
		\vspace{.8cm}
		\includegraphics[width=7cm]{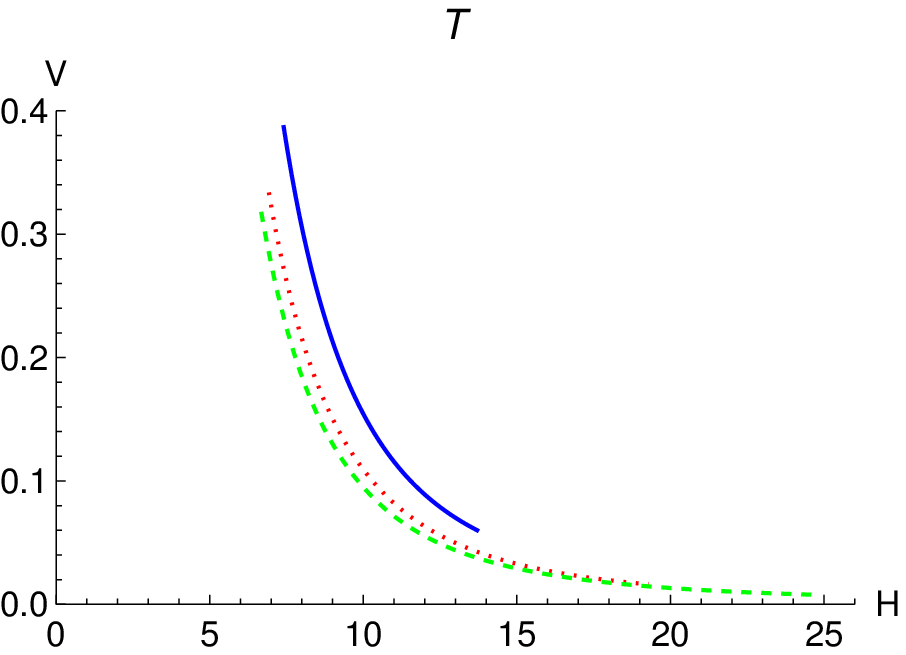}
		\vspace{.4cm}
		\caption{Angular distribution in the chiral-odd case for $ \gamma \mesonzp $ photoproduction.
			Up, left: $S_{\gamma N}=10~{\rm GeV}^2$, for $M^2_{\gamma \mesonzp}=
			3~{\rm GeV}^2$ (solid blue) and  $M^2_{\gamma \mesonzp}=
			4~{\rm GeV}^2$ (dotted red).  Up, right:
			$S_{\gamma N}=15~{\rm GeV}^2$, for $M^2_{\gamma \mesonzp}=
			3.5~{\rm GeV}^2$ (solid blue), $M^2_{\gamma \mesonzp}=
			5~{\rm GeV}^2$ (dotted red) and 
			$M^2_{\gamma \mesonzp}=
			6.5~{\rm GeV}^2$ (dashed green).
			Down: 
			$S_{\gamma N}=20~{\rm GeV}^2$, for $M^2_{\gamma \mesonzp}=
			4~{\rm GeV}^2$ (solid blue), $M^2_{\gamma \mesonzp}=
			6~{\rm GeV}^2$ (dotted red) and 
			$M^2_{\gamma \mesonzp}=
			8~{\rm GeV}^2$ (dashed green).}
		\label{fig:thetacut-rho0-odd}
	\end{center}
\end{figure}

\begin{figure}[h!]
	\begin{center}
		\psfrag{T}{}
		\psfrag{H}{\hspace{-1.5cm}\raisebox{-.6cm}{\scalebox{.7}{$\theta$}}}
		\psfrag{V}{\raisebox{.3cm}{\scalebox{.7}{$\hspace{-.4cm}\displaystyle\frac{1}{\sigma^{ \mathrm{odd} }_{\gamma \mesonzn}}\frac{d \sigma^{ \mathrm{odd} }_{\gamma \mesonzn}}{d \theta}$}}}
		\includegraphics[width=7cm]{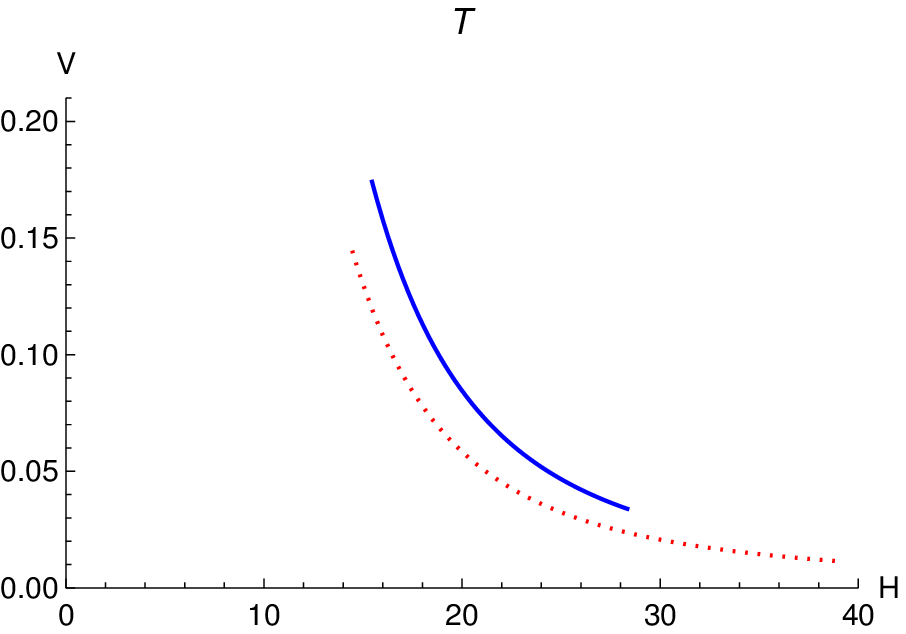}
		\includegraphics[width=7cm]{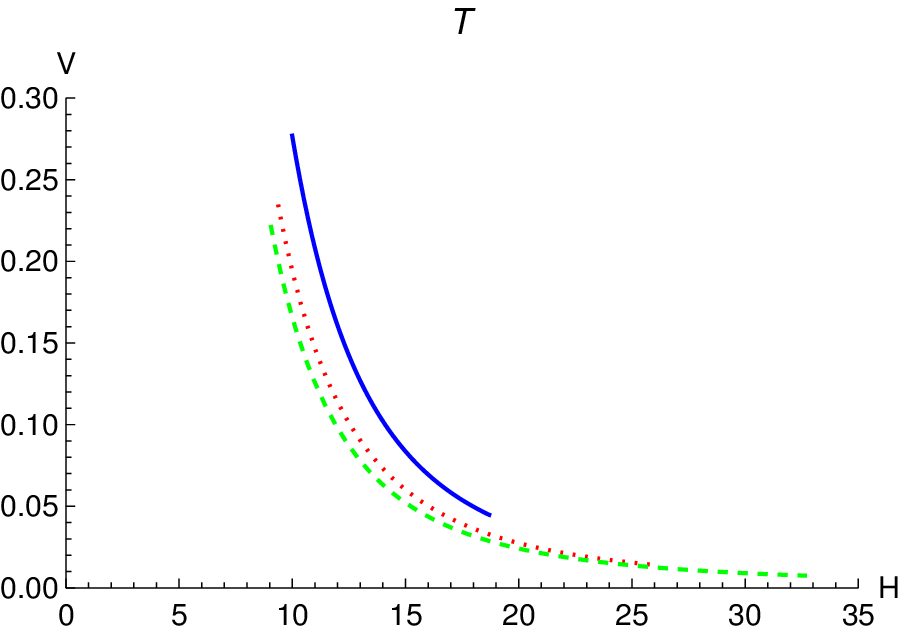}
		
		\vspace{.8cm}
		\includegraphics[width=7cm]{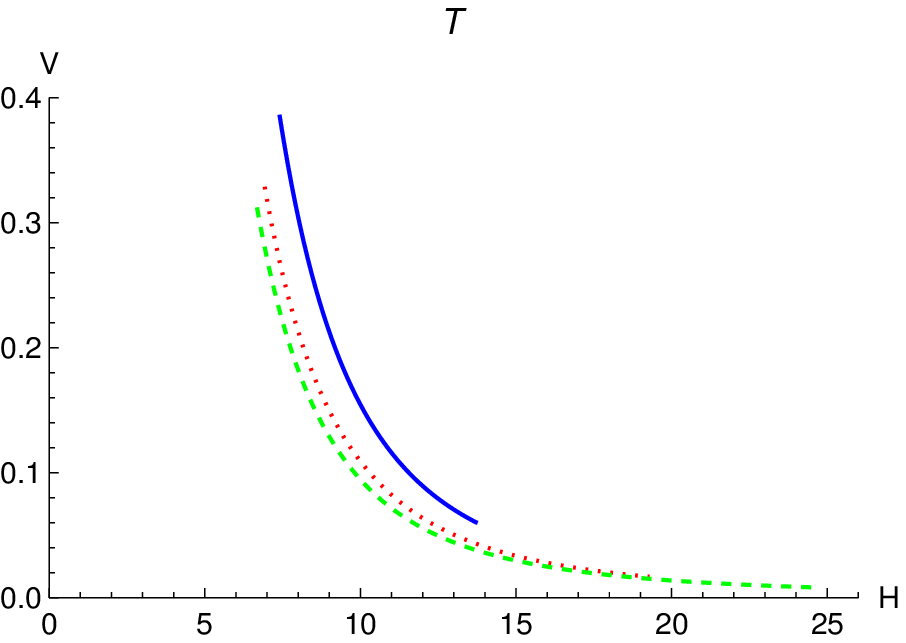}
		\vspace{.4cm}
		\caption{Angular distribution in the chiral-odd case for $ \gamma \mesonzn $ photoproduction.
			Up, left: $S_{\gamma N}=10~{\rm GeV}^2$, for $M^2_{\gamma \mesonzn}=
			3~{\rm GeV}^2$ (solid blue) and  $M^2_{\gamma \mesonzn}=
			4~{\rm GeV}^2$ (dotted red).  Up, right:
			$S_{\gamma N}=15~{\rm GeV}^2$, for $M^2_{\gamma \mesonzn}=
			3.5~{\rm GeV}^2$ (solid blue), $M^2_{\gamma \mesonzn}=
			5~{\rm GeV}^2$ (dotted red) and 
			$M^2_{\gamma \mesonzn}=
			6.5~{\rm GeV}^2$ (dashed green).
			Down: 
			$S_{\gamma N}=20~{\rm GeV}^2$, for $M^2_{\gamma \mesonzn}=
			4~{\rm GeV}^2$ (solid blue), $M^2_{\gamma \mesonzn}=
			6~{\rm GeV}^2$ (dotted red) and 
			$M^2_{\gamma \mesonzn}=
			8~{\rm GeV}^2$ (dashed green).}
		\label{fig:thetacut-rho0neutron-odd}
	\end{center}
\end{figure}

\begin{figure}[h!]
	\begin{center}
		\psfrag{T}{}
		\psfrag{H}{\hspace{-1.5cm}\raisebox{-.6cm}{\scalebox{.7}{$\theta$}}}
		\psfrag{V}{\raisebox{.3cm}{\scalebox{.7}{$\hspace{-.4cm}\displaystyle\frac{1}{\sigma^{ \mathrm{odd} }_{\gamma \mesonpp}}\frac{d \sigma^{ \mathrm{odd} }_{\gamma \mesonpp}}{d \theta}$}}}
		\includegraphics[width=7cm]{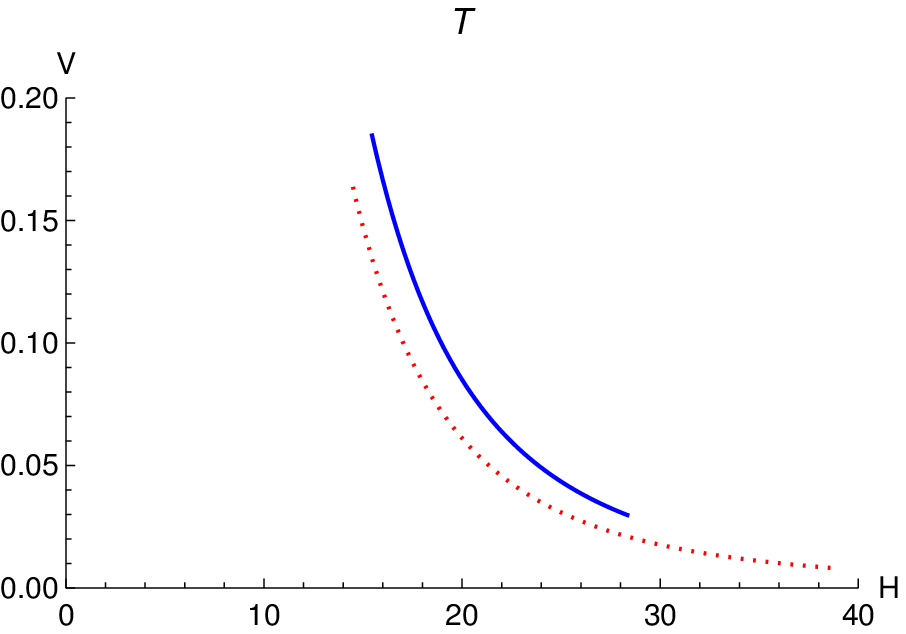}
		\includegraphics[width=7cm]{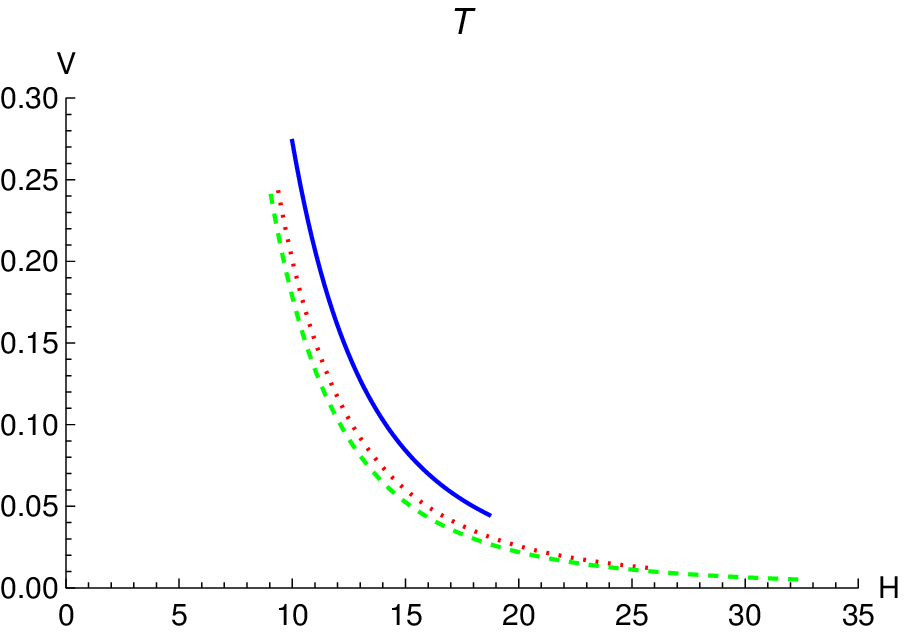}
		
		\vspace{.8cm}
		\includegraphics[width=7cm]{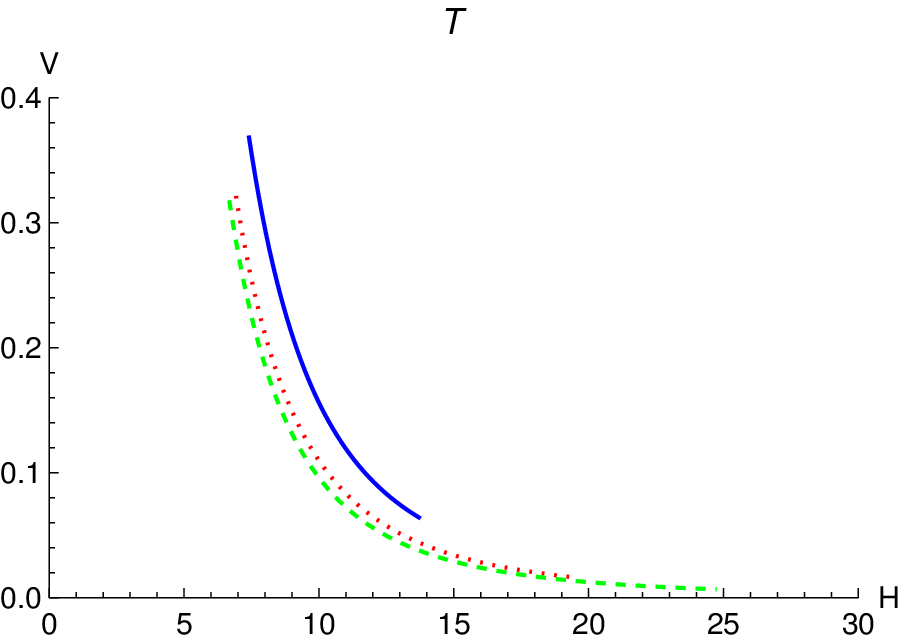}
		\vspace{.4cm}
		\caption{Angular distribution in the chiral-odd case for $ \gamma \mesonpp $ photoproduction.
			Up, left: $S_{\gamma N}=10~{\rm GeV}^2$, for $M^2_{\gamma \mesonpp}=
			3~{\rm GeV}^2$ (solid blue) and  $M^2_{\gamma \mesonpp}=
			4~{\rm GeV}^2$ (dotted red).  Up, right:
			$S_{\gamma N}=15~{\rm GeV}^2$, for $M^2_{\gamma \mesonpp}=
			3.5~{\rm GeV}^2$ (solid blue), $M^2_{\gamma \mesonpp}=
			5~{\rm GeV}^2$ (dotted red) and 
			$M^2_{\gamma \mesonpp}=
			6.5~{\rm GeV}^2$ (dashed green).
			Down: 
			$S_{\gamma N}=20~{\rm GeV}^2$, for $M^2_{\gamma \mesonpp}=
			4~{\rm GeV}^2$ (solid blue), $M^2_{\gamma \mesonpp}=
			6~{\rm GeV}^2$ (dotted red) and 
			$M^2_{\gamma \mesonpp}=
			8~{\rm GeV}^2$ (dashed green).}
		\label{fig:thetacut-rhoplus-odd}
	\end{center}
\end{figure}

\begin{figure}[h!]
	\begin{center}
		\psfrag{T}{}
		\psfrag{H}{\hspace{-1.5cm}\raisebox{-.6cm}{\scalebox{.7}{$\theta$}}}
		\psfrag{V}{\raisebox{.3cm}{\scalebox{.7}{$\hspace{-.4cm}\displaystyle\frac{1}{\sigma^{ \mathrm{odd} }_{\gamma \mesonmn}}\frac{d \sigma^{ \mathrm{odd} }_{\gamma \mesonmn}}{d \theta}$}}}
		\includegraphics[width=7cm]{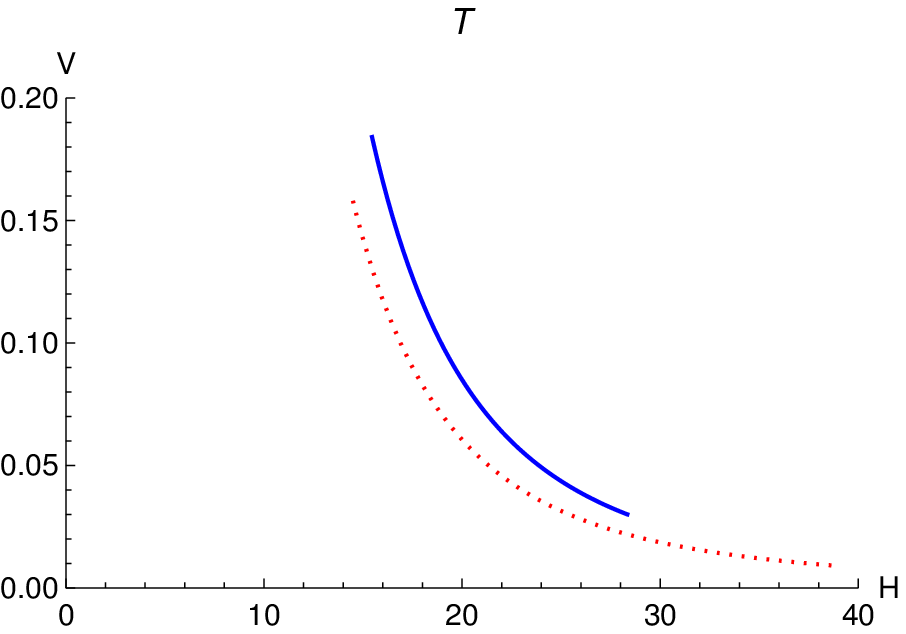}
		\includegraphics[width=7cm]{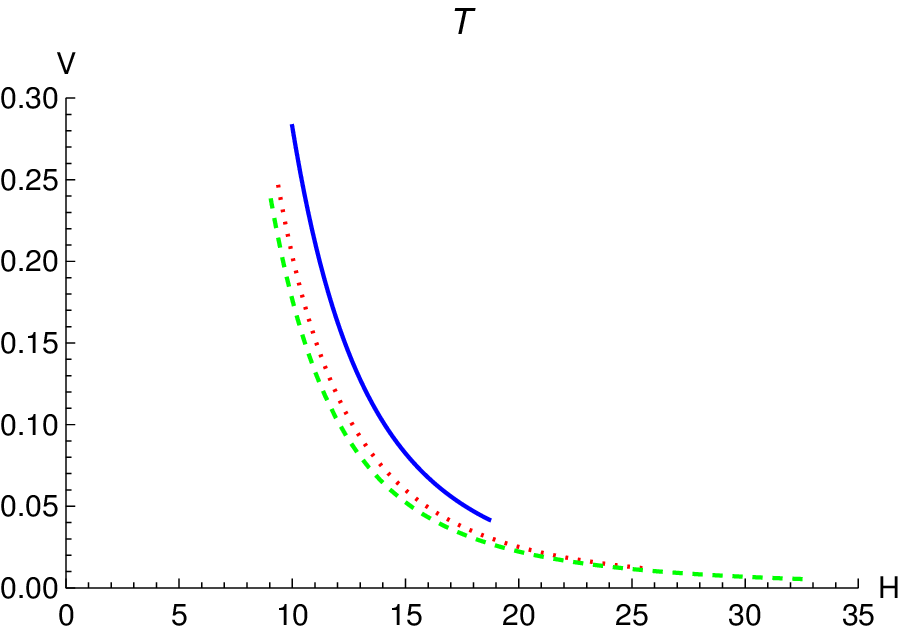}
		
		\vspace{.8cm}
		\includegraphics[width=7cm]{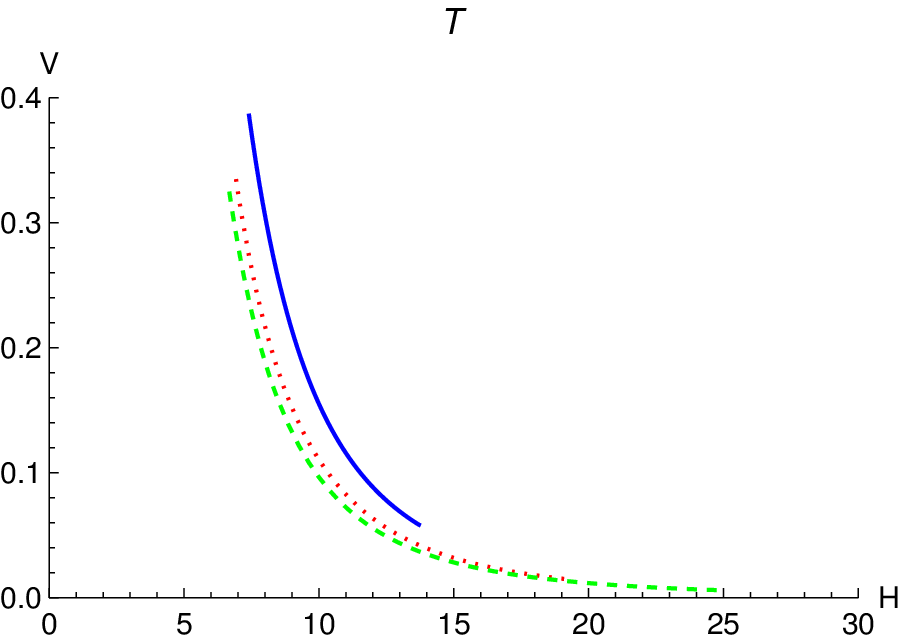}
		\vspace{.4cm}
		\caption{Angular distribution in the chiral-odd case for $ \gamma  \mesonmn $ photoproduction.
			Up, left: $S_{\gamma N}=10~{\rm GeV}^2$, for $M^2_{\gamma \mesonmn}=
			3~{\rm GeV}^2$ (solid blue) and  $M^2_{\gamma \mesonmn}=
			4~{\rm GeV}^2$ (dotted red).  Up, right:
			$S_{\gamma N}=15~{\rm GeV}^2$, for $M^2_{\gamma \mesonmn}=
			3.5~{\rm GeV}^2$ (solid blue), $M^2_{\gamma \mesonmn}=
			5~{\rm GeV}^2$ (dotted red) and 
			$M^2_{\gamma \mesonmn}=
			6.5~{\rm GeV}^2$ (dashed green).
			Down: 
			$S_{\gamma N}=20~{\rm GeV}^2$, for $M^2_{\gamma \mesonmn}=
			4~{\rm GeV}^2$ (solid blue), $M^2_{\gamma \mesonmn}=
			6~{\rm GeV}^2$ (dotted red) and 
			$M^2_{\gamma \mesonmn}=
			8~{\rm GeV}^2$ (dashed green).}
		\label{fig:thetacut-rhominus-odd}
	\end{center}
\end{figure}

The derivation of the angular distribution was performed in \APP{}E in \cite{Boussarie:2016qop}, and we do not repeat the details here. However, for the sake of completeness, we reproduce here the relevant results.

We require the outgoing photon scattering angle in the rest frame of the nucleon target. This angle $  \theta  $ is defined wrt to the direction of the incoming photon (\ie{}$ \theta $ is the angle that the outgoing photon makes with the $ -z $ axis in the nucleon rest frame). The angle $  \theta  $ satisfies
\begin{eqnarray}
	\label{tan_theta_rf-Delta}
	\tan \theta = - \frac{2M s(1+\xi)\alpha| \pv - \frac{\dv}{2}|}{  -\alpha^2 (1+\xi)^2 s^2 +(\pv - \frac{\dv}{2})^2 M^2 }\;.
\end{eqnarray}
From the relation $\alpha = M^2_{\gamma \rho}/(-u')$, see \eqref{eq:alpha}, one can express $\tan \theta$ as a function of $-u'$. To solve for $  \theta  $ in \eqref{tan_theta_rf-Delta}, one should take
\begin{alignat}{4}
	\label{theta_rf1}
&\theta &\;=\;& \arctan(\tan \theta),\quad &{\rm for} \ \tan \theta&\;>\;&0, \\
\label{theta_rf2}
&\theta &\;=\;& \pi + \arctan(\tan\theta),&\quad{\rm for} \ \tan \theta&\;<\;&0  \,,
\end{alignat}
since $\theta$ is positive. Setting $\dv=0$, \eqref{tan_theta_rf-Delta} simplifies to
\begin{eqnarray}
	\label{tan_theta}
	\tan \theta = - \frac{2M s(1+\xi) \alpha \, | \vec{p} _t| }{  -\alpha^2 (1+\xi)^2 s^2 +\pv^{\,2} M^2 }\,,
\end{eqnarray}
and using the definition of the kinematical variables in \SEC\ref{sec:kinematics}, one can obtain $  \alpha  $ in terms of $  \theta  $,
\beqa
\label{alpha-theta1}
\alpha &=& \frac{(1+\xi+ \tilde{\tau}) \, \tilde{\tau} \, \tan^2 \theta + a\left(1+\sqrt{1+\tan^2 \theta}\right)}{(1+\xi+ \tilde{\tau})^2 \tan^2 \theta+2a} ,\qquad{\rm for} \ \tan \theta>0,  \\
\label{alpha-theta2}
\alpha &=& \frac{(1+\xi+ \tilde{\tau}) \, \tilde{\tau} \, \tan^2 \theta + a\left(1-\sqrt{1+\tan^2 \theta}\right)}{(1+\xi+ \tilde{\tau})^2 \tan^2 \theta+2a} ,\qquad {\rm for} \ \tan \theta<0, 
\eqa
where
\beqa
\label{def:a_G_H}
a = \frac{4 M_{\gamma \rho}^2}s \,,\qquad
\tilde{\tau} = \frac{2 \xi}{1+\xi} \frac{M_{\gamma \rho}^2}s = \tau \frac{M_{\gamma \rho}^2}s\,.
\eqa
This thus allows us to obtain $  \left( -u' \right)  $ as a function of  $\theta$ using $-u'= \alpha M^2_{\gamma \rho}$, see \eqref{eq:alpha}. Writing
\begin{align}
	\tan  \theta = f(-u')\,,
\end{align}
the angular distribution can be obtained from the fully differential cross section through
\beqa
\label{dsigma-dtheta}
\frac{1}{\sigma}\frac{d \sigma}{d \theta}=\frac{1}{\sigma}
\frac{d \sigma}{d (-u')} \frac{1+f^2(-u'[\theta])}{f'(-u'[\theta])}\,.
\eqa

The obtained angular distribution in the chiral-even case is shown in \FIGs\ref{fig:thetacut-rho0-even}, \ref{fig:thetacut-rho0neutron-even}, \ref{fig:thetacut-rhoplus-even} and \ref{fig:thetacut-rhominus-even} for $ \mesonzp,\,\mesonzn,\,\mesonpp$ and $\mesonmn$ respectively. Similarly, the obtained angular distribution in the chiral-odd case is shown in \FIGs\ref{fig:thetacut-rho0-odd}, \ref{fig:thetacut-rho0neutron-odd}, \ref{fig:thetacut-rhoplus-odd} and \ref{fig:thetacut-rhominus-odd}
for $\mesonzp,\, \mesonzn,\,\mesonpp$ and $\mesonmn$ respectively. Each figure has 3 plots, corresponding to 3 different values of $ \SgN $, namely $ 10,\,15,\,20 \GeV^2 $. Finally, on each plot, 2 or 3 different curves are shown, which correspond to different $ \Msq $. The asymptotic DA with the standard GPD scenario are used to generate the plots.

{In the chiral-even case, the obtained angular distribution increases with $\theta$ for $ \mesonzp $ and $ \mesonzn $, while in the chiral-odd case, it decreases with $\theta$ for all of $ \mesonzp,\,\mesonzn,\,\mesonpp,\,\mesonmn$. In all cases, the distributions are dominated by moderate values of $\theta$.} In practice, at JLab Hall B, 
the outgoing photon could be detected with an angle between $5\degree$ and $35\degree$ from the incoming beam. Therefore, we find that relatively few events will be lost at JLab due to the angular cut on the outgoing photon.

\FloatBarrier

\subsection{Single-differential cross section}

\begin{figure}[h!]
	\begin{center}
		\psfrag{T}{}
		\psfrag{H}{\hspace{-1.5cm}\raisebox{-.6cm}{\scalebox{.7}{$\Msq~({\rm GeV}^{2})$}}}
		\psfrag{V}{\raisebox{.3cm}{\scalebox{.7}{$\hspace{-.4cm}\displaystyle\frac{d \sigma^{ \mathrm{even} }_{\gamma \mesonzp}}{d M^2_{\gamma \mesonzp}}~({\rm pb} \cdot {\rm GeV}^{-2})$}}}
		\includegraphics[width=7cm]{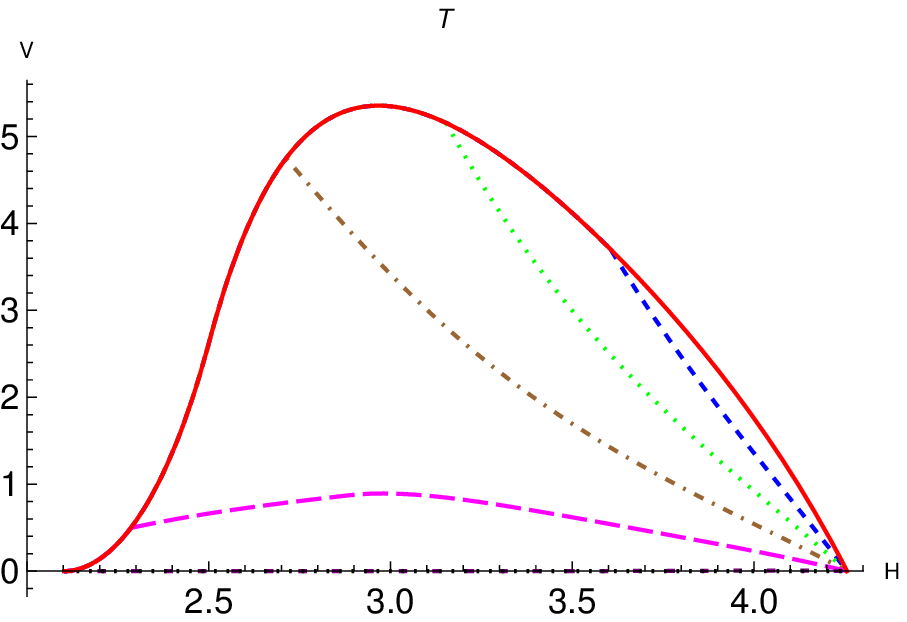}
		\includegraphics[width=7cm]{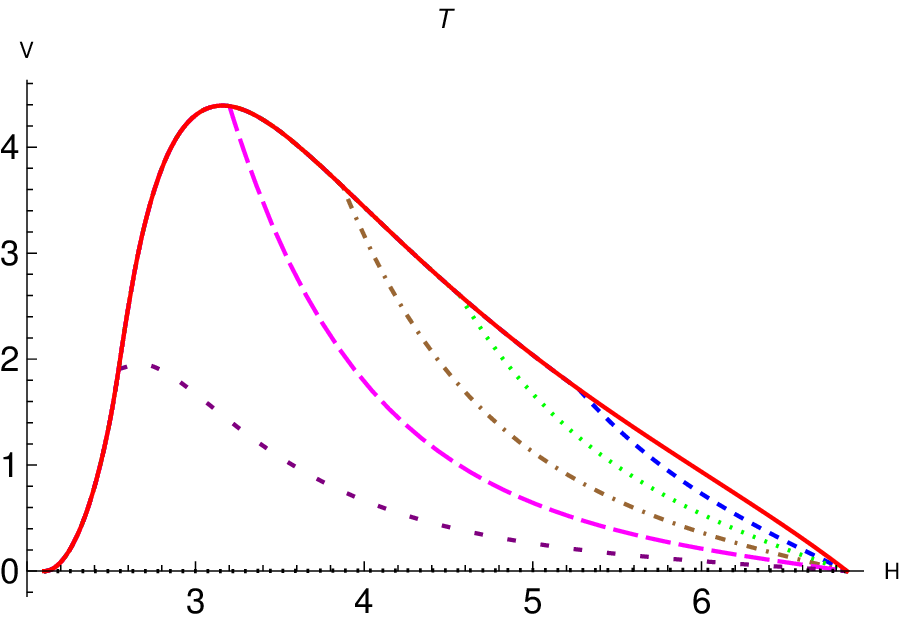}
		
		\vspace{.8cm}
		\includegraphics[width=7cm]{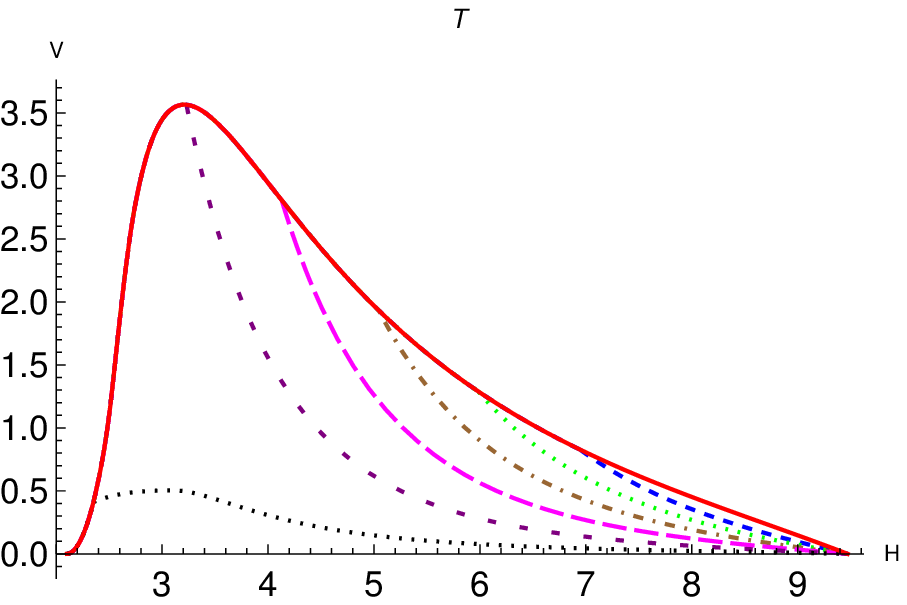}
		\vspace{.4cm}
		\caption{The chiral-even single-differential cross section as a function of $ \Msq $ for $ \gamma \mesonzp $ photoproduction. Solid red: no angular cut. Other curves show the effect of an upper angular cut $\theta$ for the outgoing $\gamma$: $35\degree$ (dashed blue), $30\degree$ (dotted green),  $25\degree$ (dashed-dotted brown), $20\degree$ (long-dashed magenta),  $15\degree$ (short-dashed purple) and $10\degree$ (dotted black).
			Up, left: $S_{\gamma N}=10~{\rm GeV}^2$.  
			Up, right:
			$S_{\gamma N}=15~{\rm GeV}^2$.
			Down: 
			$S_{\gamma N}=20~{\rm GeV}^2$.
		}  
		\label{fig:dsigmathetacut-rho0-even}
	\end{center}
\end{figure}

\begin{figure}[h!]
	\begin{center}
		\psfrag{T}{}
		\psfrag{H}{\hspace{-1.5cm}\raisebox{-.6cm}{\scalebox{.7}{$\Msq~({\rm GeV}^{2})$}}}
		\psfrag{V}{\raisebox{.3cm}{\scalebox{.7}{$\hspace{-.4cm}\displaystyle\frac{d \sigma^{ \mathrm{even} }_{\gamma \mesonzn}}{d M^2_{\gamma \mesonzn}}~({\rm pb} \cdot {\rm GeV}^{-2})$}}}
		\includegraphics[width=7cm]{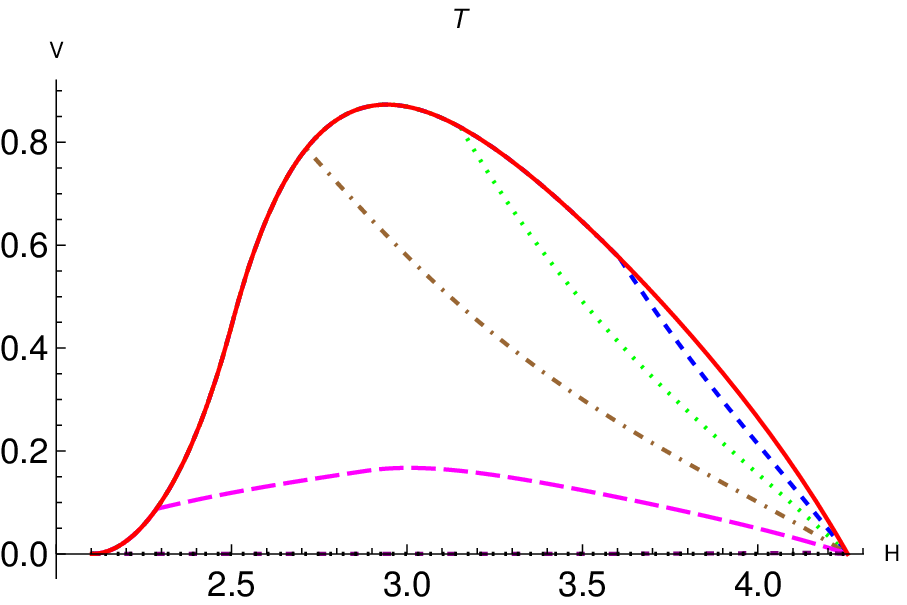}
		\includegraphics[width=7cm]{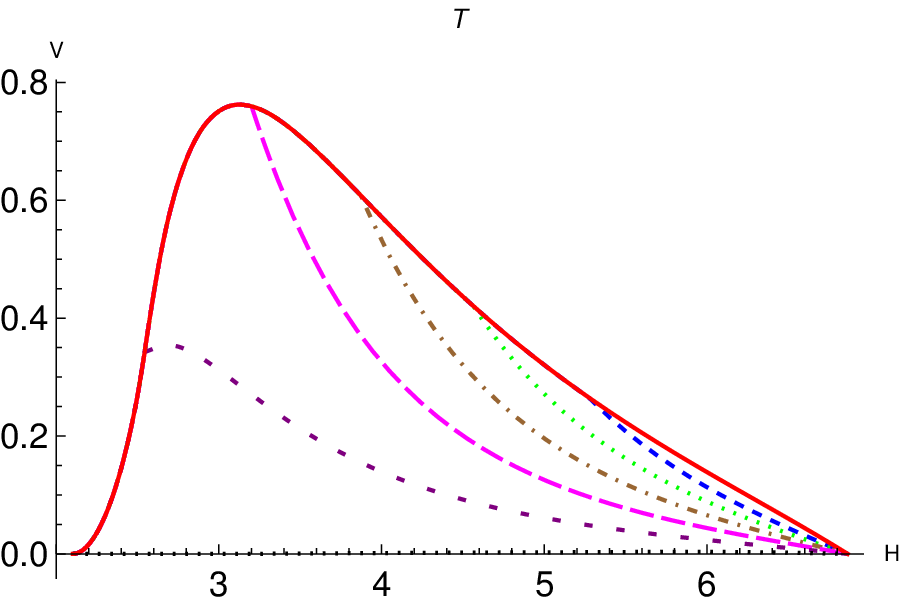}
		
		\vspace{.8cm}
		\includegraphics[width=7cm]{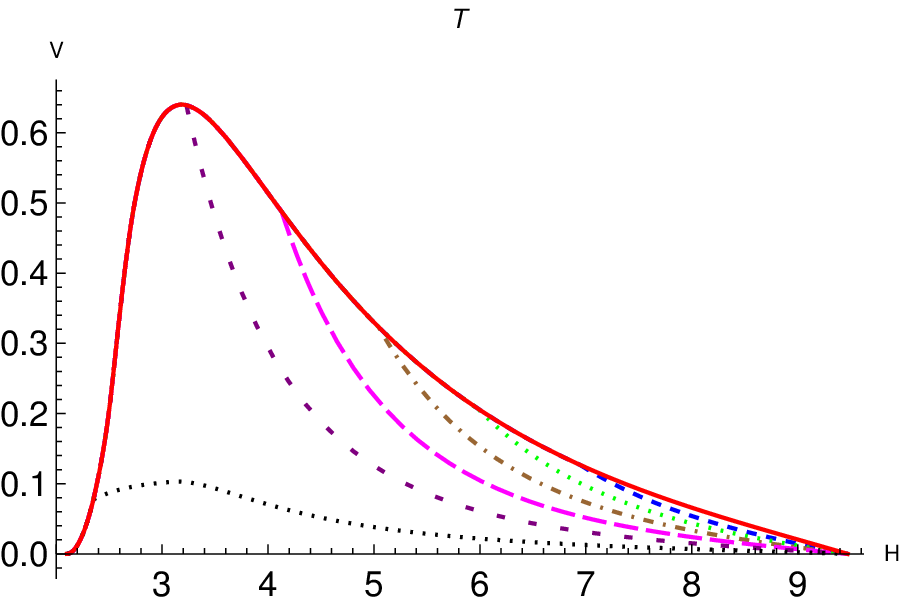}
		\vspace{.4cm}
		\caption{The chiral-even single-differential cross section as a function of $ \Msq $ for $ \gamma \mesonzn $ photoproduction. Solid red: no angular cut. Other curves show the effect of an upper angular cut $\theta$ for the outgoing $\gamma$: $35\degree$ (dashed blue), $30\degree$ (dotted green),  $25\degree$ (dashed-dotted brown), $20\degree$ (long-dashed magenta),  $15\degree$ (short-dashed purple) and $10\degree$ (dotted black).
			Up, left: $S_{\gamma N}=10~{\rm GeV}^2$.  
			Up, right:
			$S_{\gamma N}=15~{\rm GeV}^2$.
			Down: 
			$S_{\gamma N}=20~{\rm GeV}^2$.
		}  
		\label{fig:dsigmathetacut-rho0neutron-even}
	\end{center}
\end{figure}

\begin{figure}[h!]
	\begin{center}
		\psfrag{T}{}
		\psfrag{H}{\hspace{-1.5cm}\raisebox{-.6cm}{\scalebox{.7}{$\Msq~({\rm GeV}^{2})$}}}
		\psfrag{V}{\raisebox{.3cm}{\scalebox{.7}{$\hspace{-.4cm}\displaystyle\frac{d \sigma^{ \mathrm{even} }_{\gamma \mesonpp}}{d M^2_{\gamma \mesonpp}}~({\rm pb} \cdot {\rm GeV}^{-2})$}}}
		\includegraphics[width=7cm]{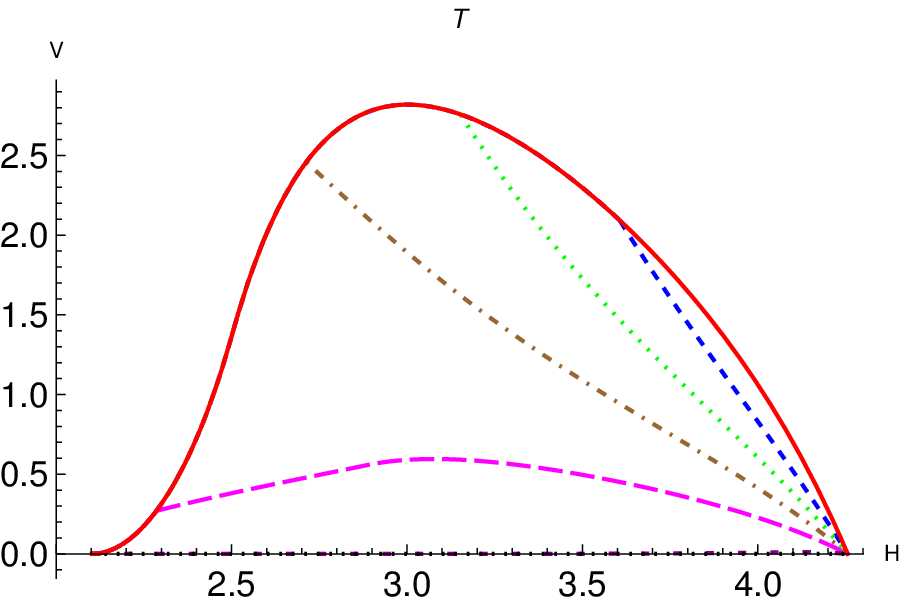}
		\includegraphics[width=7cm]{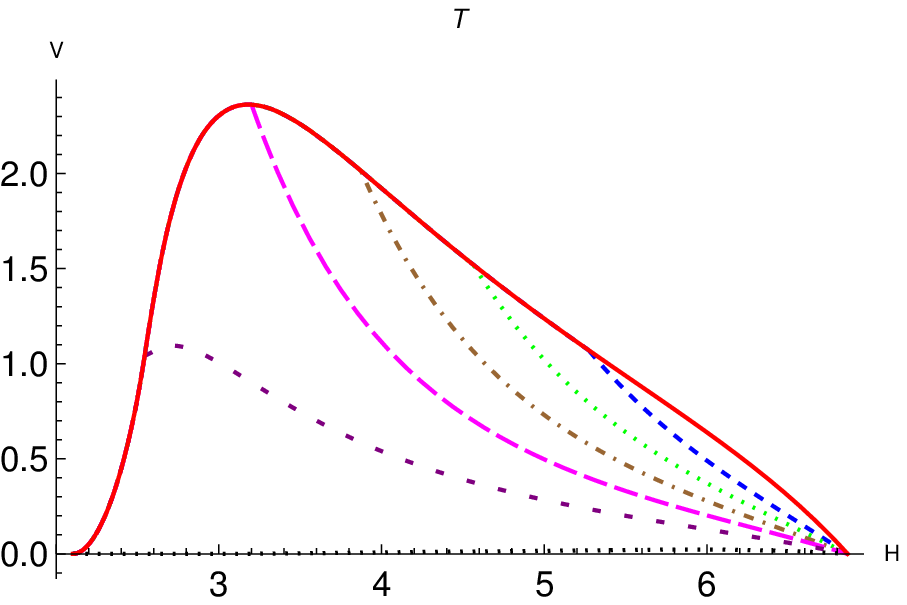}
		
		\vspace{.8cm}
		\includegraphics[width=7cm]{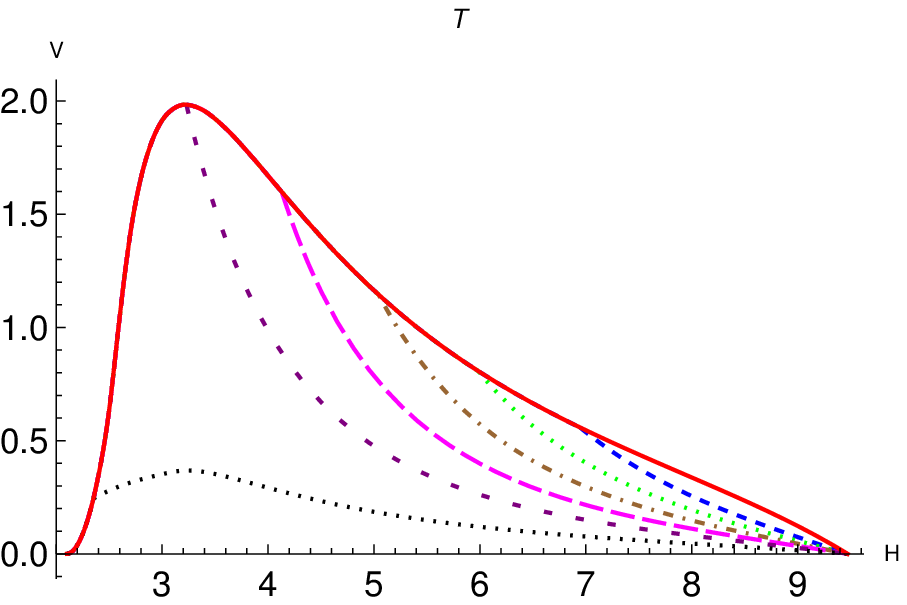}
		\vspace{.4cm}
		\caption{The chiral-even single-differential cross section as a function of $ \Msq $ for $ \gamma \mesonpp $ photoproduction. Solid red: no angular cut. Other curves show the effect of an upper angular cut $\theta$ for the outgoing $\gamma$: $35\degree$ (dashed blue), $30\degree$ (dotted green),  $25\degree$ (dashed-dotted brown), $20\degree$ (long-dashed magenta),  $15\degree$ (short-dashed purple) and $10\degree$ (dotted black).
			Up, left: $S_{\gamma N}=10~{\rm GeV}^2$.  
			Up, right:
			$S_{\gamma N}=15~{\rm GeV}^2$.
			Down: 
			$S_{\gamma N}=20~{\rm GeV}^2$.
		}  
		\label{fig:dsigmathetacut-rhoplus-even}
	\end{center}
\end{figure}

\begin{figure}[h!]
	\begin{center}
		\psfrag{T}{}
		\psfrag{H}{\hspace{-1.5cm}\raisebox{-.6cm}{\scalebox{.7}{$\Msq~({\rm GeV}^{2})$}}}
		\psfrag{V}{\raisebox{.3cm}{\scalebox{.7}{$\hspace{-.4cm}\displaystyle\frac{d \sigma^{ \mathrm{even} }_{\gamma \mesonmn}}{d M^2_{\gamma \mesonmn}}~({\rm pb} \cdot {\rm GeV}^{-2})$}}}
		\includegraphics[width=7cm]{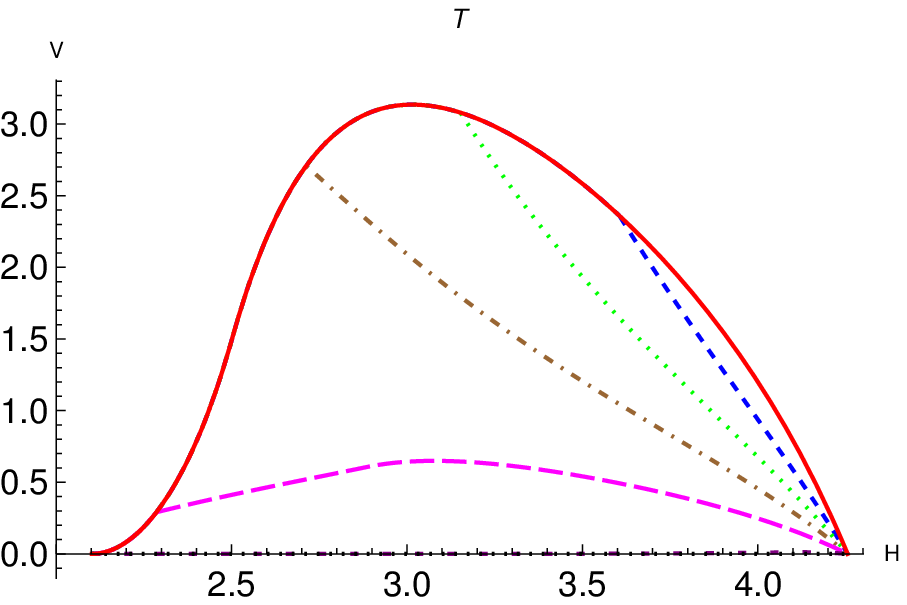}
		\includegraphics[width=7cm]{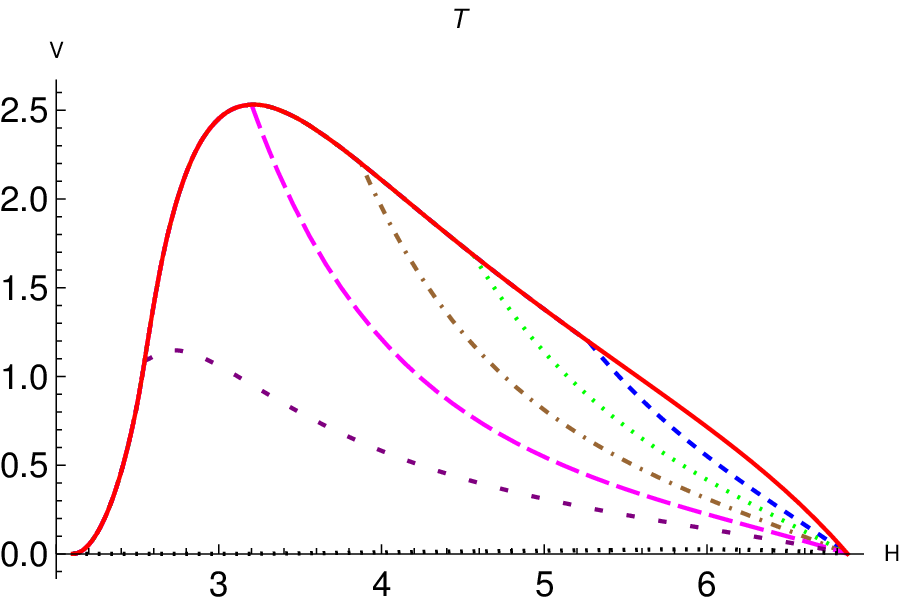}
		
		\vspace{.8cm}
		\includegraphics[width=7cm]{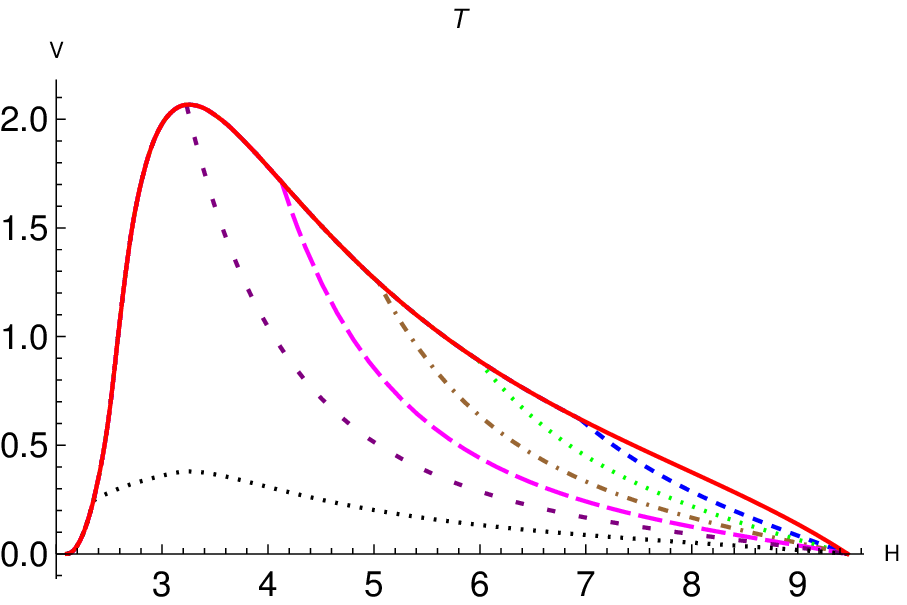}
		\vspace{.4cm}
		\caption{The chiral-even single-differential cross section as a function of $ \Msq $ for $ \gamma \mesonmn $ photoproduction. Solid red: no angular cut. Other curves show the effect of an upper angular cut $\theta$ for the outgoing $\gamma$: $35\degree$ (dashed blue), $30\degree$ (dotted green),  $25\degree$ (dashed-dotted brown), $20\degree$ (long-dashed magenta),  $15\degree$ (short-dashed purple) and $10\degree$ (dotted black).
			Up, left: $S_{\gamma N}=10~{\rm GeV}^2$.  
			Up, right:
			$S_{\gamma N}=15~{\rm GeV}^2$.
			Down: 
			$S_{\gamma N}=20~{\rm GeV}^2$.
		}  
		\label{fig:dsigmathetacut-rhominus-even}
	\end{center}
\end{figure}

\begin{figure}[h!]
	\begin{center}
		\psfrag{T}{}
		\psfrag{H}{\hspace{-1.5cm}\raisebox{-.6cm}{\scalebox{.7}{$\Msq~({\rm GeV}^{2})$}}}
		\psfrag{V}{\raisebox{.3cm}{\scalebox{.7}{$\hspace{-.4cm}\displaystyle\frac{d \sigma^{ \mathrm{odd} }_{\gamma \mesonzp}}{d M^2_{\gamma \mesonzp}}~({\rm pb} \cdot {\rm GeV}^{-2})$}}}
		\includegraphics[width=7cm]{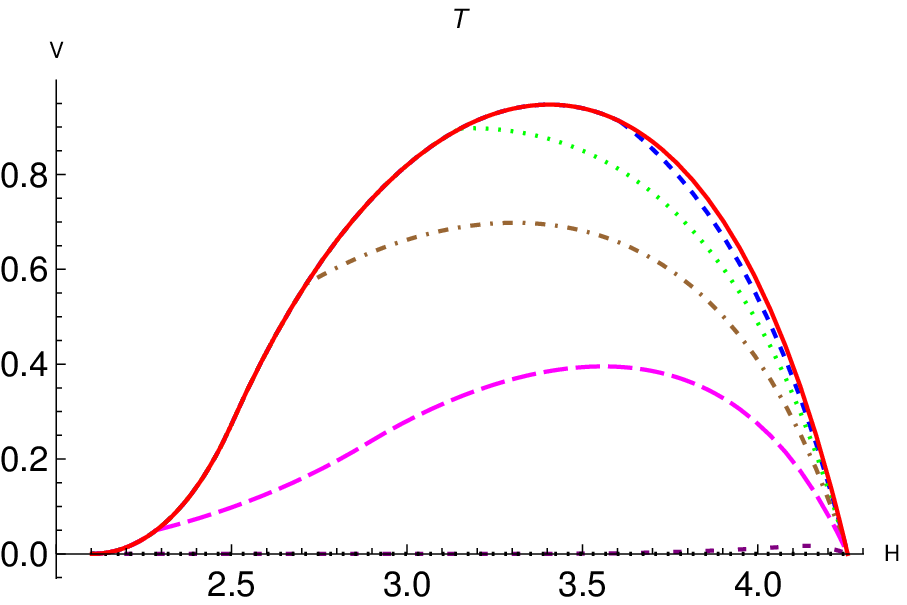}
		\includegraphics[width=7cm]{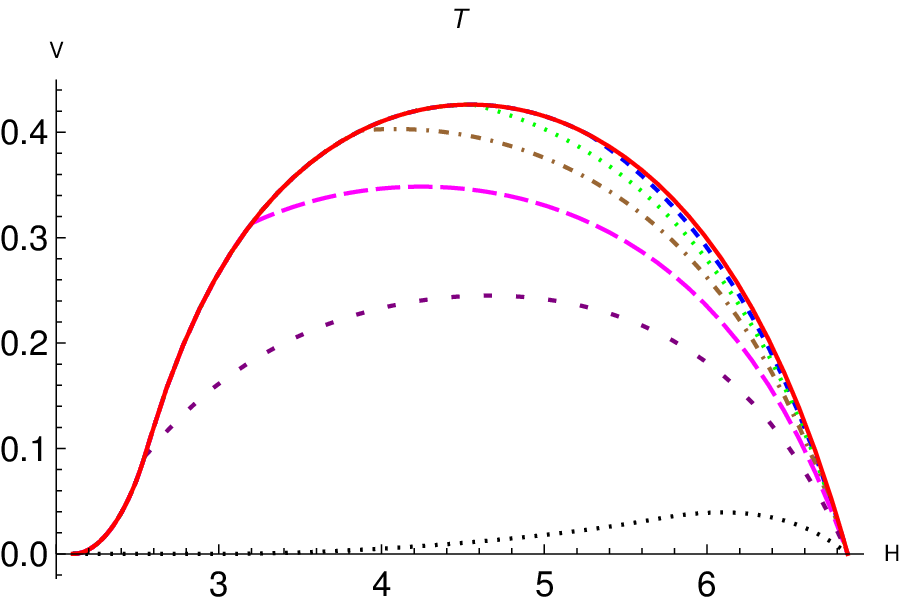}
		
		\vspace{.8cm}
		\includegraphics[width=7cm]{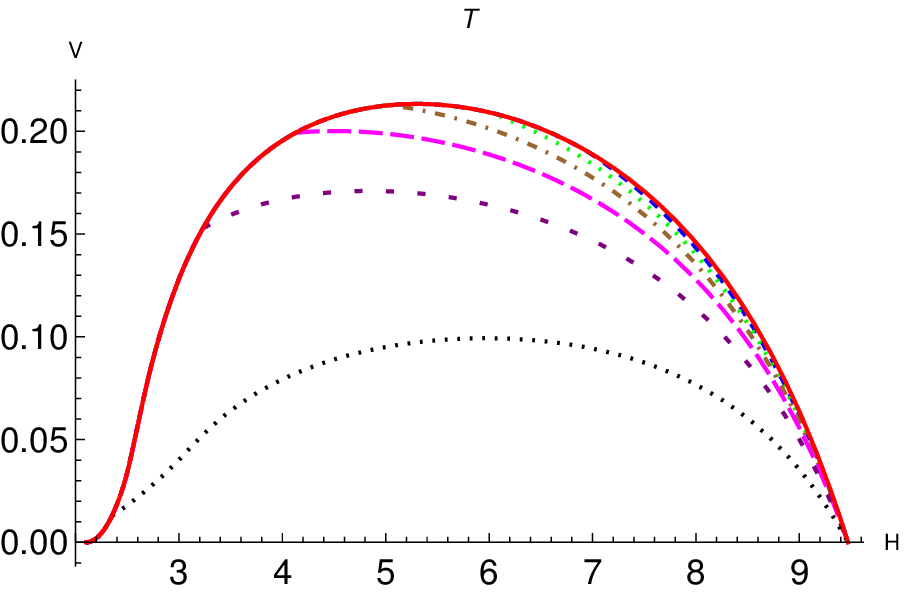}
		\vspace{.4cm}
		\caption{The chiral-odd single-differential cross section as a function of $ \Msq $ for $ \gamma \mesonzp $ photoproduction. Solid red: no angular cut. Other curves show the effect of an upper angular cut $\theta$ for the outgoing $\gamma$: $35\degree$ (dashed blue), $30\degree$ (dotted green),  $25\degree$ (dashed-dotted brown), $20\degree$ (long-dashed magenta),  $15\degree$ (short-dashed purple) and $10\degree$ (dotted black).
			Up, left: $S_{\gamma N}=10~{\rm GeV}^2$.  
			Up, right:
			$S_{\gamma N}=15~{\rm GeV}^2$.
			Down: 
			$S_{\gamma N}=20~{\rm GeV}^2$.
		}  
		\label{fig:dsigmathetacut-rho0-odd}
	\end{center}
\end{figure}

\begin{figure}[h!]
	\begin{center}
		\psfrag{T}{}
		\psfrag{H}{\hspace{-1.5cm}\raisebox{-.6cm}{\scalebox{.7}{$\Msq~({\rm GeV}^{2})$}}}
		\psfrag{V}{\raisebox{.3cm}{\scalebox{.7}{$\hspace{-.4cm}\displaystyle\frac{d \sigma^{ \mathrm{odd} }_{\gamma \mesonzn}}{d M^2_{\gamma \mesonzn}}~({\rm pb} \cdot {\rm GeV}^{-2})$}}}
		\includegraphics[width=7cm]{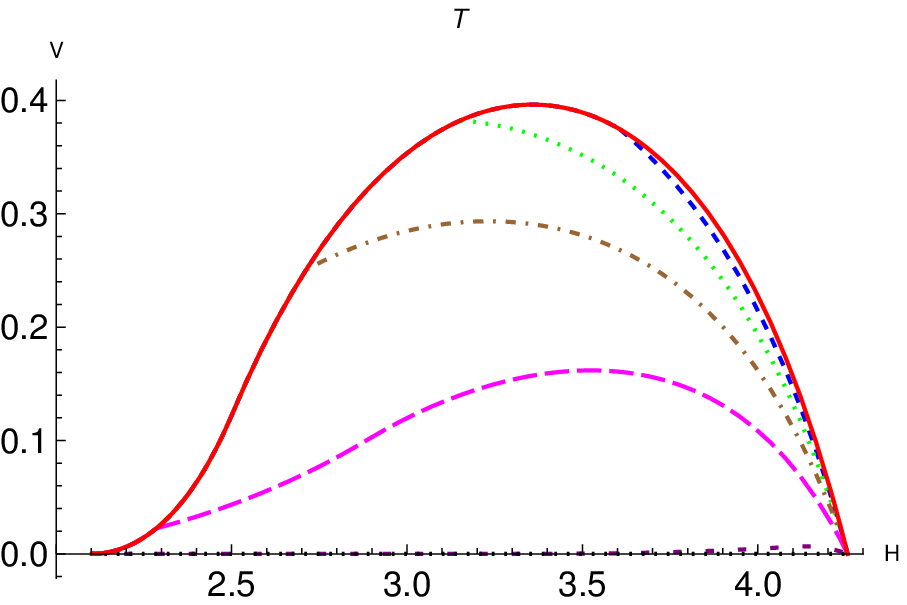}
		\includegraphics[width=7cm]{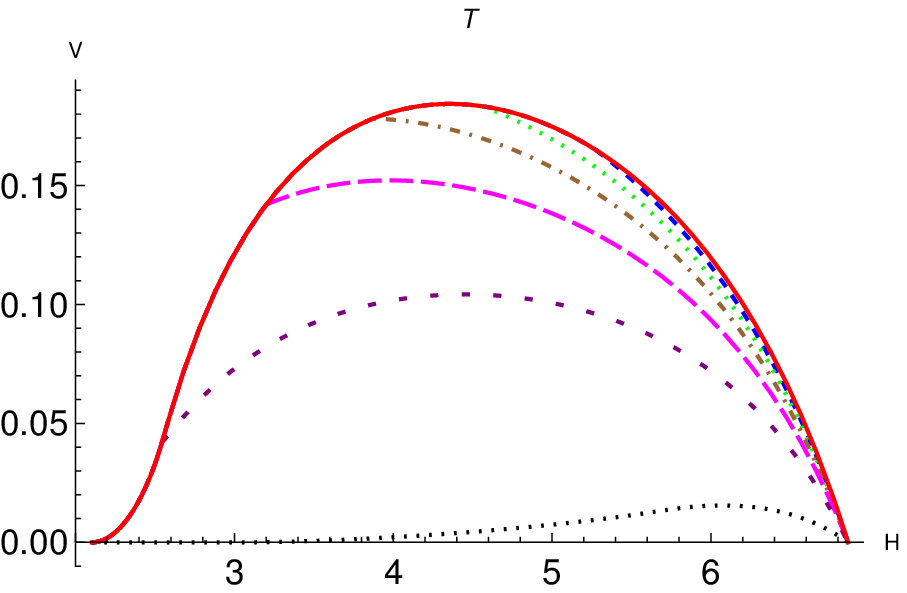}
		
		\vspace{.8cm}
		\includegraphics[width=7cm]{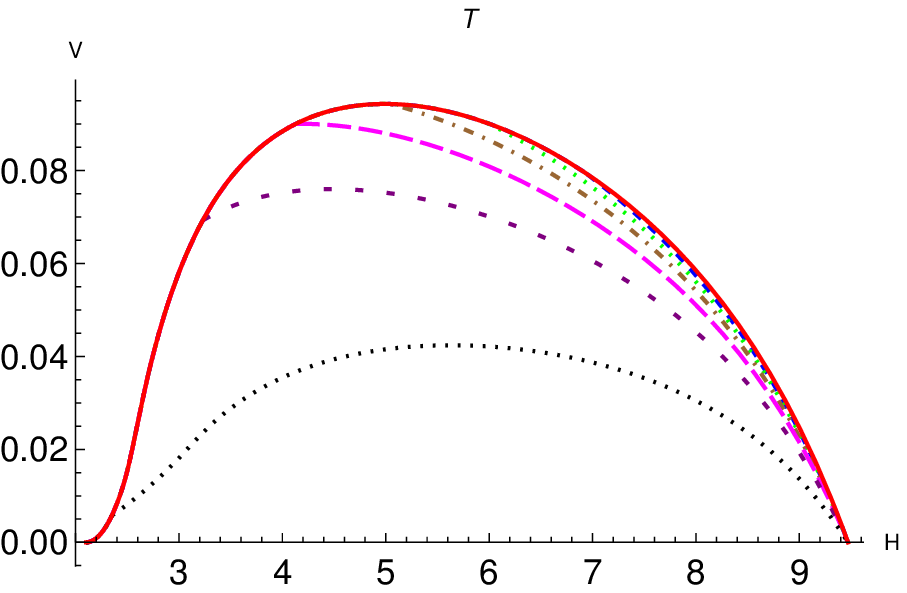}
		\vspace{.4cm}
		\caption{The chiral-odd single-differential cross section as a function of $ \Msq $ for $ \gamma \mesonzn $ photoproduction. Solid red: no angular cut. Other curves show the effect of an upper angular cut $\theta$ for the outgoing $\gamma$: $35\degree$ (dashed blue), $30\degree$ (dotted green),  $25\degree$ (dashed-dotted brown), $20\degree$ (long-dashed magenta),  $15\degree$ (short-dashed purple) and $10\degree$ (dotted black).
			Up, left: $S_{\gamma N}=10~{\rm GeV}^2$.  
			Up, right:
			$S_{\gamma N}=15~{\rm GeV}^2$.
			Down: 
			$S_{\gamma N}=20~{\rm GeV}^2$.
		}  
		\label{fig:dsigmathetacut-rho0neutron-odd}
	\end{center}
\end{figure}

\begin{figure}[h!]
	\begin{center}
		\psfrag{T}{}
		\psfrag{H}{\hspace{-1.5cm}\raisebox{-.6cm}{\scalebox{.7}{$\Msq~({\rm GeV}^{2})$}}}
		\psfrag{V}{\raisebox{.3cm}{\scalebox{.7}{$\hspace{-.4cm}\displaystyle\frac{d \sigma^{ \mathrm{odd} }_{\gamma \mesonpp}}{d M^2_{\gamma \mesonpp}}~({\rm pb} \cdot {\rm GeV}^{-2})$}}}
		\includegraphics[width=7cm]{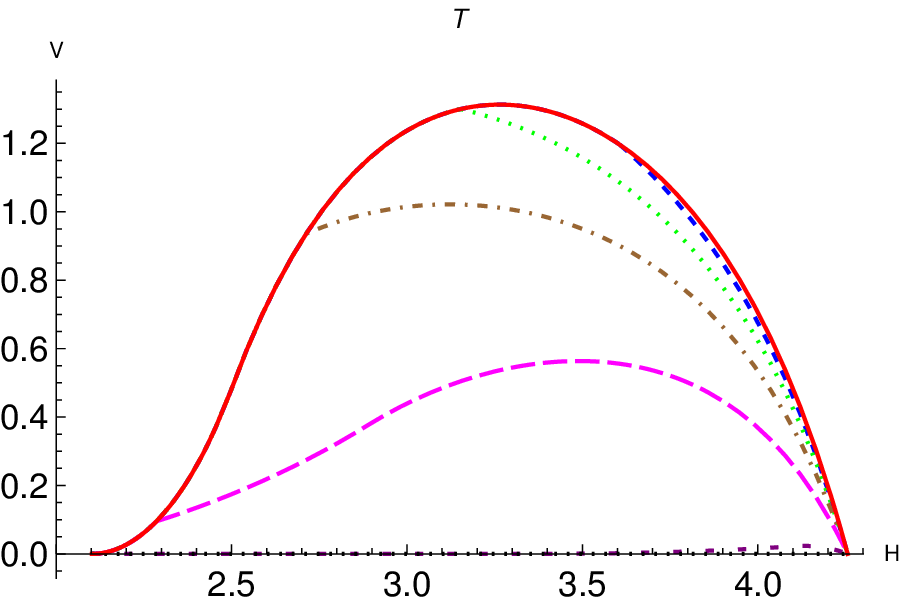}
		\includegraphics[width=7cm]{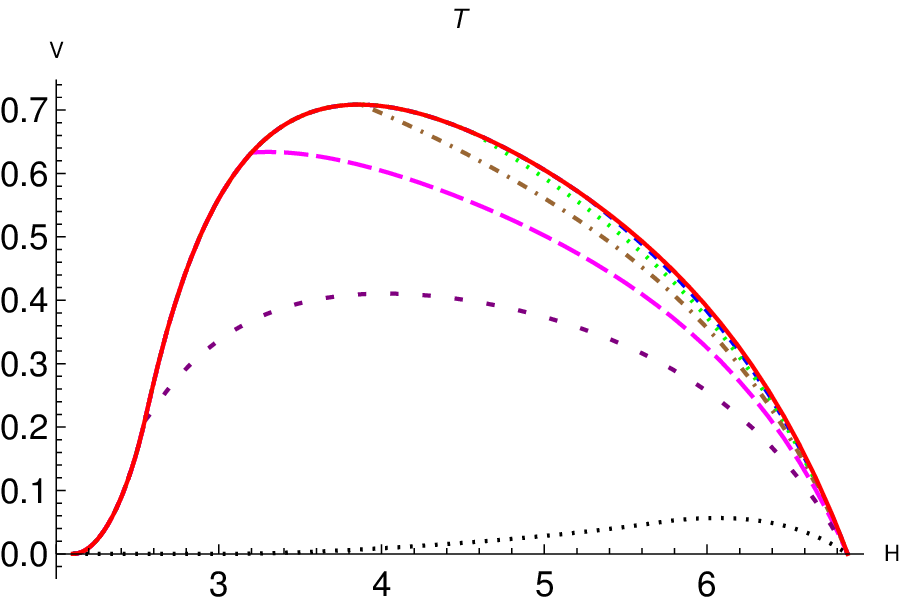}
		
		\vspace{.8cm}
		\includegraphics[width=7cm]{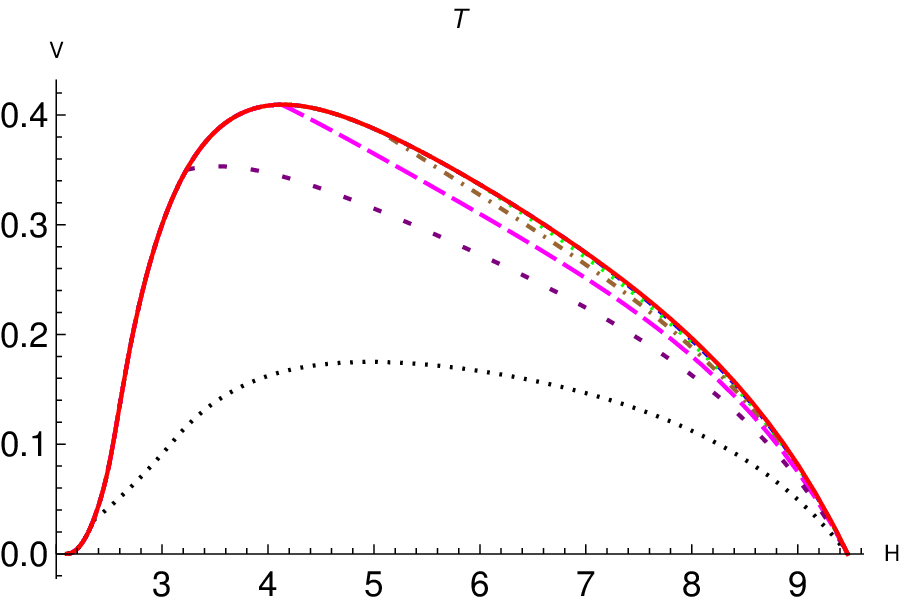}
		\vspace{.4cm}
		\caption{The chiral-odd single-differential cross section as a function of $ \Msq $ for $ \gamma \mesonpp $ photoproduction. Solid red: no angular cut. Other curves show the effect of an upper angular cut $\theta$ for the outgoing $\gamma$: $35\degree$ (dashed blue), $30\degree$ (dotted green),  $25\degree$ (dashed-dotted brown), $20\degree$ (long-dashed magenta),  $15\degree$ (short-dashed purple) and $10\degree$ (dotted black).
			Up, left: $S_{\gamma N}=10~{\rm GeV}^2$.  
			Up, right:
			$S_{\gamma N}=15~{\rm GeV}^2$.
			Down: 
			$S_{\gamma N}=20~{\rm GeV}^2$.
		}  
		\label{fig:dsigmathetacut-rhoplus-odd}
	\end{center}
\end{figure}

\begin{figure}[h!]
	\begin{center}
		\psfrag{T}{}
		\psfrag{H}{\hspace{-1.5cm}\raisebox{-.6cm}{\scalebox{.7}{$\Msq~({\rm GeV}^{2})$}}}
		\psfrag{V}{\raisebox{.3cm}{\scalebox{.7}{$\hspace{-.4cm}\displaystyle\frac{d \sigma^{ \mathrm{odd} }_{\gamma \mesonmn}}{d M^2_{\gamma \mesonmn}}~({\rm pb} \cdot {\rm GeV}^{-2})$}}}
		\includegraphics[width=7cm]{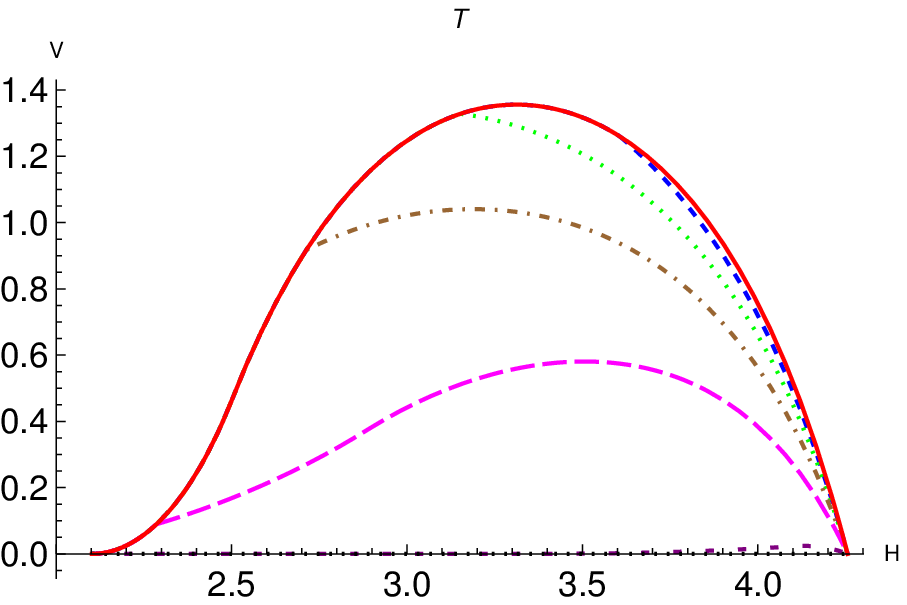}
		\includegraphics[width=7cm]{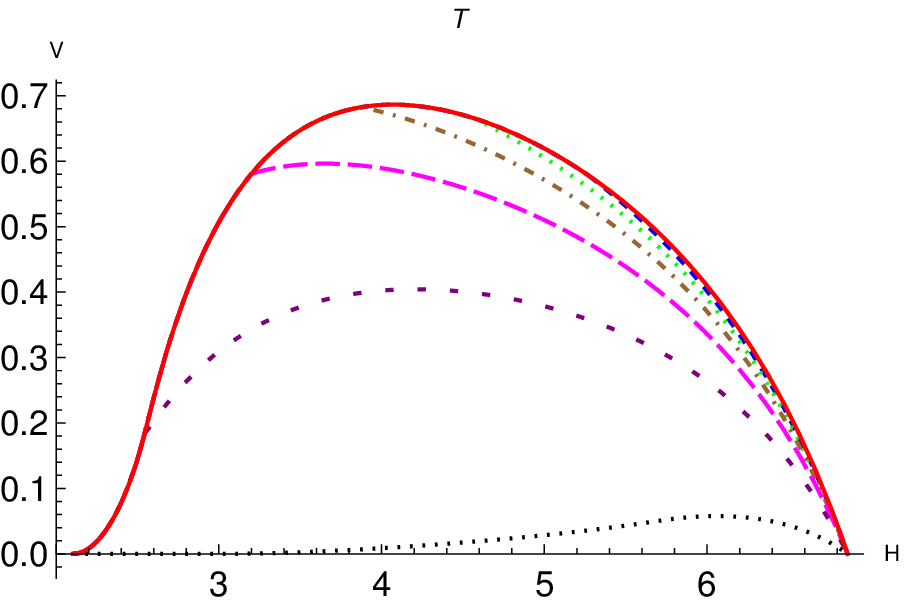}
		
		\vspace{.8cm}
		\includegraphics[width=7cm]{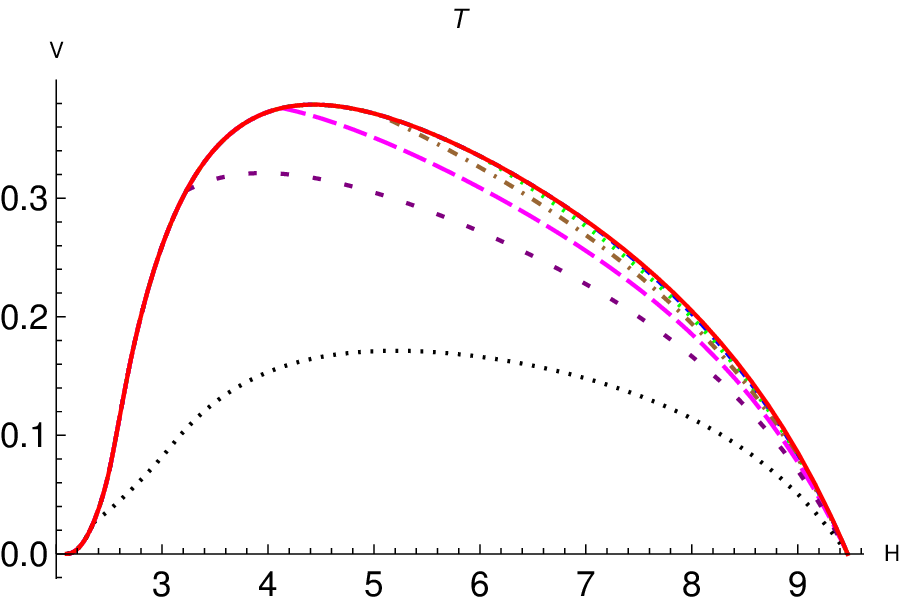}
		\vspace{.4cm}
		\caption{The chiral-odd single-differential cross section as a function of $ \Msq $ for $ \gamma \mesonmn $ photoproduction. Solid red: no angular cut. Other curves show the effect of an upper angular cut $\theta$ for the outgoing $\gamma$: $35\degree$ (dashed blue), $30\degree$ (dotted green),  $25\degree$ (dashed-dotted brown), $20\degree$ (long-dashed magenta),  $15\degree$ (short-dashed purple) and $10\degree$ (dotted black).
			Up, left: $S_{\gamma N}=10~{\rm GeV}^2$.  
			Up, right:
			$S_{\gamma N}=15~{\rm GeV}^2$.
			Down: 
			$S_{\gamma N}=20~{\rm GeV}^2$.
		}  
		\label{fig:dsigmathetacut-rhominus-odd}
	\end{center}
\end{figure}

In this subsection, we show the effect of choosing different angular cuts on the single-differential cross section, as a function of $ \Msq $. For the chiral-even case, this is shown in \FIGs\ref{fig:dsigmathetacut-rho0-even},  \ref{fig:dsigmathetacut-rho0neutron-even}, \ref{fig:dsigmathetacut-rhoplus-even} and \ref{fig:dsigmathetacut-rhominus-even} for $ \mesonzp,\,\mesonzn,\,\mesonpp$ and $\mesonmn$ respectively, while for the chiral-odd case,  this is shown in \FIGs\ref{fig:dsigmathetacut-rho0-odd},  \ref{fig:dsigmathetacut-rho0neutron-odd}, \ref{fig:dsigmathetacut-rhoplus-odd} and \ref{fig:dsigmathetacut-rhominus-odd} for $ \mesonzp,\,\mesonzn,\,\mesonpp$ and $\mesonmn$ respectively. Each figure consists of 3 plots, which correspond to 3 different values of $ \SgN $, namely $ 10,\,15 $ and $ 20 \GeV^2 $. Each plot consists of 7 curves, which correspond to 6 different angular cuts of $ 10\degree,\,15\degree,\,20\degree,\,25\degree,\,30\degree,\,35\degree $, and 1 with no angular cuts.  The asymptotic DA with the standard GPD scenario are used to generate the plots.

From the figures, we find that the angular cuts mainly affect the low $S_{\gamma N}$
domain. For the specific case of the JLab $35\degree$ upper cut (dashed-blue), the effect is negligible  both for the chiral-even and chiral-odd cases.

Moreover, we note that using cuts on $\theta$, it is possible to significantly reduce the contribution of the chiral-even contribution, in particular in the high $S_{\gamma N}$ region, while moderately reducing the chiral-odd contribution. Putting additional cuts on $M^2_{\gamma \rho}$, like  $M^2_{\gamma \rho} > 6~{\rm GeV}^2,$
would allow for an increase in the ratio of odd versus even cross section.

\FloatBarrier

\section{Vanishing of the circular asymmetry in the chiral-even case}

\label{app:vanishing-circular-asymmetry}

In this appendix, we discuss the vanishing of the circular asymmetry for the chiral-even case. For the circularly-polarised amplitudes, the analogues of \eqref{eq:pol-amp-sq-x} and \eqref{eq:pol-amp-sq-y} are given by
\begin{align}
\sum_{ \lambda _{k}}| {\cal M }_{+}|^ 2&=\frac{1}{2} \bigg[ 2|C_{A}|^2 + | \pt|^{4} |C_{B}|^2+\frac{s^2}{4}| \pt|^4 |C_{A_{5}}|^2+\frac{s^2}{4}| \pt|^4 |C_{B_{5}}|^2-2 |\pt|^2  \mathrm{Re}(C_{A}^{*}C_{B}) \;\nonumber\\[5pt]
&\phantom{aaaaa} + s |\pt|^2    \mathrm{Im} \left( C_{A} \left( C_{A_{5}}^{*}+C_{B_{5}}^{*} \right)+|\pt|^2 C_{A_{5}} C_{B}^{*}  \right) \bigg]\,,\\[5pt]
\sum_{ \lambda _{k}}| {\cal M }_{-}|^ 2&=\frac{1}{2} \bigg[ 2|C_{A}|^2 + | \pt|^{4} |C_{B}|^2+\frac{s^2}{4}| \pt|^4 |C_{A_{5}}|^2+\frac{s^2}{4}| \pt|^4 |C_{B_{5}}|^2-2 |\pt|^2  \mathrm{Re}(C_{A}^{*}C_{B}) \;\nonumber\\[5pt]
&\phantom{aaaaa} - s |\pt|^2    \mathrm{Im} \left( C_{A} \left( C_{A_{5}}^{*}+C_{B_{5}}^{*} \right)+|\pt|^2 C_{A_{5}} C_{B}^{*}  \right) \bigg]\,.
\end{align}
So,
\begin{align}
\label{eq:circular-asymmetry}
\sum_{ \lambda _{k}}| {\cal M }_{+}|^ 2 - \sum_{ \lambda _{k}}| {\cal M }_{-}|^ 2=s |\pt|^2    \mathrm{Im} \left( C_{A} \left( C_{A_{5}}^{*}+C_{B_{5}}^{*} \right)+|\pt|^2 C_{A_{5}} C_{B}^{*}  \right) \,.
\end{align}
An interesting feature of the circular asymmetry is that it only contains terms that mix vector GPD and axial GPD contributions ($ A $ and $ B $, with $ A_{5} $ and $ B_{5} $). Thus, when averaging over the target helicity, it can be shown that all terms on the RHS of \eqref{eq:circular-asymmetry} vanish. Indeed, using \EQs\eqref{eq:CA} to \eqref{eq:CB5}, we obtain, after averaging and summing over the target helicities,
\begin{align}
\frac{1}{2} \sum_{\lambda_1,\lambda_2}\left(\sum_{ \lambda _{k}}| {\cal M }_{+}|^ 2 - \sum_{ \lambda _{k}}| {\cal M }_{-}|^ 2 \right)&= \frac{|\pt|^2}{  \left( n  \cdot p \right) } \left[  {{\cal 
		H}} _{\meson A}\left(\tilde{\cal 
		H}_{\meson A_5}^{*}+\tilde{\cal 
		H}_{\meson B_5}^{*} \right) + |\pt|^2{\tilde{\cal 
		H}} _{\meson A_5} {{\cal 
		H}}_{\meson B}^{*}   \right]  \mathrm{tr} \left[  \slashed{p}_{2} \slashed{n}\gamma ^{5} \slashed{p}_{1} \slashed{n}     \right] \nonumber \\[5pt]
		&= 0\,.
\end{align}
This shows that for an unpolarised target, the circular asymmetry is identically zero. From a more physical point of view, the vanishing of the circular asymmetry is a consequence of parity invariance of QED and QCD. In particular, from \cite{Bourrely:1980mr}, one deduces that the amplitude for our process, ${\cal M } _{ \lambda _2  \lambda _k \,;\, \lambda _1  \lambda _q}$, has to obey the relation
\begin{equation}
	{\cal M } _{ \lambda _2  \lambda _k \,;\, \lambda _1  \lambda _q}= \eta\, (-1)^{ \lambda_1- \lambda_q- \left(  \lambda _2 -  \lambda _k \right) } {\cal M } _{- \lambda _2  -\lambda _k \,;\, -\lambda _1  -\lambda _q}\,,
\end{equation}
where $  \eta  $ represents a phase factor related to intrinsic spin. From this, we can deduce that
\begin{align}
	\sum_{ \lambda _i ,\,i\neq q}| {\cal M } _{ \lambda _2  \lambda _k \,;\, \lambda _1  +}|^2 = \sum_{ \lambda _i ,\,i\neq q}| {\cal M } _{ \lambda _2  \lambda _k \,;\, \lambda _1  -}|^2\,,
\end{align}
which implies that the circular asymmetry vanishes identically for an unpolarised target.

\bibliographystyle{utphys}

\bibliography{masterrefs.bib}

\end{document}